%% file: main.tex
\documentclass[twoside,a4paper,11pt]{book}
\usepackage[utf8]{inputenc}
\usepackage{latexsym} 
\usepackage{amssymb} 
\usepackage{amsmath} 

\usepackage{fancyhdr} 

\usepackage[nottoc]{tocbibind} 
\usepackage{epsfig,graphics}
\usepackage{graphicx}
\usepackage{epsfig}
\usepackage[hang,center]{subfigure}

\usepackage{multirow}
\usepackage{booktabs}  

\usepackage{enumerate}
\usepackage{amsfonts}
\usepackage{mathrsfs}

\usepackage {tabularx} 

\usepackage{amsbsy}
\usepackage{rotate}
\usepackage{bbm}
\usepackage{color}

\definecolor{MyDarkGreen}{rgb}{0,0.5,0} 
\definecolor{MyDarkBlue}{rgb}{0.23,0.21,0.69} 
\definecolor{MyLightBlue}{rgb}{0.22,0.51,0.86}

\usepackage[colorlinks=true,linkcolor=MyDarkBlue,citecolor=MyDarkGreen,urlcolor=MyLightBlue,bookmarksnumbered=true,bookmarksopen]{hyperref}

\usepackage{cite} 

\usepackage[times]{quotchap}

\newcommand{\be}{\begin{equation}}
\newcommand{\ee}{\end{equation}}
\newcommand{\bea}{\begin{eqnarray}}
\newcommand{\eea}{\end{eqnarray}}

\newcommand{\bmat}{\left(\begin{array}}
\newcommand{\emat}{\end{array}\right)}

\def\yzero{\smash{\hbox{$y\kern-4pt\raise1pt\hbox{${}^\circ$}$}}}
\def\p{\partial}
\def\a{\alpha}
\def\b{\beta}
\def\g{\gamma}
\def\d{\delta}
\def\beq{\begin{equation}}
\def\eeq{\end{equation}}
\def\beqa{\begin{eqnarray}}
\def\eeqa{\end{eqnarray}}

\def\om{\omega}
\def\th{\theta}

\def\-{\hphantom{-}}

\def\s2{\frac{1}{\sqrt2}}

\def\oh{\frac{1}{2}}
\def\beq{\begin{equation}}
\def\eeq{\end{equation}}
\def\beqa{\begin{eqnarray}}
\def\eeqa{\end{eqnarray}}
\def\tr{{\rm tr \,}}

\def\IF{\relax{\rm I\kern-.18em F}}
\def\II{\relax{\rm I\kern-.18em I}}
\def\IP{\relax{\rm I\kern-.18em P}}
\def\IC{\relax\hbox{\kern.25em$\inbar\kern-.3em{\rm C}$}}
\def\IR{\relax{\rm I\kern-.18em R}}

\def\Dsl{\,\raise.15ex\hbox{/}\mkern-13.5mu D} 
\def\IZ{Z\kern-.4em  Z}

\def\ca{{\cal A}}
\def\lam{\lambda}
\def\raw{\rightarrow}

\def\tgb{\tan{\beta}}

\def\asusy{a^{\rm SUSY}_\mu}
\def\bsg{b\to s\gamma}
\def\bmumu{B_s^0\to\mu^+\mu^-}

\def\lsim{\raise0.3ex\hbox{$\;<$\kern-0.75em\raise-1.1ex\hbox{$\sim\;$}}}
\def\gsim{\raise0.3ex\hbox{$\;>$\kern-0.75em\raise-1.1ex\hbox{$\sim\;$}}}

\def\yzero{\smash{\hbox{$y\kern-4pt\raise1pt\hbox{${}^\circ$}$}}}
\def\p{\partial}
\def\a{\alpha}
\def\b{\beta}
\def\g{\gamma}
\def\d{\delta}
\def\beq{\begin{equation}}
\def\eeq{\end{equation}}
\def\beqa{\begin{eqnarray}}
\def\eeqa{\end{eqnarray}}

\def\om{\omega}
\def\th{\theta}

\def\-{\hphantom{-}}

\def\s2{\frac{1}{\sqrt2}}

\def\oh{\frac{1}{2}}
\def\beq{\begin{equation}}
\def\eeq{\end{equation}}
\def\beqa{\begin{eqnarray}}
\def\eeqa{\end{eqnarray}}
\def\tr{{\rm tr \,}}

\def\IF{\relax{\rm I\kern-.18em F}}
\def\II{\relax{\rm I\kern-.18em I}}
\def\IP{\relax{\rm I\kern-.18em P}}
\def\IC{\relax\hbox{\kern.25em$\inbar\kern-.3em{\rm C}$}}
\def\IR{\relax{\rm I\kern-.18em R}}

\def\Dsl{\,\raise.15ex\hbox{/}\mkern-13.5mu D} 
\def\IZ{Z\kern-.4em  Z}

\def\ca{{\cal A}}
\def\lam{\lambda}
\def\raw{\rightarrow}

\def\crosssec{\sigma_{\tilde{\chi}_1^0-p}}
\def\lsim{\raise0.3ex\hbox{$\;<$\kern-0.75em\raise-1.1ex\hbox{$\sim\;$}}}
\def\gsim{\raise0.3ex\hbox{$\;>$\kern-0.75em\raise-1.1ex\hbox{$\sim\;$}}}

\def\met{\slash\hspace*{-1.5ex}E_T}

\def\p{\partial}
\def\a{\alpha}

\def\b{\beta}
\def\g{\gamma}
\def\d{\delta}
\def\th{\theta}
\def\om{\omega}

\def\-{\hphantom{-}}

\def\s2{\frac{1}{\sqrt2}}

\def\oh{\frac{1}{2}}
\def\beq{\begin{equation}}
\def\eeq{\end{equation}}
\def\beqa{\begin{eqnarray}}
\def\eeqa{\end{eqnarray}}

\def\tr{{\rm tr \,}}

\def\ca{{\mathcal A}}

\def\Dsl{\,\raise.15ex\hbox{/}\mkern-13.5mu D} 

\def\e{\epsilon}

\def\CN {{\cal N}}

\def\CL {{\cal L}}

\def\CO {{\cal O}}

\def\tr{\mbox{Tr}}
\def\str{\mbox{STr}}

\def\be{\begin{equation}}
\def\ee{\end{equation}}
\def\bea{\begin{eqnarray}}
\def\eea{\end{eqnarray}}
\def\raw{\rightarrow}

\def\IC{\mathbb{C}}
\def\IN{\mathbb{N}}
\def\IZ{\mathbb{Z}}
\def\IR{\mathbb{R}}
\def\IP{\mathbb{P}}
\def\Id{{\mathbb{I}}}

\def\oh{\frac{1}{2}}
\def\a{{\alpha}}
\def\b{{\beta}}
\def\d{{\delta}}
\def\eps{{\epsilon}}
\def\th{{\theta}}

\def\lam{{\lambda}}

\def\om{{\omega}}
\def\sig{{\sigma}}

\def\g{{\gamma}}

\def\p{{\partial}}

\newcommand{\Eq}[1]{Eq.~(\ref{#1})}
\newcommand{\Fig}[1]{Fig.~\ref{#1}}

\newcommand{\Ref}[1]{Ref.~\cite{#1}}
\newcommand{\Sec}[1]{Sec.~\ref{#1}}
\newcommand{\Ch}[1]{Ch.~\ref{#1}}
\newcommand{\App}[1]{App.~\ref{#1}}
\newcommand{\Tab}[1]{Tab.~\ref{#1}}

\title{Esa}
\author{Luis Aparicio}
\makeindex

\begin{document}

\pagenumbering{roman}

\include{portada}

\newpage
\thispagestyle{empty}
\phantom{blabla}

\newpage
\thispagestyle{empty}
\phantom{blabla}

\newpage 
\thispagestyle{empty}
\phantom{blabla}




\pagestyle{fancy}
\renewcommand{\chaptermark}[1]{%
\markboth{#1}{}}
\renewcommand{\sectionmark}[1]{%
\markright{\thesection\ #1}}
\fancyhf{} 
\fancyhead[LE,RO]{\bfseries\thepage}
\fancyhead[LO]{\bfseries\rightmark}
\fancyhead[RE]{\bfseries\leftmark}
\renewcommand{\headrulewidth}{0.5pt}
\renewcommand{\footrulewidth}{0pt}
\addtolength{\headheight}{0.5pt}
\fancypagestyle{plain}{
\fancyhead{}
\renewcommand{\headrulewidth}{0pt}}


\tableofcontents

\newpage
\thispagestyle{empty}
\mbox{}
\newpage

\pagenumbering{arabic}
\setcounter{page}{1}


\include{intro}

\include{TypeII_vacua}

\include{STDM}

\include{Yukawas}

\newpage
\thispagestyle{empty}
\mbox{}

\include{conclusionsfin}


\newpage
\thispagestyle{empty}
\phantom{blabla}
\newpage
\thispagestyle{empty}
\phantom{blabla}

\appendix

\include{appendixA}

\newpage
\thispagestyle{empty}
\phantom{lala}




\end{document}

%% file: portada.tex
\pagestyle{headings}
\pagestyle{empty}

\newcommand{\HRule}{\rule{\linewidth}{1mm}}
\setlength{\parindent}{1cm}
\setlength{\parskip}{1mm}
\noindent
\HRule
\begin{center}
\Huge{\textbf{Some phenomenological aspects \\ of Type IIB/F-theory string compactifications}}
 \\[5mm]
\end{center}
\HRule

\vspace{1.2cm}

\begin{center}

	   Memoria de Tesis Doctoral presentada por\\
	      \textbf{Luis Aparicio de Santiago} \\
       para optar al título de \textbf{Doctor en F\'isica Te\'orica} por la\\
              \textbf{Universidad Aut\'onoma de Madrid}\\
\vspace{1cm}
Tesis doctoral dirigida por \textbf{Dr. D. Luis Ibañez Santiago} \\
catedrático del Departamento de F\'isica Te\'orica de la \\ Universidad Autónoma de Madrid

\end{center}

\vspace{0.1cm}

\begin{center}
{\Large {Departamento de F\'isica Te\'orica\\ Universidad Aut\'onoma de Madrid }}\\
\vspace{0.3cm}
{\Large {Instituto de F\'isica Te\'orica\\ \vspace{0.1cm} UAM-CSIC}}\\

\end{center}

\vspace{.5cm}

\begin{figure}[ht]
\centering
\begin{tabular}{lr}

\includegraphics[scale=0.1]{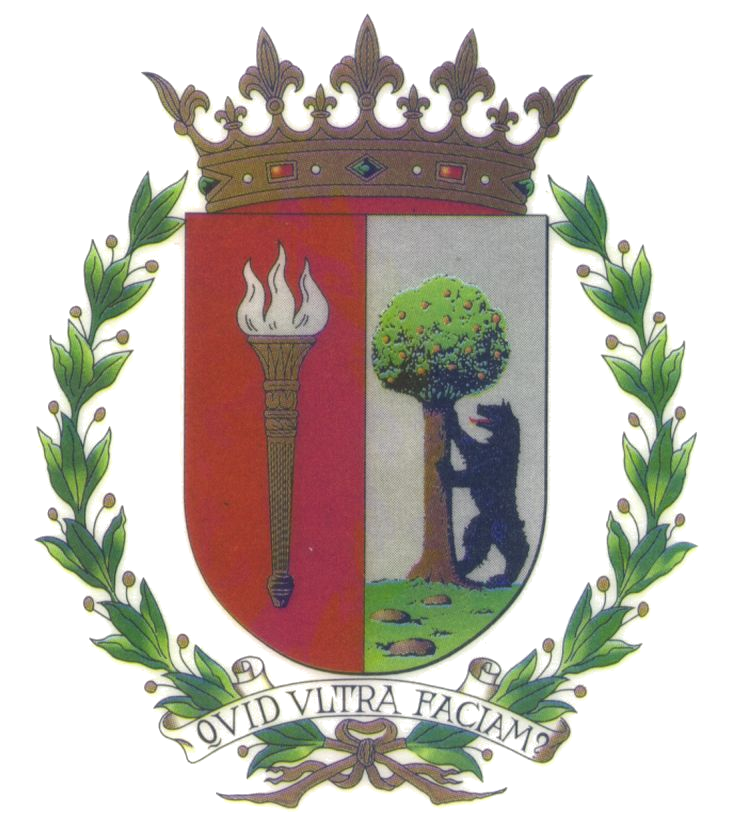}\qquad
&
\qquad\includegraphics[scale=0.5]{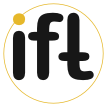}

\end{tabular}
\end{figure}

\vspace{1cm}

\begin{center}
 Junio de 2012
\end{center}

%% file: intro.tex
\chapter{Introduction}

\section{Holes in the standard paradigms}
At present, the laws describing how our Universe works seem to be irreconciliably split into two different worlds: the one corresponding to the physics of microscopic phenomena and the one which explains the (very) large scale structure. The first world is governed by quantum physics while the second one is dominated by gravity. Both approaches are respectively very well understood in terms of two mathematical frameworks: Quantum Field Theory \cite{weim1} and General Relativity \cite{wald}. The fact that General Relativity is a classical theory implies that gravity does not feel the quantum effects of the microscopic world. On the other hand, the existence of physical singularities at small space-time scales like e.g. Black Holes or the Universe initial singularity, suggests the necessity of a theory of Quantum Gravity. Nevertheless there is a problem, when one tries to quantize gravity in the usual fashion the theory turns inconsistent. \\

\renewcommand{\arraystretch}{1.4}
\begin{table}[h]
\begin{center}
\begin{tabular}{|c|c|c|c|c|c|}
\hline
\multicolumn{2}{|c|}{Fields} 
 & spin 1/2 & spin 0& spin 1 & $SU(3)_C \times SU(2)_L \times U(1)_Y$ \\  \hline \hline
 Quarks & $Q_L^i$ & $(u^i_L\>\>\>d^i_L)$ & & &  $(\>{\bf 3},\>{\bf 2}\>,\>{ 1\over 6})$ \\
($i\ = 1,\ 2,\ 3$) & $U_R^i$ & $(u^i_R)^c$ & & & $(\>{\bf \overline 3},\> {\bf 1},\> -{2\over 3})$\\
& $D^i_R$ & $(d^i_R)^c$ & & & $(\>{\bf \overline 3},\> {\bf 1},\> {1\over 3})$\\
  \hline
 Leptons & $L^i$ & $(\nu^i_L\>\>\>e^i_L)$ & & &  $(\>{\bf 1},\>{\bf 2}\>,\>-{1\over 2})$\\
($i\ = 1,\ 2,\ 3$) & $E_R^i$ & $(e^i_R)^c$ & & & $(\>{\bf 1},\> {\bf 1},\>1)$\\
  \hline
Higgs  &$\phi$ & &  $(\phi^-\>\>\>\phi^0)$ & & $(\>{\bf 1},\>{\bf 2}\>,\>-{1\over 2})$\\
\hline
Gluon & g & & &$(g^1\ ...\ g^8)$ & $(\>{\bf 8},\>{\bf 1}\>,\> 0)$ \\
\hline
W bosons &  W & & & $(W^+\>\>\> W^- \>\>\>W^0)$ & $(\>{\bf 1},\>{\bf 3}\>,\> 0)$ \\
\hline
B boson   & B & & &$B^0$ & $(\>{\bf 1},\>{\bf 1}\>,\> 0)$
\\
\hline
\end{tabular}
\caption{Standard Model particles and quantum numbers for the different representations of the gauge group.\label{StMod}}
\vspace{-0.6cm}
\end{center}
\end{table}

From the phenomenological point of view, the fundamental theories in nature are compiled in two paradigms with a succesfull predictive power: the Standard Model of particle physics and the Standard Cosmological Model. The Standard Model of particle physics \cite{hokim} is a succesfull Quantum Field Theory which may even be renormalizable if the Higgs boson is eventually found at the LHC. It is based on the gauge principle with a $SU(3)_C \times SU(2)_L \times U(1)_Y$  gauge invariance. As is shown in \Tab{StMod}, the fundamental blocks of the Standard Model are chiral Weyl fermions (quarks and leptons) representing matter fields and vector bosons describing the electroweak and strong interactions.\\ 

The Standard Cosmological Model \cite{mukanov} is supported in the theory of General Relativity under the assumption of the Cosmological Principle which states the Universe is spatially homogeneous and isotropic at large scales. This fact constraints the form of the metric of the Universe which takes Friedmann-Robertson-Walker form
\begin{equation}\label{FRW1}
ds^2 = -dt^2 + a^2(t)\left\lbrace\frac{dr^2}{1-kr^2} + r^2(d\theta^2 + \sin^2\theta d\varphi^2)\right\rbrace
\end{equation}
where $\left(r,\theta,\varphi\right)$ are the comoving coordinates, $a(t)$ the scale factor, $t$ the cosmic time, and $k$ the spatial curvature ($k > 0, k = 0, k < 0$ for a close, flat or open Universe, respectively). With this metric, Einstein equations relate the behaviour of the scale factor as a function of the curvature and the energy $\rho$ and pressure $p$ densities of the matter within the Universe  showing the dynamical evolution of the Universe with the form of the well known Friedmann equations
\begin{eqnarray}\label{Friedman1}
\frac{\ddot a}{a}& =& -\frac{4\pi}{3}(\rho + 3p) \\
\left(\dot a\over a\right)^2 &=& \frac{8\pi}{3}\rho - {k\over a^2} 
\end{eqnarray}

Despite of the experimental evidence supporting these two standard models, there exist some theoretical and phenomenological issues which do not accommodate inside these two paradigms and shows their incompleteness:
\begin{itemize}
\item Holes in Standard Model of Particle Physics: 
  \begin{itemize}
  \item \textit{Gravity}: as was commented before, we do not know how to combine consistenly quantum field theory with gravity.
  
  \item \textit{Coupling constants}: due to the behaviour of its beta-function, the QED coupling diverges for arbitrarily small distances. This implies that the Standard Model cannot be a fundamental theory. 

  \item \textit{Gauge group}: we know that the gauge group of the Standard Model is $SU(3)_C \times SU(2)_L \times U(1)_Y$ and matter fields transform in the fundamental representations of the group. However these choices are put by hand and there is no fundamental motivation. 
  
  \item \textit{Flavour problem}: why the Yukawa couplings have the structure they have? is there any kind of flavour symmetry explaining such hierarchical values or any dynamical mechanism for generating them? In the Standard Model there is no apparent explanation about neither the number of particle families nor the value of mixing paremeters or particle mass patterns.
  
  \item \textit{Spontaneous electroweak symmetry breaking}: a spontaneous breakdown of electroweak symmetry allows generating masses through the Higgs mechanism. However there is no understanding concerning the origin of the form of the scalar potential and the value of its parameters.
  
  \item \textit{Neutrino masses}: in the Standard Model there is an absence of $\nu_R$ and as a consequence neutrinos would not able to acquire masses through a direct Yukawa term, but only a non-renormalizable term of the form $LL\overline{H}\ \overline{H}$. The observed neutrino oscillation \cite{neutrinoosc} pattern can only be explained if at least two of the neutrino masses are non-vanishing and therefore, it is necessary to implement a mechanism to generate them.

  \item \textit{Hierarchy problem}: there is no protection mechanism in Standard Model to keep the scalar masses small. This implies that loop corrections to the Higgs mass are 17 orders of magnitude larger than the physical mass. As a consequence, during the renormalization process, we should make severe readjustments (1 part in $10^{26}$) at every loop order in order to get a value of order the TeV scale.  
  
  \item \textit{The question about unification}: the possibility of unifing the gauge coupling constants if one runs up their renormalization group equations is a aesthetical criterium which has been very well valued as a sign of an underlying fundamental theory. In the case of an unified Standard Model, this possibility is not realized. 
  \end{itemize}
  
\item Holes of Standard Cosmological Model
  \begin{itemize}
   \item \textit{Flatness fine-tunning}: one second after the Big Bang the $\Omega_{tot}$ must have been equal to 1 with a precision of 15 decimals in order to coincide with the recent observations of $\Omega_{tot}\sim 1$. Extrapolating in time back to the Planck scale $10^{-43}$, it would implies $\Omega_{tot}$ must have been 1 with a precission of 58 decimal places.  

  \item \textit{Horizon problem}: the Cosmic Microwave Background (CMB), which was released when the universe was $\sim 3 \cdot 10^5$ years old, is uniform in temperature to one part in $10^5$. A mechanism for stablishing this uniformity would need to transmit information at about 100 times the speed of light.

 
  \item \textit{CMB anisotropies origin}: the Standard Cosmological Model does not answer the question about the origin of anisotropies observed in the Cosmic Microwave Background, which are expected to be responsible also for the structures at large scale formation in the Universe.  
 \end{itemize}
  
\item Astroparticle and cosmological challenges: 
\begin{itemize}
 \item \textit{Dark energy}: observations  of type Ia supernovae at large red-shifts \cite{snova1} as well as the CMB radiation \cite{snova2} suggest that most part of the energy resides in an unknown new kind of negative-pressure energy called dark energy. These observations conclude that this dark energy is apparently accelerating the universe. The most popular explanations for this observation relie on one hand on the so called quintaessence models and also in Einstein's cosmological constant idea.
  
 \item \textit{Dark matter}: observations of cluster dynamics suggest that the amount of matter that can cluster in the Universe is about $\Omega_M \sim 0.3$ , while analysis of nucleosynthesis indicates that the amount of baryonic matter is much smaller, $\Omega_M \sim 0.045$. The MACHO abundance found by the Eros collaboration \cite{DM} reveals that MACHO's cannot contribute more than about $20\%$ of the galactic halo and thus cannot be used to explain the rotational curves of the galaxy. Thus, there should exist some non-baryonic dark matter. The minimal standard model does not provide any candidate for the non-baryonic dark matter and, therefore, cosmological observations again point in the direction of physics beyond the standard model.

  \item \textit{Baryogenesis}: observations indicate that the number of baryons in the Universe is grossly unequal to the number of antibaryons \cite{cosmray}. Since various considerations suggest that the Universe has started from a state with equal numbers of baryons and antibaryons, the observed baryon asymmetry must have been generated dynamically. Moreover, although the minimal standard model has the means of fulfilling the three Sakharov's conditions, it falls short to explaining the making of the baryon asymmetry of the universe. In particular, it is demonstrated that the phase of the CKM mixing matrix is an insufficient source of CP violation. It would be necessary to enlarge the symmetry breaking sector and adding a new source of CP violation to the Standard Model.

  \item \textit{Strong CP problem}: current upper bounds of the neutron electric dipole moment constrain the physically observable quantum chromodynamic (QCD) vacuum angle $|\overline{\theta}|\lesssim 10^{-11}$. In principle there is no reason for such a smallness, but there is a solution beyond the Standard Model which leads to the existence of the very light particle called axion which as a matter of fact is also a candidate for dark matter.

  \item \textit{Scale fine-tunning}: there exist a number of parameters and ratios whose smallness does not look like a result of a random choice, but are very small without any theoretical or phenomenological motivation. Some examples of these apparent fine-tunnings are: $\frac{M_e}{M_{top}}\approx 10^{-6}$, $\frac{M_W}{M_{Plank}} \approx 10^{-17}$, $\Lambda_c \approx 10^{-120}M_{Plank}^4$ or $|\theta_{QCD}|\lesssim 10^{-11}$. 

\end{itemize}

 \item The question of the origin: we can estimate that the Universe has about $10^{90}$ particles living on a space-time texture which also has physical dynamics. Where did they all come from?

\end{itemize}

\section{New data for new physics}
A pletora of new data coming from different observations and experiments have been arising in the last years shedding some light on the possible solutions of some of the problems and challeges above presented. LHC is running and its experiments: CMS, ATLAS, LHCb and ALICE are beginning to constraint the parameter space of different models. In the other hand, the Plank experiment will soon give us new data from CMB, and a big amount of other experiments related with neutrinos, flavour physics, dark matter detection, dark energy analysis, gravitational waves etc. exist or are planned.\\

\begin{figure}[h!]
\hspace*{-0.6cm}
\centering
\includegraphics[width=8cm, angle=0]{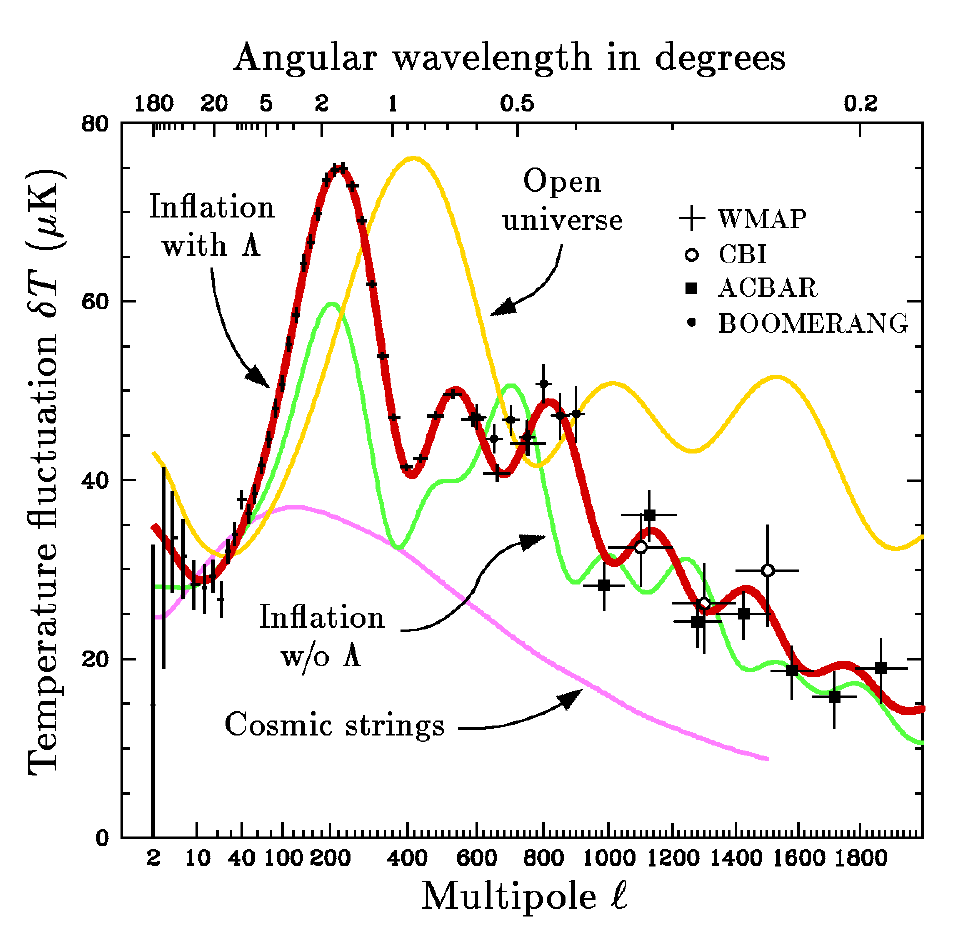} 
\caption{Comparison of the latest observational measurements of the temperature fluctuations in the CMB with several theoretical models. The solid lines show the predictions of the simplest inflationary models. The best fit corresponds to $\Lambda$CDM model \cite{etinf}.
}
\label{inflation}
\end{figure}

Concerning the Standard Cosmological Model the most convicing hypothesis seems to be the inflationary one \cite{inflation}. The $10^{-34}$ seconds period of exponential expansion at the beginning of the Universe solves the horizon and flatness problems. Moreover, thanks to inflation, quantum effects are visible in the sky, because during this process quantum mechanical fluctuations are engraved under the form of density fluctuations which constitutes the first scaffolding for the large scale structure formation and today are observed as the CMB temperature fluctuations. Inflation is also the best option from the observational point of view, its predictions for CMB (adiabatic, Gaussian and scale invariant spectrum) fits almost perfectly with data (see e.g. \Fig{inflation} \cite{etinf}). Nowadays, no other natural extension of the Big Bang cosmology is able to reproduce the CMB temperature fluctuations data so well.\\

\begin{figure}[!b]
  \includegraphics[width=8cm, angle=0]{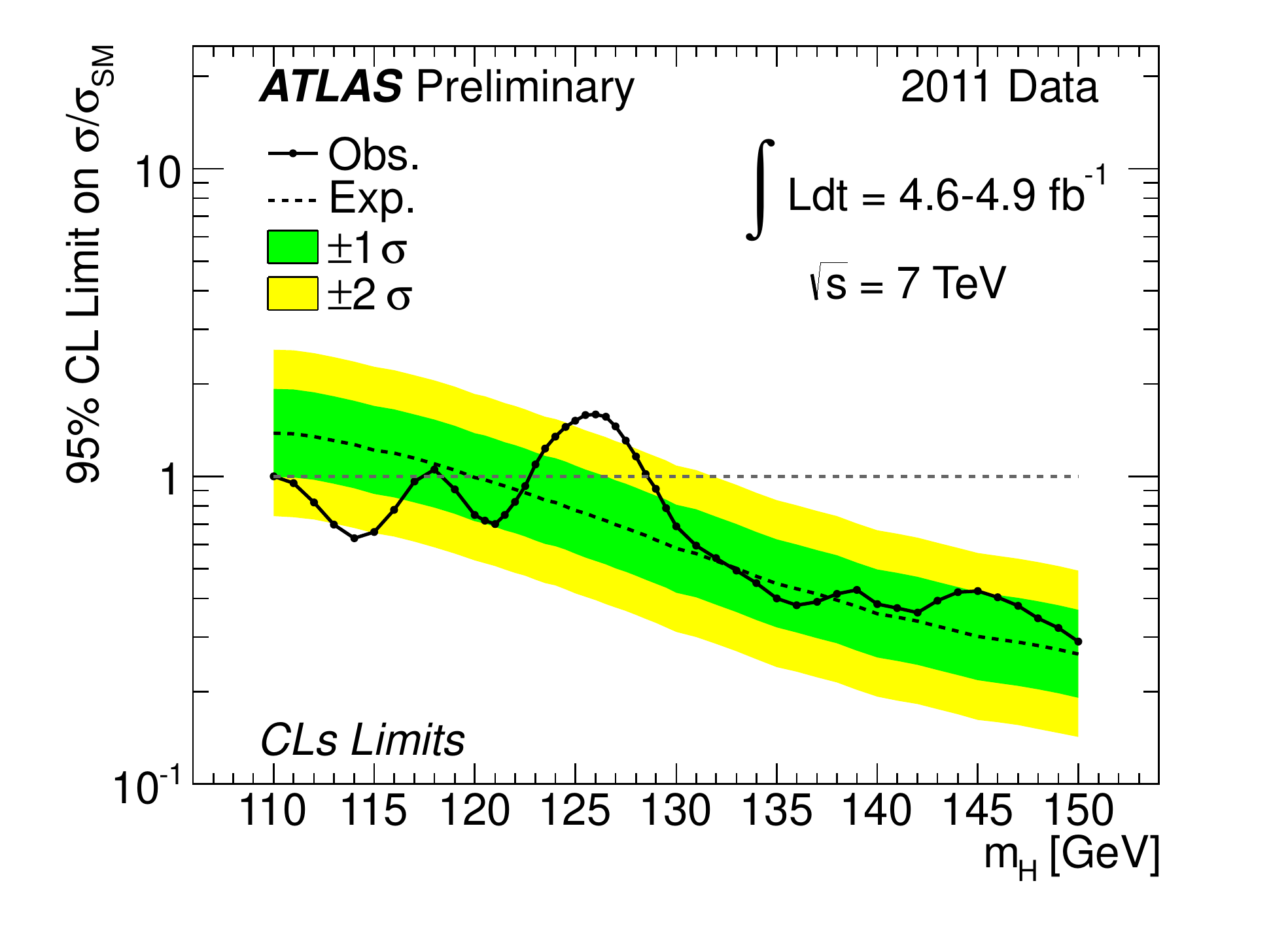}
  \hspace*{-1cm}
  \raisebox{2ex}{\includegraphics[width=8cm, angle=0]{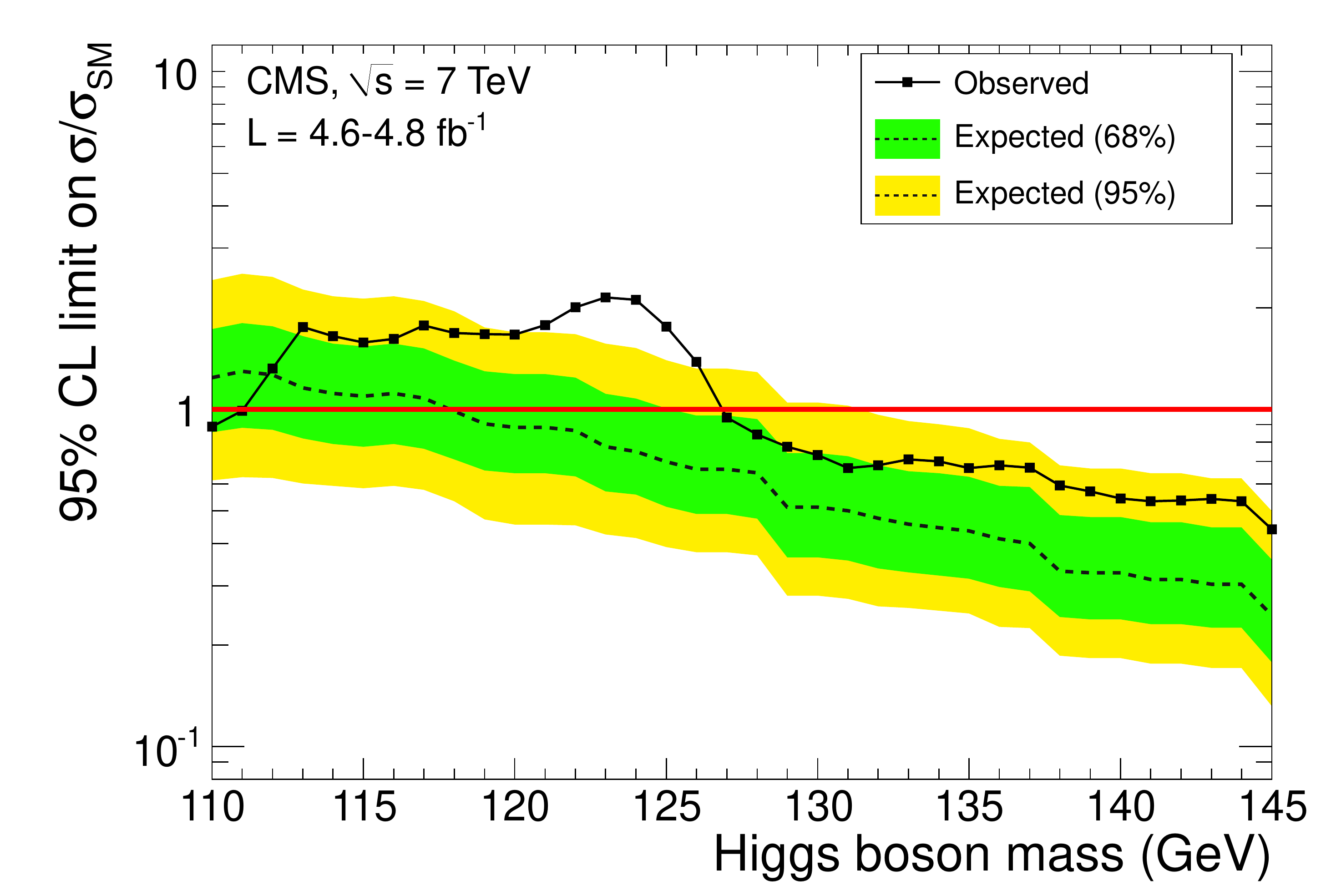}}
  \caption{SM Higgs exclusion limit at $95\%$ confidence level for $\sim4.7 fb^{-1}$ proton-proton data collected by ATLAS (left) and CMS (right) in 2010 and 2011, showing the lower Higgs mass region. Hints of a Higgs with mass around 125 GeV are observed.}
  \label{higgs}
\end{figure}

For the case of particle physics, there is vast number of models proposing new physics beyond Standard Model (SM): supersymmetry, composite models, little Higgs, extra dimensions, different kind of GUT theories, Higgsless models... Things are not so easy as in the cosmological case in order to discriminate among all possibilities. However CMS and ATLAS have published recently the first analysis on a (possible) Higgs boson signal with a mass $m_H \approx 124 - 126$ GeV (see \Fig{higgs}).

The results from CMS and ATLAS were achieved by combining searches in a number of predicted Higgs decay channels including: pairs of W or Z bosons, which decay to four leptons; pairs of heavy quarks; pairs of tau leptons; and pairs of photons. These preliminary results exclude SM Higgs boson in a range of masses of 127 – 600 GeV at $95\%$ CL. They do not exclude a SM Higgs boson with a mass between 115 GeV and 127 GeV at $95\%$ CL.

Assuming that these first results are giving us hints of the very existence of a Higgs boson with a mass around 125 GeV we can analyze the implications for physics beyond SM. First, some recent works in vacuum stability bounds \cite{spinozastrumia, Isisdori} shows that for $m_{top} \simeq 173$ GeV the $\lambda$ parameter of the Higgs potential becomes negative for a energy scale of $\Lambda \sim 10^{11}$, destabilizing the electroweak vacuum. In addition, the Higgs mass parameter $\mu^2$ blow up if we start to probe arbitrarily small distances (high energies) due to the loop corrections inducing a naturalness problem. This could suggest that at high energies the SM Higgs model is no longer valid and therefore cannot be understood as a fundamental theory of the electroweak sector i.e. there should be new degrees of freedom whose effects will appear at high energies and which contains electroweak sector as a low energy effective theory.\\

If $M_H\ =\ 125$ GeV, this constraints very much (even rules out) the simplest realizations of many models proposed for solving the weak points of the electroweak Higgs sector. Nevertheless the MSSM predicts a Higgs mass value $M_H \lesssim 130$ GeV which is in striking agreement with the not excluded range of masses given by LHC data. Supersymmetry (SUSY) \cite{martin, tatabaer} has a number of attractive theoretical and phenomenological aspects. SUSY is the most general symmetry of the S-matrix. It enlarges the Poincaré group to a so called super-Poincaré group that relates bosons and fermions through a symmetry transformation. From the phenomenological point of view some attractive features are:
\begin{itemize}
 \item[-] SUSY protects scalars from having large radiative corrections due to a loop by loop cancellation between fermion and boson quadratic contributions, and hence avoids problems related with naturalness and hierarchies.

\item[-] SUSY also protects the stability of the scalar potential from radiative corrections.

\item[-] By promoting supersymmetry transformations from global to local, in analogy with gauge theories, it arises a spin 2 massless gauge field: the graviton. It also appears its superpartner, the gravitino, which has the role of a gauge field of local diffeomorphisms. Therefore, the resulting theory contains General Relativity. This $N=1$ SUSY version of gravity is called Supergravity (SUGRA).

\item[-] Electroweak symmetry breaking is realized in a radiative way and hence with a natural dynamical mechanism. In this way the scalar potential parameters arbitrariness is much improved.

\item[-] The minimal supersymmetric extension of the SM (MSSM) allows the gauge coupling unification of SM couplings at a scale $M_{GUT} \sim 10^{16}$ GeV.

\item[-] The MSSM with a conserved R-parity contains a massive, electric and color neutral particle which constitues a very good candidate for dark matter.
\end{itemize}
   
In spite of all these attractive features, it remains for LHC to test whether supersymmetry is actually realized in nature.

\section{String Theory and its phenomenology}
Supersymmetry and inflationary theories try to address some of the problems of the standard models of particle physics and cosmology respectively. But those extensions of traditional paradigms are not yet tested and many questions remain unanswered. For instance, concerning inflation one can ask himself about the nature of inflaton: is it a scalar simple field or is composite?, is there a single field or are several?, what is the origin of this field?, is there any mechanism to produce the inflation potential which is put by hand? etc. In the case of SUSY, one important feature is that none of the superpartners of the Standard Model particles has been yet discovered. Therefore supersymmetry should be a broken symmetry of Nature. However in order to mantain the good properties above considered it should be broken in a \textit{soft} way. One consequence of soft SUSY breaking is the arising of a hundred of free parameters which bring new CP violations phases, mass patterns, mixing angles etc. In other words, that introduces a big flavour problem due to appearance of FCNC's, SUSY contribution to very constrained SM processes or large amounts of CP violation. All this parameters should be fixed by hand in order to avoid the data bounds so that one can ask himself if there is a more fundamental theory in which these particular choices have a raison d'être.\\

Consequently, both cosmology and particle physics point out the need for an underlying theory which can explain several apparently fanciful choices of their fundamental parameters and also the origin of their degrees of freedom. Besides, there are still several questions unsolved like the origin of dark energy, the strong CP problem, neutrino masses, cosmological constant and so on and so forth. String Theory \cite{polchinskis, gsw} is the most serious candidate for a fundamental theory which incorporates in principle the ingredients of the SM of particle physics and cosmology. It has a number of attractive features to describe a unified theory of all interactions \cite{book}. On the other hand, String Theory is also the most promising candidate for a consistent theory of quantum gravity and has several theoretical achievements such as the string theory computation of the Bekenstein-Hawking black hole entropy \cite{BH} or the explicit realization of the holographic principle in Maldacena’s AdS/CFT conjecture \cite{maldacena}.\\

\begin{figure}[h!]
\hspace*{-0.3cm}
\begin{center}
\includegraphics[scale=0.7]{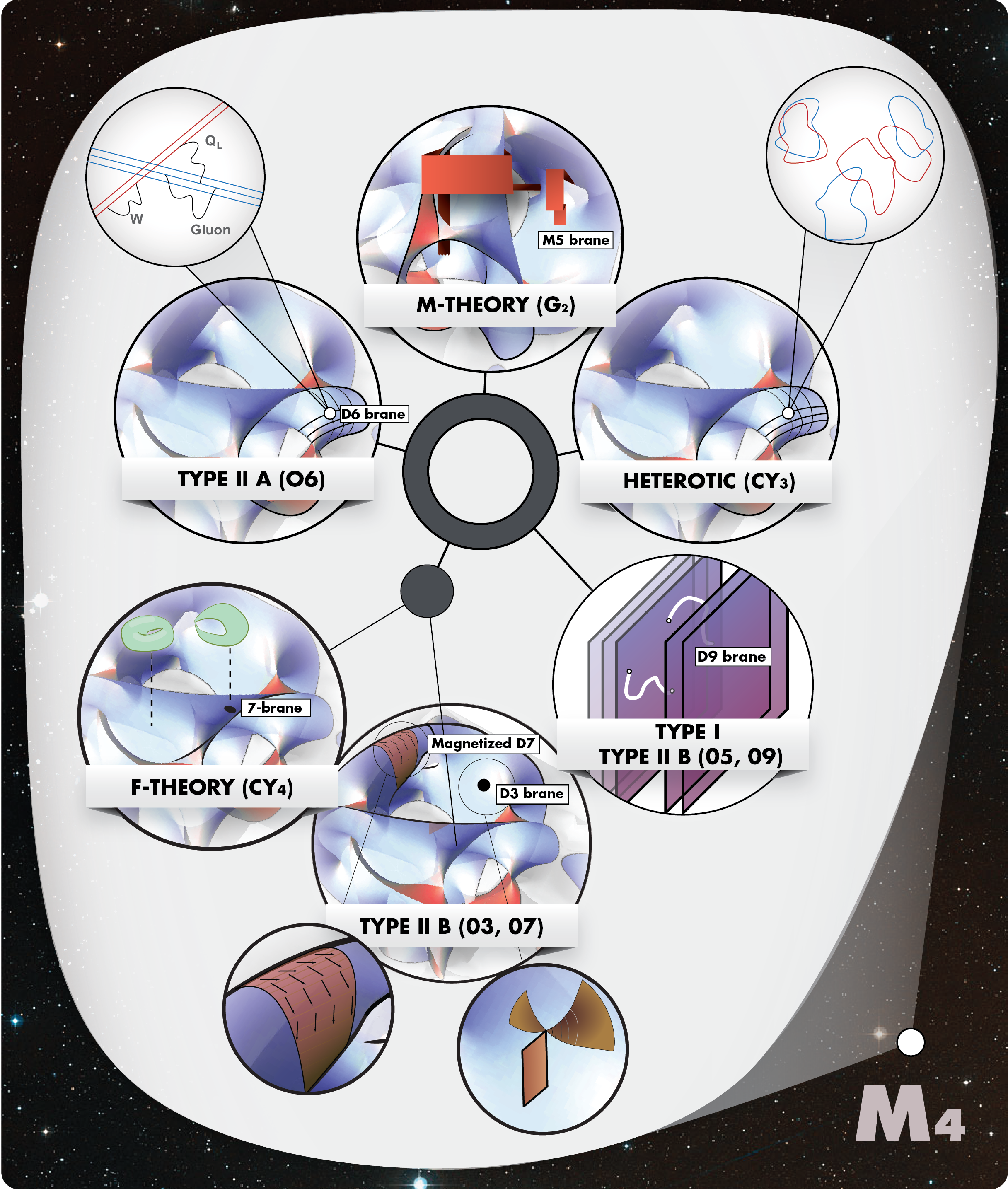}
\caption{Artistic view of the 10d (11d) string theories compactified as $M_4 \times \mathcal{X}_6 $ leading to the five large classes of 4d chiral compactifications.}
\label{star}
\end{center}
\end{figure}

From the point of view of unification, the essetial aim is to understand how the SM or the MSSM may be obtained as a low energy limit of String Theory. There exist different options to achive the construction of 4d compactifications which lead to a chiral spectrum of massless fermions at low energies. In \Fig{star} we can find a summary of the possible chiral vacua classified according to the underlying 10d or 11d theory taken as starting point. All the branches of the circle in \Fig{star} are connected through different dualities so that the different vacua should be considered as a single underlying theory but reflected in different mirrors. 
    
\section{Plan of the thesis}

We are going to focus on Type II compactifications because of the potential for the construction of realistic MSSM-like compactifications. In particular we will concentrate in Type IIB Calabi-Yau orientifolds and its non-perturbative realization: F-theory. These sort of models, have attracted a lot of attention during recent years due to their phenomenological interest. The plan of the thesis is as follows:
\begin{itemize}
 \item \Ch{ch1} will be devoted to an introductory survey of some concepts and aspects of Type II vacua like e.g. the low energy effective action or soft terms. It is also included a brief presentation of F-theory stressing the phenomenological interest of local models.  

\item In \Ch{stdm} we present an analysis of the theoretical and phenomenological issues of modulus dominated SUSY breaking \cite{aci}. In addition it is examined its status in comparison with recent LHC data \cite{aciII}. 

\item \Ch{capyuk} is devoted to the analysis of flux and instanton effects on local F-theory models. Yukawas and matter fields wave functions corresponding to these models are calculated. The results may allow for an understanding of the problem of fermion hierarchies in the Standard Model \cite{afim}.

\item Finally, an appendix and some conclusions. 
\end{itemize}

%% file: TypeII_vacua.tex
\chapter{Some aspects of Type II vacua}\label{ch1}

\section{D-branes in Type II string theory}
Beyond the perturbative sector in String Theory, there exist some non-perturbative states which play a fundamental role in the theory: D\textit{p}-branes. They can be interpreted in a double way: in one hand they are soliton-like solutions \cite{Ortin:2004ms} (topological defects) of the Type II low energy supergravity equations of motion with $p+1$ extended dimensions where $p$ is the number of spatial dimensions. In the other hand, they admit a fully stringy perturbative description as $(p+1)$-dimensional subspaces on which open strings can end \cite{Polchinski:1995mt}. The interactions with closed string sector show some of the physical properties of D$p$-branes. In particular D$p$-branes are sources of RR and NSNS fields \cite{Polchinski:1995mt}. They are sources of $C_{p+1}$ RR fields and of graviton and dilaton NSNS fields, and due to these couplings they carry charge under RR $(p+1)$-forms and tension (because the interaction with the 10-dimensional graviton). Since in Type IIB the RR fields have rank $(2p+1)$, the only posible (BPS) D-branes will be D$(2p+1)$-branes and from a similar argument in Type IIA we will have D$(2p)$-branes.\\

Another important feature of D$p$-branes in flat spacetime is that they preserve a linear combination of left and right moving supersymmetries of the 10-dimensional theory
\begin{equation}
 Q = \epsilon_R Q_R + \epsilon_L Q_L
\end{equation}
where $Q_L$ and $Q_R$ are the 16-component spinor Type II supercharges and $\epsilon_{L,R}$ are spinor coefficients satisfying
\begin{equation}
 \epsilon_L = \Gamma^0...\Gamma^p\epsilon_R
\end{equation}
This implies that D$p$-branes are $\frac{1}{2}$BPS states and therefore there exists a relation between the RR charges and their tension.\\

The tension-charge relation above metioned has a very important consequence. We can consider dynamically stable configurations of several parallel D$p$-branes  because there is no net force among parallel branes due to the cancellation of the gravitational attraction by the Coulombian RR charge repulsion. In this kind of configurations, if we label each one of the $n$ branes with an index $a \ =\ 1,\ ... ,\ n$ (Chan-Paton degrees of freedom, see below) \cite{Paton:1969je}, we see that we have $n^2$ open string sectors labelled $ab$ for an open string starting at the $a-th$ D-brane and ending at $b-th$ D-brane.\\
 
\noindent The mass formula for the open string states is given by:    
\begin{equation}
 M_{ab}^2 = \sum_{i=p+1}^9 \left(\frac{x_a^i-x_b^i}{2\pi\alpha'}\right)^2 + \frac{1}{\alpha'}(N_B + N_F + E_0)
\end{equation}
where $x^0,\ ...,\ x^p$ are the spanning directions of the D-branes which are located at $x_a^i$ in the $(9-p)$ transverse directions and where $E_0 = -1/2,\ 0$ in the NS and R sectors. $N_B$ and $N_F$ are the bosonic and fermionic number operators and $\alpha'$ is the inverse of the string tension squared. The massless sector corresponds to a $(p+1)$-dimensional theory with $U(n)$ gauge symmetry bosons in the sector $ab$ and $U(1)^n$ in the $aa$ sector . This stack of coincident D-branes promote the abelian $U(1)$ gauge symmetry of single branes to a non abelian gauge group \cite{Witten:1995im}. The rest of the massless spectrum for D-branes in flat space is given by $(9-p)$ adjoint scalars and adjoint fermions, filling a $U(n)$ vector multiplet with respect to the 16 unbroken supersymmetries.\\

The Chan-Paton factors mentioned above are labeled by $\lambda^a$ and are interpreted as a set of degrees of freedom for every open string end. These Chan-Paton labels can be represented by matrices that satisfy a Lie algebra as a symmetry group of open string interactions, so that $\lambda^a$ can be chosen as hermitian matrix generators where $a$ is the adjoint index. The symmetries of open string scattering amplitudes turn out to be compatible with symmetry algebras $U(N)$ (also $\ SO(N),\ \mathrm{or}\ Sp(N)$ in the presence of orientifold planes \cite{Schwarz:1982md, Marcus:1982fr}). This symmetry is a global symmetry from the point of view of the worldsheet sigma-model, but local (gauge) in the ten-dimensional target space-time.\\

Using these configurations of D$p$-brane stacks we can obtain a general gauge group of the form:
\begin{equation}
 \mathcal{G} = \prod_a U(N_a) \times \prod_b SO(N_b) \times \prod_c Sp(N_c) 
\end{equation}
and open strings ending on the branes will transform in the adjoint or bi-fundamental $(\square_a,\ \overline{\square}_b)$ representations\footnote{For the case of $SO(N)$ or $Sp(N)$ we will have symmetric, anti-symmetric and $(\square_a,\ \square_b)$ representations because of the $\Omega$ orientifold action.}. These may correspond to some of the essential blocks of Standard Model and that is the reason why D$p$-branes could be useful for connecting particle physics and String Theory \cite{Uranga:2005wn}.

\section{Type IIA orientifolds}
\subsection{Intersecting branes}
\begin{figure}[tb]
\begin{center}
\includegraphics[scale=0.65]{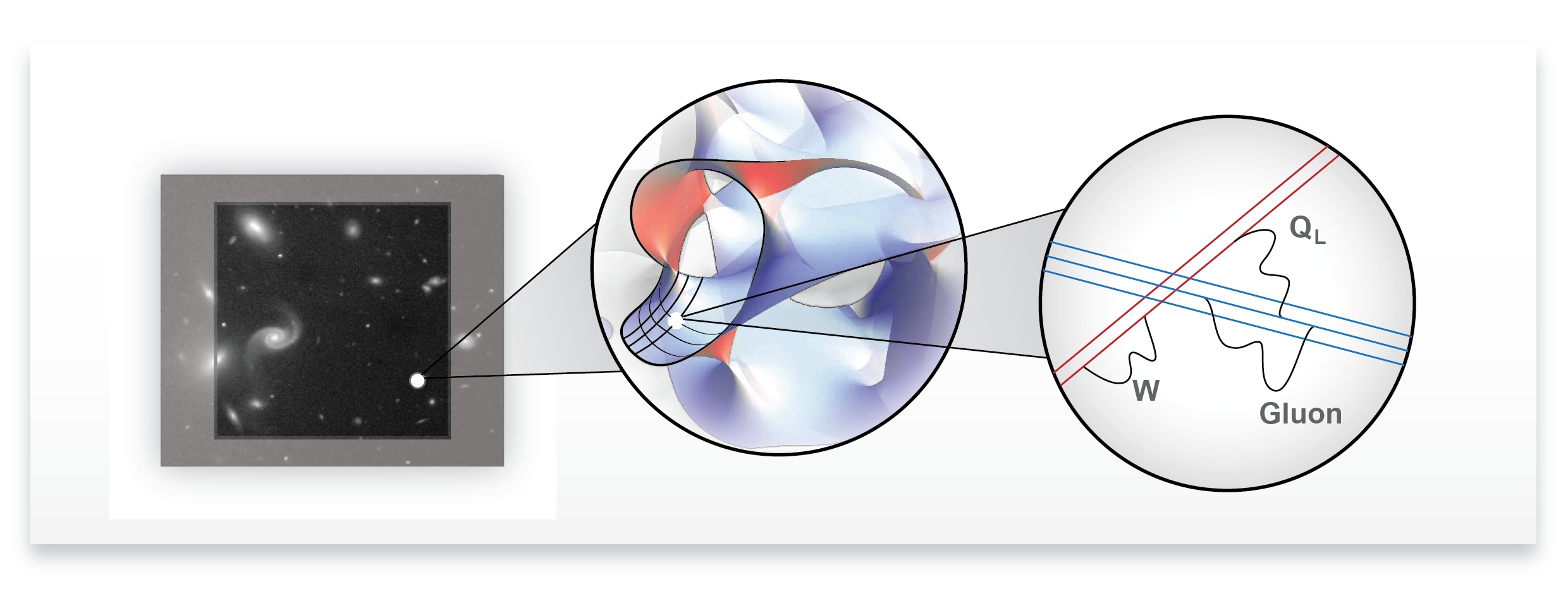}
\caption{\small{Pictorial representation of the idea of intersecting branes.}}\label{fig2Marchart}
\end{center}
\end{figure}

\begin{figure}[tb]
\begin{center}
\includegraphics[scale=0.6]{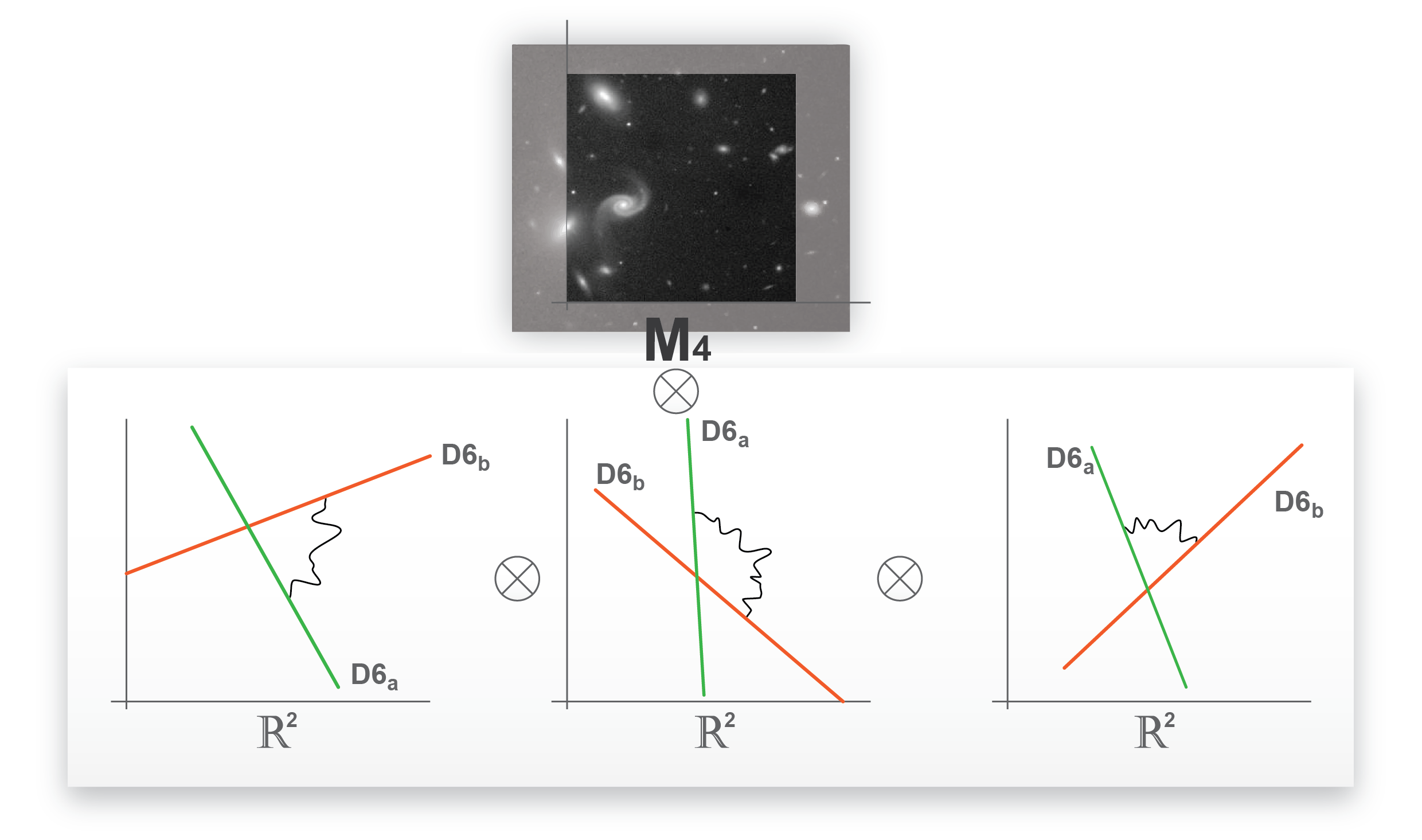}
\caption{\small{Diagram of two D-branes locally intersecting at angles in $M_{10} =M_4 \times \mathbb{R}^2 \times \mathbb{R}^2 \times \mathbb{R}^2$. The branes intersect each other at angles in the extra dimensions, going from brane $D6_a$ to $D6_b$ in the $ab$ sector.}}\label{fig23.eps}
\end{center}
\end{figure}

Up to now we have studied the D$p$-branes in the 10-dimensional space-time. However we need to make compact 6 of the 10 dimensions in order to recover the four dimensional physics. The different stacks of D-branes may intersect each other and at these intersections live the quarks and leptons in realistic models motivated by the Standard Model. A schematic picture of this idea is shown in \Fig{fig2Marchart}. Before dealing with the intersecting branes in compactified dimensions let us consider them as hyperplanes in $M_{10}$ where open strings are attached in and let us make them to intersect in flat space \cite{Berkooz:1996km, Arfaei:1996rg, Balasubramanian:1996uc, SheikhJabbari:1997cv}. Then we can say that two of these $p+1$-planes are related by an isometry belonging to $(SO(8)/SO(p-1))\times T_{9-r}$ where $T_s$ is the group of translations in $s$ dimensions and $r$ the dimension of the direct sum of both tangent spaces. Considering a vector basis where a rotation $R \in SO(8)/SO(p-1)$ can be diagonalized and rewriting the coordinates in terms of complex coordinates  $z^m = x^{2m} + i x^{m+1}$ we can express such a rotation without loss of generality as:
\begin{equation}
 R : z^m \rightarrow e^{i\pi\theta^m}z^m 
\end{equation}
so that the rotation preserves the complex structure. $R$ can be seen as an element of the subgroup $U(4) \subset SO(8)$. One example of this situation is illustrated in \Fig{fig23.eps} for $p=6$, i.e. D$6$-branes.\\

The mass operator for the open string sector beginning at $a$ ($\sigma=0$) and ending at $b$ ($\sigma = \pi$) is \cite{Berkooz:1996km}:
\begin{equation}
\alpha'M_{ab}^{2}=\frac{Y^{2}}{4\pi^{2}\alpha'}+ N_{\nu}+\left(\sum_{\mu=1}^{4}\theta_{ab}^{m}-1\right)\nu \ ,
\end{equation}
where $\alpha'$ is the string tension, $Y^{2}$ is the squared length of the transverse separation of two D-branes (just in case they are parallel in some of the dimensions), and $N_{\nu}$ stands for the number operator of the oscillations, which on the Ramond $(\nu=0)$ and Neveu-Schwarz $(\nu=\frac{1}{2})$ sectors is given by
\begin{eqnarray}
N_0 &  =  &
\sum_{\mu=1}^4 \left[\sum_{n>0}\left( \alpha^{\mu}_{-n_{+}} \alpha^{\mu}_{n_+} + \alpha^{\mu}_{-n_-} \alpha^{\mu}_{n_-} \right) + \alpha^{\mu}_{-\theta^{i}} \alpha^{\mu}_{\theta^{\mu}}\right] \nonumber \\
& + & \sum_{\mu=1}^4 \left[\sum_{r>0}\left( r_+\psi^\mu_{-r_+} \psi^\mu_{r_+} + r_-\psi^\mu_{-r_-} \psi^\mu_{r_-} \right) + \theta^\mu\psi^\mu_{-\theta^\mu} \psi^\mu_{\theta^\mu}\right] \label{numero1} \\
N_{\frac{1}{2}} &  =  & 
\sum_{\mu=1}^3 \left[\sum_{n>0}\left( \alpha^\mu_{-n_+} \alpha^\mu_{n_+} + \alpha^\mu_{-n_-} \alpha^\mu_{n_-} \right) + \alpha^\mu_{-\theta^\mu} \alpha^\mu_{\theta^\mu}\right] \nonumber \\
& + & \sum_{\mu=1}^4 \left[\sum_{r>0}\left( r_+\psi^\mu_{-r_+} \psi^\mu_{r_+} + r_-\psi^\mu_{-r_-} \psi^\mu_{r_-} \right)\right]. 
\end{eqnarray}
As usual in string theory, $\alpha_{\pm n}^{\mu}$ and $\psi_{\pm r}^{\mu}$ come from expansion of the bosonic $X^{\mu}$ and fermionic $\Psi^{\mu}$ world-sheet operators but keeping in mind that the boundary conditions are the ones corresponding to branes intersecting at angles.  This boundary conditions can be shown considering two D6-branes in flat space intersecting in the $(x^i,x^j)$-planes with $i=4,6,8$ and $j=5,7,9$, therefore the boundary conditions can be expressed as a mixing of Neumann and Dirichlet:
\begin{eqnarray}
\left .\partial_\sigma X^i  \right|_{\sigma=0}=0 &  &\left .\cos\theta_{ij}\partial_\sigma X^i + \sin\theta_{ij}\partial_\sigma X^j\right|_{\sigma=\mathcal{\ell}} = 0 \label{intbr} \\
\left .\partial_\tau X^j  \right|_{\sigma=0}=0 &  &\left .-\sin\theta_{ij}\partial_\tau X^i + \cos\theta_{ij}\partial_\tau X^j\right|_{\sigma=\mathcal{\ell}} = 0 \label{intbr2}
\end{eqnarray}
and subindices stand for the corresponding oscillation mode: $n\in \mathbb{Z}$ for bosonic oscillators and $r\in \mathbb{Z}+\nu$ for fermionic oscillators with $\nu$ integer for the Ramond sector and half-integer for Neveu-Schwarz sector.\\

A more efficient way for treating the string oscillations (at least the lightest ones) is provided by the bosonize formulation \cite{gsw, Berkooz:1996dw}. In this formulation the rotation is encoded in a four dimensional shift vector $v_{\theta}=(\theta_{ab}^{1},\theta_{ab}^{2},\theta_{ab}^{3},\theta_{ab}^{4})$ and oscillating states are described by the vector sum $r+v_{\theta}$, where $r_i \in (\mathbb{Z} + \frac{1}{2}+ \nu)$. The GSO projection \cite{SheikhJabbari:1997cv, Gliozzi:1976jf} is implemented through the condition $\sum_{i}r_{i}=odd$, and the mass of any state is given by
\begin{equation}
\alpha'M_{ab}^{2}=\frac{Y^{2}}{4\pi \alpha'}+ N_{bos}(\theta)+\frac{(r+v_{\theta})^{2}}{2}-\frac{1}{2}+E_{ab} \label{massbrane}
\end{equation}
where $N_{bos}(\vartheta)$ stands for the bosonic oscillator contribution and $E_{ab}$ is the vacuum energy
\begin{equation}
E_{ab}=\sum_{\mu=1}^{4}\frac{1}{2}|\theta^{m}|(1-|\theta^{m}|).
\end{equation}

Notice that each of the $ab$ states is localized at the D-branes intersections, so both the chiral fermions and the scalars are stuck at a single point in the dimensions additional to $M_{4}$. Concerning the rest of the sectors (closed strings, open strings attached to a single D-brane) appearing in this kind of configurations are similar to other string constructions \cite{Aldazabal:2000sa}.\\

\subsection{D6-branes in toroidal and orientifold compactifications}
In Type IIA  we have only D4, D6 and D8-branes which can contain $M_4$ inside. They wrap respectively 1, 3 and 5-cycles in $\mathbf{X}_6$ compact dimensions, but $\bf{CY}_3$ compact spaces do not admit 1 or 5 cycles \cite{aspinwall} because those generically have Betti numbers $b_1 = b_5 = 0$ and hence only D6-branes are available for this analysis.\\

One of the main problems of 4d Type II theory is the large amount of supersymmetry preserved in the D-brane configurations. This can be heuristically understood in the following way \cite{Uranga:2005wn}: the 4d chirality is a violation of 4d parity and because of the correlation between the 4d chirality and the chirality in the 6d extra-dimensions (implied by the GSO projection) it is needed to have a preferred six dimensional orientation in order to violate the 6d parity. Now, let us impose the identification of some of the D-brane worldvolume dimensions with $M_4$. This implies configurations of D$(3+n)$-branes and therefore each pair of D-branes will be related not by a general $SO(8)$ rotation but by an $SO(6)$ element or a $SU(3)$ element in a particular complex basis if we choose properly the angle relations. Hence we must set $\theta^4 \equiv 0$. As a consequence, the fourth entry of our vector $r + v_\theta$, which will be either an integer ($\nu = 0$) or half integer ($\nu = \frac{1}{2}$) number, will provide the four-dimensional Lorentz quantum number of our state. Thus, a string oscillation on the Ramond (R) sector will yield four-dimensional fermions, whereas Neveu-Schwarz (NS) excitations will yield bosons from the four dimensional point of view.\\ 

As chirality appears from the fact that two D$p$-branes introduce a preferred orientation in the transverse 6d extra-dimensions, then we need a $p$ big enough for defining such orientation. For example two stacks of D6-branes in flat 10d space $M_4 \times \mathbb{R}^2 \times \mathbb{R}^2 \times \mathbb{R}^2$ spanning $M_4$ times a line in each of the three 2-planes have enough dimensions to define an orientation in the 6d transverse space, but that is not possible in the case of two stacks of D5-branes in Type IIB. For the case of D6-branes it is easy to see that the only possible massless state of the $R$-sector is 
\begin{equation}
 r + v_{\theta} = \left(-\frac{1}{2} + \theta_1 ,\ -\frac{1}{2} + \theta_2,\ \frac{1}{2} + \theta_3 ,\ -\frac{1}{2} \right) 
\end{equation}
where the orientation has been set with $\theta_1, \theta_2 \geq 0 , \theta_3 \leq 0$ and allowing $\sum_i \theta_i = 0$. This means that the combination of 10d GSO projection and orientation kills one of the two possible 4d spinors leaving us with chirality in our 4d field theory.\\

The angles relation condition we need for having a $SU(3)$ subgroup of $SO(6)$ rotations beween branes is given by  
\begin{equation}
\sum_i \theta_i = 0 
\end{equation}
which depending on the angles choice has a more general form 
\begin{equation}
 \pm\theta_1 \pm \theta_2 \pm \theta_3 = 0 \ \mathrm{mod}\ 2\pi
\end{equation}

Notice that configurations satisfying this kind of condition generically preserve 4d $\mathcal{N}=1$ supersymmetry because there exist Killing spinors in the transverse dimensions $\mathbf{X}_6 = \mathbb{R}^2 \times \mathbb{R}^2 \times \mathbb{R}^2$, i.e there are $SU(3)$ singlets under the spinor decomposition from $SO(6)$ to the $SU(3)$ subgroup:
\begin{equation}
\begin{array}{ccccc}
SO(10)  & \rightarrow & SO(6)\times SO(1,3)  & \rightarrow & SU(3) \times SO(1,3)  \\
\mathbf{16} &  \rightarrow & (\mathbf{4},\mathbf{2}) + (\mathbf{\bar{4}},\mathbf{2'}) & \rightarrow & (\mathbf{3},\mathbf{2}) + (\mathbf{\bar{3}},\mathbf{2'}) + (\mathbf{1},\mathbf{2}) + (\mathbf{1},\mathbf{2'})  
\end{array}
\end{equation}
where $\mathbf{2}$ or $\mathbf{2'}$ are representing the chiral handedness in 4d.\\  
  
Notice also that, given that the volume wrapped by the D-branes in the flat case ($M_4 \times \mathbb{R}^2 \times \mathbb{R}^2 \times \mathbb{R}^2$) is  infinite, the gauge coupling constants would be infinitely suppresed and the gauge symmetry would look in an effective way as a global one. It is necessary, in order to get phenomenologically viable situations, to consider constructions in which the D$p$-branes wrap compact cycles of six-dimensional manifolds.

\subsubsection{Toroidal compactifications} 
The simplest constructions are the compactifications of the corresponding Type IIA string theory on a $T^{6}$ manifold, with the D$6$-branes wrapping compact $3$-cycles on the $T^{6}$. Actually, such constructions were considered in \cite{Aldazabal:2000dg, Blumenhagen:2000wh}, where it was found that only for the $p=6$ case chirality and $N =1$ SUSY was achieved. Aside of these simple toroidal-like settings, other approaches have been considered in the literature, such as orbifold compactifications \cite{Blumenhagen:2001mb}, and compactifications on more general Calabi-Yau spaces \cite{Uranga:2002pg, Blumenhagen:2002wn}. The latter kind of models are in general less useful from the point of view of model building since the curved geometry is notoriously difficult for explicit computations.\\

\begin{figure}[tb]
\begin{center}
\includegraphics[scale=0.6]{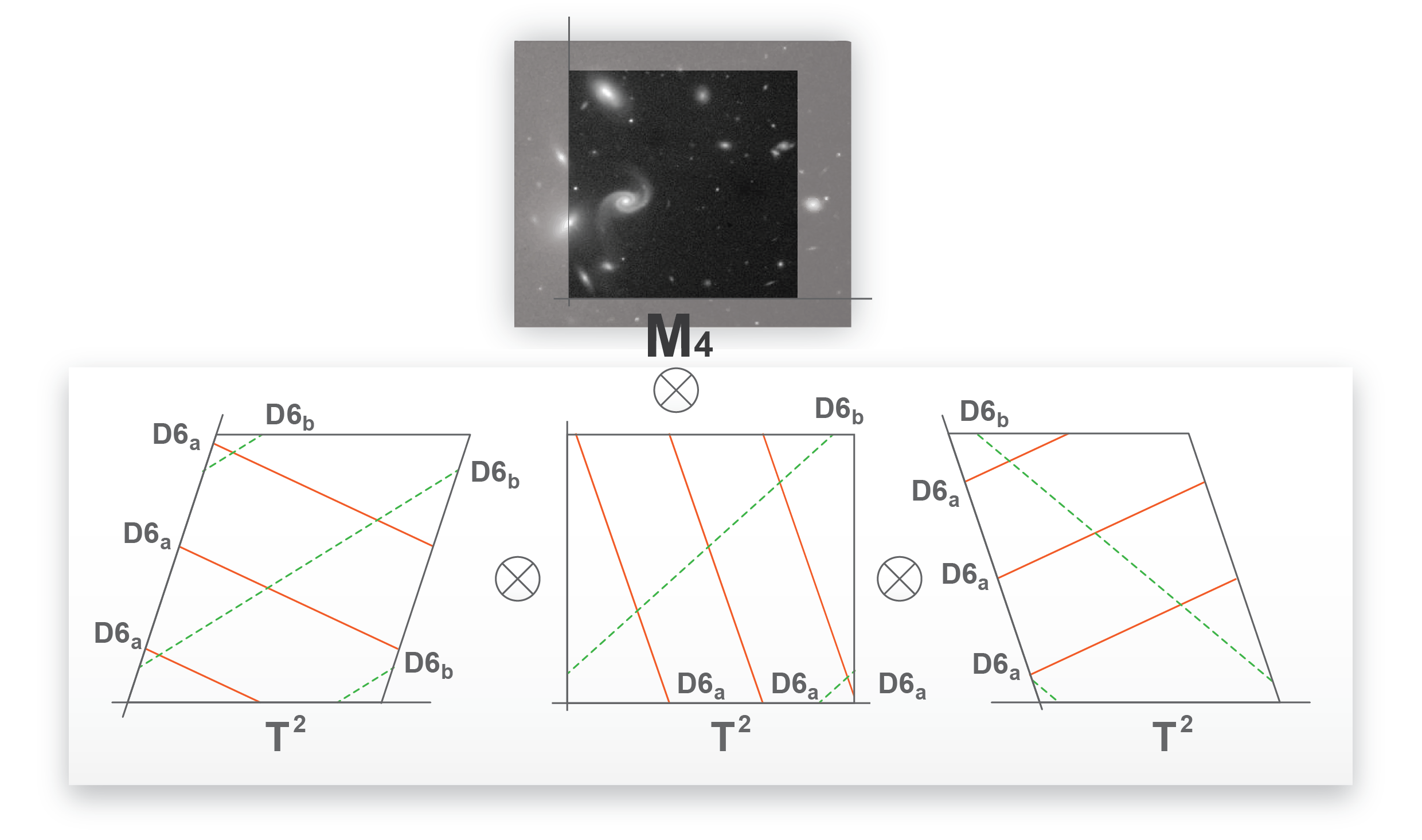}
\caption{\small{D6-branes wrapped on factorisable 3-cycles on compact $T^6 = T^2 \times T^2 \times T^2$ and filling the four non-compact dimensions $M_4$. In comparison with the picture shown in figure \ref{fig23.eps}, ona can realize that D6-branes are intersecting several times.}}\label{fig24.eps}
\end{center}
\end{figure}

Let us suppose the case of intersecting D6-branes wrapping 3-cycles on a $T^{6}$. Let us assume that the $T^{6}$ has a factorisable metric as a product of 3 two-tori, $T^{6}=T^{2}\times T^{2} \times T^{2}$. We will also assume that the homology 3-cycles of the $T^{2}$ where each brane is wrapped on can be written simultaneously for each brane as a direct product of three 1-cycles, each one wrapping a 
different $T^{2}$, i.e., of the form
\begin{equation}
\Pi_{a}=\bigotimes_{r=1}^{3}(n_{a}^{r}[a_r] + m_{a}^{r}[b_r]) \label{homologyclass}
\end{equation}
where $[a]$ and $[b]$ are the basis elements of the homology group of 1-cycles. This kind of 3-cycles are called \textit{factorisable} 3-cycles. By this choice, the world-volume of each brane is given by $M^{4}\times T^{3}$. Branes of this kind are $\frac{1}{2}BPS$ objects as well. An example of such configurations involving two D6-branes (i.e., $n=3$) is shown in the \Fig{fig24.eps}.\\

There are a couple of comments to make about this kind of constructions. First, note the fact that the gauge coupling constants are finite and given by
\begin{equation}
 \frac{1}{g_{a}^{2}}=\frac{M_{s}^3}{(2\pi)^4\lambda_{II}}\mathrm{Vol}(\Pi_a)
\end{equation}
where $M_s = \alpha'^{-\frac{1}{2}}$ is the string scale and $\lambda_{II}$ is Type IIA coupling constant. The second one is that automatically a given pair of branes will typically intersect a finite number of times. This number is a topological quantity named the \textit{intersection number} and is given by
\begin{equation}
I_{ab}=\prod_{r=1}^{3}(n_{a}^{r}m_{b}^{r}-m_{a}^{r}n_{b}^{r}) \label{intnumb}
\end{equation}
for a given pair of branes $a$ and $b$. Note that the quantum numbers of the particles in different intersections of the same pair of branes are identical. Thus, the particle generations have a nice interpretation in this setup: it is the number of times two given D-branes intersect each other. This is a generic feature of all intersecting brane models, regardless of the compactification space or the degree of complexity cycles may acquire.\\

For the general case of $\mathbf{CY_3}$ compactification, D6-branes may be wrapping special 3-cycles called special Lagrangian 3-cycles. Those 3-cycles are volume-minimizing submanifolds and hence the stability of D6-branes is ensured. These special Lagrangian 3-cycles are defined by
\begin{eqnarray}
(\mathcal{F}+iJ)\arrowvert_{\Pi} = 0\\
\mathrm{Im}(e^{-i\phi}\Omega_3)\arrowvert_\Pi = 0
\end{eqnarray}
where $\mathcal{F}$ is the usual gauge invariant field strength in a D-brane worldvolume while $J$ and $\Omega_3$ are respectively the CY Kähler 2-form and holomorphic 3-form obtained from the internal Killing spinor of the compactification. Notice that the above condition implies $\mathcal{F}=0$, i.e. not allowed flux through the cycle. Finally $e^{-i\phi}$ is a constant phase which only depends on the homology class of $\Pi_3$. The meaning of these two conditions is that tangent spaces at different points of 3-cycle $\Pi_3$ are related to each other by SU(3) rotations and, recalling the discusion in the previous subsection, it means that the D6-branes wrapped in the special Lagrangian cycles preserve a 4d $\mathcal{N}=1$ supersymmetry. However, far away from the brane, on the underlying $\mathbf{CY_3}$, still there is $\mathcal{N}=2$ supersymmetry.\\  

String theories with open string sectors in configurations with no infinite transverse directions, i.e. in compact extra-dimensions or in n-cycles where D-branes are wrapped in, have to satisfy a consistency condition called $RR$ tadpole cancellation. In the D6-branes case this is translated to the fact that they are sources of charge under the RR 7-form. The RR tadpole cancellation in 10d can be interpreted as a generalization of the Gauss law which implies that for compact spaces the total charge must add up to zero. It is encoded in the language of 3-cycle homology in the following condition:
\begin{equation}\label{eqRRa}
[\Pi_{tot}]\equiv \sum_a N_a [\Pi_a] = 0 
\end{equation}
From the field theory poiny of view, RR tadpoles are equivalent to the presence of anomalies, and the condition of RR cancellation implies the anomaly cancellation as well.\\  

\subsubsection{Orientifold compactifications}
In order to verify \Eq{eqRRa} it is necessary the presence of anti-D$p$-branes. The configurations that contain systems of branes and anti-branes are however no longer stable because D$p$-branes and $\overline{\mathrm{D}p}$-branes have opposite charge and then attract each other and eventually annihilate. Moreover, the supersymmetries preserved by the $\overline{\mathrm{D}p}$-brane are precisely the supersymmetries broken by a D$p$-brane along the same directions. This implies that the whole system breaks all the supersymmetries.\\
     
The problem of RR tadpole cancellation in SUSY may be solved by performing an orientifold projection. An orientifold of a given string theory (generically a Type II) is obtained by performing a projection of that theory that involves the world-sheet orientation reversal $\Omega$. This $\Omega$ flips the orientation of fundamental strings.\\ 

An orientifold projection consists in combining the world-sheet parity operator $\Omega$ with a $\mathbb{Z}_2$ discrete symmetry of the background, typically an isometry of the metric in the internal space $\mathbf{X}_6$, for making the projection on Type II string theory. In this class of quotients the action of $\Omega$ is not of geometric nature (i.e., it does not act on the target space) but is a symmetry of the string theory whose action is taken on the worldsheet. The prototype example is the case of Type I string theory in 10d, which is obtained as a Type IIB projection with $\Omega$ and where the discrete symmetry is the identity.\\

In Type IIA string theory on a 10d flat space $M_{10}$ the orientifold action is given by $\Omega\mathcal{R}(-1)^{F_L}$. Here $\mathcal{R}$ is a $\mathbb{Z}_2$ geometric action, acting locally as $(x^5, x^7, x^9) \rightarrow (-x^5, -x^7, -x^9)$ and $(-1)^{F_L}$ is a left-moving world-sheet fermion number projection with $F_L$ introduced for technical reasons. There are some effects that are important when we are dealing with this kind of constructions in Type IIA string theory:
\begin{itemize}
\item The quotient theory has special subspaces in spacetime, fixed under the geometric part of $\mathcal{R}$ of the orientifold projection. Such subspaces are 7d-planes defined by $x^5=x^7=x^9=0$, and spanning $x^i,\ i=0,1,2,3,4,6,8$ coordinates. These spaces fixed under the orientifold action are called orientifold 6-planes or O6-planes (it has 6 spatial dimensions and 1 for time). Physically it corresponds to a region of spacetime where the orientation of a string can flip.  

\item There exist some similarities among O$p$-planes and D$p$-planes with the same $p$. O$p$-planes are also charged under the RR $(p+1)$-form $C_{p+1}$. This corresponds to a RR tadpole of a crosscap close string diagram. For instance, the O6-plane is charged under the RR 7-form, and its charge is given by $Q_{O6}=\pm4$, in units where the D6-brane charge is +1. From the phenomenological point of view the only interesting kind of O-planes are those with negative charge in order to cancel the D-branes charge\footnote{Incidentally this would correspond to a $SO$ projection choice.}. The O-planes also preserve supersymmetry and therefore there is a relation between its charge and its tension and due to the sign choice made above we will talk about negative tension. Furthermore, systems of D-branes and O-planes can be stable because of the charge cancellation amongst each other.  

\item Despite of having charge and tension, the O-planes are not physical objects because they have no open strings attached in and therefore they do not carry world-volume degrees of freedom.
 
\item Moreover, for each brane added to a given configuration one has to add its image under $\Omega\mathcal{R}$, in order for the whole configuration to stay $\Omega\mathcal{R}$-invariant. In the particular case of branes wrapping factorisable 3-cycles this amounts to add, for each D-brane $a$ wrapping a $\otimes(n_{a}^{(i)},m_{a}^{(i)})$ cycle, another brane $a^{*}$ wrapping a $\otimes(n_{a}^{(i)},-m_{a}^{(i)})$ cycle. An example of this case is illustrated in \Fig{fig25.eps}.

\begin{figure}[tb]
\begin{center}
\includegraphics[scale=0.65]{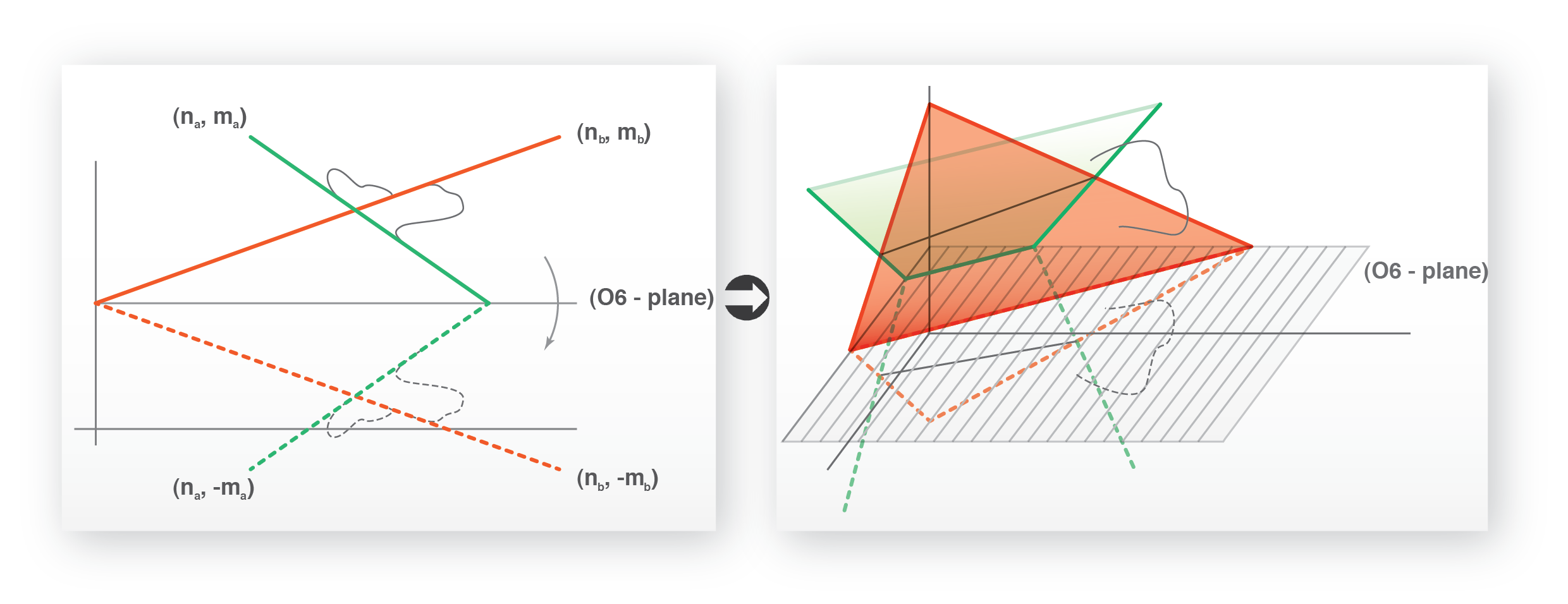}
\caption{\small{Sketch of a orientifold projection where intersecting D-branes are wrapped on a $(n,m)$-cycle with wrapping numbers $(n,m)$. Natice that the geometrical action of $\mathcal{R}$ on the internal space is inducing an action on the space of homology equivalent clases as $\mathcal{R} : [(n, m)] \mapsto [(n, -m)]$.}}\label{fig25.eps}
\end{center}
\end{figure}

\item As a consequence, depending on the choice of D-brane/O-plane configuration, new physical scenarios can be engineered. Some interesting examples are: 
\begin{itemize}
 \item For the case of coincident O6/D6 the Chan-Paton projection leads to $SO(N)$ or $Sp(N)$ gauge groups instead of the usual $U(N)$. They also corresponds repectively to negative and positive charged O-planes.

\item For the case of intersecting D6-branes it is possible to obtain new kind of chiral fermion representations like $(\square_a,\ \square_b)$ in addition to $(\square_a,\ \overline{\square}_b)$ or symmetric and antisymmetric representations.
\end{itemize}
  
\item Regarding the closed string sector, the systems of D6-branes and O6-planes preserve 4d $\mathcal{N}=1$ supersymmetry. Conversely, D6-branes themselves preserve $\mathcal{N}=1$ supersymmetry in the open string sector only if there is a common supersymmetry with the O6-planes.

\item Finally, an important effect will be the modification of tadpole conditions for the case of orientifold compactifications with D6-branes to:
\begin{equation}\label{eqRR}
[\Pi_{tot}]\equiv \sum_a N_a [\Pi_a] +  \sum_a N_a [\Pi_a'] -4 \times [\Pi_{O6}]= 0 
\end{equation}
\end{itemize}

\section{Type IIB orientifolds with magnetized D7-branes}\label{D7magn}
In the previous section we have summarized some aspects of Type IIA Calabi-Yau orientifolds, namely D6-branes wrapped on intersecting (special Lagrangian) 3-cycles $\Pi_3$. In Type IIB string theory we can consider D3, D5, D7 and D9-branes as sources of RR $C_{2p+1}$ IIB forms. Unlike Type IIA string theory, the method of intersecting the branes does not solve the lack of chirality present in the theory and additional ingredients are required, like e.g. singular (e.g. orbifold) spaces \cite{Aldazabal:2000sa} and or gauge backgrounds.\\ 

For the case of intersecting D3 and D5-branes, there are no spanning dimensions enough in the internal space $\mathbf{Y}_6$ in order to define a preferred orientation and hence, chirality does not arise. Regarding intersecting D7-branes, one has the opposite problem, i.e. they have too many dimensions spanning in the $\mathbf{Y}_6$ and therefore they have to intersect in 2-cycles. In such case we have one of the $\theta_i$ angles equal to zero and hence, in comparison with intersecting D6-branes, appears an indetermination enabling another massless fermionic state with opposite chirality. For instance, recalling \Eq{massbrane} and the choice of the orientation made in last section, if we set $\theta_1=0$ then the state 
\begin{equation}
 r + v_{\theta} = \left(-\frac{1}{2} ,\ -\frac{1}{2} + \theta_2,\ \frac{1}{2} + \theta_3 ,\ \frac{1}{2} \right) 
\end{equation}
would be also massles but with the opposite 4d handedness. Nevertheless, it is possible to obtain chiral fermions from wrapped non-intersecting D-branes, if the geometry of the wrapped manifold or the topology of the internal world-volume gauge bundle are non-trivial.  

\subsection{Type IIA - Type IIB duality}
There exists a relation between Type IIA and Type IIB compactifications through the so called mirror symmetry. This symmetry can be understood as a duality establishing that for each Type IIA on a threefold $\mathbf{X}_6$ there exists a Type IIB on a different ``mirror'' threefold $\mathbf{Y}_6$. It has a particular property, actually the one that gives his name, it exchanges the roles of the Kähler and complex moduli in such a way that the Hodge diamonds of both CY manifolds are related by a mirror relation:
\begin{eqnarray}
h_{1,1}(\mathbf{X}_6) = h_{2,1}(\mathbf{Y}_6) \\
h_{2,1}(\mathbf{X}_6) = h_{1,1}(\mathbf{Y}_6)
\end{eqnarray}

In addition, mirror symmetry exchanges the role of Type IIA D$(2p)$-branes and Type IIB D$(2p+1)$-branes. It means that D6-branes wrapped in (special Lagrangian) 3-cycles are T-dual (or mirror symmetric) to D3, D5, D7 or D9-branes  wrapping even-dimensional holomorphic cycles $\mathcal{S}_a$.\\

The BPS conditions of Type IIB D-branes on a compact manifold $\mathbf{Y}_6$ are given by 
\begin{eqnarray}
 \mathcal{F}^{(2,0)} = 0 \\
\mathrm{Im} \left(e^{-i\phi}e^{\mathcal{F}+iJ}\sqrt{\hat{A}(T_{\mathcal{S}_a})}\right)=0
\end{eqnarray}
where $\mathcal{F}$ is a non-trivial holomorphic gauge bundle, $J$ is the Kähler 2-form, $\hat{A}$ is the A-roof genus encoding the D-brane curvature couplings and $T_{\mathcal{S}_a}$ is the tangent bundle of $\mathcal{S}_a$.\\

The phase $e^{i\phi}$ depends on the homology class of holomorphic cycles. This phase will be fixed by the orientifold projection. Moreover, the way in which the geometric orientifold action $\mathcal{R}$ acts in the internal space, establishes two possibilities:
\begin{eqnarray}
\mathrm{For} \ \phi = \frac{\pi}{2}: \; \;&&  \mathcal{R}(J) = J, \; \; \mathcal{R}(\Omega) = \Omega \\   
\mathrm{For} \ \phi = 0: \; \; && \mathcal{R}(J) = J, \; \; \mathcal{R}(\Omega) = - \Omega
\end{eqnarray}
These two possibilities of $\mathcal{R}$ over the holomorphic and Kähler forms are given by a particular kind of Type IIB orientifold compactifications, more specifically:
\begin{eqnarray}
\mathrm{For} \ \phi = \frac{\pi}{2}: \; \;&&  O3 \ \mathrm{and/or} \ O7 \ \mathrm{planes} \label{eqO3} \\   
\mathrm{For} \ \phi = 0: \; \; && O5 \ \mathrm{and/or} \ O9 \ \mathrm{planes} \label{eqO5}
\end{eqnarray}
  
As in the case of Type IIA, orientifold projections in Type IIB are very useful in order to obtain globally supersymmetric Calabi Yau compactifications, but the origin of chirality is again related to the open string sector and hence to the presence of D-branes in these compactifications.\\ 

Like in Type IIA, in Type IIB there exists a cancellation of RR tadpoles among D-branes and O-planes and as a consequence, for each compactifications with O$(2p+1)$-planes there must be the corresponding D$(2p+1)$-branes in order to cancel their RR charges. Namely, for O7-planes, O3-planes, O5-planes and O9-planes there will be respectively be D7-branes, D3-branes, D5-branes and D9-branes. In this context and following \Eq{eqO3} and \Eq{eqO5} we can conclude that systems of D7/D3 and D9/D5 branes fulfill BPS conditions.\\

A useful tool to understand the chirality in Type IIB is looking to the mirror or T-dual map of Type IIA:
\begin{equation}
 \begin{array}{ccc}
 \rm{Type\ IIA\ in\ O6/D6\ models} & \rightarrow &  \rm{Type\ IIB\ O3/D3\ and/or\ O7/D7 \ models} \\
 \rm{Type\ IIA\ in\ O6/D6\ models} & \rightarrow &  \rm{Type\ IIB\ O5/D5\ and/or\ O9/D9 \ models}  
\end{array}
\end{equation}

\noindent There are two simple examples of chiral Type IIB orientifolds:
\begin{itemize}
 \item Magnetized D-branes on toroidal compactifications, i.e. with non zero $\mathcal{F}$ gauge bundle background (the so called open string fluxes). Prototypical examples of that are the models of magnetized D9 or D7-branes.   

  \item Systems of D3/D7 (O3/O7) or D5/D9 (O5/O9) in orbifold (orientifold) sigularities. A particularly attractive class of models for these systems corresponds to D3-branes at singularities \cite{Aldazabal:2000sa}. 
\end{itemize}

\subsection{Magnetized D-branes}  
We are now going to focus in the case of magnetized Type IIB D-branes \cite{Bachas:1995kx, Angelantonj:2000hi, Blumenhagen:2001te}. Let us consider the bosonic part of the two-dimensional open string sector action on the worldsheet
\begin{eqnarray}
 S &=& \frac{1}{4\pi\alpha'}\int_{\Sigma}{d^2\sigma \sqrt{-g}\left[\left(g^{ab}G_{\mu\nu}(X)+\epsilon^{ab}B_{\mu\nu}(X)\right)\partial_a X^\mu \partial_b X^\nu + \alpha'R\Phi(x) \right]} \nonumber \\
 &+& \int_{\partial\Sigma}{d\tau A_i(x)\partial_\tau X^i} \label{sigmamod} 
\end{eqnarray}
where $g$ and $R$ are the world-sheet $\Sigma$ internal metric and the associated Ricci scalar, and $G_{\mu\nu}$, $B_{\mu\nu}$, $\Phi$ and $A_i$ are the target-space background fields that we are going to consider constant. If we vary \Eq{sigmamod} respect to $X^\mu$ fields, then we will find that those fields satisfy the usual two-dimensional wave equation. \\

Now let us suppose a D$p$-brane where the $X^\mu$ fields act as coordinates such that worldvolume coordinates are $X^i$ with $i=0,...,p$ and transverse coordinates $X^m$ where $m=p+1, ..., 9$.  Then, the boundary conditions of the wave equation for $X^i$ fields are the boundary conditions at $\sigma = 0, \pi$ on the D-brane. However in this case this conditions are a little bit different from the usual Neumann ($\partial_\sigma X^i=0$) boundary conditions:
\begin{eqnarray}
 \partial_\sigma X^i + \mathcal{F}_{j}^{i} \partial_\tau X^j &=& 0 \\
X^m &=& x_0^m
\end{eqnarray}
where $\mathcal{F}$ is the invariant flux on the brane in terms of the antisymmetric tensor $B$ and the gauge field strengh $F=dA$ expressed in the following way
\begin{equation}
 \mathcal{F}=\frac{B}{\sqrt{G}}+2\pi\alpha'F .
\end{equation}

If we consider, for simplicity, that all the components are zero excepting those of the $(4,5)$-plane, one has $\mathcal{F}\equiv \mathcal{F}_{12} = -\mathcal{F}_{21}$ and the boundary conditions for this plane will be:
\begin{eqnarray}
\partial_\sigma X^4 + \mathcal{F}\partial_\tau X^5 &=& 0 \nonumber \\
\mathcal{F}\partial_\tau X^4 - \partial_\sigma X^5 &=& 0  \label{magbc}
\end{eqnarray}

In order to see the explicit T-duality with Type IIA we can parametrize $\mathcal{F}=\tan\theta_{4,5}$ and hence \Eq{magbc} can be reexpressed as
\begin{eqnarray}
\left .\partial_\sigma X^4  \right|_{\sigma=0}=0 &  &\left .\cos\theta_{45}\partial_\sigma X^4 + \sin\theta_{45}\partial_\tau X^5\right|_{\sigma=\mathcal{\ell}} = 0 \label{intbrM} \\
\left .\partial_\sigma X^5  \right|_{\sigma=0}=0 &  &\left .-\sin\theta_{45}\partial_\tau X^4 + \cos\theta_{45}\partial_\sigma X^5\right|_{\sigma=\mathcal{\ell}} = 0 \label{intbr2M}
\end{eqnarray}
These conditions coincide with Eqs.(\ref{intbr}, \ref{intbr2}) exchanging Neumann and Dirichlet boundary conditions $ \partial_\tau X^5 \leftrightarrow \partial_\sigma X^5$, which corresponds to T-duality along the $x^5$ direction. A oscillator mode expansion with that boundary condition will correspond to a system of D-branes intersecting at $\theta_{45}$ angles in the (4,5)-plane.\\

This argument can be generalized to an $ab$ open string stretched between two D-branes with magnetic fluxes $\mathcal{F}_a$ and $\mathcal{F}_b$ respectively, in this case the parametrization will be given by
\begin{equation}
 \theta_{ab}= \arctan\mathcal{F} - \arctan\mathcal{F}_b \label{arctg}
\end{equation}

\subsection{Magnetized D-branes in toroidal orientifold compactifications}
One of the interesting properties of T-duality between Type IIA and Type IIB is the fact that all the features of intersecting branes in toroidal and orientifold compactifications can be recovered for the case of magnetized branes.\\

In order to understand the duality for the case of toroidal compactifications with magnetized D-branes we are going to suppose a vanishing $B$-field so that $\mathcal{F} = 2\pi\alpha'F$. The total magnetic flux $F$ in the $\mathbf{T^2}$ obeys a flux Dirac quantization condition given by
\begin{equation}
 m \int_{\mathbf{T^2}}{F}=2\pi n \label{diraccond}
\end{equation}
where $n\in\mathbf{Z}$ is the quantum of $F$ flux and $m$ represents the multiplicity of times that a D-brane is wrapped in one of the cycles. Actually, a single D-brane wrapped $m$ times generates a $U(m)$ gauge group, i.e., each time is a $U(1)$ factor. \Eq{diraccond}  can be generalized to a $N_a$ stack of D-branes in a factorizable $\mathbf{T^6}$
\begin{equation}
 m_a^i \int_{\mathbf{T^2_i}}{F_a^i}=2\pi n_a^i \label{diraccond2}
\end{equation}
where now $m_a^i$ indicates the i-th 2-torus $\mathbf{T^2}$ in which D-branes are wrapped and with $n_a^i$ units of $U(1)_a$ magnetic flux.\\

It is interesting to realize that a stack $N_a$ of D-branes wrapping a cycle $m_a$ times will generate a $U(N_a m_a)$ gauge group. Notice that both $n$ and $m$ are topological numbers because they do not depend on the geometry and describe completely the system. This pair of numbers $(n,m)$ characterizes a magnetized D-brane in Type IIB and are T-dual to the wrapping numbers on non-trivial cycles on the dual $\mathbf{T^2}$ in Type IIA.\\

As in Type IIA the open string sector is constructed among different stacks of D-branes. As it was argued above, the gauge group produced by each stack of D-branes is $U(N_a m_a)$ and $U(N_b m_b)$ respectively and therefore the open string sector will have a $U(N_a m_a)\times U(N_b m_b)$ gauge group with $ab$ open string fermions transforming in $(\square_a,\ \overline{\square}_b)$ bi-fundamental representations of this group. In particular for two stacks of D-branes wrapped $m_a$ and $m_b$ times  and supposing the large $\mathbf{T^2}$ volume so that the magnetic fields are diluted \cite{Cascales:2003pt, book}, this group is broken down by the $n_a$ and $n_b$ monopole quanta of open string flux $F$ through
\begin{equation}
 U(N_a m_a)\times U(N_b m_b)\rightarrow U(N_a)^{m_a}\times U(N_b)^{m_b} \rightarrow U(N_a)\times U(N_b)
\end{equation}
and hence there will be an splitting of the original bi-fundamental in the following way
\begin{equation}
 (\square_a,\ \overline{\square}_b)\rightarrow(\underline{\square_a ,...}\ ,\ \underline{\overline{\square}_b ,...})\rightarrow m_a m_b  (\square_a,\ \overline{\square}_b)
\end{equation}

The first breaking can be interpreted as passing from the local to the global point of view, i.e. locally a D-brane wrapped several times ($m$ times) is like having a stack of $m$ branes whose positions conmmute (due to $U(N_a)^{m_a}$). From the field theory point of view it is like we would be seeing KK modes but when we change to global perspective we realize that among these KK modes, there is no KK zero mode. The second breaking is nothing but realizing that all the $m$ stacks are actually the same brane and therefore it is like dealing with a branch-cut of a multivalued function in a complex plane in such a way that the wrapping of the D-brane around a non-trivial cycle could be interpreted as non-trivial monodromy in the brane topology.\\  

Regarding chirality, the reason why Type IIB string theory after dimensional reduction contains chiral fermions arising from the KK reduction of 10d chiral fermions from $(\square_a,\ \overline{\square}_b)$ is due to the non zero index of the Dirac operator coupled to the gauge bundle. In this case the index for $\mathbf{T^2}$ is given by
\begin{equation}
 \mathrm{ind}\ \displaystyle{\not}D_{ab} = \frac{1}{2\pi}\int_{\mathbf{T^2}}{(F_a-F_b)}=\frac{n_a}{m_a}-\frac{n_b}{m_b}
\end{equation}

Following this line we can realize that the chirality for the case of magnetized D-branes is field-theory-like and similar to the one used in heterotic models. In this sense, the origin is different from the case of intersecting branes where it comes from the geometrical aspects of the D-brane system. Moreover, the number of these lower dimensional chiral fermions in the $(\square_a,\ \overline{\square}_b)$ is given by $m_a m_b$ times the index in the following form
\begin{equation}
 I_{ab}=m_a m_b\frac{1}{2\pi}\int_{\mathbf{T^2}}{(F_a-F_b)}=n_a m_b - m_a n_b \label{Iab}
\end{equation}
which is totally equivalent to the T-dual case of intersecting D6-branes in Type IIA, see \Eq{intnumb}.\\

One interesting property that can be observed from the developed Chern-Simons part of D-brane action for vanishing B-field
\begin{eqnarray}
 S_{cs}&=&\mu_p\left(\int_{W_{p+1}}{C_{p+1}} + (2\pi\alpha')\int_{W_{p+1}}{C_{p-1}\wedge \rm{tr}\,F} \right. \nonumber \\
 &+& \left. \frac{1}{2}(2\pi\alpha')^2\int_{W_{p+1}}{C_{p-3}\wedge \rm{tr}\,F^2}-\frac{1}{24(8\pi^2)}\int_{W_{p+1}}{C_{p-3}\wedge \rm{tr}\,R^2} +... \right) \label{CS}
\end{eqnarray}
is that the description for a given D$p$-brane implies automatically other kinds of lower dimensional D-branes. For example if we consider $N_a$ D$9_a$-branes in $\mathbf{T^6}$ such that they wrap $m_a^i$ times on the i-th 2-torus $(\mathbf{T^2})_i$ with $n_a^i$ units of magnetic flux in $(\mathbf{T^2})_i$ therefore the complete system is given by
\begin{equation}
 \begin{array}{ccc}
  D9 & \rightarrow & (n_a^1,m_a^1)(n_a^2,m_a^2)(n_a^3,m_a^3) \nonumber \\
 D7 & \rightarrow & (1,0)(n_a^2,m_a^2)(n_a^3,m_a^3) \nonumber \\
 D5 & \rightarrow & (n_a^1,m_a^1)(1,0)(1,0) \nonumber \\
 D3 & \rightarrow & (1,0)(1,0)(1,0) 
 \end{array}
\end{equation}
where $n_a^i = (1,0)$ means that those lower dimensional branes are localized at a point in the corresponding $(\mathbf{T^2})_i$. Therefore a D$p$-brane with non-trivial gauge bundle can be seen as a D$p$-brane where the lower dimensional D-branes allowed by the CS couplings have been dissolved and hence the chiral spectrum can be interpreted as the sum of intersection numbers between their elementary components.\\

Given that intersection numbers are defined by topological properties associated to the homology, it is interesting to define the homology class of the magnetized D-brane. It can be done in a similar way as we defined $[\Pi_a]$ in \Eq{homologyclass} by introducing the homology classes $[0]_i$, which corresponds to the point where lower dimensional branes are seated, and $[\mathbf{T^2}]_i$ for the each $i$-th 2-torus class such that
\begin{equation}
 [\mathbf{Q}_a]=\prod_{i=1}^3\left(m_a^i [\mathbf{T^2}]_i + n_a^i [0]_i \right) 
\end{equation}
From this expression we realize that the number of chiral fermions \Eq{Iab} can be also expressed as the product of homology classes:
\begin{equation}
 I_{ab}=[\mathbf{Q}_a]\cdot[\mathbf{Q}_b]
\end{equation}

Another consequence of the CS action in \Eq{CS}, is the fact that D-branes with open string fluxes are sources for the RR even-degree forms. In particular, for the case of D9-branes $C_{10}$, $C_{8}$, $C_{6}$ and $C_{4}$. This implies the existence of RR tadpoles that, as in the case of Type IIA, signal an inconsistency of the theory. Like in Type IIA, the consistency condition of RR cancellation is given by
\begin{equation}
 \sum_a N_a [\mathbf{Q}_a]=0
\end{equation}
 
This condition will be modified by the action of orientifold projections. A particularly interesting example of that is the case of magnetized D7-branes on the $\mathbf{T^6}/(\mathbf{Z}_2\times\mathbf{Z}_2)$ orientifold which is constructed with the $\mathbf{T^6}/(\mathbf{Z}_2\times\mathbf{Z}_2)$ orbifold modded out by  $\Omega \mathcal{R}_1\mathcal{R}_2\mathcal{R}_3(-1)^{F_L}$, where $\mathcal{R}_i$ acts as $z_i \rightarrow -z_i$. In this model the gauge sectors of the model are localized on intersecting D7-branes with magnetic fluxes. The model contains 64 O3-planes and 4 O7-planes and in order to cancel the RR tadpoles there are D$9_a$-branes and their orientifold mirrors D$9_{a'}$-branes. In this case the tadpole cancellation conditions will be given by 
\begin{equation}
 \sum_a N_a [\mathbf{Q}_a] +\sum_a N_a [\mathbf{Q}_{a'}]-32 [\mathbf{Q}_{O_p}]=0
\end{equation}

Other interesting models which include also realizations of GUT constructions are developed in e.g. \cite{Blumenhagen:2008zz}.\\

\section{F-theory unification}\label{ftheory}
The $SL(2,\mathbf{Z})$ self-duality symmetry in Type IIB string theory \cite{Schwarz:1995du} has as a consequence the appearance of new degrees of freedom, the so called $(p,q)$-branes. We need a good framework to describe these new objects in Type IIB, in particular the $(p,q)$7-branes or simply 7-branes. A useful way to describe Type IIB compactifications with 7-branes in a generic situation is called F-theory \cite{Vafa:1996xn}. In the other hand $SL(2,\mathbf{Z})$ connects the perturbative and non-perturbative regimes of Type IIB string theory, allowing us to go further in the knowledge of non-perturbative compactifications of Type IIB string theory \cite{Vafa:1996xn, Morrison:1996pp}. In this sense F-theory is for Type IIB like M-theory is for Type IIA.\\

Nevertheless, F-theory, unlike M-theory, does not correspond to a Lorentz invariant higher dimensional theory. In other words, we are not dealing with a fundamental theory. F-theory should be thought of as a non-perturbative description of a class of string vacua which is accessible from other string descriptions in certain limits \cite{weigandRev}:
\begin{itemize}
 \item F-theory as (strongly coupled) Type IIB theory with 7-branes and varying dilaton.
\item F-theory as dual to $E_8 \times E_8$ heterotic theory.
\item F-theory as dual to M-theory on a vanishing $\mathbf{T}^2$.
\end{itemize}

\noindent In this section we will concentrate on third and the first ones following very closely \cite{weigandRev}, \cite{Sen:1998kr} and \cite{heckmanRev}. We want to understand the relation between Type IIB orientifold compactifications and F-theory in order to apply them to phenomenological issues \cite{bhv1, bhv2, dw1, dw2}. Moreover the F/M-theory duality is the one that captures the dynamics in the most general way, as we will see, because it gives a geometric meaning to the $SL(2,\mathbf{Z})$ self-duality in Type IIB string theory.

\subsection{$\mathbf{SL(2,Z)}$ in Type IIB}
If we consider Type IIB string theory at strong coupling $g_s \rightarrow \infty$ and we recall that the mass scale for D$p$-branes is given by $M\simeq \alpha'^{-1/2}g_s^{-1/(p+1)}$ then it is inmmediate to see that the fundamental object of the theory at strong coupling is the D1-brane. But a D1-brane is a one dimensional object, i.e. a string, showing that the dual theory in strong regime is governed by strings as well so that we can say that it is also a string theory. Moreover, at strong coupling, the supersymmetry of the theory is still given by a 10d $\mathcal{N}=2$ chiral supergravity multiplet, and therefore we are again talking about a Type IIB string theory but where the fundamental object is a D1-brane.\\

This connection between strong and weak regimes is giving us hints of an underlying invariance $g_s \rightarrow \frac{1}{g_s}$. This is in agreement with the fact that the Type IIB effective supergravity action written in Einstein frame displays an invariance under the so called Type IIB S-duality \cite{Schwarz:1995du}:
\begin{equation}
e^{-\phi} \rightarrow \frac{1}{e^{-\phi}}, \; \; \; \; \left(\begin{array}{c} B_2 \\ C_2 \end{array}\right) \rightarrow \left(\begin{array}{cc} 0 & -1 \\ 1 & 0 \end{array} \right) \left(\begin{array}{c} B_2 \\ C_2 \end{array}\right). \label{SLsimpp}
\end{equation}

In Type IIB supergravity the axion $C_0$ and dilaton $\phi$ interchanges with the dilatino $\lambda$ under supersymmetry transformations. Then, it is interesting to construct the axio-dilaton field by complexifying the pair axion-dilaton in the supergravity action through $\tau = C_0 + i e^{-\phi} = C_0 + \frac{i}{g_s}$. Using this field composition, S-duality can be promoted to a larger $SL(2,\mathbf{R})$ invariance for the Einstein frame Type IIB supergravity action 
\begin{equation}
\tau \rightarrow \frac{a\tau+b}{c\tau +d}, \; \; \; \; \left(\begin{array}{c} B_2 \\ C_2 \end{array}\right) \rightarrow \left(\begin{array}{cc} a & b \\ c & d \end{array} \right) \left(\begin{array}{c} B_2 \\ C_2 \end{array}\right) \label{SL}
\end{equation}
where $ad-bc=1$. This symmetry is broken at non-perturbative level to $SL(2,\mathbf{Z})$. Notice that it would not be able to be a symmetry of the theory because if $a$, $b$, $c$ and $d \in \mathbb{R}$ then the charge Dirac quantization condition would be broken. But actually, quantization is preserved by the subgroup $SL(2,\mathbf{Z})$ which is the one that continues as invariance of the theory.\\

The $SL(2,\mathbf{Z})$ symmetry exchanges F1 (fundamental strings) and D1 branes (see \Eq{SLsimpp}) due to the fact that fundamental strings couple to $B_2$ NSNS 2-form and D1-branes couple electrically to $C_2$ RR forms. It seems appropiate to combine F1 and D1-brane into a doublet of $SL(2,\mathbf{Z})$ such that F1-string is a $(1,0)$ vector and D1-brane is a $(0,1)$ one. However it is possible to make more general transformations with \Eq{SL} transforming a $(1,0)$ vector into a general $(p,q)$-vector. This general $(p,q)$-strings carries $p$ units of $B_2$ electric charge and $q$ units of $C_2$ electric charge. Such objects exist as supersymmetric bound states for $p$, $q$ coprime \cite{Witten:1995im}. Finally, it is interesting to be aware that $SL(2,\mathbf{Z})$ has the same formal properties than the modular group. Actually the arising of F-theory comes from trying to give a modular-like meaning to this $SL(2,\mathbf{Z})$.

\subsection{F-theory from the geometry of elliptic fibrations}
Let us recall that T-duality relates Type IIA compactified to 9d on a $\mathbf{S}^1$ of radius $R$ with Type IIB on a dual $\mathbf{S}^1$ of radius $R'=\alpha'/R$. On the other hand Type IIA at strong coupling limit is equivalent to a 11d M-theory compactified in $\mathbf{S}^1$ in the decompactification limit. Therefore there is a duality between M-theory on a $\mathbf{T}^2$ and Type IIB on a $\mathbf{S}^1$. In this equivalence one can identify the Type IIB axio-dilaton $\tau$ with the complex structure parameter $\tau=R_2/R_1 e^{i\theta}$ of the $\mathbf{T}^2$ on which M-theory is compactified.\\

There is also a relation between the radius $R$ of $\mathbf{S}^1$ and the area of $\mathbf{T}^2$, $A=R_1 R_2 \sin \theta$ given by $M_{11}^3 A \simeq 1/R$. Therefore, in the decompactification limit $R \rightarrow \infty$ the $\mathbf{T}^2$ area vanishes. That corresponds to collapsing one of the 1-cyles of the $\mathbf{T}^2$. In this limit we have a duality between Type IIB in 10d and M-theory in $\mathbf{S}^1$ and we can give a geometrical meaning to $SL(2,\mathbf{Z})$ self-duality of Type IIB string theory: it corresponds to the modular group of M-theory compactified on $\mathbf{T}^2$ in the vanishing area limit $A \rightarrow 0$. This idea is valid not only for two-torus but in general:
\begin{equation}
 \mathrm{F-theory}\; \mathrm{on}\; \mathcal{K}\times S^1 \leftrightarrow \mathrm{M-theory}\; \mathrm{on}\; \mathcal{K} 
\end{equation}

F-theory is nothing but this idea of giving a geometric meaning to $SL(2,\mathbf{Z})$ as a modular group where the axio-dilaton is the modular parameter. Therefore it can be interpreted as a limit ($A \rightarrow 0$) of a M-theory compactification in a manifold expressed as a $\mathbf{T}^2$ fibration (elliptic fibration, see later). For the simplest case of 10d Type IIB orientifold, this geometrization would correspond to constructing a theory such that when is compactified in a torus fibration (identifying the axio-dilaton with the torus complex structure) we recover Type IIB string theory. This 12d theory is what we understand by F-theory and in this case the torus fibration is a trivial fibration given by direct product of 10d Minkowsky with $\mathbf{T}^2$ as fiber
\begin{equation}
\begin{array}{ccc}
 \mathbf{T}^2 \rightarrow &\mathcal{M}_{10} \times  \mathbf{T}^2& \\ 
				  &   \downarrow   &               \\ 
				   & \mathcal{M}_{10} &

\end {array}
\end{equation}
Because the fibration (actually the fiber bundle) is trivial, the fiber is the same for each point of the base space, i.e. the axio-dilaton is constant for all the 10d Minkowski space. As a matter of fact, this 12d theory on $\mathbf{T}^2$ is equivalent to 11d M-theory on a $\mathbf{T}^2$ in the vanishing area limit in which a new dimension grows. Notice that the extra 2d of F-theory is an artifact for achieving the 2-torus which is giving the geometric meaning to $SL(2,\mathbf{Z})$, but they are not physical space dimensions.\\

Now we can generalize the previous idea to other manifolds which are not trivial fiber bundles such that fiber is able to have a different value in each point of the base space. This implies: 
\begin{equation}
\begin{array}{ccc}
 \mathcal{F} \rightarrow &\mathcal{M}_{p} \times  \mathcal{X}_{10-p+2}& \\ 
				  &  \downarrow   &               \\ 
				   & \mathcal{M}_{p} &

\end {array}
\end{equation}
 where  $\mathcal{F} =\mathbf{T}^2$ such that
\begin{equation}
\begin{array}{ccc}
 \mathbf{T}^2 \rightarrow & \mathcal{X}_{10-p+2}& \\ 
				  &   \downarrow   &               \\ 
				   & \mathcal{B}_{10-p} &

\end {array}
\end{equation}

In our case the fiber is a two torus, if we denote $z$ as the base of complex coordinates, then the complex structure will depend on them as $\tau =\tau(z)$. Following this argument we can conclude that F-theory is a way of compactifying type IIB avoiding the restriction of a constant complex dilaton.\\

This kind of fibrations are called elliptic fibrations because the $\mathbf{T}^2$ can be described algebraically as:
\begin{equation}
 y^2 = x^3 + fx +g \label{weistras}
\end{equation}
where $x$ and $y$ are complex variables and $f$ and $g$ are complex parameters. This equation is called Weierstrass equation and is a general form to describe an elliptic curve. Elliptic curves are therefore, topologically, two-torus. The way to describe the elliptic fibration on a certain manifold is promoting $f$ and $g$ to functions of the base coordinate $z$. It is interesting to emphasize that the zeros of the discriminant of the cubic \Eq{weistras}
\begin{equation}
 \Delta =27g^2+4f^3
\end{equation}
are describing the singularities of the elliptic curve. This corresponds to the points where $\tau(z)$ diverges to $i\infty$, which geometrically can be viewed as one of the $(p,q)$-cycles of the $\mathbf{T}^2$ collapsing to zero.


\subsection{F-theory from D7-brane backreaction}
Another way to access to F-theory is through the backreaction of D7-branes. In general, the description of perturbative Type II orientifolds with D-branes is done neglecting the backreaction of the branes and orientifold planes on the background geometry because asymptotically away from them it is really negligible. It can be explained heuristically through the Poisson equation for the background fields sourced by the brane \cite{weigandRev}. These sourced fields are asymptotically zero excepting the critical case of branes with codimension 2. This is precisely the case of D7-branes in Type IIB string theory. D7-branes or more generically $(p,q)$7-branes expanded along $0, .., , 7$  look like charged point particles localised in the two normal directions. 7-branes are $C_8$ electrically charged and magnetically charged with $C_0$, which combines with $g_s$ to form the complex dilaton. If we define a complex coordinate with 7-brane normal coordinates $z= x^8 + i x^9$, it turns out that the axio-dilaton is a function of $(z,\bar{z})$ but due to supersymmetry conditions, it must be a holomorphic function of $z$ $\tau(z)$. It is possible to see that the Poisson equation solution  is given by
\begin{equation}
 \tau(z) = \tau_0 + \frac{1}{2\pi i} \mathrm{ln}(z-z_0)+... \label{tau}
\end{equation}
for a region very close to the brane $z=z_0$. \\

\Eq{tau} shows a branch cut in $z_0$ and hence there exists a monodromy associated. Schematically it means that passing around a circle surrounding the D7-brane in $z$-plane the complex dilaton transforms as 

\begin{equation}
 \tau(z)\rightarrow \tau(z) +1
\end{equation}

\noindent which is equivalent to a $SL(2,\mathbf{Z})$ transformation (see \Eq{SL}) with $a=d =1$, $c=0$ and $b=1$ and hence monodromies leave invariant the theory. In terms of brane sourced fields it means that the relation $F_1=dC_0$ is not globally defined because $C_0$ shifts $C_0\rightarrow C_0 +1$. That allows us to conect with the F/M-theory point of view because on $z_0$ (on top of the 7-brane) $\tau$ diverges. In terms of F/M-theory duality this would be equivalent to $\Delta(z_0)=0$, and that is the reason why we say that generically in the $z$'s where the elliptic fibre pinches off there exist a 7-brane seated at this point.

\subsection{GUT's in local F-theory models}\label{fthguts}
One of the most powerful features of F-theory is the fact that the precise way in which an elliptic curve is pinching off can admit a classification in terms of the ADE Dynkin diagrams which is known as the Kodaira classification of singular fibres. Depending on this classification, the gauge groups of the 7-branes have $A_n$, $D_n$ or exceptional $E_{6,7,8}$ algebras. This corresponds to gauge groups $SU(n+1)$, $SO(2n)$, $E_6$, $E_7$ and $E_8$. This is an important property since allows for the existence of $SU(5)$ or $SO(10)$ GUT symmetries.  \\

A second useful property is that, just as in perturbative Type IIB string theory, spacetime filling 7-branes occupy the same subspace in the internal directions, and then we can use the already known methods in the perturbative theory for engineering higher dimensional gauge theories. Let us consider for instance the case of realistic 4d compactifications of F-theory. We will use an elliptic fibration of a CY fourfold  which locally is a complex threefold with a $\mathbf{T}^2$ fibered on top of each point. This would be equivalent to compactify Type IIB string theory in a $B_6$ but with a non trivial complex string coupling. In the non-perturbative regime we will have 7-branes in the points of base space in which the two-torus have a collapsed 2-cycle. \\

\begin{figure}[h!]
\begin{center}
\includegraphics[scale=0.65]{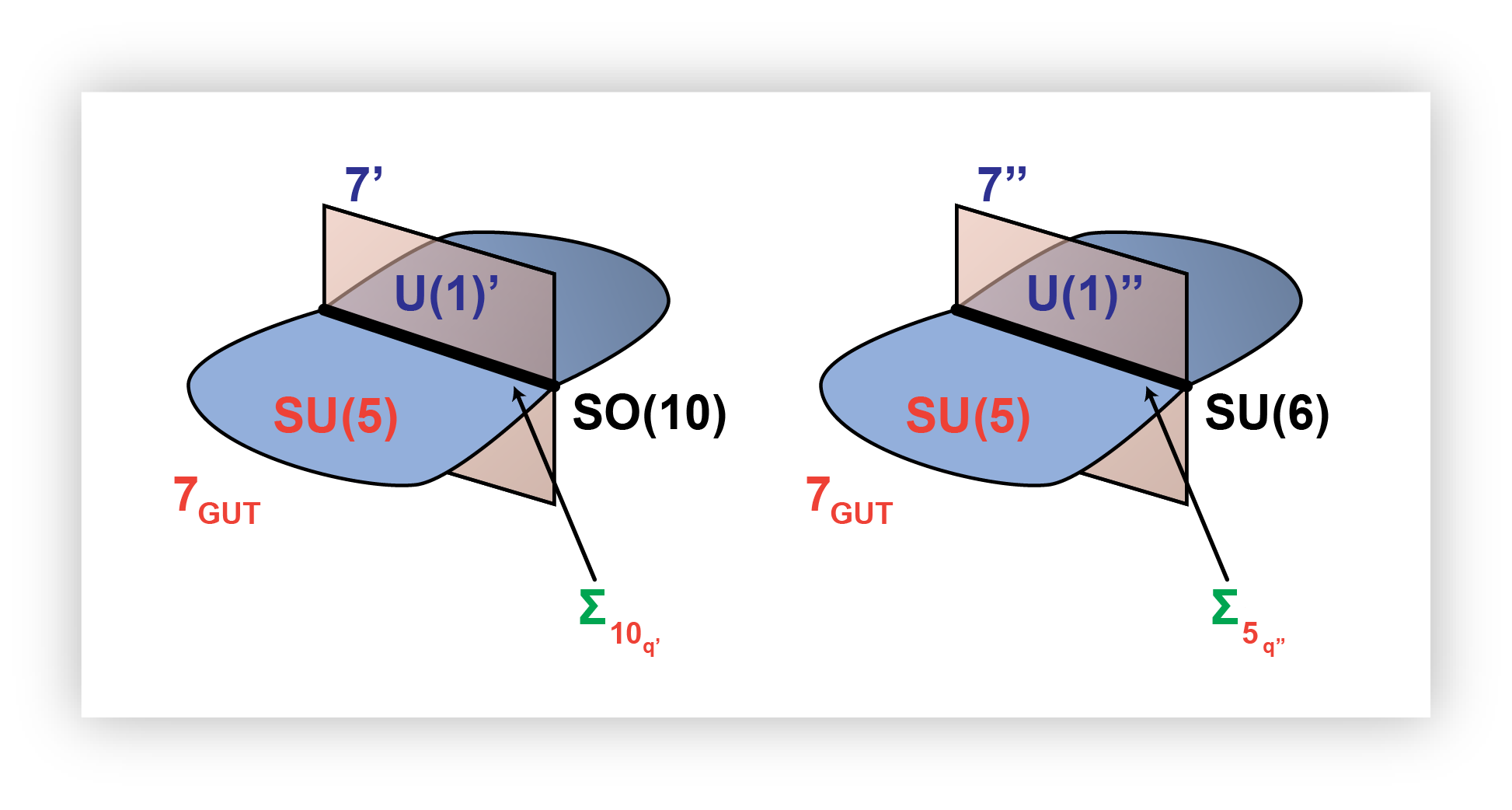}
\caption{\small{Matter fields live at matter curves. These matter curves ($10$ and ${\bar 5}$)  correspond to the intersection of the 7-branes wrapping a 4-cycle $S$ with other $U(1)$ 7-branes.}}\label{flocal1}
\end{center}
\end{figure}

Both D7-branes in the perturbative level and 7-branes in the non-perturbative language will fill 4d Minkowski and will be wrapped on 4-cycles (or 2-folds in the complex space). Moreover those 4-cycles will generically be intersecting in a 2-cycle, i.e. a complex curve, which in F-theory jargon is often called matter curve $\Sigma$. In these kind of constructions, chiral matter lives at the intersection of 7-branes corresponding to an enhanced degree of the sigularity. In \Fig{flocal1} is shown an scheme for the case of a $SU(5)$ F-theory GUT, for which the enhancement is unfolded as
\begin{eqnarray}
  SU(6) &\rightarrow& SU(5) \times U(1)'\\
  \mathbf{35} &\rightarrow& \!\!\! \mathbf{24}_0 + \mathbf{1}_0 + [ \mathbf{5}_1 + c.c.] \\
  SO(10) &\rightarrow& SU(5) \times U(1)''\\
  \mathbf{45} &\rightarrow& \!\!\! \mathbf{24}_0 + \mathbf{1}_0 + [ \mathbf{10}_4 + c.c.] 
\end{eqnarray}

The GUT group lives on 7-branes whose 4 extra dimensions beyond Minkowski wrap a 4-cycle $S$. This $S$ manifold is inside a 3 complex dimensional manifold $B_3$ where the 6 extra dimensions are compactified. There is also the possibility of two matter curves $\Sigma_1$ and $\Sigma_2$ in the base $S$ to intersect in a point. This would correspond to a triple intersection of 7-branes. At the intersection of $\Sigma_1$ and $\Sigma_2$ the rank of the singularity type would be increased in two units. \\

\begin{figure}[t!]
\begin{center}
\includegraphics[scale=0.7]{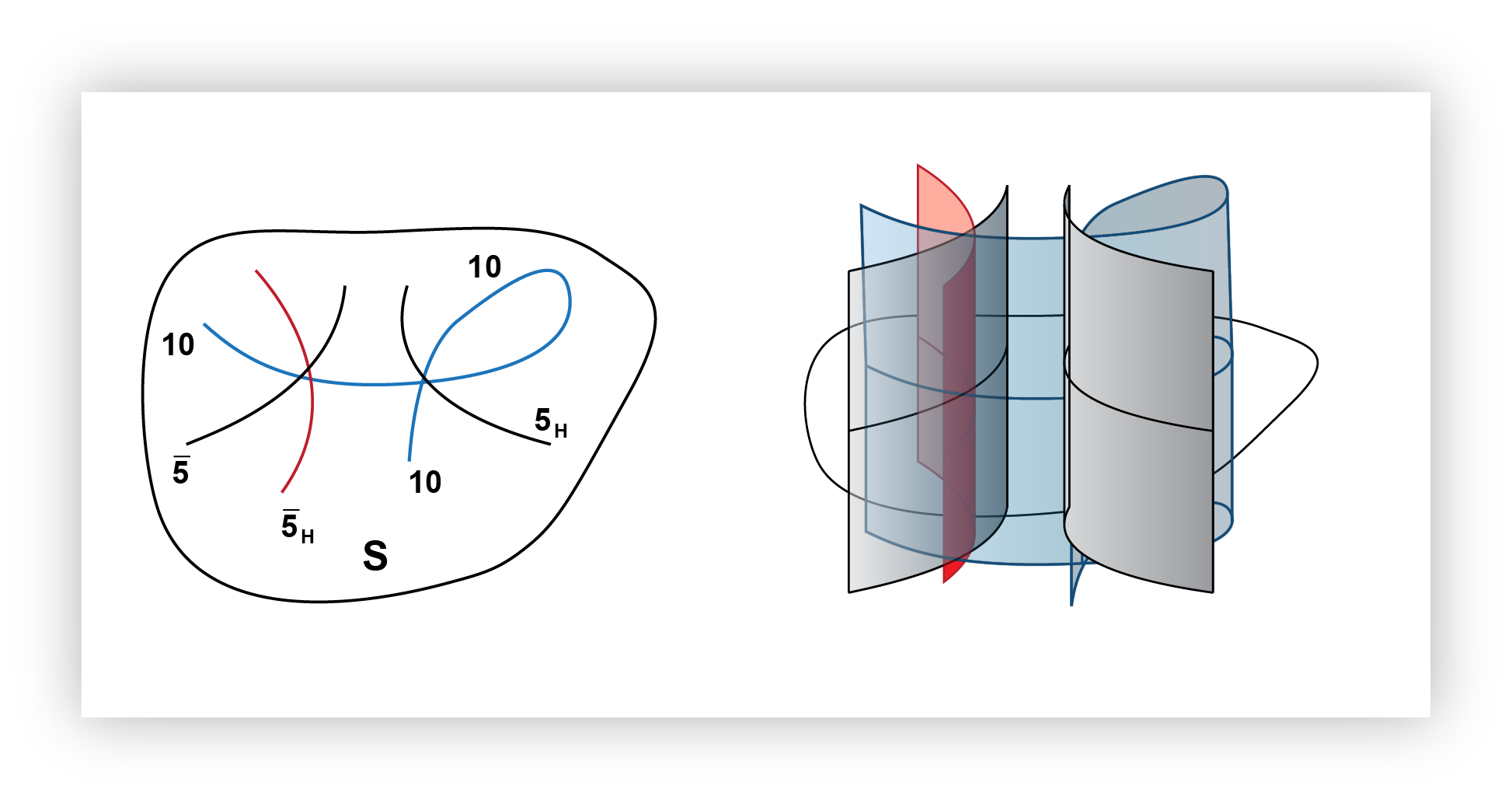}
\caption{\small{Intersection matter curves of chiral and Higgs multiplets of a $SU(5)$ GUT model with the intersection points leading to Yukawa couplings}}\label{flocal2}
\end{center}
\end{figure}

For the case of local F-theory models involving a $SU(5)$ GUT symmetry, there is one matter curve for each $SU(5)$ rep. and at the intersection of matter curves with Higgs curves $H_u, H_d$ we obtain Yukawa couplings, as is illustrated in \Fig{flocal2}. The gauge bosons live in the bulk of $S$ whereas quarks, leptons, and Higgsses are localized in complex curves inside $S$. Chirality is obtained by adding a $U(1)$ magnetic background in the underlying 7-branes with gauge group $U(1)$.\\

Thus the gauge group in F-theoretical 7-branes goes beyond what one can get in perturbative Type IIB orientifold $D7$-branes in which only $SU(n+1)$ and $SO(2n)$ gauge groups may be obtained. Furthermore in F-theory the matter content in models with $SO(2n)$ gauge symmetry may include spinorial representations which are not present in perturbative IIB orientifold compactifications. This is important since e.g. it allows for an underlying $SO(10)$ GUT structure which contains SM fields in 16-dimensional spinorial representations. In addition, it also allows the construction of $SU(5)$ GUT's with a $\mathbf{10} \cdot \mathbf{10} \cdot \mathbf{5}$ Yukawa coupling. This coupling is forbidden in D-brane models at perturbative level since a perturbative $U(5)$ coupling of the form $\mathbf{10}_2 \cdot \mathbf{10}_2 \cdot \mathbf{5}_1$ is not invariant under the $U(1) \subset U(5)$. It can however arise in F-theory from a point of an enhanced exceptional $E_6$ symmetry
\begin{eqnarray}
  E_6 &\rightarrow& SU(5) \times U(1) \times U(1)'\\
  \mathbf{78} &\rightarrow& \!\!\! Adjoints\ + [ (\mathbf{10},\ -1,\ -3) + (\mathbf{10},\ 4,\ 0) + (\mathbf{5},\ -3,\ 3) + (\mathbf{1},\ 5,\ 3) + c.c.] 
\end{eqnarray}
Note that a $\mathbf{10} \cdot \mathbf{10} \cdot \mathbf{5}$ coupling is now allowed by the $U(1)$ symmetries. This solves the top-Yukawa problem which appears in Type IIB orientifolds and is realized at the intersection of four 7-branes as is shown in \Fig{flocal2}. Concerning the case of leptonic and down-quark Yukawas, they are contained in a coupling $\mathbf{10} \cdot \overline{\mathbf{5}} \cdot \overline{\mathbf{5}}$ which arises in an $SO(12)$ enhancement:
\begin{eqnarray}
  SO(12) &\rightarrow& SU(5) \times U(1) \times U(1)'\\
  \mathbf{66} &\rightarrow& \!\!\! Adjoints\ + [ (\mathbf{10},\ 4,\ 0) + (\overline{\mathbf{5}},\ -2,\ 2) + (\overline{\mathbf{5}},\ -2,\ 2) + c.c.] 
\end{eqnarray}
which is allowed both in the perturbative and non-perturbative realizations of Type IIB orientifolds through the intersection of branes (see \Fig{flocal2}).\\

Finally, F-theory gives us a natural framework for GUT's. This is a very nice characteristic of heterotic compactifications but it is not so easy to find in Type II orientifold compactifications. Conversely, and unlike the heterotic case, all the tools of local Type II Calabi-Yau orientifolds can be used in F-theory and therefore one can adopt a bottom-up approach. Hence, in a first analysis, one can ignore the global characteristics of the Calabi-Yau, the fixing of the moduli etc. and one can simply assume that the local F-theory model will be eventually embedded in a global model. In conclusion, F-theory is inheriting the good features of Heterotic and Type II orientifold compactification.

\section{Type II 4d effective action for O3/O7 orientifolds}
We are going to focus now our attention on the 4d effective action of the Type IIB Calabi-Yau orientifold\footnote{For reviews on orientifold models see  \cite{interev} .}. The bosonic $N=1$ supergravity action will contain terms as 
\begin{equation}
S =-\int{\frac{1}{2} R + K_{I\bar{J}}DM^I \wedge \ast D\bar{M}^{\bar{J}}+ \frac{1}{2}\rm{Re}f_{\kappa\lambda}F^{\kappa}\wedge \ast F^{\lambda} + \frac{1}{2}\rm{Im}f_{\kappa\lambda}F^{\kappa}\wedge F^{\lambda} + V} + ...   \label{IIBaction}
\end{equation}
where the scalar potential $V$ is given by
\begin{equation}
 V = e^K \left(K^{I\bar{J}}D_{I}WD_{\bar{J}}\bar{W}-3|W|^2 + \frac{1}{2}\left(\frac{1}{\mathrm{Re} f}\right)^{\kappa\lambda}D_\kappa D_\lambda\right) \label{scalarpot}
\end{equation}
and where $M^I$ denote all complex scalars in chiral multiplets present in the theory and $K_{I\bar{J}}$ is a Kähler metric satisfying $K_{I\bar{J}}=\partial_I \partial_{\bar{J}} K(M,\bar{M})$. The scalar potential is expressed in terms of the Kähler-covariant derivative $D_I W= \partial_I W + (\partial_I K)W$. Finally $f$ is the gauge kinetic function. 


\subsection{Kähler potential}\label{subsection kahler}
As we said for the case of Type IIB orientifold Calabi-Yau compactifications, one projects the string spectrum over $\Omega \mathcal{R}$ invariant states, $\Omega$ being  the worldsheet parity operator and $\mathcal{R}$ some particular order-2 geometric action on the compact CY coordinates. That operation leaves invariant certain submanifolds of the compact CY variety corresponding to orientifold planes. We further concentrate on the case in which these orientifold planes are $O(3)$ and $O(7)$ planes. In the first case the $O(3)$ planes are volume filling (i.e. contain Minkowski space) and are pointlike on compact space. The $O(7)$ planes will also fill Minkowski space and wrap orientifold invariant 4-cycles in the CY space. Global consistency of the compactification will in general require the presence of volume filling $D3$- and $D7$-branes, the latter wrapping 4-cycles $\Sigma_4$ in the CY. MSSM fields will appear from open strings exchanged among these D-branes.\\

The closed string spectrum will include complex scalars, the axidilaton $S$, Kahler moduli $T_i$ and complex structure $U_n$ fields defined as \cite{book}
\beq
S\ =\ \frac {1}{g_s}\ +\ iC_0 \ \ ;
T_j\ =\ \frac {1}{g_s (\alpha ')^2}{\rm Vol}(\Sigma_j)\ +\ iC_4^j \ ;\ U_n\ =\ \int_{\sigma_n} \Omega \,,
\label{losmoduli}
\eeq
where $g_s$ is the $D=10$ IIB dilaton, $\Sigma(\sigma)$ are 4(3)-cycles in the CY and $\Omega $ is the holomorphic CY 3-form. Here $C_p$, $p=0,4$ are Ramond-Ramond (RR) scalars coming from the dimensional reduction of antisymmetric p-forms. In terms of these the $N=1$ supergravity Kahler potential of the moduli is given by
\beq
K(S,T_i,U_n) \ =\ -\log(S+S^*)\ -\ 2\log({\rm Vol}[{\rm CY}])\ -\ 
\log[-i\int \Omega \wedge {\overline \Omega}]\,.
\label{elkaler}
\eeq
To get intuition it is useful to consider the case of the (diagonal) moduli in a purely toroidal $T^2\times T^2\times T^2$ orientifold. There are 3 diagonal $T_i$ Kahler moduli with Re$T_i=\frac {1}{g_s(\alpha ')^2} A_jA_k$, $i\not=j\not=k$, $A_i$ being the area of the i-th 2-torus. The complex structure moduli $U_n$, n=1,2,3  coincide with the three  geometric complex structure moduli $\tau_i$ of the three  2-tori. One then obtains
\beq
K\ =\ -\log(S+S^*) \ -\ \log(\Pi_i (T_i+T_i^*)) \ -\log (\Pi_n
(\tau_n+\tau_n^*))\,. \label{elkahler2}
\eeq 
In equations \Eq{elkaler} and \Eq{elkahler2} $N=1$, $D=4$ string effective actions one recognizes the standard log (no-scale) structure of the Kahler potential. The gauge interactions and matter fields to be identified with the SM reside at the $D7$, $D3$-branes and/or their intersections.\\

\begin{figure}[t!]
\begin{center}
\includegraphics[height=6cm]{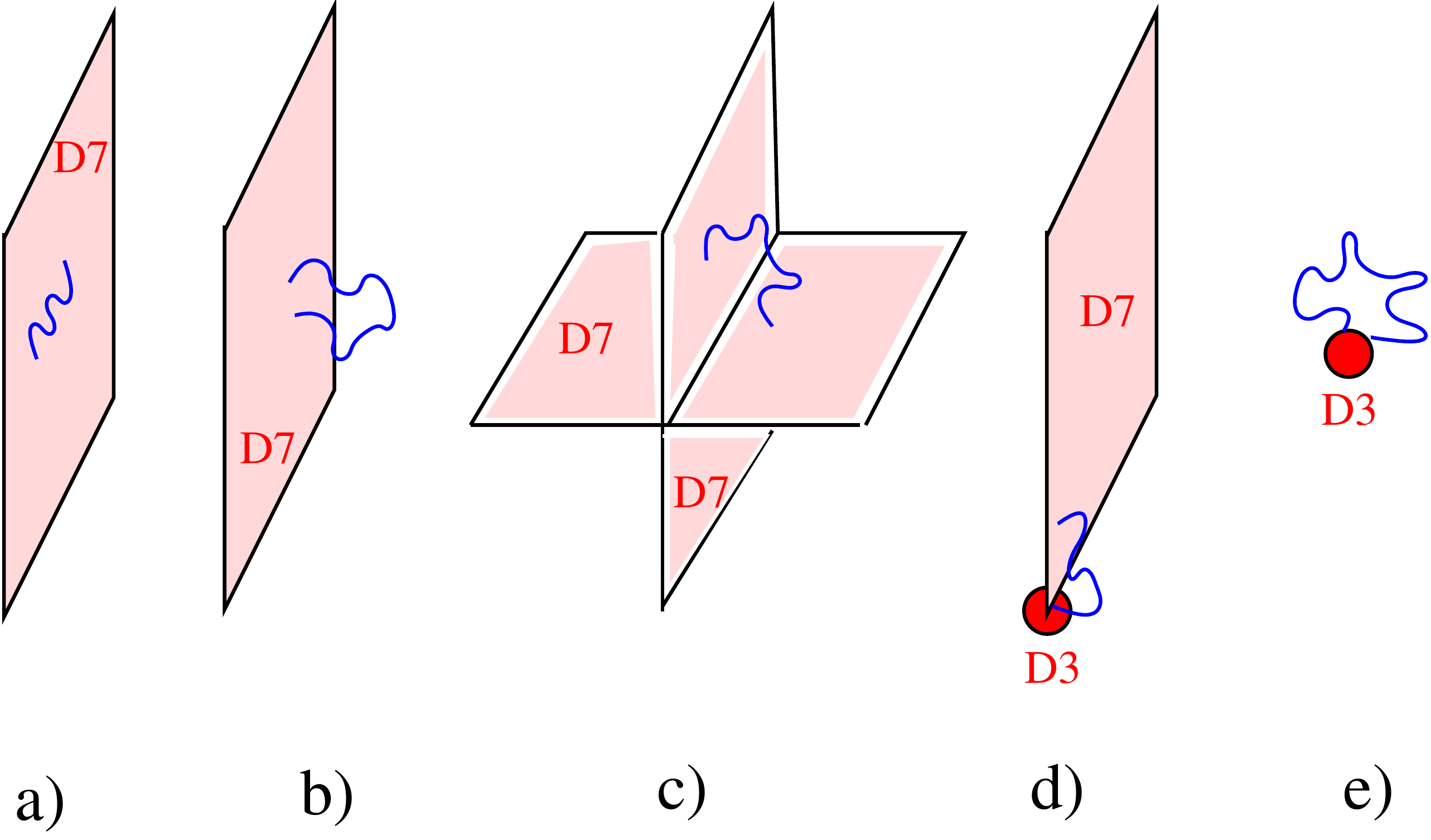}
\caption{Different origin of matter fields in $D7/D3$ configurations.
a) States from reduction of gauge fields A within the $D7$ worldvolume;
b) From reduction of D=8 fields $\phi$ parametrizing the $D7$ position;
c) From two intersecting $D7$-branes; d) From open strings between a
$D3$ and a $D7$; e) From open strings starting and ending on $D3$ branes.}
\label{soft1}
\end{center}
\end{figure}

In order to get a semirealistic model we need D-brane configurations giving rise to chiral gauge theories. In the classes of IIB models here considered one can classify the origin of chiral matter fields in terms of five possibilities which are pictorially summarized in \Fig{soft1}. The worldvolume theory of $D7$-branes is 8-dimensional supersymmetric Yang-Mills and contains both vector  A and matter scalar  $\phi$ multiplets (plus fermionic partners) before compactification to $D=4$. The first class of matter fields a) come from massless modes of the gauge multiplet fields inside the $D7$ worldvolume (plus fermionic partners). The case  b) corresponds to massless modes coming from the $D=8$ scalars $\phi$  which parametrize the position of $D7$-branes in transverse space (plus fermionic partners). The case c) corresponds to massless fields coming from the exchange of open strings between intersecting $D7$-branes. Open strings between $D3$ and $D7$ fields give rise to matter fields of type d) whereas matter fields living fully on $D3$ branes correspond to type e). Generically, in order to make those zero modes chiral in $D=4$ additional ingredients are required. In particular, there should be  some non-vanishing magnetic field in the worldvolume of the  $D7$-branes. In the case of $D3$-branes one may obtain chirality if they are located at some (e.g. orbifold) singularity in the compact CY space.\\

In principle SM fields could come from any of these five configurations but, as we will see, there are a number of constraints which reduce considerably the possibilities. For the moment let us state that there are specific semirealistic constructions in which SM fields live in any of these five configurations. For example, in section 9.1 of \cite{ciu2} there are examples in which SM particles reside in the a,b and d type of open strings above. Models with SM particles in $(33)$ branes at singularities have been discussed  in \cite{Aldazabal:2000sa, bjl, vw}. Concerning the c) configurations, they appear in magnetized $D7$-brane models which are mirror to IIA intersecting D6-brane models \cite{cim1, cim2, ms, ms04, coiso, interev}.

\subsubsection{Kahler metrics for toroidal orientifolds}\label{kaler toroidal}
For illustration, we consider now the form of the Kähler metrics in a factorized 6-torus, keeping only the diagonal moduli. These are the untwisted moduli present in $\mathbf{Z}_2 \times \mathbf{Z}_2$ orientifold compactifications. Let us consider a factorized 6-torus $T^2\times T^2\times T^2$, for this case the complex structure moduli\footnote{Note that the toroidal complex structure $\tau_j$ is equal to the moduli field $U_j$ that appears in the $D=4$ Type IIB supergravity action.} is given by

\begin{equation}\label{ecm2.22}
\tau_j=\frac{1}{e^2_{jx}}(A_j+ie_{jx}\cdot e_{jy})
\end{equation}

\noindent where $e_{jx}$ and $e_{jy}$ are the two-torus $T^2_j$ lattice vectors. On the other hand from \Eq{losmoduli} we have the Kähler moduli \cite{fi} 
\begin{equation}\label{ecam2.23}
T_j=e^{-\phi_{10}}\frac{A_jA_k}{\alpha'^2}+ia_j\ ; \; j\neq k \neq i
\end{equation}

\noindent where $a_j$ are the axions that arise from the RR 4-form $C_4^j$. And finally we recalll that the complex dilaton $S$ is given by
\begin{equation}\label{ecm2.21}
S=e^{-\phi_{10}}+ia_0
\end{equation}

\noindent where $a_0$ is a zero mode from the RR $C_0$ form and $e^{-\phi_{10}}$ is the string coupling constant $g_s=e^{-\phi_{10}}$. For convenience we define
\begin{equation}\label{ecm2.24}
s=S+\bar{S} \ ; \;t_i=T_i + \bar{T}_i \ ; \;u_i=U_i + \bar{U}_i.
\end{equation}

After the compactification to four dimensions the gravitational coupling is $G_N=\kappa^2/8\pi$ where
\begin{equation}\label{ecm2.27}
\kappa^{-2}=\frac{{M_{Pl}}^2}{8\pi}=e^{-2\phi_{10}} \frac{A_1A_2A_3}{\pi\alpha'^4}=\frac{(st_1t_2t_3)^{1/2}}{4\pi\alpha'},
\end{equation}
and the four-dimensional dilaton is given by
\begin{equation}\label{ecm2.28}
\phi_4=\phi_{10}-\frac{1}{2}\log(A_1A_2A_3/\alpha'^3)
\end{equation}
so that in terms of \Eq{ecm2.24}
\begin{equation}\label{ecm2.29}
e^{\phi_4}=\frac{(st_1t_2t_3)^{1/4}}{2}.
\end{equation}

We are going to consider compactifications with a MSSM sector with a Kähler potential developed in terms of chiral matter fields $C_{\alpha}$ as 
\begin{equation}\label{ecm2.30}
K=\hat{K}(M,\bar{M})+\sum_{I,J}\tilde{K}_{I\bar{J}}(M,\bar{M})C_I\bar{C}_J+\frac{1}{2}\sum_{I,J}[Z_{IJ}(M,\bar{M})C_IC_J+c.c.]+...
\end{equation}
where $M={S,T_i,U_i}$ are the string moduli and the Kähler potential for these string moduli is given by
\begin{equation}\label{ecm2.31}
\kappa^2\hat{K}(M,\bar{M})=-\log s-\sum_{i=1}^3\log t_i -\sum_{i=1}^3\log u_i.
\end{equation}
 
Concerning the chiral matter fields $C_{\alpha}$, they belong to the open string sector. If we consider the five types of matter fields discussed above (see \Fig{soft1}), they correspond in the toroidal/orbifold case to states with a brane structure
\beq
a=(7^i7^i)_j \ ;\ b=(7^i7^i)_i \ ;\ c=(7^i7^j)  \ ; \ d=(37^i) \ ;\
e=(33)_i \ \ \ \ , \ \ \ i\not=j\,,  \label{abcde}
\eeq 
where the superindices label the three types of $D7^i$-branes whereas the subindex in the 1-st, 2-nd and 5-th cases correspond to the complex planes $j=1,2,3$. Following the notation of \Ref{fi}, we will label them as $C_I={C_j^{7_i},\ C_i^{7_i},\ C^{7_i7_j},\ C^{37_i}}$ and $C_i^{3}$ respectively.\\

We will therefore distinguish between 'untwisted' states, those corresponding to open strings beginning and ending on the same stack of branes and 'twisted', the ones which correspond to open strings living at the intersection of different stacks of D-branes. The untwisted sector gives 3 massless chiral multiplets in a toroidal setting (it also gives a gauge multiplet as usual in D-branes) which transform in the adjoint of the group of the stack of branes. In particular, $C_i^{7_i}$ and $C_i^{3}$ describe the position of the brane in the transverse $T_i^2$ and $C_j^{7_i}$ describes the Wilson lines on the two internal complex dimensions.  The twisted sector corresponds to massless chiral multiplets transforming in the bi-fundamentals of the D-brane gauge groups. In addition we consider a diagonal Kähler metrics for the matter fields $\tilde{K}_{I\bar{J}} = \tilde{K}_{\alpha}$ i.e. the metrics vanish when $J\neq I$.\\ 

The general expresions for the Kähler metrics for the untwisted sector are given in \cite{lmrs, lrs} and in our notation have the following form
\begin{eqnarray}
\label{kalunt1} && \kappa^2\tilde{K}_{C_i^{7_i \rm{\ or\ 3} }}=\frac{e^{\phi_4}}{\alpha'^2u_i}\sqrt{\frac{\alpha'A_i}{A_lA_k}}\left|m_i^lA_l+i\alpha'n_i^l\right|\left|m_i^kA_l+i\alpha'n_i^k\right| \quad \mathrm{if} \quad i=j\\
\label{kalunt2} && \kappa^2\tilde{K}_{C_j^{7_i}}=\frac{e^{\phi_4}}{u_j}\sqrt{\frac{\alpha'A_j}{A_iA_l}}\left|\frac{m_i^lA_l+i\alpha'n_i^l}{m_i^jA_j+i\alpha'n_i^j}\right| \quad \mathrm{if} \quad i\neq j
\end{eqnarray}
with $ j \neq k \neq l \neq j=1,2,3 $ the wrapping numbers and $i=a,b,c$ the stack of D7-branes. These correspond to cases a) and b) of \Eq{abcde}. These expressions are valid for the case $\mathbf{Z}_2 \times \mathbf{Z}_2$ orbifold orientifolds. Similar expressions may be obtained for other orbifolds.\\

In order to calculate the metric of twisted sector of magnetized D-branes on toroidal compactifications (cases c) and d) of \Eq{abcde}) we need to remember that $F^i$ is the magnetic flux going through the i-th 2-torus which may be written as
\beq
F^i \ =\ n^i  (\frac{s t_i}{t_jt_k})^{1/2} \,, 
\eeq
with $n^i$ quantized integer fluxes, and that 
\beq
\theta _{ab}^j \ =\ \arctan(F_b^j)-\arctan(F_a^j) \ .
\eeq 
We also define
\begin{equation}\label{ecm2.39}
\hat{\nu}_l=\frac{\theta_{ij}^l}{\pi}
\end{equation}
which $\hat{\nu}_1+\hat{\nu}_2+\hat{\nu}_3=0$. With this we can construct $\nu_l$ such that $0\leq\nu_l<1$ and $\nu_1+\nu_2+\nu_3=2$ in the following way
\begin{equation}\label{ecm2.40}
\nu_l=\left\{\begin{array}{ll}
1+\bar{\nu}_l & \textrm{if $\bar{\nu}_l<0$}\\
\bar{\nu}_l  & \textrm{if $\bar{\nu}_l\geq0$}
\end{array}\right.
\end{equation}

With these definitions one obtains for the general expression for the Kähler metrics for twisted sector\footnote{Notice that the definition of the parameter $\nu_l$ in the way as we have introduced it, is in order to have positive gamma function arguments.}
\begin{eqnarray}
\label{kalt1} && \kappa^2\tilde{K}_{C^{7_i7_j \rm{\ or\ 3}7_i}}=e^{\phi_4}\prod_{l=1}^3u_l^{-\nu_l}\sqrt{\frac{\Gamma(1-\nu_l)}{\Gamma(\nu_l)}}\\
\nonumber
\end{eqnarray}
where $u_j$ are the real parts of the complex structure moduli, $\Gamma $ is the Euler Gamma function. These will correspond to states coming from open strings in between (magnetized) branes  $D7^a$, $D7^b$ or $D7^a$ with $D3^b$ embedded into, wrapping different 4-tori. See \App{appA} for an application to a new MSSM-like model.\\

As discussed in \cite{imr, lmrs, lrs}, using first order expansion terms without fluxes of  \Eq{kalunt1}, \Eq{kalunt2} and \Eq{kalt1} (see also the example in \App{appA}), the  Kahler metrics corresponding to \Eq{abcde} brane configurations are
\beq
K_{C_j^{7_i}}= \frac {1}{t_k} \ ;\
K_{C_i^{7_i}}= \frac {1}{s} \ ;\
K_{C^{7_i7_j}}= \frac {1}{s^{1/2}t_k^{1/2}} \ ;\
K_{C^{37_i}}= \frac {1}{t_j^{1/2}t_k^{1/2}} \ ;\
K_{C_i^{3}}= \frac {1}{t_i}\,.
\eeq
Notice that we have not included here the dependence on the complex structure fields $U_n$, which will play no role in modulus dominated SUSY-breaking that we will develope in \Sec{stdm}. Notice also that all metrics are flavour diagonal at this level. In terms of the overall Kähler modulus $t=t_i$ we will thus have for the the chiral field metrics in this toroidal case \cite{aci}:
\beq
K_\alpha\ =\ \frac {1}{s^{1-\xi}t^{\xi}}
\ ;\  \ \xi =0,\ 1/2,\ 1  \ .
\eeq
where fields in \Eq{abcde} of type a), d) and e) have modular weight $\xi=1$, and those of type c) have $\xi=1/2$ and type b) has $\xi=0$.

\subsubsection{Kahler metrics beyond the toroidal setting}
\begin{figure}
\begin{center}
\includegraphics[height=4cm]{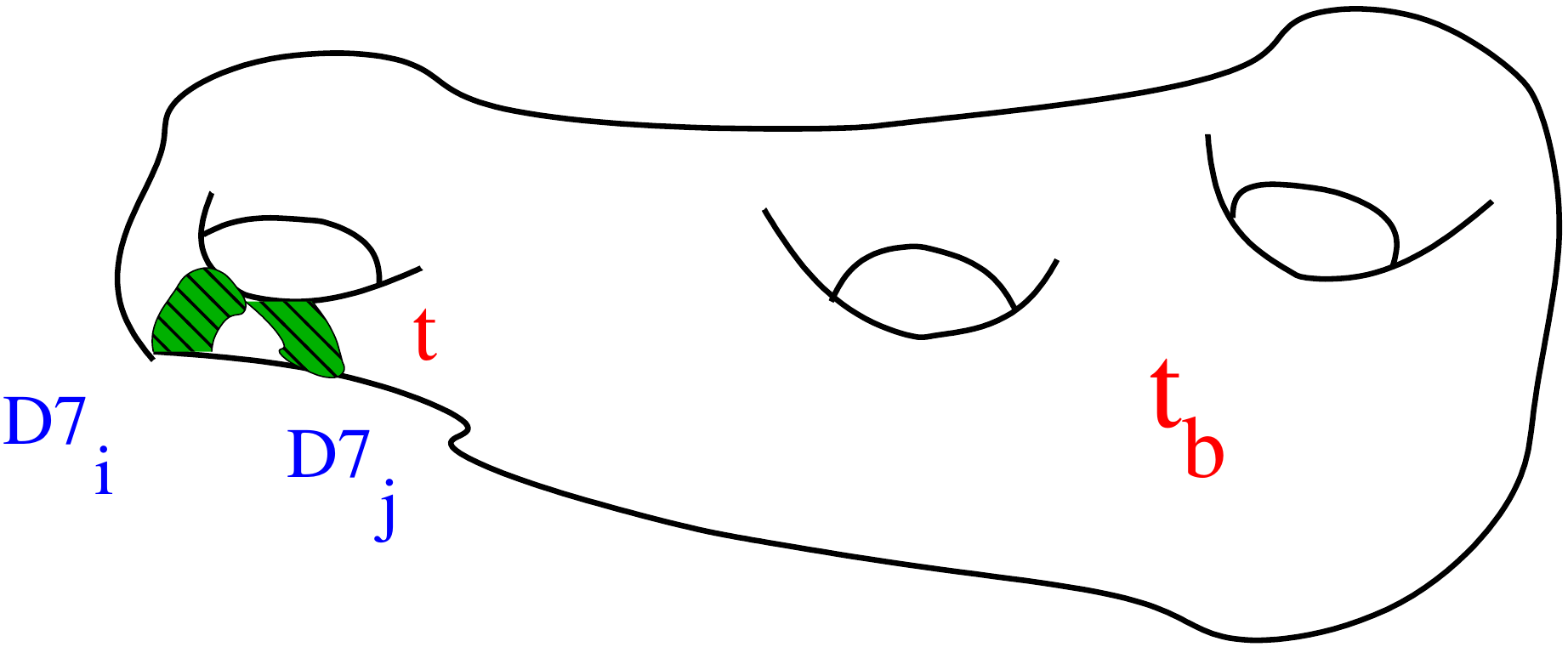}
\caption{The SM $7$-branes wrap local 4-cycles of size $t$ in a CY whose
(larger) volume is controlled  by $t_b$.}
\label{instanton4}
\end{center}
\end{figure}

Toroidal  orientifolds are very special in some ways and we would like to see to what extent the above findings generalize to more general IIB CY orientifolds. In particular $D7$-branes wrap 4-tori whose volume are directly related to the overall volume of the compact manifold and then the Kahler moduli appearing in the MSSM effective action are directly related to the overall volume of the tori. This is unnecessarily restrictive. One would expect more general situations in which the $D7$-branes wrap local cycles whose volume need not be directly related  to the overall volume of the CY. An example of this is provided by the 'swiss cheese' type of compactifications discussed in \cite{conlon1, conlon2, ccq, bhp, acqs, ccq2} (see also \cite{bmp}). In this class of  more general models one  assumes that the SM resides  at stacks of $D7$-branes wrapping  'small cycles' in a CY whose overall volume is controlled by a large modulus $t_b$ (see fig.\,\ref{instanton4}) \cite{conlon1, conlon2}:
\beq
{\rm Vol}[{\rm CY}] \ =\ t_b^{3/2} \ -\ h(t_i)\,, 
\label{volcy}
\eeq
where $h$ is a homogeneous function of the 'small' Kahler moduli $t_i$ of degree 3/2. The simplest example of CY with these characteristics is the manifold ${\bf P}^4_{[1,1,1,6,9]}$ which has two Kahler moduli with \cite{ddf} \footnote{For discussions for the multi Kahler moduli case see \cite{bhp, ccq2, bmp}.}
\beq
{\rm Vol}[{\rm CY}] \ =\ t_b^{3/2} \ - \ t^{3/2}\,.
\label{dosmoduli}
\eeq

It is important to emphasize that, in contrast to the assumptions in \cite{conlon1, conlon2}, we will not necessarily assume that $t_b$ in \Eq{dosmoduli} is hierarchically larger than the $t_i$, rather we will just assume that $t_b$ is big enough so that an expansion in powers of $t_i/t_b$ makes sense. In the SUSY-breaking analysis in the next chapter we will consider for simplicity the dependence on a single small Kahler modulus $t$ and take both $t_b$ and $t$ sufficiently large so that the supergravity approximation is valid. In this  case one can write a large volume expansion for the Kahler metrics of the form  \cite{conlon1, conlon2, ccq}
\beq
K_{\alpha}\ = \frac {t^{(1-\xi_\alpha )}}{t_b} \,,  
\label{metrilla}
\eeq
with $\xi_\alpha$ the modular weights discussed above. Note that for $t_b=t$ one would recover the form for the metric we found in toroidal cases (ignoring have the s-dependence, irrelevant in our SUSY-breaking analysis). The precise values of $\xi_\alpha$ and hence the Kähler metrics for matter fields may be computed using scaling arguments as is shown in \Sec{yukweightand}.

\subsection{Modular weight possibilities and Yukawa couplings}\label{yukweightand}

We are going to consider now the case of modular weights $\xi_\alpha$ beyond the toroidal setting. Although for \Eq{metrilla} the explicit computation of the $t$-dependence used in toroidal models of \Sec{kaler toroidal} is not directly applicable, one can use arguments based on the scaling of Yukawa couplings to indirectly compute the $\xi_\alpha$'s. The idea \cite{ccq} is to recall that the physical Yukawa couplings  ${\widehat Y}_{\alpha \beta \gamma}$ among three fields $\Phi_\alpha $ after normalization is given in $N=1$ supergravity by
\beq {\widehat Y}_{\alpha \beta \gamma} \ =\ e^{K/2} \frac {Y^{(0)}_{\alpha \beta \gamma }}
{(K_\alpha K_\beta K_\gamma )^{1/2} }\,, 
\eeq

\noindent where $Y^{(0)}$ is the unnormalized Yukawa coupling which in Type IIB orientifolds is known to be holomorphic on the complex structure fields $U_n$ and independent of the Kahler moduli $T_i$ (at least at the perturbative level we are considering). From that we conclude (using eqs.(\ref{volcy}-\ref{metrilla})) that the physical Yukawa coupling scales with $t$ like
\beq
{\widehat Y}_{\alpha \beta \gamma }\ \simeq \   t^{ 1/2 (\xi_\alpha + \xi_\beta +\xi_\gamma\ -\ 3)}\,.
\label{estay}
\eeq

Note that the dependence on the large modulus $t_b$ has dropped to leading order in $t/t_b$. This is expected since the wave functions of the matter fields are localized at the small cycles and the Yukawa couplings are local quantities, expected to be independent of the overall volume. On the other hand one can estimate the $t$-scaling of the physical Yukawa couplings by noting that in Type IIB string theory those  may be found by computing overlap integrals of the form

\beq
{\widehat Y}_{\alpha \beta \gamma }\ \simeq \ 
\int dx^6\ \Psi_\alpha \Psi_\beta \Psi_\gamma \,, 
\eeq 
where the integration is dominated by the overlap region of the three extra dimension zero mode states $\Psi$. This assumes those internal wave functions to be normalized:

\beq
\int |\Psi_\alpha|^2\ = \ \int |\Psi_\beta|^2\ =\ \int
|\Psi_\gamma|^2\ =\ 1\,.
\label{normal}
\eeq

There are several possible Yukawa coupling structures. However, we are going to focus on the intersecting D7-branes possibilities because it is the one motivated from F-theory models as we saw in \Sec{ftheory}. Consider first the case of three sets of $D7_i$ branes i=1,2,3, each overlapping with each other on  a complex curve of real dimension two. Then the matter fields at each intersection $C^{7_i7_j}C^{7_j7_k}C^{7_k7_i}$ overlap over a 2-dimensional space and, given \Eq{normal}, their wave function would scale like $1/R\simeq t^{-1/4}$. For a trilinear coupling to exist the three $D7$ branes overlap in this case at a point so that the  physical Yukawa should scale like ${\widehat Y}\simeq   (1/t^{1/4})^3 = t^{-3/4}$. This agrees with (\ref{estay}) if the three chiral fields at intersecting $D7$-branes have $\xi_\alpha=1/2$. \\

Using  a similar argument, one can calculate the Yukawa coupling of type $C^{7_i7_j}C^{7_j7_i}C_j^{7_i}$. One finds that bulk fields of type $C_j^{7_i}$ have $\xi_\alpha =1$. Finally from the existence of couplings of type $C_j^{7_i}C_j^{7_i}C_i^{7_i}$ one gets that the zero modes from fields of type $C_i^{7_i}$ have $\xi_\alpha =0$. In summary, there are essentially only three Yukawa coupling options corresponding to the modular weight distributions\footnote{There also exist other type of couplings, e.g. $(1,1,1)$, but they lead to no SUSY-breaking soft terms to leading order.} $(1/2,1/2,1/2)$, $(1/2,1/2,1)$  or $(1,1,0)$. From now on we will denote these three options by (I-I-I), (I-I-A)
and (A-A-$\phi$) respectively.\\

Note that for $t_b\propto t$ the Kahler metrics from (\ref{metrilla}) will then have the same $t$-scaling that we found for the different type of zero modes A, $\phi$, I in the toroidal case. Notice also that the above results are rather general and independent of the particular geometry as long as one can approximate the scaling behaviour with a single local Kahler modulus. Note also that one can only extract information on the sum of the three modular weights (not the individual ones) and obviously only in models in which those trilinear couplings actually exist.

\subsection{Gauge kinetic function}\label{secgaugek}
The gauge kinetic function in Type II orientifolds may be obtained from the D-brane action. More specifically from de DBI part of the action
\begin{equation}
 S_{DBI}=-\mu_p\int{e^{-\phi}\sqrt{\mathrm{det}(G+B-2\pi\alpha'F)} }, \; \; \mu_p = \frac{(\alpha')^{(p+1)/2}}{(2\pi)^p} 
\end{equation}

\noindent and expanding to quadratic order in gauge field strength $F$ one obtains the inverse of the gauge coupling constant for a D$p$-brane
\begin{equation}
 \frac{1}{g^2_i}=e^{-\phi} \frac{(\alpha')^{(3-p)/2}}{(2\pi)^{p-2}} \mathrm{Vol}(\Pi_{p-3})
\end{equation}

\noindent this expression is particularly simple in the case of a factorized $\mathbf{T}^6$. The real parts of the gauge kinetic functions for the groups arising in $D7_i$ and $D3$  are calculated from the expression \cite{CIMQuivers, lmrs, lrs}
\begin{equation}
\mathrm{Re}\ f_i=\frac{e^{-\phi_{10}}}{(2\pi)^5\alpha'^2}\prod_{l\neq i}|m_l^iA_i+i\alpha'n_l^i| = \frac{1}{g^2_i} \label{ecm2.56}
\end{equation}

\noindent For a $D3$-brane and  a $D7$ brane wrapping a 4-cycle $\Sigma_j$ one gets  the simple result
\beq 
f_{D7^j} \ =\ T_j\ \ \ \ \ ;\ \ \ \ f_{D3} \ =\ S\,.
\eeq
We apply all these formulae on a toroidal MSSM-like example in \App{appA}.\\

\subsection{Subleading effects from magnetic gauge fluxes}\label{sublmageff}

In previous sections we have mentioned that getting chirality of the massless $D=4$ spectrum in general requires the presence of magnetic fluxes\footnote{These open string gauge  fluxes should notbe confused with antisymmetric RR and NS fluxes which may break SUSY
and come from the closed string sector.} on the worldvolume of $D7$ or F-theory 7-branes. So a natural question is how these fluxes may modify the Kahler metrics of matter fields and the gauge kinetic functions associated to 7-branes. In the large $t$ limit the magnetic fluxes are diluted and 
one expects this effect to be subleading. To flesh out this expectation we study in this subsection how the presence of magnetic fluxes affects the 
effective action in the well studied case of toroidal/orbifold perturbative IIB orientifolds. The presence of fluxes affects in the same manner
perturbative and F-theory compactifications so one expects similar corrections to appear in more general CY orientifolds or F-theory compactifications.\\

As was discussed in \Sec{D7magn}, in configurations with sets of overlapping $D7$-branes in generic 4-cycles, chirality is obtained by the addition of magnetic fluxes. In the case of toroidal/orbifold IIB orientifold models the corrections to the Kahler metrics coming from magnetic fluxes have been computed in \cite{lmrs, lrs}. The results for states coming from strings within $D7$s wrapping the same cycle is
\beq
K_{C_j^{7_i}} \ =\ \frac {1}{t^k}\ |\frac{1+iF^k}{1+iF^j}| \ \ ;\ \ 
K_{C_i^{7_i}} \ =\ \frac {1}{s}\ (1+|F^jF^k|)\,, 
\eeq
where $i\not=j\not=k$ label the 3 2-tori and $F^i$ is the magnetic flux going through the i-th 2-torus which may be written as
\beq
F^i \ =\ n^i  (\frac{s t_i}{t_jt_k})^{1/2} \,, 
\eeq
with $n^i$ quantized integer fluxes. For states coming from open strings in between (magnetized) branes  $D7^a$, $D7^b$ wrapping different 4-tori one has from \Eq{kalt1}
\beq
K_{C^{7_a7_b}}\ =\ \frac {1}{(s t_1t_2t_3)^{1/4}}
\Pi_{j=1}^3 \ u_j^{-\theta_{ab}^j}\
\sqrt{ \frac {\Gamma (\theta_{ab}^j)} {\Gamma (1-\theta_{ab}^j) } }\,, 
\label{metrica}
\eeq
where $u_j$ are the real parts of the complex structure moduli, $\Gamma $ is the Euler Gamma function and 
\beq
\theta _{ab}^j \ =\ \arctan(F_b^j)-\arctan(F_a^j) \ .
\eeq 
The gauge kinetic functions are also modified in the presence of magnetic fluxes as
\beq
Ref^a_i \ =\  T_a \ ( 1 \ +\ |F_a^jF_a^k|) \ .  
\eeq

These results apply for the case of a toroidal and/or orbifold setting. However we can try to model out what could be the effect of fluxes in a more general setting following the above structure. To model out the possible effect of fluxes we consider again the limit with $t_i = t$ and diluted fluxes $|F_i|= F$, i.e. large $t$ and ignore the dependence on the complex structure $u_i$ fields. Then from the above formulae one obtains  the dilute flux results 
\beqa
K_{C_j^{7_i}}^i \ &=& \ \frac {1}{t}\  \ \ ;\ \
K_{C_i^{7_i}} \ =\ \frac {1}{s}\ (1 + a_i \frac {s}{t})\ ,  \\
K_{C^{7_a7_b}}\ & =& \ \frac {1}{(s^{1/2}t^{1/2}) } (1\ +\  c_{ab}\ \frac
{s^{1/2}}{t^{1/2}})\ \label{ult}, \\
{\rm Re}f_i \ &=&\   t \ +\  a_i  s \ ,
\label{metricaaprox}
\eeqa
where $a_i,c_{ab}$ are constants (including the flux quanta) of order one. The limit of \Eq{metrica} has been obtained by expanding the Gamma functions for dilute flux\footnote{A complete expression in terms of Gamma and Beta functions was discussed in the example of the appendix.} and ignoring the dependence through the complex structure fields $u_i$. The above three formulae may be summarised by \cite{aci}:
\beq 
K_{matter} \ =\ \frac {1}{s^{(1-\xi)}t^\xi } \times (1\ +\ c_\xi (s/t)^{1-\xi}) \ =
\ \frac {1}{s^{(1-\xi)}t^\xi} 
\ +\ \frac{c_\xi}{t}\, ,
\label{esta}
\eeq
with $c_{\xi}$ some flux-dependent constant coefficient whose value will depend on the modular weight $\xi$ and the magnetic quanta. This means that the addition of fluxes has the effect in all cases to add a term which goes like $1/t$. Heuristically,  we know that with the addition of large fluxes the $D7$-branes turn into localized branes, very much like $D3$-branes and we know that matter fields for $D3$-branes have kinetic terms which go like $1/t$. This would explain the modulus dependence of the last term in eq.(\ref{esta}).\\ 

Note that in the diluted flux limit ($s/t\rightarrow 0$) one recovers the metrics we discussed above. Note also that the metric for the matter fields of type $(7^i7^i)_j, j\not=i$ does not get corrections to leading order. As we will see, this will imply that fields with modular weight $\xi=1$ will remain massless even after SUSY-breaking. This will be phenomenologically relevant, as we will discuss in the numerical analysis.\\ 

For a non-toroidal compactification one expects analogous flux corrections to exist with $\xi =0,1/2,1$ corresponding to chiral fields of types $\phi $, I and A respectively. One can easily generalize this  to the case in which $t$ is the local modulus coupling to the MSSM in a CY whose volume is controlled by a larger modulus $t_b$, as we discussed in section (2.3). From the above expressions one then finds a Kahler moduli dependence of the matter metrics of the form \cite{aci}
\beq
K_{matter} \ =\ \frac {t^{1-\xi}}{t_b} \ (1\ +\ c_\xi t^{\xi-1}) \ =\ 
\frac {t^{(1-\xi)}}{t_b} \ +\ \frac {c_\xi}{t_b} \ .
\label{flujillas}
\eeq
as a generalization of eq.(\ref{metrilla}). We have ignored here the explicit dependence on $s$ and the complex structure, not relevant for the computation of soft terms. Note that the flux correction actually will depend on the large modulus $t_b$ rather than on the local modulus $t$.

\section{Soft terms}
With the knowledge of the matter metrics and gauge kinetic function we can address the computation of SUSY-breaking soft terms. The MSSM superpotential $W$ has the general form
\beq
W\ =\ W_0 \ +\ \frac {1}{2} \mu H_uH_d \ +\ \frac {1}{6}
Y^{(0)}_{\alpha \beta \gamma } C^\alpha C^\beta C^\gamma \ +\ \ldots
\label{super}
\eeq
Here $W_0$ is a moduli dependent superpotential (including typically a flux-induced constant term which controls the scale of SUSY breaking). The $C^\alpha $ are  the SM chiral superfields and $H_{u,d}$ the minimal Higgs multiplets. In our computations below we are going to assume that there is an explicit $\mu$-term in the Lagrangian which can be taken (at least in some approximation) as independent of the Kahler moduli $t_b,\ t$. A simple origin for such a term  is again RR and NS fluxes. Indeed, it is known that certain SUSY preserving fluxes may give rise to  explicit supersymmetric mass terms to chiral fields, $\mu$-terms (see e.g. \cite{ciu1, fluxed, grana, lrs, ciu2}). If in addition there are SUSY-breaking fluxes a $B$-term and the rest of soft terms appear. These two kind of fluxes are  in principle  independent so that one can simply consider an explicit constant (moduli independent) $\mu$-term in the effective action as a free parameter. On the other hand, if both the origin of the $\mu$-term and SUSY-breaking soft terms is fluxes,  it is reasonable to expect both to be of the same order of magnitude, solving in this way the $\mu$ problem\footnote{An alternative
to get a Higgs bilinear could be the presence of an
additional SM singlet chiral field $X$  coupling to the Higgs fields
like $XH_{u}H_d$. It is trivial to generalize all the above formulae to this
NMSSM case.}. However, in the general formulae below we will  allow for  the presence of a Giudice-Masiero term  of the form $Z(T,T)H_uH_d+ h.c.$ in the Kahler potential \cite{gm}.
\\

\noindent The form of the effective soft Lagrangian obtained is given by (see e.g.\cite{bim})
\begin{eqnarray}
{\cal L}_{soft} &=& \frac{1}{2}(M_a 
{\widehat{\lambda}}^a {\widehat{\lambda}}^a + h.c.)
- m_{\alpha}^2 \widehat{C}^{*\overline {\alpha}}
\widehat{C}^{\alpha}
\nonumber\\ &&
-\
\left(\frac{1}{6} A_{\alpha \beta \gamma} \widehat{Y}_{\alpha \beta \gamma}
	    \widehat{C}^{\alpha} \widehat{C}^{\beta} \widehat{C}^{\gamma}
  + B {\widehat{\mu}} \widehat{H}_u \widehat{H}_d+h.c.\right)\ ,
\label{F6}
\end{eqnarray}
with
\begin{eqnarray}
M_a &=&\frac{1}{2}\left(Ref_a\right)^{-1} F^m \partial_m f_a  \; ,
\label{F}\\
{m}_{\alpha}^2 &=& 
\left(m_{3/2}^2+V_0\right) - {\overline{F}}^{\overline{m}} F^n 
\partial_{\overline{m}}\partial_n \log{ K_{\alpha}}\ ,
\label{mmmatrix}
\\
A_{\alpha\beta\gamma} &=& 
F^m \left[  { K}_m + \partial_m \log Y^{(0)}_{\alpha\beta\gamma} 
- \partial_m \log({ K_{\alpha}} { K_{\beta}}
{  K_{\gamma}}) \right]\ ,
\label{mmmatrix2}
\\
B &=& {\widehat{\mu}}^{-1}({ K}_{H_u}{\ K}_{H_d})^{-1/2}
\left\{ \frac{ { W}^*}{|{ W}|} e^{{ K}/2} \mu 
\left( F^m \left[ {\hat K}_m + \partial_m \log\mu\right.\right.\right.
\nonumber\\ && 
\left.\left.-\ \partial_m \log({ K_{H_u}}{ K_{H_d}})\right]
- m_{3/2} \right)
\nonumber\\ &&
\ + 
\left( 2m_{3/2}^2+V_0 \right) {Z} -
m_{3/2} {\overline{F}}^{\overline{m}} \partial_{\overline{m}} Z
\nonumber\\ &&
+\ m_{3/2} F^m \left[ \partial_m Z - 
Z \partial_m \log({ K_{H_u}}{ K_{H_d}})\right]
\nonumber\\ &&
\left.-\ {\overline{F}}^{\overline{m}} F^n 
\left[ \partial_{\overline{m}} \partial_n Z - 
 (\partial_{\overline{m}} Z) 
\partial_n \log({ K_{H_u}}{ K_{H_d}})
\right] \right\}
\ ,  
\label{mmmatrix3}
\end{eqnarray}
where $V_0$ is the vacuum energy which is assumed to be negligibly small from now on. Here $\widehat{C}^{\alpha}$ and $\widehat{\lambda}^a$ are the scalar and gaugino canonically
{\it normalized} fields respectively
\begin{eqnarray}
\widehat{C}^\alpha &=& { K}_{\alpha}^{1/2} C^\alpha\ ,
\label{yoquese}
\\
\widehat{\lambda}^a &=& (Re f_a)^{1/2} \lambda^a\ , 
\label{F7}
\end{eqnarray}
and the rescaled Yukawa couplings and $\mu$ parameter 
\begin{eqnarray}
{\widehat{Y}}_{\alpha \beta \gamma}    &=& Y_{\alpha \beta \gamma}^{(0)} \; 
 \frac{ { W}^*}{|{ W}|} \; e^{{ K}/2} \;
({ K}_\alpha { K}_\beta { K}_\gamma)^{-1/2}\ ,
\label{yyoquese}
\\
\widehat{\mu} &=& 
\left(  \frac{ { W}^*}{|{ W}|} e^{ K/2} {\mu}
+ m_{3/2} Z -
{\overline {F}}^{\overline{m}} \partial_{\overline{m}} Z \right)
({ K}_{H_u}{ K}_{H_d})^{-1/2}\ ,
\label{rescalado}
\end{eqnarray}
have been factored out in the $A$ and $B$ terms as usual. We will apply all these formulae on the computation of soft terms in the next chapter.

%% file: STDM.tex
\chapter{Modulus dominated SUSY breaking phenomenology in Type IIB orientifold models}\label{stdm}

In this chapter we will apply the physics of the previous chapter in order to calculate SUSY-breaking soft terms corresponding to modulus dominance in Type IIB orientifold/F-theory models. We will not assume a definite  mechanism which fixes all moduli and then compute the soft terms but rather the reverse. We will assume that the auxiliary field of a local Kahler modulus is the dominant source of SUSY-breaking in the MSSM sector. Under these assumptions we will end up with very definite patterns for the low energy SUSY spectrum. Finally we will study the phenomenological viability under certain assumptions. This kind of calculus and analysis is interesting because if low energy SUSY exists we may measure the mass of the SUSY particles (sleptons, squarks, gluinos...) at LHC, and the value of squark and gluino masses would provide information about how is the SUSY masses pattern and therefore how is the breaking of SUSY. Conversely, a measurement of the SUSY spectrum of sparticles a the LHC would also provide an experimental test for large classes of string compactifications leading us to rule out them or to go further in their analysis.\
Most of the results in this chapter are based on the papers \cite{aci} \cite{aciII}.

\section{Construction of semirealistic Type II string vacua}\label{semirealistic}
With the advent of the Large Hadron Collider (LHC) Particle Physics is entering into a new era in which a wealth of theoretical models, scenarios and ideas are being tested. One of the most prominent ideas beyond the Standard Model (SM)  is low energy supersymmetry (SUSY) and its simplest implementation, the Minimal Supersymmetric Standard Model (MSSM). Although at the moment no sign of supersymmetric particles has been seen, there is at least one recent LHC result which points in the direction of supersymmetry. The 2011 run of LHC has restricted the most likely range for the  Higgs particle mass to be  $115.5-131$~GeV (ATLAS) \cite{ATLASHiggs} and $114.5-127$ GeV (CMS) \cite{CMSHiggs}. In addition, there are  hints observed by both CMS and ATLAS of an excess of  events that might correspond to $\gamma \gamma$, $ZZ^*\rightarrow 4l$ and $WW^*\rightarrow 2l$ decays of a Higgs particle with a mass in a range close to $125$ GeV. Interestingly, such values for the Higgs mass are consistent with the expected range $<130$ GeV for the lightest Higgs in the MSSM.\\

Although in qualitative agreement with MSSM expectations, the hints of a $125$~GeV Higgs are slightly uncomfortable for  models like the Constrained Minimal Supersymmetric Standard Model (CMSSM), in which the complete SUSY spectra is determined in terms of a few universal soft supersymmetry-breaking parameters $M,\,m,\,A,\,B,\,$ and $\mu$ \cite{CMSSM}. Indeed lighter Higgs masses of order $110-115$~GeV are generic  in the CMSSM parameter space. In order to get values as large as $125$~GeV one needs to have heavy stops with a sizable LR-mixing and large values of $\tan\beta$, leading typically to a very heavy  SUSY spectrum. In fact it has been noted \cite{Baer:2011ab,Hall:2011aa,Arbey:2011ab,Akula:2011aa} (see also \cite{Gogoladze:2011aa,Carena:2011aa}) that the areas of the CMSSM parameter space compatible with $125$ GeV Higgs show a very strong preference for the region with $A\simeq -2m$ if the SUSY spectrum is not to be very heavy \footnote{Note in passing that a $125$ GeV Higgs is difficult to accommodate in the simplest gauge and anomaly mediation scenarios since $A=0$ in these schemes, see Refs.\ \cite{Arbey:2011ab,Draper:2011aa,Evans:2012hg}.}. But why should nature be centered in that peculiar corner of parameter space?\\  
  
A possible explanation for relations among soft terms like, e.g., $A$ and $m$ requires going beyond the general assumptions underlying the CMSSM scheme and being more specific about the origin of SUSY breaking. The CMSSM boundary conditions are obtained in supergravity mediation schemes with unification (GUT-like) constraints and universal kinetic terms for all the SM matter fields. In order to get relations among the  $M,\,m,\,A,\,M,\ \mu$ parameters one needs very specific classes of low energy $N=1$ supergravity models. It is here where string unification models arising from specific classes of string compactifications may be useful.\\
  
As we saw in previous chapter, in low-energy supergravity models coming from string compactifications the gauge kinetic functions as well as the kinetic (Kahler metrics) terms of the SM fields are not arbitrary and depend on the moduli of the corresponding string compactification.  If the auxiliary fields of the moduli are the source of SUSY-breaking, specific relations among the different soft terms are obtained. These have been worked out for heterotic vacua \cite{Brignole:1993dj, Cvetic:1991qm, il, Kaplunovsky:1993rd} (see e.g.  Ref.\,\cite{bim} for a review and further references) and generalized for the more recent case of Type II orientifold compactifications  \cite{fluxed, aci}. See also Ref.\,\cite{othersoft, conlon2, acharya} and references therein for explicit SUSY-breaking models in Type II orientifolds.\\

With the advent of the D-brane techniques it has been possible to construct Type II string orientifold configurations of branes yielding a massless spectrum close to that of the MSSM (see Ref.\,\cite{book} for a review). As we reviewed in the previous chapter, a particularly successful scheme is the one based on Type IIB orientifolds with the SM fields residing on intersecting  7-branes and their non-perturbative generalization, F-theory. One of the attractive aspects of this large class of compactifications is that it is well understood how the presence of antisymmetric field fluxes and possibly non perturbative effects  can give rise to a complete fixing of the moduli of the compactification \cite{kklt} (for reviews see Refs.\,\cite{Denef:2008wq, Douglas:2006es,book}). In addition, the large number of possible fluxes allows to fine-tune the vacuum energy to a small but positive value, in a way compatible with a non-vanishing positive cosmological constant. Besides fixing the moduli, such fluxes in general give rise to soft SUSY breaking terms for the MSSM fields in semirealistic compactifications \cite{grana, ciu2, ciu1, lrs, fi, softfromflux}. In particular, it has been found that certain ISD (imaginary self-dual) fluxes correspond to the presence of Kahler modulus dominated SUSY-breaking, providing an explicit realization of gravity mediation SUSY-breaking in string theory.\\

\section{Kähler moduli dominated SUSY-breaking in string theory}
It is well known that, within the mentioned scenarios, the Kahler moduli in string compactifications have a classical no-scale structure \cite{cfkn} in such a way that one can obtain SUSY breaking with a vanishing (tree level) cosmological constant. This is true both in the Heterotic as well as Type I, Type IIB  orientifold and  other compactifications and corresponds to having non-vanishing vacuum expectation values for the auxiliary fields of the Kahler moduli. In Type IIB Calabi-Yau (CY) orientifolds this corresponds to string compactifications with  spontaneously broken SUSY solving the classical equations of motion \cite{gkp}. It has also been shown \cite{gkp, ciu1, lrs, ciu2} that in those Type IIB orientifolds such SUSY-breaking corresponds to the presence of RR and NS antisymmetric tensor fluxes in the compactification. Such compactifications will {\it generically }  have non-vanishing   fluxes so it looks like a most natural source of supersymmetry breaking in string theory. The scale of SUSY breaking may be hierarchically small if the flux in the compact region where the SM sector resides is appropriately small. So in what follows we will assume the dominance of Kähler moduli in SUSY breaking.\\

We are going also to assume that all SM gauginos get a mass through this SUSY-breaking mechanism at leading order. This has a phenomenological motivation, it excludes large classes of string compactifications including perturbative heterotic compactifications, Type I string compactifications and Type IIB orientifolds with the SM residing on $D3$-branes at local singularities. In all these classes of compactifications the gauge kinetic function is independent of the (untwisted) Kahler moduli to leading order so that  modulus dominance does not give rise to gaugino masses. This argumentation leave us just with Type IIB orientifold compactifications with MSSM fields residing on $D7$ branes (or their intersections) and their non-perturbative F-theory extensions. Alternatively one may consider the mirror description in terms of Type IIA orientifolds with the MSSM fields living at D6-branes   (or their intersections)  or  their non-perturbative extensions from M-theory compactified on manifolds with $G_2$ holonomy. Being in principle equivalent we will  concentrate in the case of the MSSM chiral fields residing in the bulk of $D7$-branes or at the intersection of $D7$-branes (or F-theory 7-branes) since at present little is known about the effective action of M-theory compactifications in manifolds with $G_2$ holonomy (see \cite{acharya} for work in this direction).\\

The unification of couplings is one of the most solid arguments in favour of the MSSM \cite{gcu}. We will thus  also  impose that gauge couplings of the MSSM unify. A simple way in which this is achieved is having at some level some  $SU(5)$, $SO(10)$ or $E_6$ symmetry relating the couplings. This unification structure may be obtained most naturally in F-theory \footnote{It is also naturally obtained in  heterotic string compactifications. However in the perturbative heterotic case modulus dominated SUSY breaking does not give rise to SUSY breaking soft terms to leading order, due to the classical no-scale structure.}. It requires that the 7-branes  associated to the SM gauge group all reside in the same stack. Note that this does not imply the existence of an explicit GUT structure since the symmetry may be directly broken to the SM at the string scale. The value of the gauge coupling is related to the (inverse) size of the 4-volume  $V_7$ wrapped by the 7 branes. There is a Kahler modulus whose real part $t$ is proportional to this volume $V_7$. This is the local modulus which will be relevant for the generation of the MSSM SUSY-breaking soft terms. In particular we will assume that a non-vanishing VEV for the auxiliary field of  this {\it local}  Kahler modulus $t$ exists giving rise to gaugino masses and other SUSY-breaking soft terms.\\

The final assumption we will make, requires the existence of at least one Yukawa coupling of order $g$ (the MSSM gauge coupling constant), the one giving mass to the top quark\footnote{Masses for the rest of quarks and leptons, being much smaller, could have their origin alternatively in non-renormalizable Yukawa couplings and/or non-perturbative (e.g. string instanton) effects. That is extremely unlikely in the case of the top-quark.}. That means that there should exist a trilinear superpotential coupling a Higgs field to the left and right-handed top quarks through cubic Yukawa couplings. As it was shown in \Sec{yukweightand} such a cubic Yukawa couplings of order one may only be present for couplings of type (A-A-$\phi$) (I-I-A) or (I-I-I) in the context of Type IIB orientifold models with chiral fields living in $D7$-branes (as well as for more general F-theory configurations). It was also shown that these chiral fields come in three classes depending on their geometric origin:  A-fields (from the vector multiplets A in the $D=8$ worldvolume action of the 7-brane and fermionic partners), $\phi$-fields (from  $D=8$ scalar  $\phi $ multiplets and fermionic partners) and I-fields (fields from the intersection of different 7-branes). The existence of this large top Yukawa coupling allowed us to use the arguments of \Sec{yukweightand} to compute the corresponding modular weights $\xi_a$.

\section{Supersymmetry breaking and soft terms}\label{sectST}

Following \Sec{subsection kahler} we will assume here that the SM resides at stacks of 7-branes wrapping a 4-cycle $S$ (of size controlled by a Kahler modulus $t$) in a 6-manifold whose overall volume is controlled by a large modulus $t_b\gg t$. In the F-theory context these moduli $t,\,t_b$ would correspond to the size of $S$ and $B_3$, as it was developed in \Sec{ftheory}. As in \Sec{subsection kahler}, we can  model out this structure with a Kahler potential of the form \cite{conlon1, conlon2, ccq}
\beq
G\ =\ -2\log(t_b^{3/2} \ -\ t^{3/2}) \ +\ \log|W|^2\, ,
\eeq
with $t=T+T^*$ being the relevant local modulus. The gauge kinetic function and Kahler metrics of a MSSM matter field $C_\alpha$ will be given to leading order by\footnote{We will ignore here the dependence of the Kahler potential on the axidilaton and complex structure fields since they play no role in the computation of the soft terms under consideration, at least to leading order in $\alpha '$.}
\beq
f_a\ =\ T \ \ \ ;\ \ \  K_\alpha \ =\ \frac {t^{(1-\xi_\alpha )}
}{t_b}\, . 
\eeq
It is important to emphasize that what is relevant here is the no-scale structure of the moduli Kahler potential and that analogous results would  be obtained in a more general CY or F-theory compactification with more  Kahler moduli as long as we assume that our local modulus $t$ is smaller than the overall modulus $t_b$ so that an expansion on $t/t_b$ is consistent. As emphasized in \cite{conlon1, conlon2, ccq} the soft terms obtained for the local MSSM 7-branes will only depend on the local modulus $t$ and the value of its auxiliary field $F_t$ and will not directly depend on the large moduli. Note that, being a Kahler modulus, assuming $F_t\not=0$ corresponds to the assumption of the presence of non-vanishing SUSY-violating antisymmetric ISD fluxes \cite{gkp} in the region in compact space where the SM 7-branes reside.\\

Applying \Eq{F}, \Eq{mmmatrix}, \Eq{mmmatrix2} and \Eq{mmmatrix3} with the Kahler metrics and gauge kinetic functions implemented in last sections, and keeping in mind that in Type IIB orientifolds the holomorphic perturbative superpotential is independent of the Kahler moduli so that the derivatives of ${\widehat Y}^{(0)}$ in the expression for $A$ vanish, we obtain general  soft terms (for $t_b\gg t$ ) as follows:
\beqa
M\ & = & \frac {F_{t}}{t}\, , \\
m_\alpha ^2\ & = &\ (1-\xi_\alpha ) |M|^2
 \ \ , \ \  \alpha =Q,U,D,L,E,H_u,H_d \, ,\ \\ \nonumber
A_U\  & = & \ -M(3-\xi_{H_u}-\xi_Q-\xi_U)\, , \\ \nonumber
A_D\  & = & \ -M(3-\xi_{H_d}-\xi_Q-\xi_D)\, , \\ \nonumber
A_L\  & = & \ -M(3-\xi_{H_d}-\xi_L-\xi_E)\, , \\ \nonumber
B \   & = & \ -M(2-\xi_{H_u}-\xi_{H_d}) \ .
\label{cojosoftgen2}
\eeqa
Note that the dependence on the SUSY-breaking from the large modulus $t_b$ disappears to leading order and the size of soft terms is rather controlled by the local  modulus $t$. In particular, gaugino masses corresponding to the SM gauge groups depend only on the modulus of the local 4-cycles they wrap, rather than the large volume modulus $t_b$. These gaugino masses set the scale of the SUSY-breaking soft terms. As in \cite{conlon1, conlon2, ccq} here the gravitino mass will be given approximately by the auxiliary field of large moduli, $m_{3/2}\simeq -F_{t_b}/t_b$, with corrections suppressed by the large volume $t_b$. One can check that all  the  bosonic soft terms above may be understood as coming from the positive definite scalar potential 
\beq
\ V_{SB}\ =\ \sum_\alpha  \ (1-\xi_{\alpha}) \vert \partial_\alpha  W \ -\ M^*C_{\alpha }^*\vert ^2
\ +\ \sum_\alpha  \xi_\alpha  \vert \partial_\alpha  W \vert ^2\, .
\label{complimas}
\eeq
The positive definite structure may be seen as a consequence of the 'no-scale' structure of modulus domination and is expected to apply also in more general  F-theory compactifications.\\

Within the philosophy of gauge coupling unification we will assume unified modular weights:
\beq
\xi_f \ =\ \xi_Q =\xi_U =  \xi_D =\xi_L =\xi_E\, .
\eeq
This is also reasonable within e.g. an F-theory approach with an underlying GUT-like symmetry like $SO(10)$ in which one expects all fermions to have the same modular weight. One also expects flavour independent modular weights, since different generations in these magnetized models come from different Landau levels with identical couplings to leading order in $\alpha'$. So we will assume a universal modular weight $\xi_f$ for all quarks and leptons. Concerning the Higgs multiplets we have seen that they can have $\xi_H=0,1,1/2$ and hence there is no reason why they should have the same modular weight as chiral fermions. We will however assume that  both Higgses have the same modular weight $\xi_H=\xi_{H_u}=\xi_{H_d}$, as would also be expected in models with an underlying left-right symmetry.\\

Under these conditions the summary of the soft terms obtained for the three possibilities for brane distributions with consistent Yukawa couplings, (A-A-$\phi$) , (I-I-A) and (I-I-I) are shown in \Tab{opciones reales}.
\begin{table}[htb] \footnotesize
\renewcommand{\arraystretch}{1.50}
\begin{center}
\begin{tabular}{|c|c||c|c|c|c|c|c|}
\hline  $(\xi_L,\xi_R,\xi_H)$   &  Coupling &
   M  &  $m_L^2$
 &  $m_R^2$    &  $m_H^2$
 &   A     &  $B$     \\
\hline\hline
 $(1,1,0)$ &   (A-A-$\phi$) &
$M$   &   0  &    0     &   $ {|M|^2}$   &  $-M$  &  $-2M$
   \\
\hline
 $(1/2,1/2,1)$ &   (I-I-A)  &
$M$   &   $\frac {|M|^2}{2}$    &  $\frac {|M|^2}{2}$   
  &   0    &  $-M$  &  0    \\
\hline
$(1/2,1/2,1/2)$ &   (I-I-I)  &
$M$   &   $\frac {|M|^2}{2} $  
& $\frac {|M|^2}{2} $ &  $\frac {|M|^2}{2}$    &  -3/2M    &  $-M$      
\\
\hline \end{tabular}
\end{center} \caption{\small Modulus dominated soft terms for choices of modular weights $\xi_\alpha$  which are consistent with the existence of a large trilinear Yukawa coupling in $7$-brane systems. }
\label{opciones reales}
\end{table}

Note that in the scenarios with couplings (A-A-$\phi$) and (I-I-A) it is natural to assume that the Higgs field is identified with fields of type $\phi$ and $A$ respectively and these are the cases displayed in the table. Concerning the $B$ parameter it is obtained assuming an explicit $\mu$-term.\\

We will study the phenomenological viability of these three options in the next sections. Since the cases with couplings (I-I-A) and (I-I-I) only differ in the origin of the Higgs fields, and hence in the values of their modular weights, they will give rise to a similar phenomenology. For this reason, we will analyse the general case in which the Higgs modular weight is a free parameter, $\xi_H$, and regard examples (I-I-A) and (I-I-I) as the limiting cases with $\xi_H=1$ and $\xi_H=1/2$, respectively. Notice that this can be understood as if the physical MSSM Higgs was a linear combination of two fields with the correct quantum numbers, one of them living in the intersection of two $D7$-branes and the other one in the bulk of one of them. On the other hand, the case with couplings (A-A-$\phi$) is unrelated to the previous ones and will be studied separately.

\subsection{Corrections to soft terms from magnetic fluxes}
The Kahler metrics and gauge kinetic functions used  in the computations above correspond to the leading behaviour in $\alpha '$. It is interesting to estimate what could be the effect of subleading terms  coming from possible magnetic fluxes in the 7-branes, as discussed in \Sec{sublmageff}.
We know that the presence of such fluxes is required to get chirality. To do that we can use the results for the Kahler metrics given in the diluted flux approximation in \Eq{flujillas}. We find for the soft terms the results \cite{aci}:
\beqa
M \ &=&\ \frac{F_t}{t+as}\, ,\\
m_\alpha ^2\ & = & \  \frac {|F_t|^2}{t^2}  (1-\xi_\alpha)\ 
\frac { (1+ \frac {c_{\alpha}\xi_\alpha }{t^{(1-\xi_\alpha )}})}
{(1+ \frac {c_\alpha }{t^{(1-\xi_\alpha )}})^2}\, ,
\\
A_{\alpha \beta \gamma}\ & = & \ -\ \frac {F_t}{t} \ \sum_{i=\alpha, \beta, \gamma}
\ (1-\xi_i)(1-\frac {c_i}{t^{(1-\xi_i)}})\, ,\\
B\ &=&\ -\frac {F_t}{t} \sum_{i=H_u, H_d} \ (1-\xi_i)(1- \frac {c_i}{t^{1-\xi_i}} ) \, .
\eeqa
Note that for matter fields coming from a $D=8$ vector multiplet (modular weight $\xi=1$) the scalar terms are still zero and get no flux correction. In the computation of the $B$-term an explicit $\mu$ term independent on $t,t_b$ has been assumed. We will make use of these formulae to try to estimate the effect of fluxes on the obtained low energy physics below.

\section{Electroweak symmetry breaking}\label{ewsb old}
We are now ready to extract the low energy implications of the MSSM soft terms listed in \Tab{opciones reales}. We will use those values as boundary conditions at the string scale which we will identify with the standard GUT scale at which the MSSM gauge couplings unify. We will solve numerically the  Renormalization Group Equations (RGEs) and calculate the low energy SUSY spectrum. We will also impose standard radiative electroweak symmetry breaking \cite{ir2}.

\subsection{Radiative EW symmetry breaking and experimental constraints}\label{ewsb old method}
The minimization of the loop-corrected Higgs potential leaves the following two conditions, which are imposed at the SUSY scale, 
\begin{eqnarray}
  \mu^2 & = & 
  \frac{ - m_{H_u}^2\tan^2\beta + m_{H_d}^2}{\tan^2\beta-1} 
  - \frac{1}{2} M_Z^2 ,
  \label{muterm}\\
  \mu B & = &  \frac{1}{2} \sin 2 \beta \, (m_{H_d}^2 +m_{H_u}^2 + 2
  \mu^2)\,, 
  \label{Bterm}
\end{eqnarray}
where $\tan\beta\equiv\langle H_u \rangle/\langle H_d \rangle$ is the ratio of the Higgs vacuum expectation values and $m_{H_{u,d}}^2$ correspond to the Higgs mass parameters. It should be noted that our choice for the sign of the $\mu$ parameter, consistent with our convention for the superpotential (\ref{super}), is opposite to the usual one. \\

A usual procedure consists then in fixing the value of $\tgb$ and use the experimental result for $M_Z$ to determine the modulus of the $\mu$ parameter via eq.(\ref{muterm}). The $B$ parameter is then obtained by solving \Eq{Bterm}. In this approach, once the modular weights which describe the specific model are given, the only free parameters left are the common gaugino mass, $M$, the value of $\tan\beta$ and the sign of the $\mu$ parameter (not fixed by condition \Eq{muterm}).\\

On the other hand, given that the value of the bilinear soft term, $B(M_{GUT})$, is also a prediction in these constructions (dependent on the source of the $\mu$ term) a more complete approach consists in imposing it as a boundary condition for the RGEs at the string scale. Conditions (\ref{muterm}) and (\ref{Bterm}) can then be used to determine both $\tgb$ and $\mu$ as a function of the only free parameter, $M$. Note that this is a extremely constrained situation and it is not at all obvious a priori that solutions passing all experimental constraints exist.\\

It is not possible to derive an analytical solution for $\tan\beta$ from Eqs.(\ref{muterm}) and (\ref{Bterm}), since the Higgs mass terms depend on $\tan\beta$ in a non-linear way. Furthermore, $\tan\beta$ is needed in order to adjust the Yukawa couplings at the GUT scale so that they agree with data. Thus, to find a solution an iterative procedure has to be used where the RGEs are numerically solved for a first guess of $\tan\beta$ using the soft parameters as boundary conditions at the GUT scale. The resulting $B$ at the SUSY scale is then compared with the solution of \Eq{Bterm}. In subsequent iterations, the value of $\tan\beta$ is varied, looking for convergence of the resulting $B(M_{SUSY})$. It is not always possible to find a solution with consistent REWSB, and this excludes large areas of the parameter space. In our analysis we have implemented such iterative process through a modification of the {\tt SPheno2.2.3} code \cite{spheno}.\\

Once the supersymmetric spectrum is calculated, compatibility with various experimental bounds has to be imposed\footnote{Some of the bounds and constraints imposed in the calculation of this section haven been updated in recently. The new values of $\bsg$ and $(\bmumu)$ are presented in \Sec{smfp} and used for the study in detail of the (I-I-I) configuration phenomenonlogy. Nevertheless, concerning the conclusions of the analysis of this section the old values are sufficient.}. We have taken into account the constraints obtained by LEP on the masses of supersymmetric particles, as well as on the lightest Higgs boson \cite{pdg}. We will analyze the impact of recent LHC Higgs limits in \Sec{hsmdcmssm}. Also, the experimental limits on the contributions to low-energy observables have been included in our analysis. More specifically, we impose the experimental bound on the branching ratio of the rare $\bsg$ decay, $2.85\times10^{-4}\le\,{\rm BR}(\bsg)\le 4.25\times10^{-4}$, obtained from the experimental world average reported by the Heavy Flavour Averaging Group \cite{bsgHFAG07}, and the theoretical calculation in the Standard Model \cite{bsg-misiak}, with errors combined in quadrature.  
We also take into account the upper constraint on the $(\bmumu)$ branching ratio obtained by CDF, BR$(\bmumu)<5.8\times10^{-8}$ at $95\%$ c.l. \cite{bmumuCDF07}. As we said, more recent limits are included in the analysis in \Sec{hsmdcmssm}.\\ 

Regarding the muon anomalous magnetic moment, a constraint on the supersymmetric contribution to this observable, $\asusy$, can be extracted by comparing the experimental result \cite{g-2}, with the most recent theoretical evaluations of the Standard Model contributions \cite{g-2_SM,newg2,kino}. When $e^+e^-$ data are used the experimental excess in $a_\mu\equiv(g_\mu-2)/2$ would constrain a possible supersymmetric contribution to be $\asusy=(27.6\,\pm\,8)\times10^{-10}$, where theoretical and experimental errors have been combined in quadrature. However, when tau data are used, a smaller discrepancy with the experimental measurement is found. Due to this reason, in our analysis we will not impose this constraint, but only indicate the regions compatible with it at the $2\sigma$ level, this is $11.6\times10^{-10}\le\asusy\le43.6\times10^{-10}$.\\

Assuming R-parity conservation, and hence the stability of the LSP, we also investigate the possibility of obtaining viable neutralino dark matter. This is, in the regions of the parameter space where the neutralino is the LSP we compute its relic density by means of the program {\tt micrOMEGAs} \cite{micromegas}, and check compatibility with the data obtained by the WMAP satellite \cite{wmap5yr}, which constrain the amount of cold dark matter to be $0.1037\le\Omega h^2\le 0.1161$. The value of the mass of the top quark is particularly relevant. In our computation in this section, we have used the central value corresponding to the measurement by CDF \cite{topCDF08}, $m_t=172 \pm 1.4\,{\rm GeV}$. We will briefly comment on the effect that deviations from this quantity may have on REWSB.\\

The presence in SUSY theories of scalar fields which carry electric and colour charges can lead to the occurrence of minima of the Higgs potential where charge and/or colour symmetries are broken when these scalars take non-vanishing VEVs. If these minima are deeper than the physical (Fermi) vacuum, the latter would be unstable, i.e. unbounded from below (UFB). The different directions in the field space that can lead to this situation were analysed and classified in \cite{clm1}. It was found there that the most dangerous direction corresponds to the one labelled as UFB-3, where the stau and sneutrino take non-vanishing VEVs, since the tree-level scalar potential could even become unbounded from below. These UFB constraints were found to impose stringent constraints on the parameter space of general supergravity theories \cite{cggm03-1}, as well as in different superstring and M-theory scenarios \cite{ibarra,ckm08}. In our study we will comment on the constraints which are derived when the absence of such charge and/or colour-breaking minima is imposed\footnote{Strictly speaking, the existence of a global charge and/or colour-breaking vacuum cannot be excluded if the lifetime of the metastable physical minimum is longer than the age of the Universe, as is usually the case.}.\\

In order to understand the effect of all these experimental and astrophysical constraints we have performed a scan over the gaugino mass parameter, $M$, and $\tan\beta$ for the three different consistent models that were specified in \Tab{opciones reales}. We will discuss first the results obtained without imposing the predicted boundary condition for $B$ and analyse the effect of this constraint in the following subsection.

\subsection{The intersecting 7-brane (I-I-I)-(I-I-A) configurations}
\begin{figure}[t!]
  \includegraphics[width=8cm]{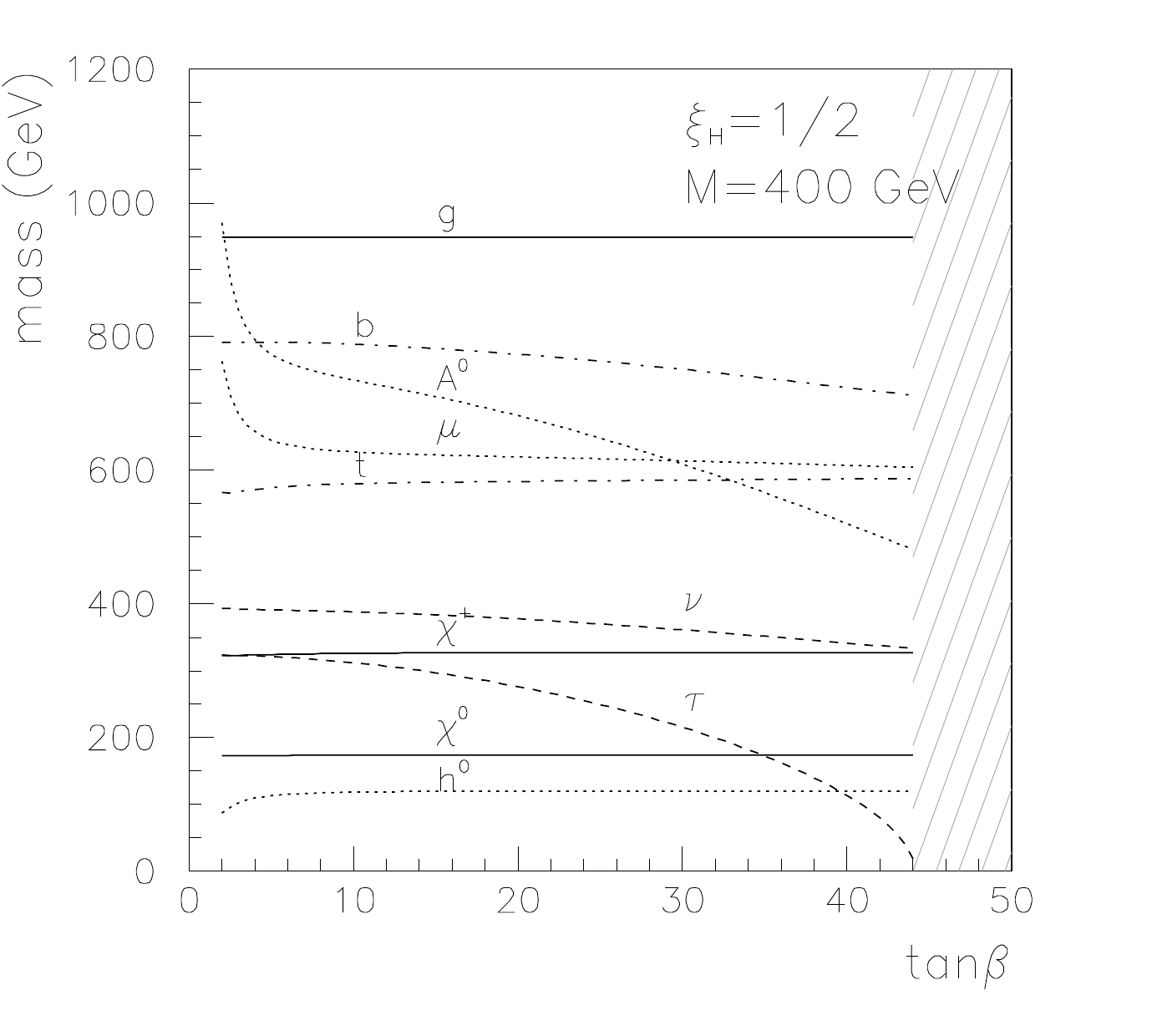}
  \includegraphics[width=8cm]{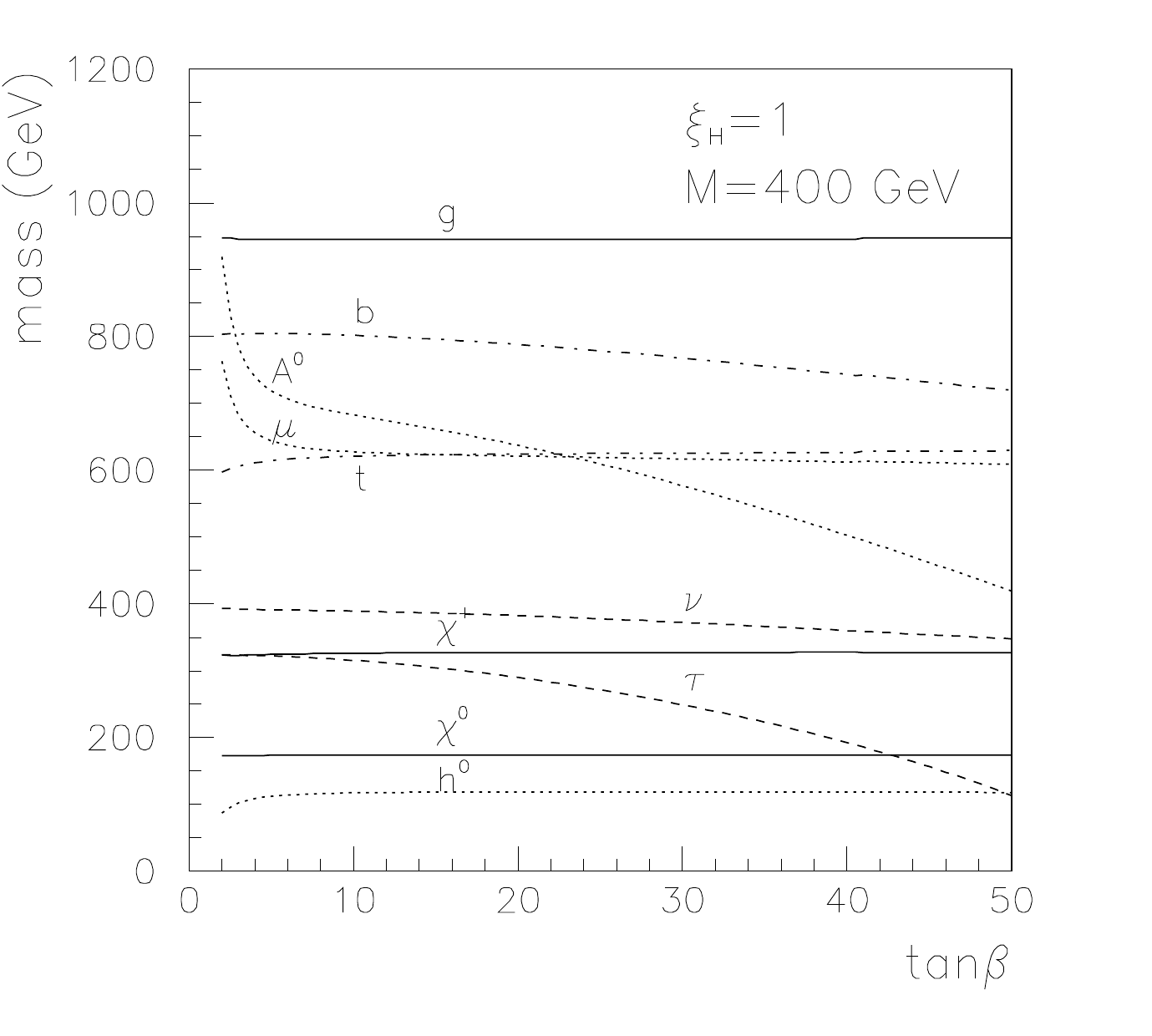}
   \vspace*{-1cm}
  \caption{
    Low-energy supersymmetric spectrum as a function of $\tan\beta$
    for $\xi_H=1/2,$ (left) and $\xi_H=1$ (right) with 
    $M=400\ {\rm GeV}$ and $\mu<0$. From bottom to top, the solid lines
    represent the masses of the lightest neutralino, the lightest
    chargino, and the gluino. Dashed lines display the masses of the
    lightest stau and lightest sneutrino. Dot-dashed lines correspond to
    the stop and sbottom masses. Finally, the lightest Higgs mass, the 
    pseudoscalar Higgs mass
    and the absolute 
    value of the $\mu$ parameter are displayed by means of
    dotted lines. The ruled area for large $\tan\beta$ is excluded by
    the occurrence of tachyons in the slepton sector. 
    \label{spectrum-iih}
  }
\end{figure}

As commented at the end of \Sec{sectST}, we will analyse these two cases together within the framework of a generic scenario in which the modular weight for the Higgs is a free parameter. In this  approach, $\xi_H$ can take any value between $\xi_H=1/2$, which corresponds to case (I-I-I), and $\xi_H=1$, as in case (I-I-A). The soft parameters for such a model can be extracted from (\ref{cojosoftgen2}) and read
\begin{eqnarray}
m_{L,E,Q,U,D} ^2\ & = &\ |M|^2/2\, ,\ \\ \nonumber
m_{H_u,H_d} ^2\ & = &\ (1-\xi_H )|M|^2\, ,\ \\ \nonumber
A_{U,D,L}\  & = & \ -M(2-\xi_{H})\, , \\ \nonumber
B \   & = & \ -2M(1-\xi_{H}) \, .
\label{soft-iih}
\end{eqnarray}
A sample of the resulting supersymmetric spectrum is represented in \Fig{spectrum-iih} as a function of the value of $\tan\beta$ for $M=400\,{\rm GeV}$ and $\mu<0$ for the two limiting cases $\xi_H=1/2$ and $\xi_H=1$. As evidenced by the plot, despite the fact that the slepton mass squared terms are always positive at the GUT scale, the running down to the EW scale can lead to the occurrence of tachyonic eigenstates for large tan$\beta$. This is typically the case of the lightest stau. The negative contribution to the RGEs governing the stau mass parameters increases with $\tan\beta$, since it is proportional to the lepton Yukawa, which varies as $1/\cos\beta$. As a consequence, the stau mass decreases towards large values of $\tan\beta$, first becoming the LSP (which in the $\xi_H=1/2$ example occurs for $\tan\beta\gsim 35$), then falling below its experimental lower bound and eventually turning tachyonic. This sets an upper constraint on the possible value of $\tan\beta$ (which obviously increases for larger values of $M$). The effect of this constraint is more important when $\xi_H$ is small, since the trilinear parameter $A_L$ is larger (more negative) and enhances the negative contribution in the RGEs of the stau mass parameters. Thus, the bound $\tan\beta\lsim45$ for $\xi=1/2$ is relaxed to $\tan\beta\lsim55$ with $\xi=1$.\\

Another interesting feature is that the resulting value for $|\mu|$, calculated from \Eq{muterm}, turns out to be relatively large, of order $1.5\,M$ (see \Fig{plotmu}). Similarly, the pseudoscalar Higgs (as well as the heavy neutral and charged Higgses) is also heavy, decreasing for large values of $\tan\beta$. When the value of $\xi_H$ increases the predicted pseudoscalar mass is slightly smaller.\\

The rest of the properties of the spectrum are less sensitive to variations in the Higgs modular weight. For small values of $\tan\beta$ the lightest neutralino is typically the LSP in this example. Since the value of the $\mu$ parameter is always large, the lightest neutralino is mostly bino-like. The universality of gaugino masses at the GUT scale and the large values of $|\mu|$ also lead to the well known low-energy relation among the masses of the lightest neutralino, the lightest chargino and the gluino, $m_{\tilde\chi^0}:m_{\tilde\chi^+}:m_{\tilde g}\approx1:2:5.5$. The squark sector is rather heavy, another consequence of the universality of the soft masses. Still squarks in the region $\sim 1$ TeV are by now almost excluded by LHC so that values $M$ larger that 500 GeV are required, see \Sec{susyspectrum}. \\

\begin{figure}[t!]
  \hspace*{-0.6cm}
  \includegraphics[width=8cm]{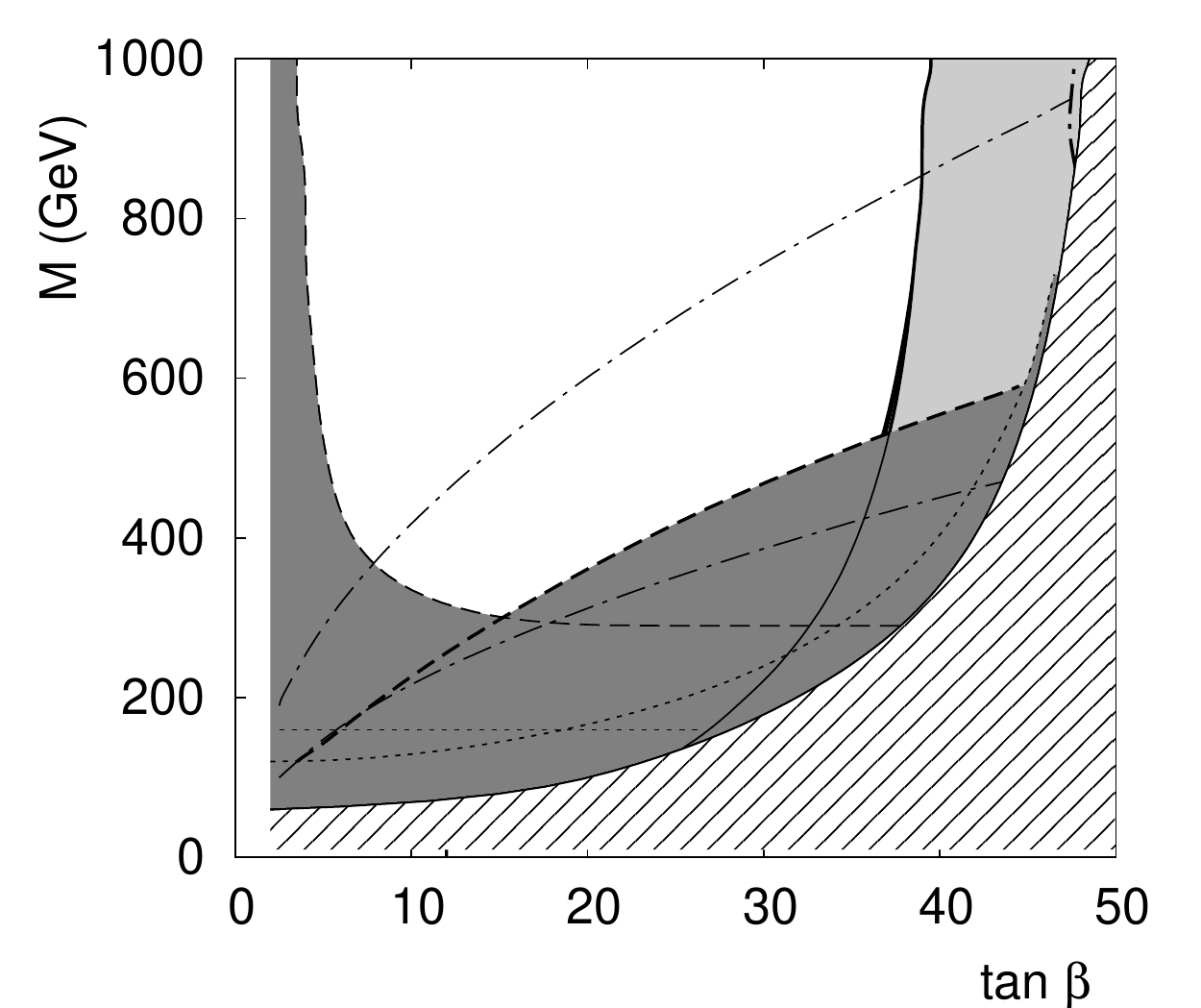}
  \includegraphics[width=8cm]{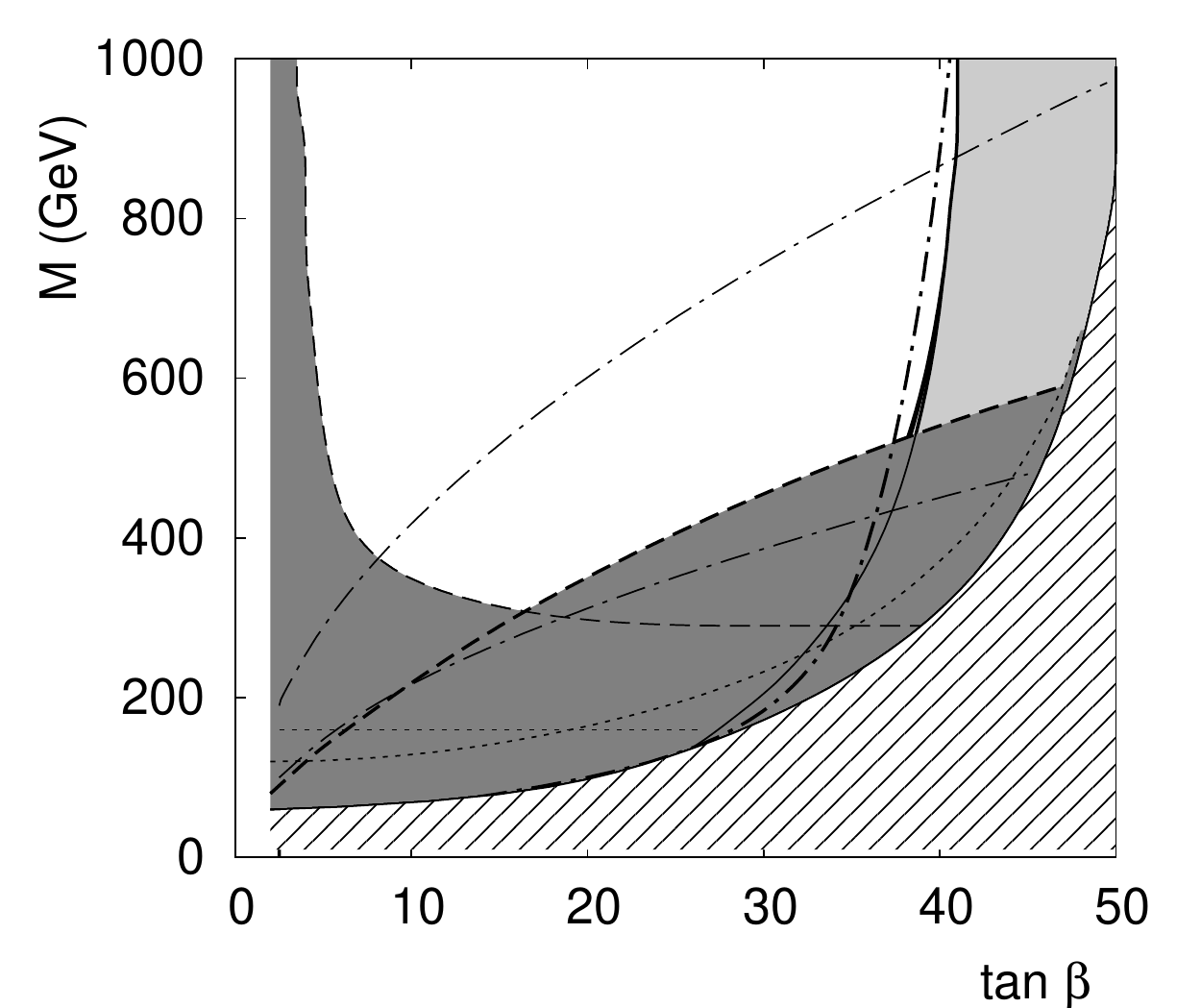}
  \hspace*{-0.6cm}
  \includegraphics[width=8cm]{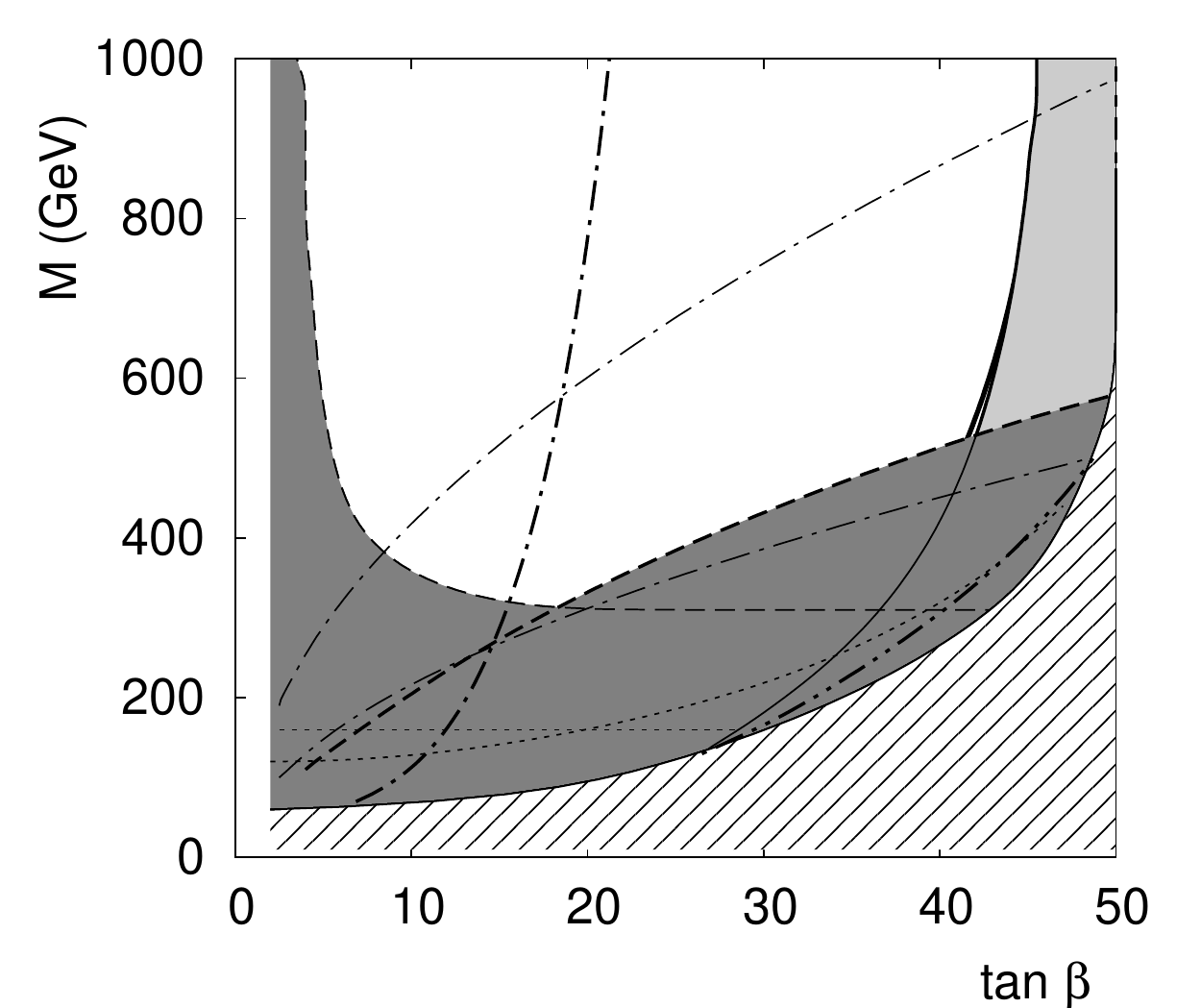}
  \includegraphics[width=8cm]{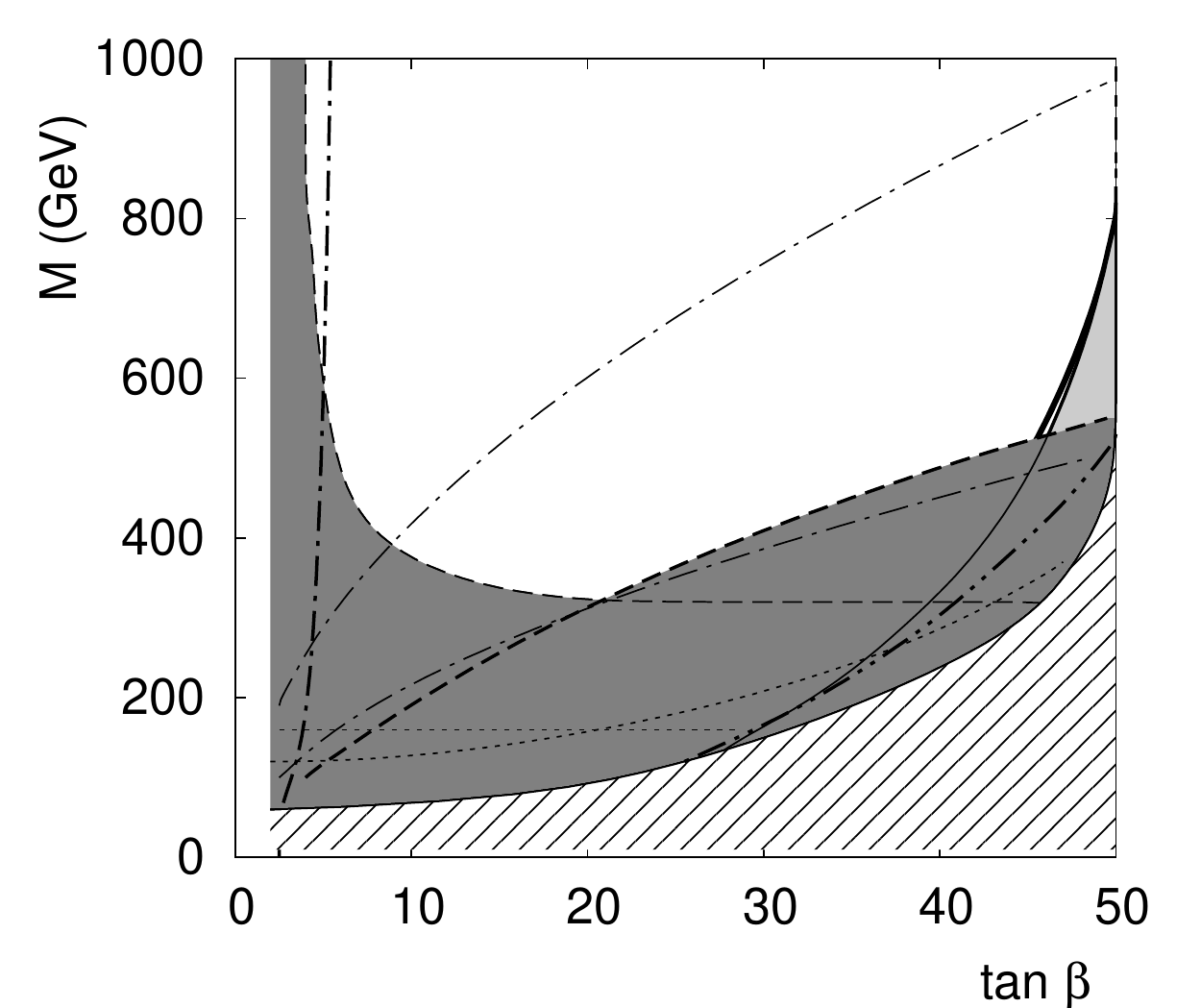}
  \vspace*{-0.7cm}
  \caption{Effect of the various experimental constraints on the $(M,\,\tgb)$ plane for cases with $\xi_H=0.5,\,0.6,\,0.8$, and $1$, from left to right and top to bottom. Dark grey regions correspond to those excluded by any experimental bound. Namely, the area below (and to the left of) the thin dashed line is ruled out by the lower constraint on the lightest Higgs mass. The region below the thin dotted lines is excluded by the lower bounds on the stau and chargino masses. The area below the thick dashed line is excluded by $b\to s\gamma$. The region below the double dot-dashed line is excluded by $\bmumu$. The thin dot-dashed lines correspond, from top to bottom, to the lower and upper constraint on $\asusy$. The area contained within solid lines corresponds to the region in which the stau is the LSP, and is depicted in light grey when experimental constraints are fulfilled. In the remaining white area the neutralino is the LSP. The thin black area, in the vicinity of the region with stau LSP, corresponds to the region where the neutralino relic density is in agreement with the WMAP bound.  
    \label{mtgb-iih}
  }
\end{figure}

For a better understanding of the effect of the various experimental constraints a full scan on the two free parameters $M$, and $\tan\beta$ (for $\mu<0$) is presented in \Fig{mtgb-iih} for four examples of Higgs modular weight\footnote{In these figures the constraint for the B-parameter has not yet been imposed. We will see below that the dark matter constrained combined with the prediction for $B$ singles out the case with $\xi_H\simeq 0.5-06$.}, $\xi_H=0.5,\,0.6,\,0.8$ and $1$. Some of the features of the supersymmetric spectrum we have just described are also clearly displayed. For example, the ruled area for large $\tan\beta$ and small $M$ corresponds to the area excluded due to tachyons in the stau eigenstates. This area becomes smaller as the modular weight for the Higgs increases as a consequence of the increase in the stau mass.   At the same time, the region with stau LSP (which is represented by light grey) is also shifted towards larger $\tan\beta$, thereby enlarging the area in which the neutralino is the LSP.\\

The region below the thin dashed line is excluded by the LEP constraint on the Higgs mass, and corresponds to $\tan\beta\lsim5$ and $M\lsim300\ {\rm GeV}$. This constitutes the stronger non-LHC lower limit on $M$ for small values of $\tan\beta$. For $\tan\beta\gsim15$, however, the experimental bound on BR($\bsg$) sets a more constraining lower limit on $M$, which increases with $\tan\beta$ and reaches $M\gsim 500\,{\rm GeV}$. Regarding the supersymmetric contribution to the muon anomalous magnetic moment, the region of the parameter space which is favoured by the experimental constraint corresponds to the area between the thin dot-dashed lines. The lower(upper) bound on $\asusy$ sets an upper(lower) limit on $M$. Since $\asusy$ increases for large $\tan\beta$ (through an enhancement of the contribution mediated by charginos and sneutrinos), both the upper and lower limits on $M$ also increase. As we already indicated, in our analysis this constraint is not imposed.\\

As a result, there are vast regions (white area) of the parameter space compatible with all the experimental constraints and in which the lightest neutralino is the LSP. In order to determine its viability as a dark matter candidate, its relic density has to be computed and compared with existing bounds on the abundance of cold dark matter.\\

As mentioned above, the neutralino is mostly bino in these constructions, and for this reason its relic density easily exceeds the recent WMAP constraint. The correct neutralino abundance is only found in those regions of the parameter space where the neutralino mass is very close to the stau mass, since then a coannihilation effect \cite{gs90} takes place in the early Universe which reduces very effectively the neutralino abundance\footnote{Another possible mechanism that could help in reducing the neutralino relic density is their resonant annihilation through an s-channel mediated by a pseudoscalar Higgs. However, that is only possible when $2m_{\tilde\chi^0}\approx m_{A^0}$. As we saw in \Fig{spectrum-iih} the pseudoscalar is too heavy in these constructions for such a resonance to take place.}. After imposing the constraint on the neutralino relic density, the only regions of the parameter space which are left correspond to very narrow bands in the vicinity of the area with stau LSP. Interestingly, these favour a very narrow range of values for $\tan\beta$, which is always large. Also, as $\xi_H$ increases, the allowed region is shifted towards larger $\tan\beta$. Thus, while $35\lsim\tan\beta\lsim40$ for $\xi_H=1/2$ (case (I-I-I)), $45\lsim\tan\beta\lsim55$ is needed for case (I-I-A) with $\xi_H=1$.\\

So far we have not commented on the effect of UFB constraints. Most of the parameter space turns out to be disfavoured on these grounds. The reason for this is the low value of the slepton masses, and more specifically, of the stau mass. Indeed, the lighter the stau, the more negative the scalar potential along the UFB-3 direction becomes, thus easily leading to a minimum deeper than the realistic (physical) vacuum. In particular, the whole $(M,\,\tan\beta)$ plane with $M<1000$~GeV is disfavoured for these reason in all the cases represented in \Fig{mtgb-iih}. On the other hand, as we will see later, a Higgs mass $\sim 125$ GeV favors large values of $M \sim 1.4$ TeV.

\subsection{B parameter constraint on the (I-I-I)-(I-I-A) configuration}  
So far we have not imposed the boundary condition on the value of the $B$ parameter at the GUT scale, which was also predicted in terms of the modular weights and related to the rest of the soft parameters by \Eq{soft-iih}. As we already explained, in this approach the REWSB condition (\ref{Bterm}) must be used in order to determine the value of $\tan\beta$ by means of an iterative procedure. This leaves only one free parameter, $M$, to describe all the soft terms and, if a solution for the REWSB equations is found, a value of $\tan\beta$ is predicted for each $M$.\\ 

\begin{figure}[t!]
    \hspace*{-0.5cm}
    \includegraphics[width=8cm]{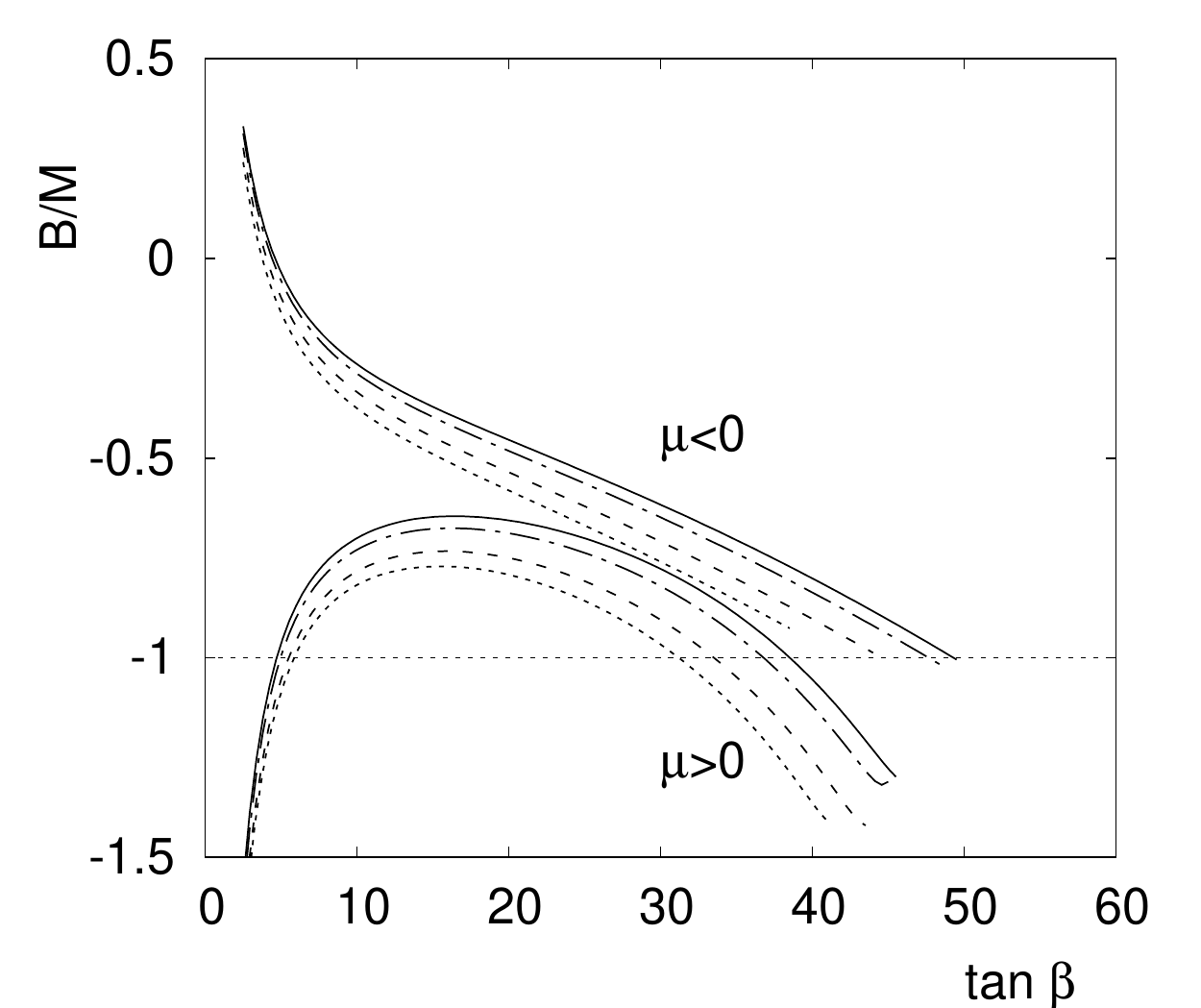}
    \includegraphics[width=8cm]{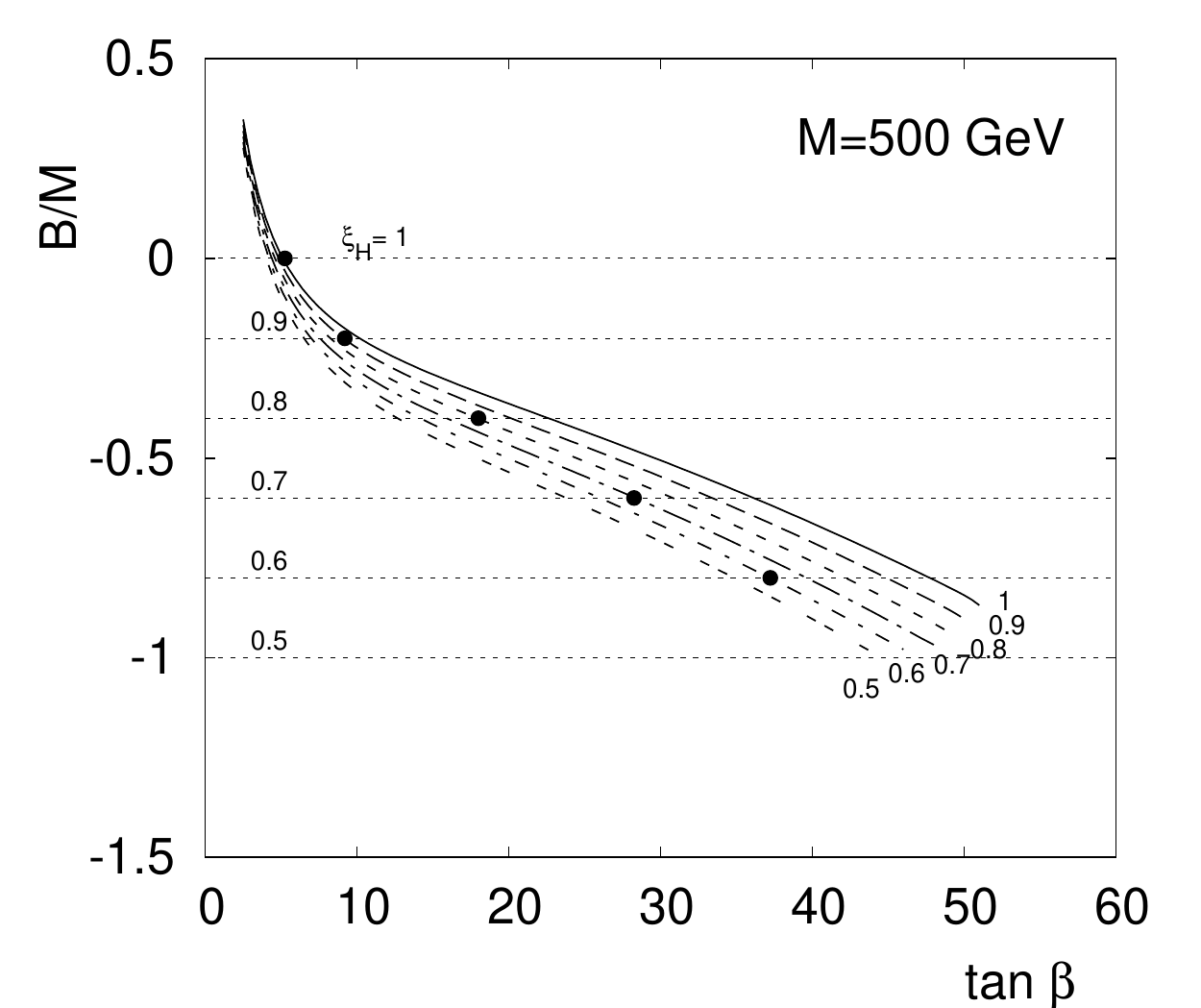}
    \caption{Left) Resulting $B(M_{GUT})/M$ as a function of
    $\tan\beta$ for the case with $\xi_H=0.5$. 
    The dotted, dashed, dot-dashed, and solid lines
    correspond to $M=300,\,500,\,1000$, and $1500\,{\rm
    GeV}$, respectively, for both
    signs of the $\mu$ parameter. 
    The boundary condition $B=-M$ 
    is indicated with a horizontal dotted
    line.  
    Right) The same, but for 
    the cases with $\xi_H=0.5,\,0.6,\,0.7,\,0.8,\,0.9$, and $1$, from
    bottom to top, with $\mu<0$ and $M=1000\,{\rm GeV}$.
    The corresponding boundary conditions for $B$ are represented with
    horizontal dotted lines, and 
    the solid circles indicate the values of $\tan\beta$ consistent
    with these. 
  }
  \label{btgb-iih}
\end{figure}

In \Fig{btgb-iih} (left panel) we display the value of $B/M$ at the GUT scale as a function of $\tan\beta$ for several values of $M$ and for the two possible choices of the sign of the $\mu$ parameter. 
The solutions of the REWSB equations correspond to the values of $\tan\beta$ where the different lines intersect the dotted line, which represents the boundary condition $B/M=-1$ in the $\xi_H=1/2$ case. As we can see,  solutions are found for $\mu<0$ only when $\tan\beta$ is very large. In the $\mu>0$ case (disfavoured by $b\rightarrow s\gamma$ limits) solutions are found  for $\tan\beta\sim4$ and  $30\lsim\tan\beta\lsim40$.\\

When the modular weight for the Higgses increases, the boundary condition for $B$ is seriously affected. For instance, when 
$\xi_H=1$ as in case (I-I-A), one obtains $B=0$. As a consequence, the ranges of $\tan\beta$ which are solutions of the REWSB conditions change significantly. This is illustrated on the right hand-side of \Fig{btgb-iih}, where $B/M$ at the GUT scale is represented as a function of
$\tan\beta$ (with $\mu<0$ and $M=500$ GeV) for the cases with $\xi_H=0.5,\,0.6,\,0.7,\,0.8,\,0.9$, and $1$, from bottom to top, respectively. The boundary conditions corresponding to these values of the Higgs modular weights are represented by means of dotted lines, also from bottom ($\xi_H=0.5$) to top ($\xi_H=1$). The solutions for $\tan\beta$ for each choice of modular weight correspond to the intersection of the $B/M$ line with the corresponding boundary condition, and are indicated with filled circles. As we see in the figure, already slightly above $\xi_H=1/2$ (e.g. for $\xi_H=0.52$) correct EW symmetry breaking is obtained with the predicted $B$-term. This happens around tan$\beta\simeq 40$. As $\xi_H$ increases from $1/2$ to $1$ correct EW symmetry breaking is obtained at the predicted $B$ with lower and lower values of $\tan\beta$. In the (I-I-A) limit with $\xi_H=1$ solutions for REWSB are  obtained for $\tan\beta\simeq 4$.\\

The resulting values for $\tan\beta$ as a function of the common scale $M$ have been also superimposed on \Fig{mtgb-iih} by means of a thick
dot-dashed line. Consistently with what we just explained, when $\xi_H=1/2$ solutions are only found for $\tan\beta\approx45$ and $M\gsim900\,{\rm GeV}$, in the region with stau LSP. However, even with a slight increase in $\xi_H$, due to the rapid change in the boundary condition for $B$, solutions of the REWSB equations are found with smaller values of $\tan\beta$. If we want to obtain successful REWSB in a region consistent with appropriate neutralino dark matter, one is lead only to the region with $\xi_H\simeq 0.6$, quite close to the boundary conditions (I-I-I) with $\xi_f=\xi_H=1/2$. This may be seen in \Fig{mtgb-iih} (upper right) in which the thick dot-dashed line is very close to the line marking the separation between neutralino and stau LSP regions, i.e., the coannihilation region. On the other hand, in the  (I-I-A) case with $\xi_H\simeq 1$ one obtains appropriate REWSB for tan$\beta\simeq 4$, far away from the coannihilation region and hence too much dark matter is predicted. Thus  insisting in getting neutralino dark matter consistent with WMAP measurements and consistent REWSB selects the region close to the (I-I-I) boundary conditions in which all MSSM matter fields live at intersecting 7-branes.\\

It is worth mentioning here that variations in the value of the top mass slightly alter the running of the $B$ parameter and, as a consequence, lead to a small shift in the solutions for $\tan\beta$. We have checked that in the previous examples this shift is $\Delta\tan\beta\approx\pm1$ when the top mass varies from $m_t=169.2\,{\rm GeV}$ to $m_t=174.8\,{\rm GeV}$ (which corresponds to a $2\sigma$ deviation from the experimental central value). Although smaller top mass also implies a more stringent constraint from the experimental bound on the lightest Higgs mass, the rest of the constraints are not significantly affected and the coannihilation region remains basically unaltered. One can therefore understand $\Delta\tan\beta$ as a small uncertainty on the trajectories for $\tan\beta$ in \Fig{mtgb-iih} to be taken into account when demanding compatibility with the regions with viable neutralino dark matter.

\subsection{The bulk 7-brane (A-A-$\phi$) configuration}
Let us consider now the other alternative left, in which the MSSM resides at the bulk of the 7-branes. As seen in table 1, in this case  the sfermion soft masses vanish at the GUT scale. This has important implications on the resulting low-energy spectrum. Although squark masses (which receive large positive contributions in the corresponding RGEs from the gluino mass parameter) easily become large enough, slepton masses remain rather light. This is particularly problematic for the lightest stau, due to the negative contribution proportional to the Yukawa in the RGEs. For this reason the stau mass-squared becomes negative even for moderate values of $\tan\beta$. In this example, this sets an upper bound of $\tan\beta\lsim25$.\\

\begin{figure}[t!]
  \includegraphics[width=8cm]{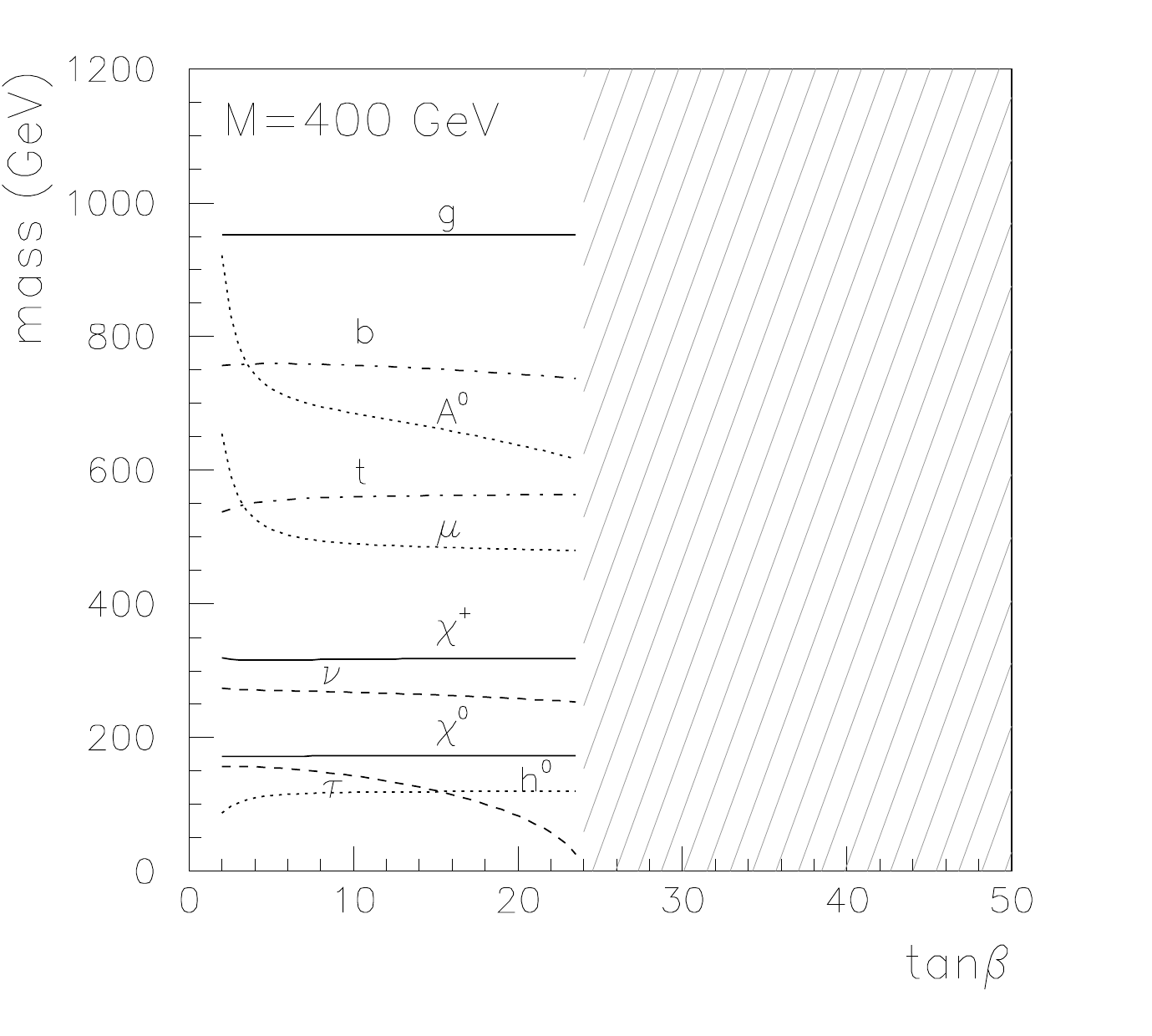}
  \hspace*{-1cm}
  \raisebox{2ex}{\includegraphics[width=8cm]{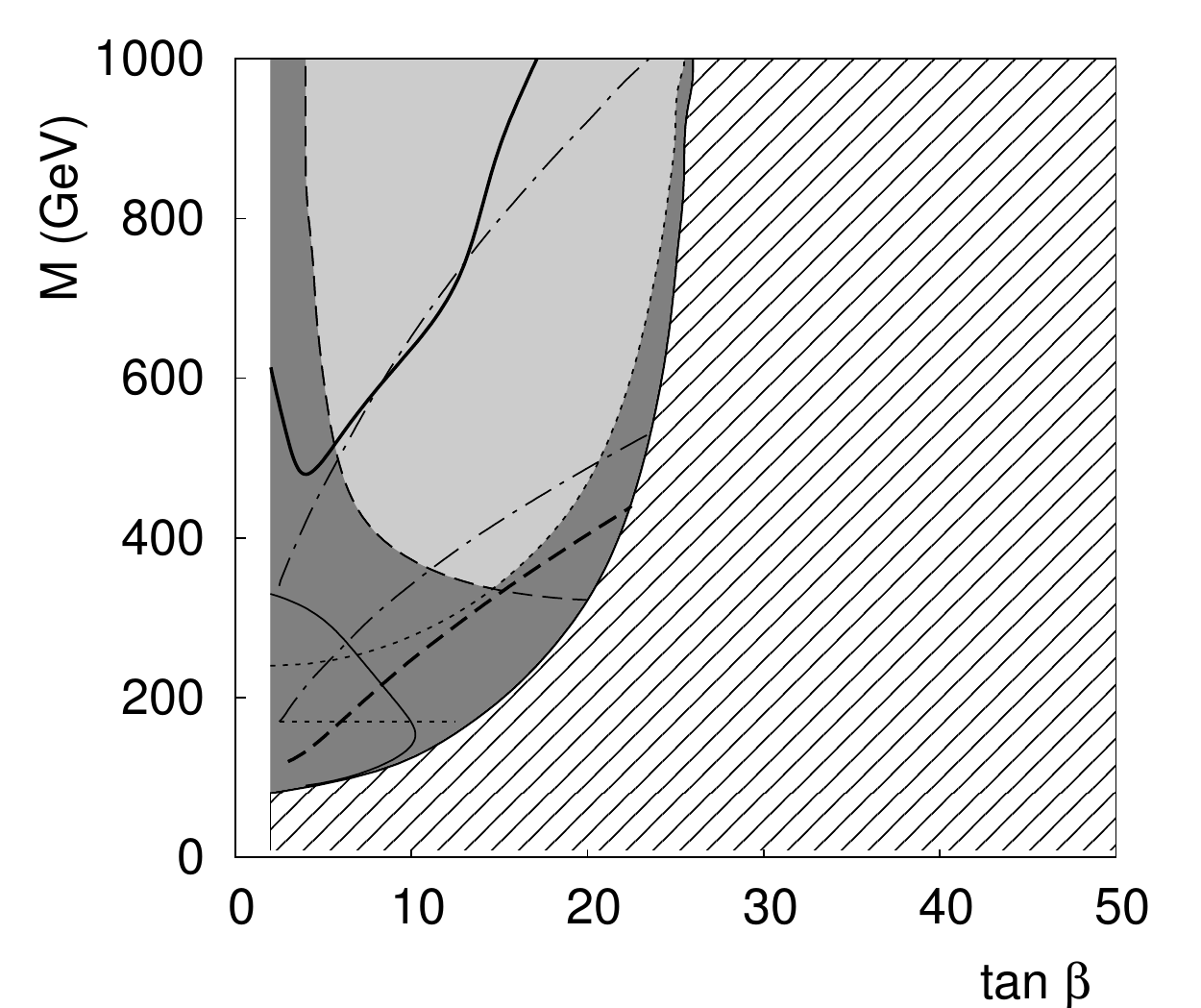}}
  \caption{Left) Low-energy supersymmetric spectrum as a function of
  $\tan\beta$ for case (A-A-$\phi$). 
    Right) Effect of the various experimental constraints on the
    $(M,\,\tgb)$ plane for case (A-A-$\phi$). 
    Colour and line conventions are the same as in
    fig.\ref{mtgb-iih}.
    In addition, the UFB constraints are fulfilled in the region above
    the thick solid line. 
    }
  \label{mtgb-aap}
\end{figure}

In the remaining allowed areas of the parameter space the stau becomes the LSP. A stable charged LSP would bind to nuclei, forming exotic isotopes on which strong experimental bounds exist. Phenomenological consistency would then require such staus to decay, a possibility which arises if R-parity was broken. There is therefore no viable supersymmetric dark matter candidate in this scenario.\\

The resulting SUSY spectrum, together with the effect of the rest of the experimental constraints on the parameter space are shown in \Fig{mtgb-aap}, clearly displaying all the above mentioned features. Interestingly, and contrary to what we observed for the intersecting $7$-brane configurations, there is a region of the parameter space which satisfies the UFB constraints, corresponding to the area above the thick solid line with $M\gsim 500$~GeV and $\tan\beta\lsim 20$. This is possible because of the increase in the Higgs mass parameters at the GUT scale (remember that now $m_{H_u,H_d}^2=M^2$), which entails a less negative contribution to the scalar potential along the UFB directions.\\

If we further impose the prediction for the B-parameter the situation is worse. Indeed in this scenario the boundary condition for $B$ at the string scale is $B=-2M$. No solutions are found for $\tan\beta$ neither for $\mu<0$, nor for $\mu>0$ satisfying this boundary condition.

\begin{figure}[t!]
  \hspace*{-0.6cm}
  \includegraphics[width=8cm]{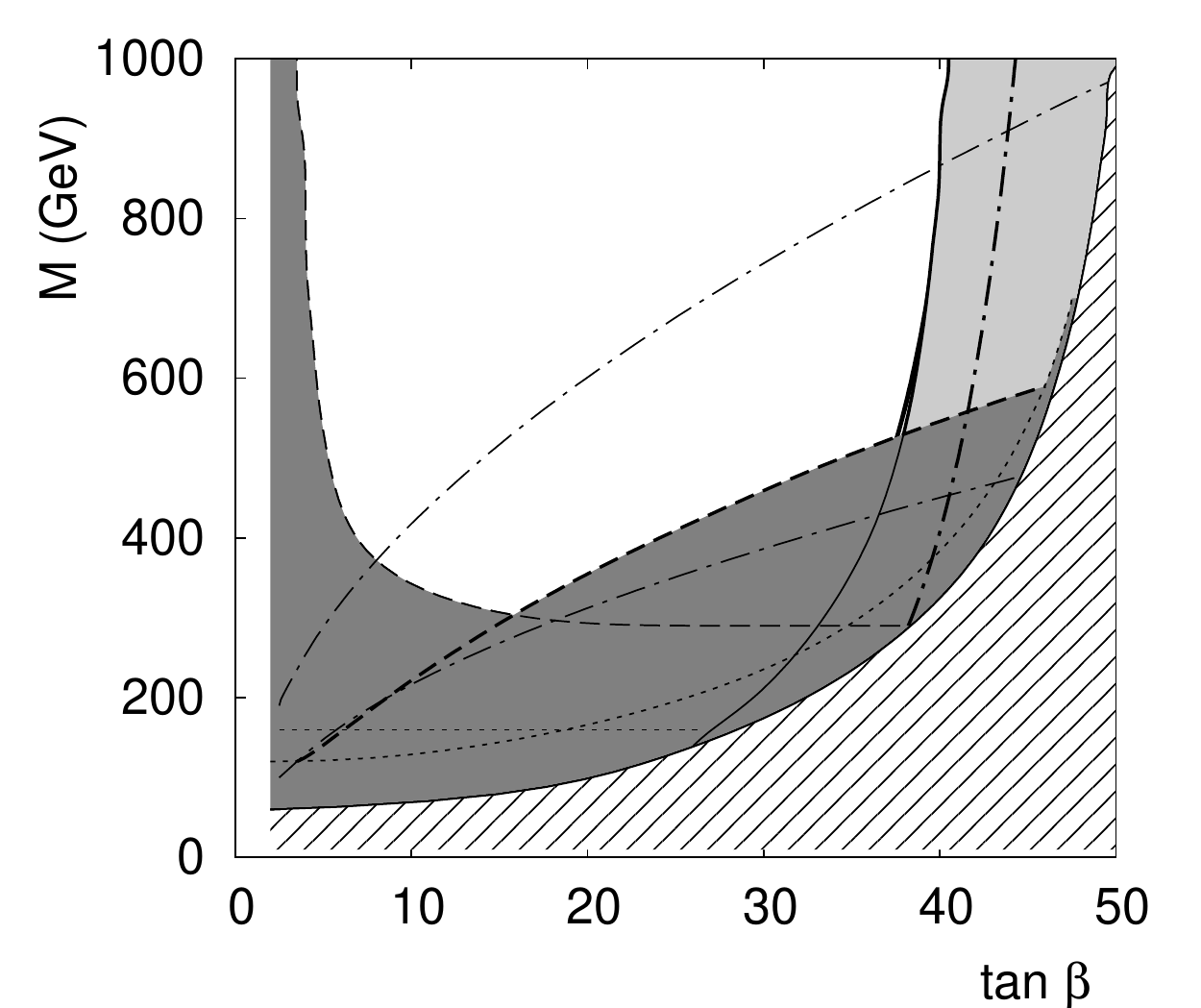}
  \includegraphics[width=8cm]{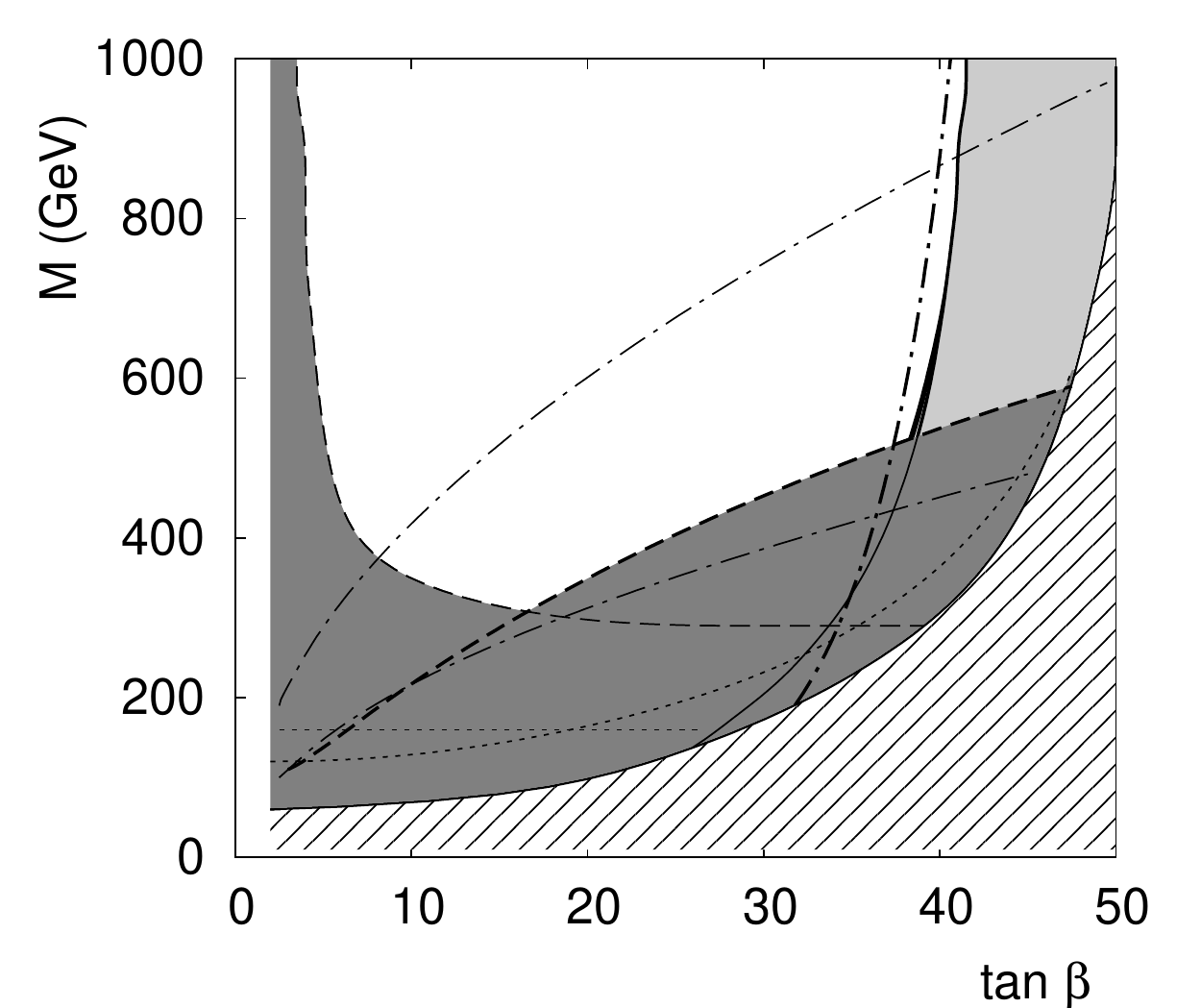}
  \caption{The same as in fig.\,\ref{mtgb-iih} but for case (I-I-I)
  where corrections coming from fluxes are included with $\rho_H=0.1$
  (left) and $\rho_H=0.2$
  (right). In both of them $\rho_f=0$.
    }
  \label{mtgb-iiiflux}
\end{figure}

\subsection{Effect of magnetic fluxes}
We have seen how the intersecting 7-brane configuration (I-I-I)-(I-I-A) with $\xi_H\simeq 0.6$  is consistent with all constraints including appropriate amount of neutralino dark matter. This 'effective modular weight' $\xi_H\simeq 0.6$ could be understood if the Higgs field is a linear combination of fields with modular weights $\xi_H=1/2$ (predominant) and $\xi_H=1$.\\

Alternatively one could think that there could be some higher order correction which could explain the small deviation from the fully intersecting configuration (I-I-I) with $\xi_H=1/2$. As we discussed, one possible source for such small corrections could be the magnetic fluxes which are anyway required for the spectrum to be chiral. Using the results in section (3.2) we can estimate what could be the structure of such corrections. In agreement with the unification hypothesis, we will assume that all sfermions have the same flux correction in their Kahler metrics proportional to some parameter $c_f$. We will then parametrize the corrections in terms of parameters defined for the different cases as:
\beq
\rho=\frac {(c_H-as)}{t} \ ;\ 
\sigma=\frac {as}{t} \ ;\ \rho_f=\frac {c_f}{t^{1/2}} \ ;\ 
\rho_H=\frac {c_H}{t^{1/2}}\, ,
\eeq
where $a$ is defined in \Eq{metricaaprox} (in our case $a_i=a$ for gauge coupling unification). The results for the soft terms are shown in \Tab{correcciones de flujos}. 
\begin{table}[htb] \footnotesize
\renewcommand{\arraystretch}{1.50}
\begin{center}
\begin{tabular}{|c||c|c|c|c|}
\hline   Coupling &
    $m_f^2$
 &   $m_H^2$
 &   A     &  $B$        \\
\hline\hline
   (A-A-$\phi$) &
    0       &   $ {|M|^2(1-2\rho)}$ &   $-M(1-\rho )$   &  $-2M(1-\rho)$  
  \\
\hline
    (I-I-A)  &
   $\frac {|M|^2}{2}(1-\frac {3}{2}\rho_f)$ 
   &    0    &  $-M(1-\rho_f)$  &  0     \\
\hline
   (I-I-I)  &
   $\frac {|M|^2}{2}(1-\frac {3}{2}\rho_f) $  &   $\frac {|M|^2}{2}(1-\frac {3}{2}\rho_H)$    
&  $-\frac {1}{2}M(3-\rho_H-2\rho_f)$    &  
$-M(1-\rho_H)$  
\\
\hline \end{tabular}
\end{center} \caption{\small  Corrections from magnetic fluxes to the 
different soft terms. The parameters $\rho,\rho_H, \rho_f$
are defined in the main text. }
\label{correcciones de flujos}
\end{table}
To get the results we have assumed that $\rho_H,\rho_f \gg \sigma$, given their different large $t$ behaviour. Looking at the table one observes as a general conclusion that the size of scalar soft terms decreases with respect to gaugino masses as fluxes are turned on. This is consistent with the known fact that as fluxes increase 7-branes localize and get boundary conditions more and more similar to 3-branes, whose  matter fields are known do not get bosonic soft terms.\\
\begin{figure}[t!]
  \hspace*{-0.6cm}
  \includegraphics[width=8cm]{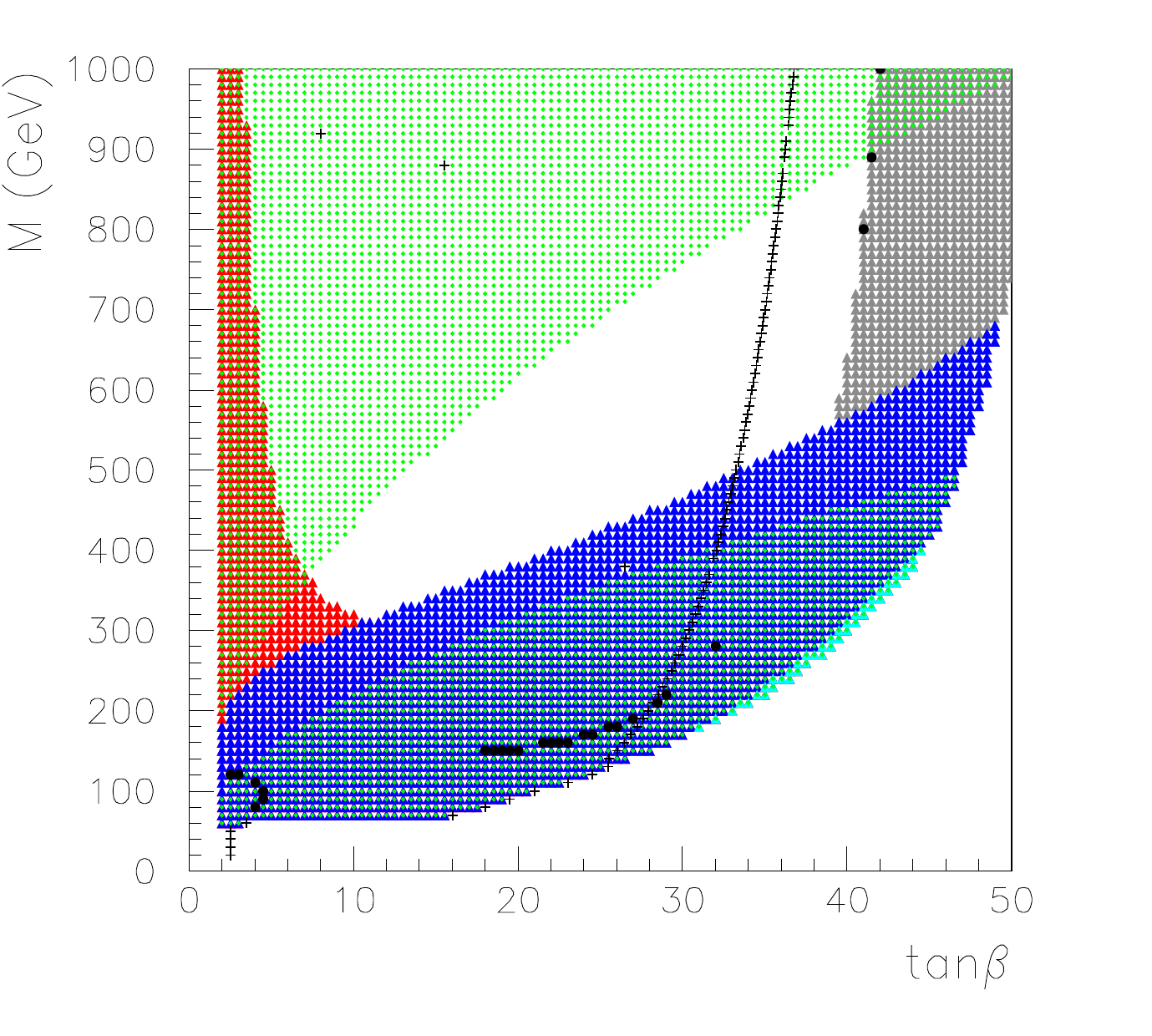}
  \includegraphics[width=8cm]{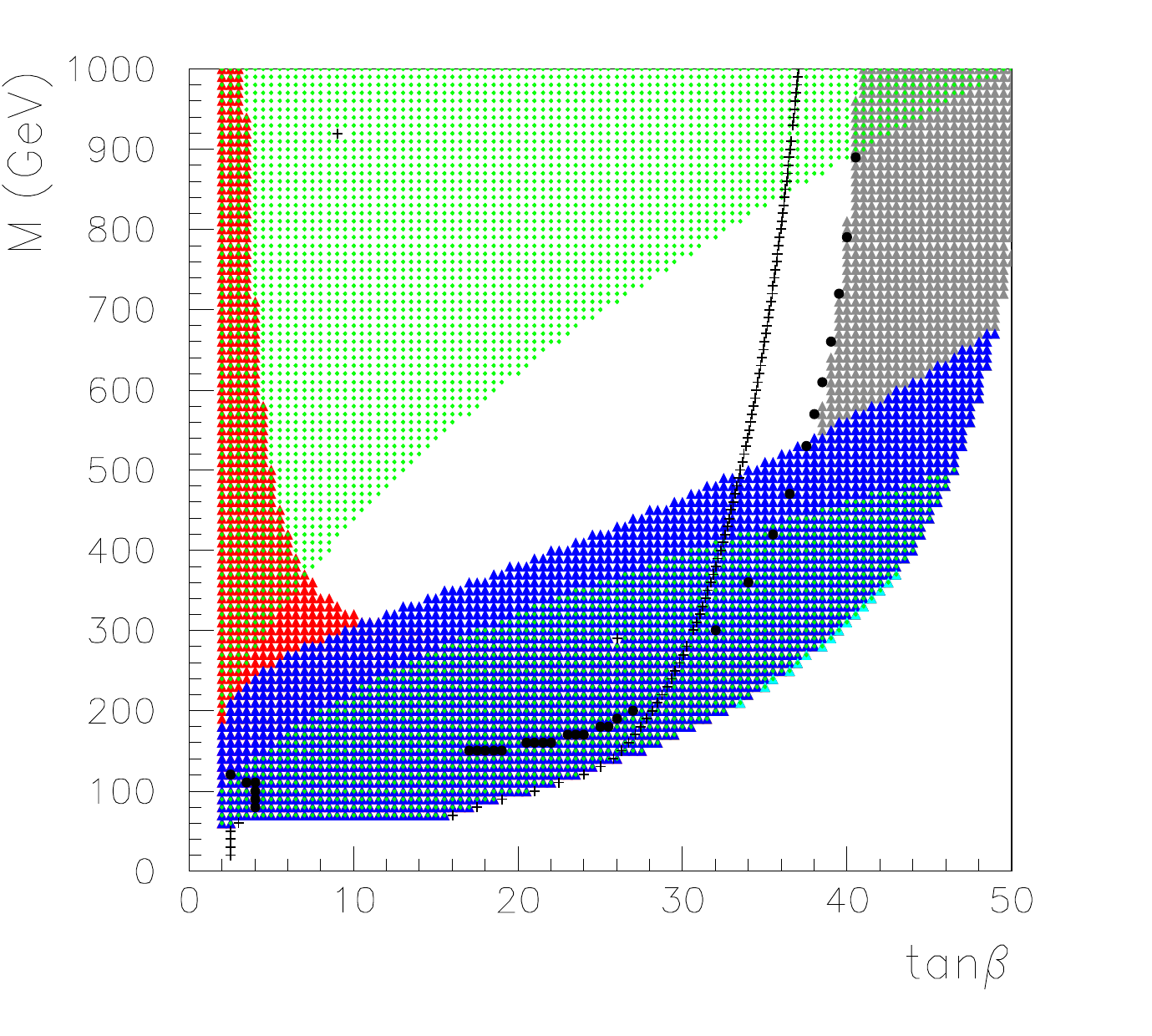}
  \hspace*{-0.6cm}
  \includegraphics[width=8cm]{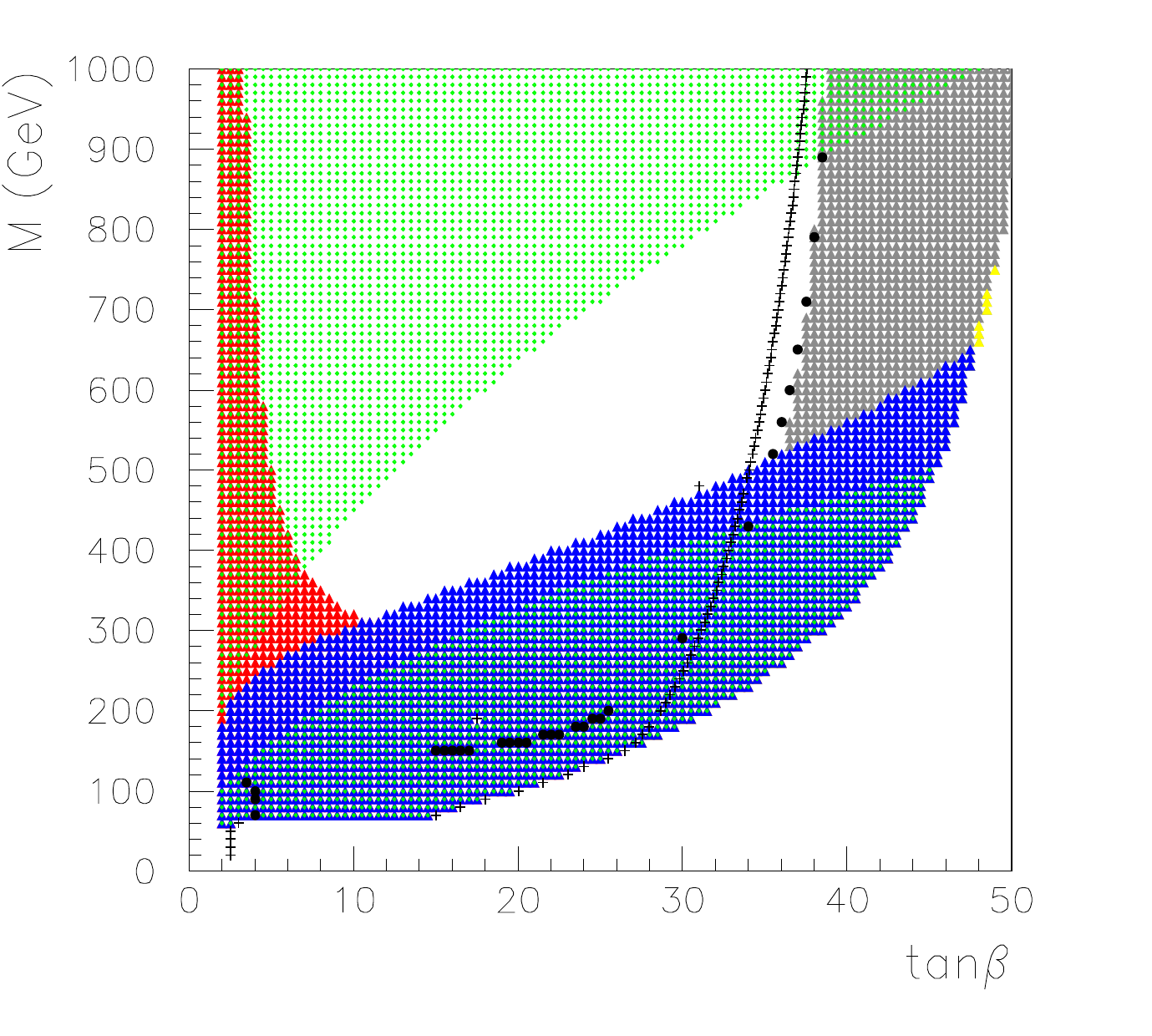}
  \includegraphics[width=8cm]{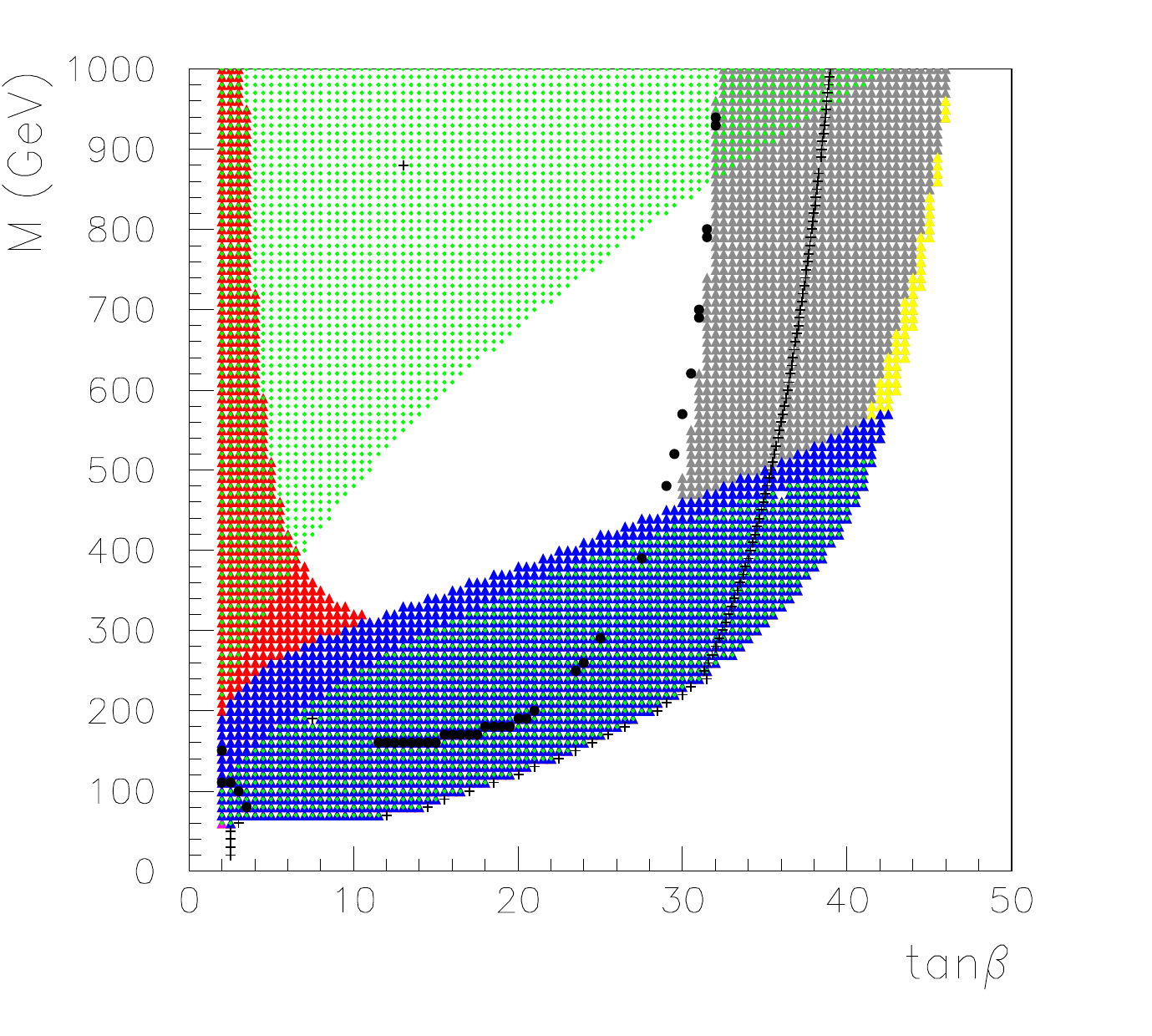}
  \vspace*{-0.8cm}
  \caption{From left to right and top to bottom flux correction of:  $\rho_f = 0.03$ with $\rho_H = 0.3$, $\rho_f =0.06 $ with $\rho_H = 0.3$, $\rho_f =0.13 $ with $\rho_H = 0.3$, and  $\rho_f =0.3 $ with $\rho_H = 0.3$.
The region in red colour is ruled out by the lower constraint on the lightest Higgs mass, the blue area correspond to values excluded by $b \rightarrow s \gamma$, the green areas correspond, from top to bottom, to the lower and upper constraint on $a_\mu^{SUSY}$ and the yellow area is ruled out by lower bound on the stau mass. The white part on the right of the graphics correspond to the area excluded due to tachyons in the stau eigenstates. The area depicted in grey  is the region in which the stau is the LSP when the above experimental constraints are fulfilled. In the remaining white area the neutralino is the LSP. The black points are the region where the neutralino relic density is in agreement with WMAP bound. Finally, the cross line is the one that represents the values of the parameter space for which there exists radiative Electro Weak symmetry breaking (REWSB). 
\label{figHFC}
  }
\end{figure}
\begin{figure}[t!]
  \hspace*{-0.6cm}
  \includegraphics[width=8cm]{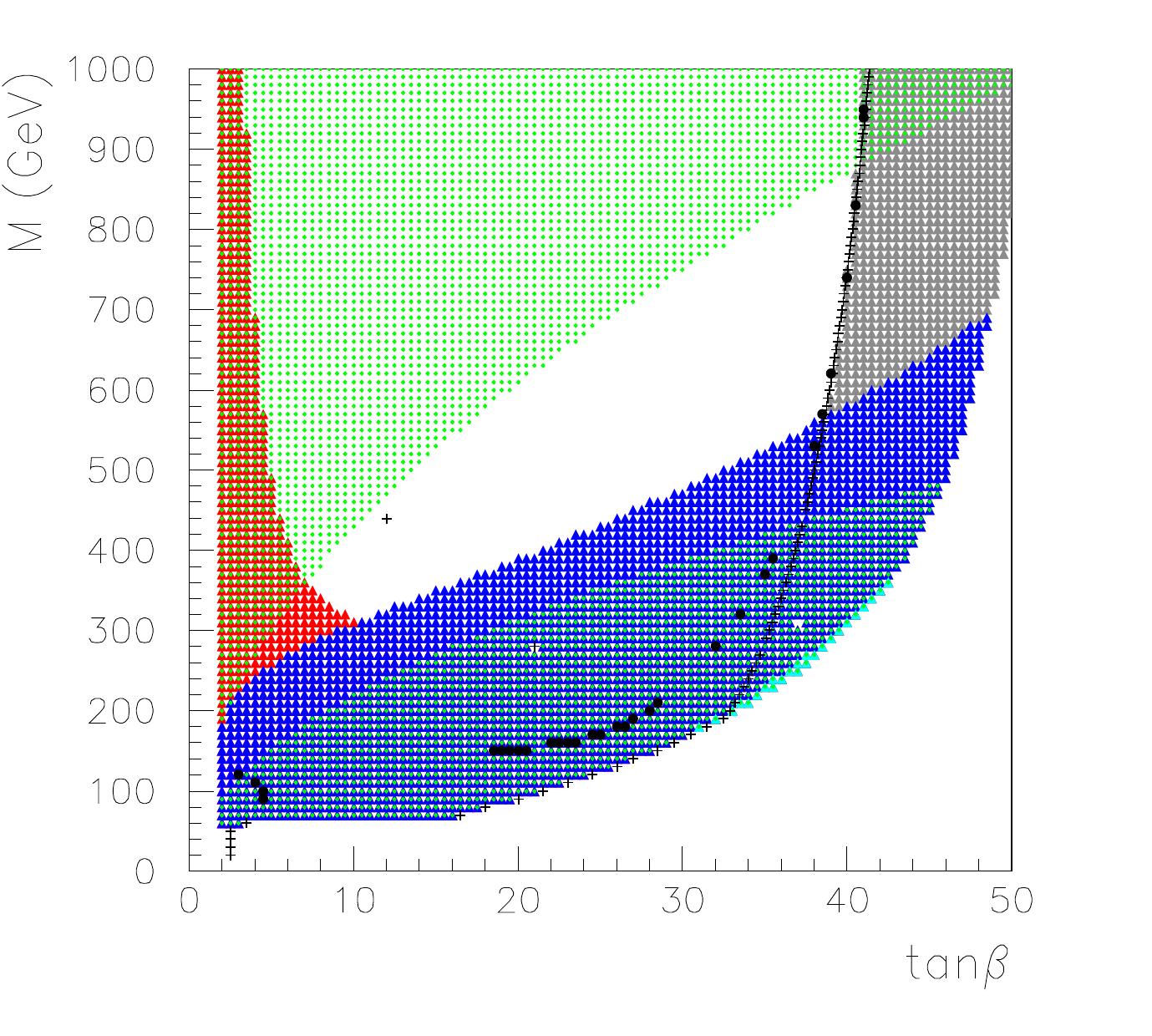}
  \includegraphics[width=8cm]{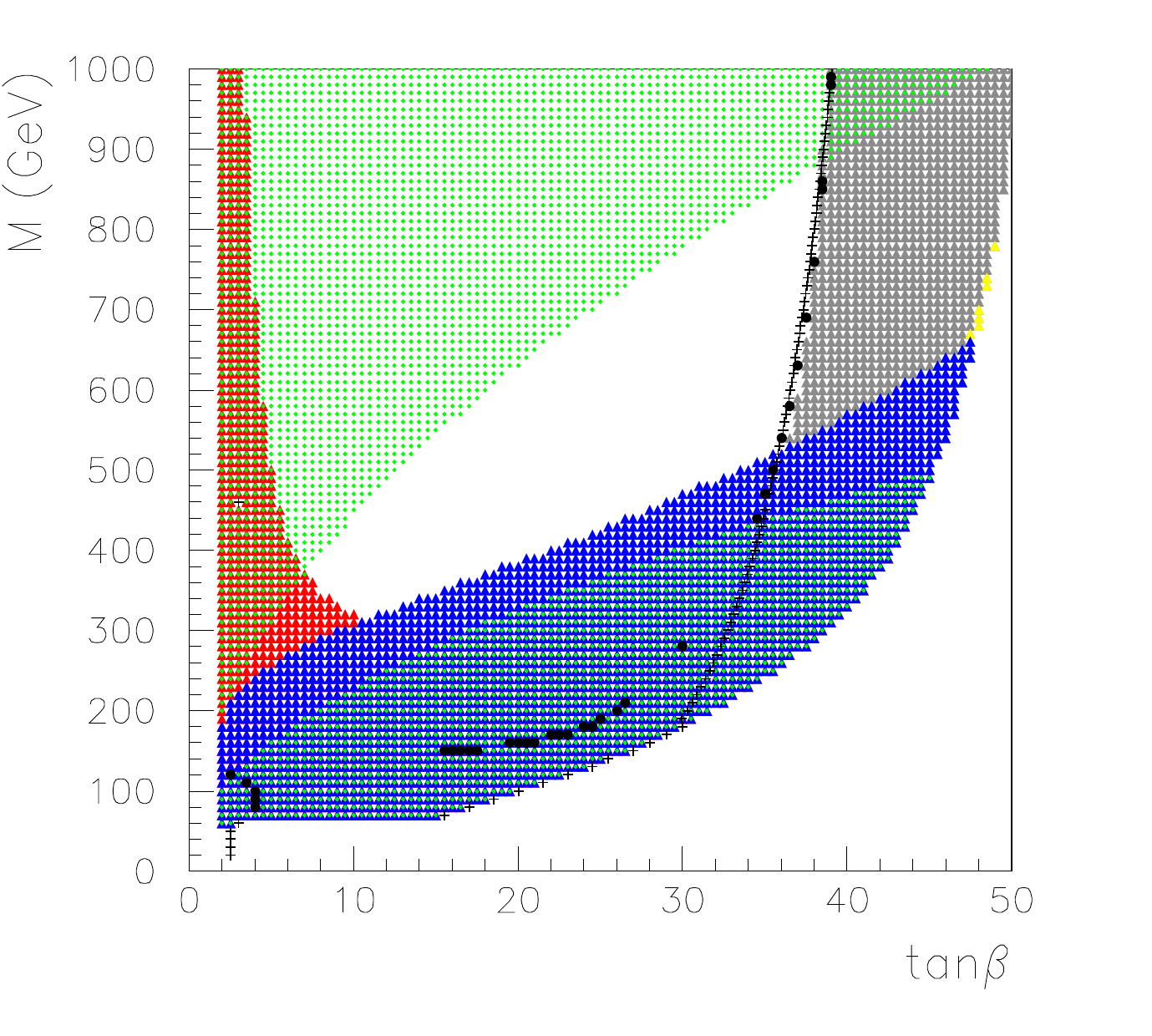}
  \hspace*{-0.6cm}
  \includegraphics[width=8cm]{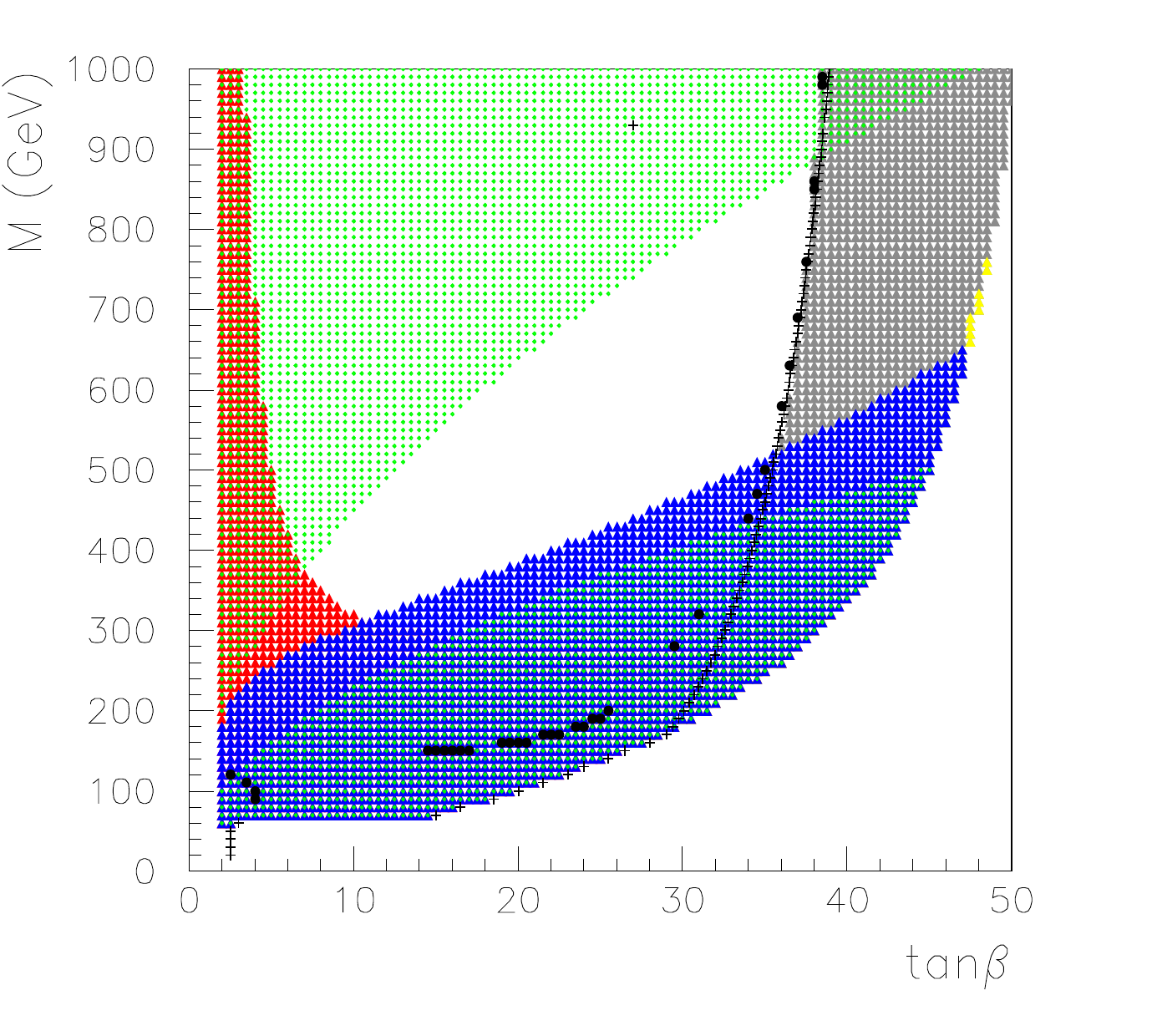}
  \includegraphics[width=8cm]{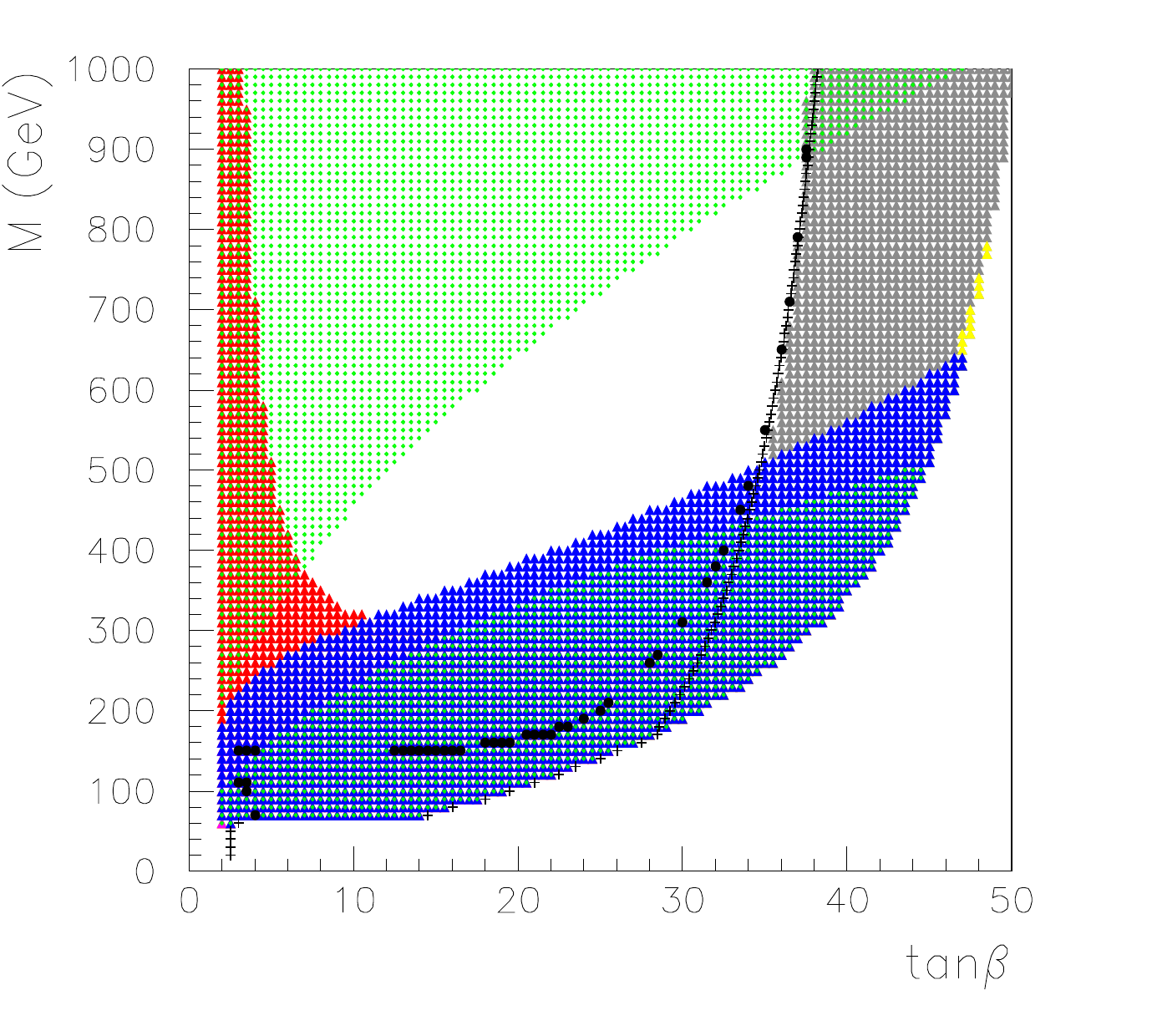}
  \vspace*{-0.6cm}
  \caption{From left to right and top to bottom flux correction of: $\rho_f = 0$ with $\rho_H = 0.18$, $\rho_f =0.13 $ with $\rho_H = 0.23$, $\rho_f =0.17 $ with $\rho_H = 0.27$, and $\rho_f =0.21 $ with $\rho_H = 0.3$. 
\label{HFCvsFFC}
  }
\end{figure}

Note  that, as we mentioned, the scalar masses of fields of A-type (modular weight $\xi =1$) remain massless even after the addition of magnetic fluxes. So the problem of the scheme (A-A-$\phi$) of having too light (or even tachyonic) right-handed sleptons cannot be solved with the addition of fluxes. Interestingly, the inclusion of magnetic fluxes also alters the boundary condition for the bilinear parameter. In particular, for case (I-I-I) $B$ at the GUT scale becomes less negative. This is welcome, as we already saw in the previous subsection, in order to obtain successful radiative electroweak symmetry breaking.\\

We have explored the effect of a small flux correction to the (I-I-I) case ($\xi_H=1/2$)  with $\rho_H=0.1,0.2$ and $\rho_f=0$. The results are shown in \Fig{mtgb-iiiflux}. The low-energy supersymmetric spectrum is not very affected by the change on the Higgs mass parameters. The only visible effect is a very slight increase in the stau mass for large $\rho_H$ as a consequence of the decrease in $|A_L|$. Hence, the allowed region with neutralino LSP and correct dark matter abundance barely changes. On the contrary, the line along which the boundary condition for $B$ is satisfied changes significantly, being shifted towards smaller $\tan\beta$ as $\rho_H$ increases. Compatibility with viable neutralino dark matter is found around $\rho_H\approx0.2$ with $\xi_H=1/2$. Thus indeed, magnetic fluxes could provide for the small correction required to get full agreement with the appropriate dark matter for the (I-I-I) intersecting 7-branes scheme.\\

Note that the corrections to sfermion masses in \Tab{correcciones de flujos}, parametrized by $\rho_f$, imply a decrease of their mass at the GUT scale. This leads to lighter staus at the EW scale, and therefore the region where the neutralino is the LSP (and obviously the region with correct relic density) is shifted towards smaller values of $\tan\beta$. This would make it more complicated to obtain compatibility of successful REWSB and neutralino dark matter (larger $\rho_H$ would be needed). However it is interesting to make a further comparative analysis between the effect of fermion and Higgs fluxes. For that purpose we present a scan of the two free parameters $M$ and $\tan \beta$ (for ($\mu < 0$) for different combinations of fermion flux correction $\rho_f$ for a fixed Higgs flux correction $\rho_H$ as one can see in figure \Fig{figHFC}.\\

We have taken fermion and Higgs flux corrections of $\rho_i = 0.3$ as maximun values because we want to consider the flux as a (perturbative) correction. From the observation of these graphics we can conclude the following:
\begin{itemize}
 \item Effects of fermion flux correction:
\begin{itemize}
 \item It enlarges the zone in which the stau is the LSP.
\item It enlarges the tachyon zone.
\item It drives the REWSB line to the zone in which the stau is the LSP. 
\end{itemize}
\end{itemize}
\begin{itemize} 
 \item Effects of Higgs flux corrections:
\begin{itemize}
 \item It enlarges the zone in which the neutralino is the LSP.
\item It reduces the zone of tachyons.
\item It drives the REWSB line to the neutralino LSP zone.
\end{itemize}
\end{itemize}

One may understand this behaviour seeing  how RGEs work and looking to the first approximation of the soft terms in \Tab{correcciones de flujos}. The enlarging of the tachyon zone due to fermion flux is related with the value of $\rho_f$: when $\rho_f$ goes large, $m_f$ becomes smaller, and therefore we will have the $m_f^2$ closer to negative values. In summary, if one includes fermion flux corrections, one needs to include Higgs flux corrections to compensate so that the combination of effects of both fluxes finally allow you to obtain REWSB. We have also made an analysis combining the values of fermion and Higgs fluxes that make REWSB to stay in the neutralino LSP zone and that are in agreement with the WMAP bound. We compute the relic densisty by means of the program micrOMEGAs, and check compatibility with the data obtained by WMAP satellite. The result is shown in the \Fig{HFCvsFFC} .\\

Note that as an order of magnitude one can make numerically a rough estimate of the effect of the fluxes contribution. From the flux quantization condition it follows that
\begin{equation}
\langle F\rangle\approx\frac{1}{R^2} \label{numapprox}
\end{equation}
We also know 
\begin{equation}
\frac{4\pi}{g^2}\sim \mbox{Wrapped volume} = R^4,
\end{equation}
where $\frac{g^2}{4\pi}$ is $\alpha_{GUT}$ in this case and therefore from \Eq{numapprox} one can estimate the order of the expected flux density corrections
\begin{equation}
\rho_i \sim \alpha_{GUT}^{1/2},
\end {equation}

Note that as an order of magnitude one thus numerically expects $\rho_H\simeq 1/t^{1/2}\simeq \alpha_{GUT}^{1/2}\simeq 0.2$. In what follows we will only consider the case with a $\rho_H$ flux, although we have done an analogous analysis with $\rho_f\not=0$ which yields completely analogous results (although requiring slightly larger $\rho_H$).

\section{Higgs and SUSY  spectrum  in the  Modulus Dominated CMSSM}\label{hsmdcmssm}
We have found in the previous sections that imposing correct radiative electroweak symmetry breaking (REWSB) \cite{ir2} and viable neutralino dark matter, essentially only the (I-I-I) configuration with magnetic fluxes survives. We emphasize that this scenario corresponds to models in which the SM fields live at the intersection of 7-branes and hence $\rho_\alpha=1/2$ for all $\alpha$. It is very much like the recent F-theory GUT constructions that we reviewed in \Sec{fthguts}. In the latter, quarks and leptons live confined in complex matter curves embedded in the bulk 7-brane in which the SM gauge group lives (see \Fig{flocal2}).\\

\begin{figure}[h!]
\hspace*{-0.6cm}
\centering
\includegraphics[width=8.cm, angle=0]{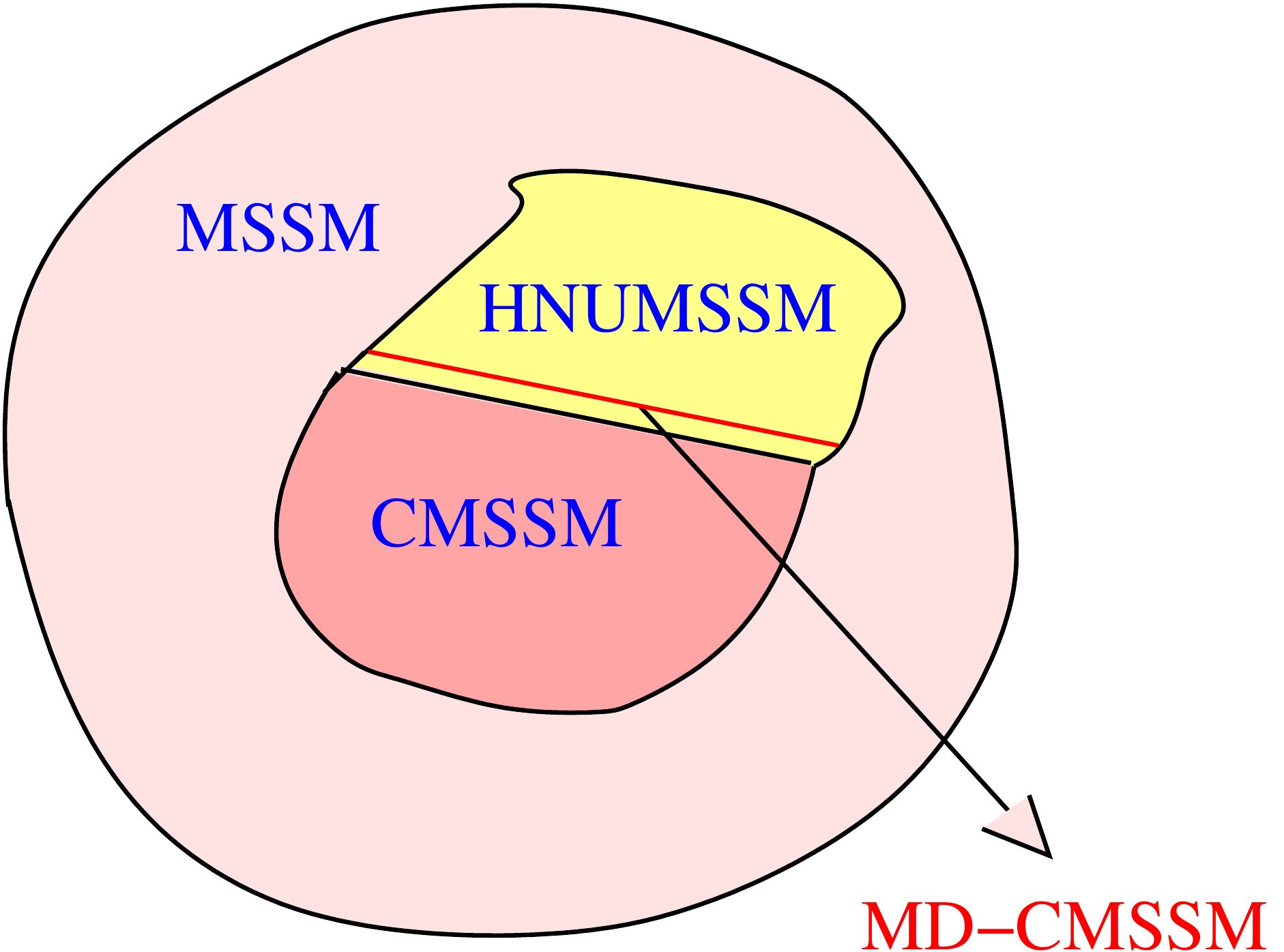}
\caption{Pictorial view of the modulus dominance constrained MSSM as a
slice of the Higgs non-universal HNUMSSM which is a slight deformation 
(due to the small flux parameter) of the CMSSM.}
\label{MDCMSSM}
\end{figure}

As shown in \Tab{correcciones de flujos}, in this scheme we have soft terms at the string unification scale with the relations
\beqa
m_{\tilde f}^2 \  & = & \ \frac{1}{2}|M|^2\,, \\  
m_{H}^2 \ & = & \   \frac{1}{2}|M|^2  (1- \frac{3}{2}\rho_H)\,,  \\
A \ & = & \ -\frac{1}{2}M(3\ -\ \rho_H)\,, \\ 
B\ & = & \ -M(1-\rho_H)\,, 
\label{boundconditionsfinal}
\eeqa
where $\rho_H$ parametrizes the effect of magnetic fluxes on the Higgs Kahler metrics. Note that this set of soft terms constitutes a deformation of a slice of the CMSSM with slightly non-universal Higgs masses. We will call it here Modulus Dominated CMSSM (MD-CMSSM, \Fig{MDCMSSM}).\\

Consistency of the scheme requires this parameter to be small so that indeed the interpretation of $\rho_H$ as a small flux correction makes sense. Note that we thus have essentially two free parameters, $M$ and $\mu$, with a third parameter $\rho_H$ restricted to be small. From now on we are going to carry out a more through analysis of this string theory configuration beyond the results of \Sec{ewsb old}, covering the full parameter space and studying its phenomenological consequences, including a study in detail of Higgs and SUSY spectra.  We will also analyze the impact of the 2011 LHC data on our results and explore the eventual LHC reach in testing the model. For that purpose we are going to impose again these two constraints: 1) consistent REWSB and 2) correct neutralino dark matter abundance. These two constraints are very stringent and it is non-trivial that both conditions may be simultaneously satisfied in such a constrained system as it was shown in \Sec{ewsb old}.

\subsection{REWSB and dark matter constraints: a model with a single free parameter}\label{smfp}

We are going to follow the same steps used in \Sec{ewsb old method}. For this further analysis, we have implemented the iterative process through a series of changes in the public code {\tt SPheno\,3.0} \cite{spheno, Porod:2011nf}. As we commented in \Sec{ewsb old method}, this code solves numerically the renormalization group equations of the MSSM and provides the SUSY spectrum at low energy and also calculates the theoretical predictions for low-energy observables such as the branching ratios of rare decays ($b\to s\gamma$, $B_s\to\mu^+\mu^-$) and the muon anomalous magnetic moment. The results are sensitive to the value of the top quark mass, particularly for the Higgs mass, see below. In the computation we use the central value in $m_t=173.2\pm 0.9$~GeV \cite{Lancaster:2011wr}. In addition to correct REWSB we also impose the presence of viable neutralino dark matter, assuming R-parity conservation. The relic density of the neutralino is calculated numerically using the MSSM module of the code {\tt MicrOMEGAs 2.4} \cite{micromegas, Belanger:2010gh} and check for the compatibility with the data obtained from the WMAP satellite, which constrain the amount of cold dark matter to be  $0.1008 \leq \Omega h^2\leq 0.1232$ at the $2\,\sigma$ confidence level \cite{snova2}.\\ 

\begin{figure}[t!]
\hspace*{-0.6cm}
\includegraphics[width=8.5cm, angle=0]{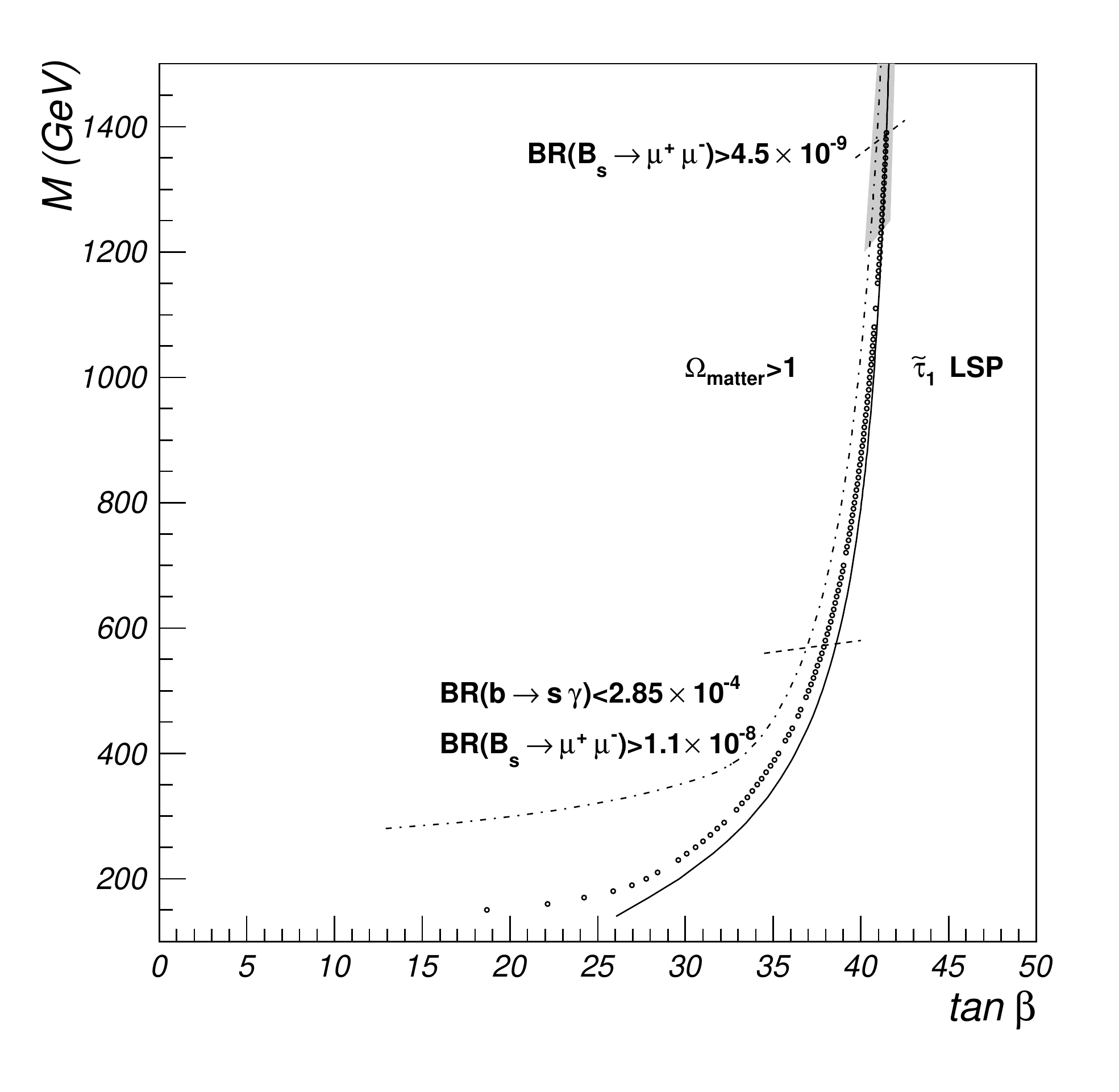}
\includegraphics[width=8.5cm, angle=0]{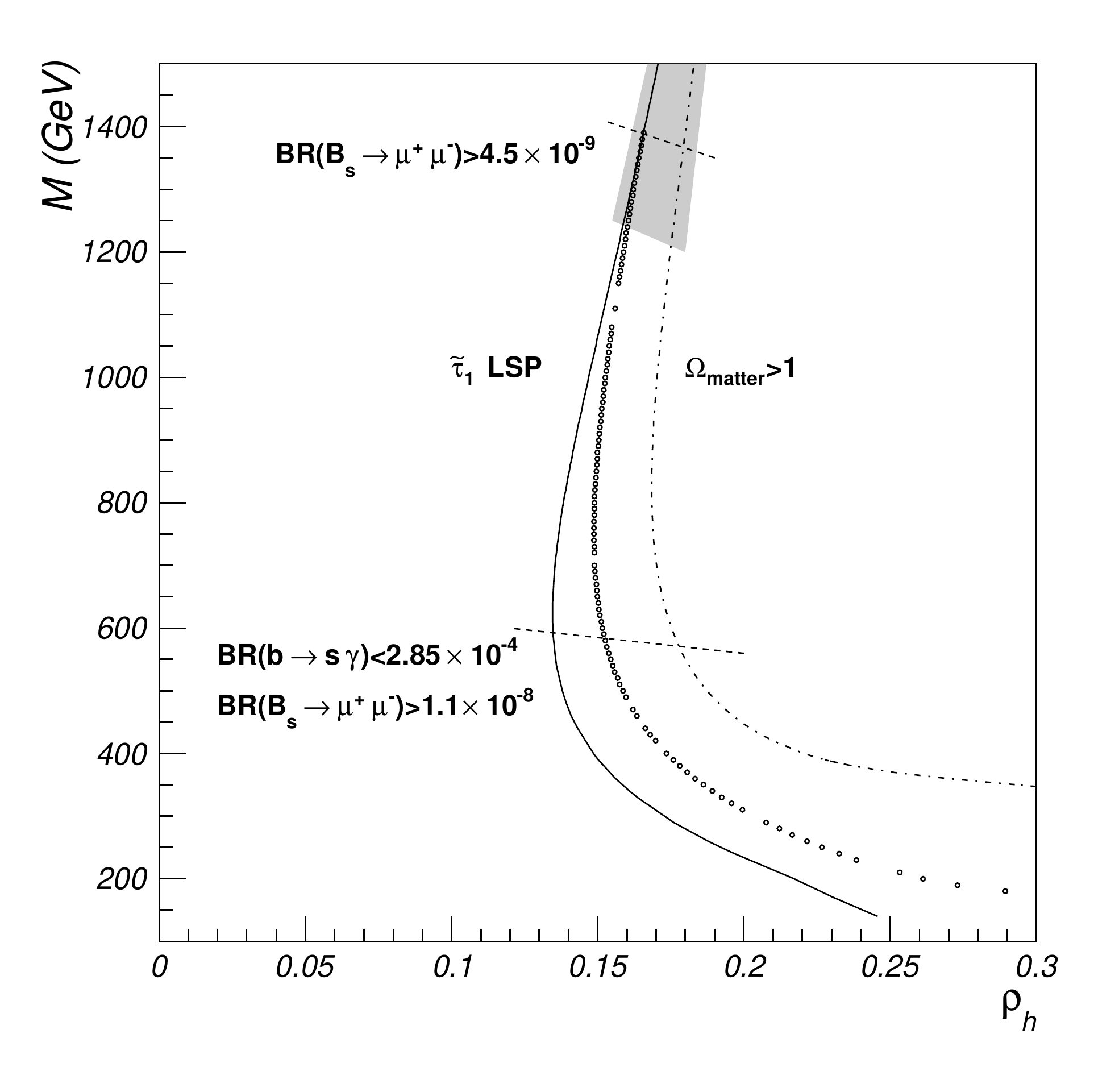} 
\caption{Left) Trajectory in the $(M,\tan\beta)$ plane for which the REWSB conditions are fulfilled and the correct amount of 
dark matter is obtained. Right) Corresponding values of the flux, $\rho_H$. In both cases, dots correspond to points fulfilling the central value in WMAP result for the neutralino relic density. The dot-dashed line denotes points along which the matter density is critical, $\Omega_{matter}=1$, whereas the solid line indicates the points for which the stau becomes the LSP. The points below the dashed line are excluded by the lower bound on BR$(b\to s\gamma)$ and the upper bound on BR$(B_s\to\mu^+\mu^-)$ from Ref.\,\cite{cmslhcb} and the recent LHCb result \cite{Aaij:2012ac}. The gray area indicates the points compatible with the latter constraint when the $2\sigma$ error associated to the SM prediction is included.}
\label{Mtanbeta}
\end{figure} 
Imposing both conditions we are left with a model with a single free parameter or, equivalently, lines in the  
$(M,\,\tan\beta)$ and  $(M,\,\rho_H)$ planes.   In Fig.\,\ref{Mtanbeta} we show the trajectories consistent with both REWSB and viable
 neutralino dark matter. The left hand-side of Fig.\,\ref{Mtanbeta} shows how the viable values for $\tan\beta$ are confined to a large value region,
 $\tan\beta \simeq 36-41$. The maximum values for $M$ and $\tan\beta$ occur for $M\simeq 1.4$~TeV, $\tan\beta\simeq 41$. 
 The existence of these maximal values are due to the dark matter condition. Indeed, as we said, the LSP in
 this scheme is mostly pure Bino and generically its abundance exceeds the WMAP constraints. However along the line in the figure 
 the lightest neutralino $\chi_1^0$ is almost degenerate in mass with the lightest stau ${\tilde \tau}_1$  (see Fig.\,\ref{plotHiggs1})  and a coannihilation effect takes place
 in the early universe reducing very effectively its abundance. Above the point $M\simeq 1.4$ TeV, 
 coannihilation is not sufficiently efficient in depleting neutralino abundance and 
  $\chi_1^0$  ceases to be a viable dark matter candidate. Thus viable dark matter 
 gives rise to a very strong constraint on the $M$ value, $|M|\leq 1.4$~TeV, which in turn implies an upper bound on the SUSY and Higgs spectrum, see below. 
Notice that for small values of the gaugino mass the predicted $\tan\beta$ can also be smaller. In principle one could get to values of $\tan\beta$ as low as $~10$ while still fulfilling REWSB and the neutralino relic density with $M\gsim150$~GeV. However, the resulting SUSY spectrum is extremely light and already well below the current experimental bounds. First, demanding $m_h>115.5$~GeV leads to a lower bound on the common scale $M\gsim340$~GeV with $\tan\beta\gsim34$. 
Similarly, current LHC lower bounds on the masses of gluinos and second and third generation squarks imply $M\gsim400$~GeV and $\tan\beta\gsim35$. 
There is a  more stringent lower bound coming from the BR($b\to s\gamma$) constraint, which implies $M\gsim570$~GeV and $\tan\beta\gsim38$. 
\\

\begin{figure}[t!]
\centering
\includegraphics[width=8.5cm, angle=0]{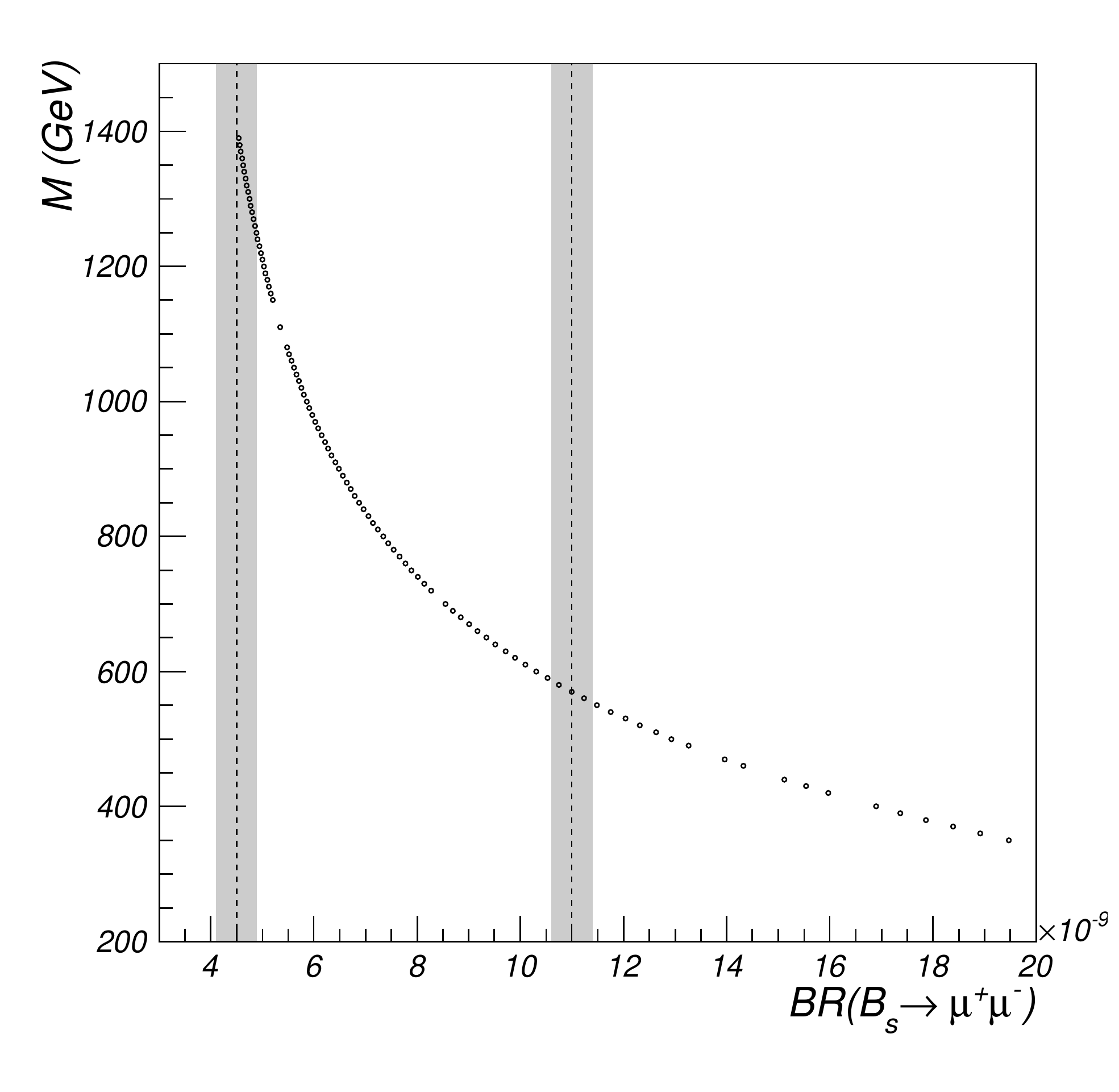}
\caption{Universal gaugino mass versus the theoretical prediction for BR$(B_s\to\mu^+\mu^-)$. The dashed lines denote the experimental upper bound on this observable from Ref.\,\cite{cmslhcb} and the recent LHCb result \cite{Aaij:2012ac}. The $2,\sigma$ theoretical error on the SM prediction is indicated by means of a shaded region in both cases.}
\label{fig:bmumu}
\end{figure} 

Finally, the experimental upper limit on BR($B_s\to\mu^+\mu^-$) has a profound impact on the allowed parameter space.
A combination of CMS \cite{Chatrchyan:2011kr} and LHCb \cite{lhcb} data recently set a bound as low as BR$(B_s\to\mu^+\mu^-)<1.1\times10^{-8}$ \cite{cmslhcb}. This would lead to $M\gsim560$~GeV, thus having a similar effect as the other constraints mentioned above.
However, very recently, the experimental bound was significantly improved by the LHCb collaboration \cite{Aaij:2012ac}, leading to the unprecedented constraint BR$(B_s\to\mu^+\mu^-)<4.5\times10^{-9}$. This is in fact very close to the SM prediction BR$(B_s\to\mu^+\mu^-)=(3.2\pm0.2)\times10^{-9}$ \cite{Buras:2010mh,Buras:2010wr} and thus
has important implications in our parameter space. Given that our model entails large values of $\tan\beta$ and a significant mixing in the stop mass matrix, the resulting BR$(B_s\to\mu^+\mu^-)$ is relatively large. 
Fig.\,\ref{fig:bmumu} represents the theoretical predictions for this observable as a function of the corresponding universal gaugino mass $M$, showing that BR$(B_s\to\mu^+\mu^-)\gsim4.4\times10^{-9}$. We display in the plot the experimental bound from Ref.\,\cite{cmslhcb} and Ref.\,\cite{Aaij:2012ac}, explicitly showing the effect of the improved measurement. For each case, we take into account the $2\sigma$ theoretical uncertainty of the SM contribution.
It is in fact expected that this upper bound improves in the near future with new data from CMS and LHCb. This has the potential to disfavour our construction if no deviation from the SM value is observed.
\footnote{It should be pointed out in this respect that the inclusion of non-vanishing flux correction $\rho_f$  for sfermions in Eq.\,(\ref{boundconditionsfinal})
can slightly alter the allowed regions in the parameter space, shifting the viable points towards smaller values of $\tan\beta$, thereby decreasing the SUSY contribution to BR$(B_s\to\mu^+\mu^-)$.}\\

On the right hand-side of Fig.\,\ref{Mtanbeta} we display the line in the $(M,\,\rho_H)$ plane that is consistent with REWSB and viable neutralino dark matter.
Interestingly enough, after applying experimental constraints, the value of $\rho_H$ is indeed small, of order $0.15-0.17$ and is very weakly dependent on $M$.
This is consistent with the interpretation of $\rho_H$ as a small correction arising from gauge fluxes, as discussed in the previous chapter.
Indeed the values for $\rho_H$  obtained are of the expected order of magnitude, $\rho_H\propto \alpha_{GUT}^{1/2}\simeq 0.2$.
 
The viable points of the parameter space lie along a narrow area of the parameter space. In fact, small deviations in any of the parameters, $M$, $\tan\beta$ or $\rho_h$ have catastrophic consequences, since either the relic density becomes too large (it very rapidly overcloses the Universe) or the stau becomes the LSP. We illustrate this in Fig.\,\ref{Mtanbeta}, where the dashed and solid lines represent the points for which $\Omega_{matter}=1$ and $m_{{\tilde \tau}_1}=m_{\chi_1^0}$, respectively.
The line with critical density extends to $M\approx2.5$~TeV, but the region fulfilling WMAP $2\,\sigma$ region stops at $M=1.4$~TeV. 
Interestingly, the flux $\rho_h$ cannot vanish (since the stau becomes the LSP), this is, even though small, a deviation from the CMSSM is necessary. Also, it cannot be too large or we would have an excessive amount of dark matter. \\

\begin{figure}[t!]
\hspace*{-0.6cm}
\centering
\includegraphics[width=11.cm, angle=0]{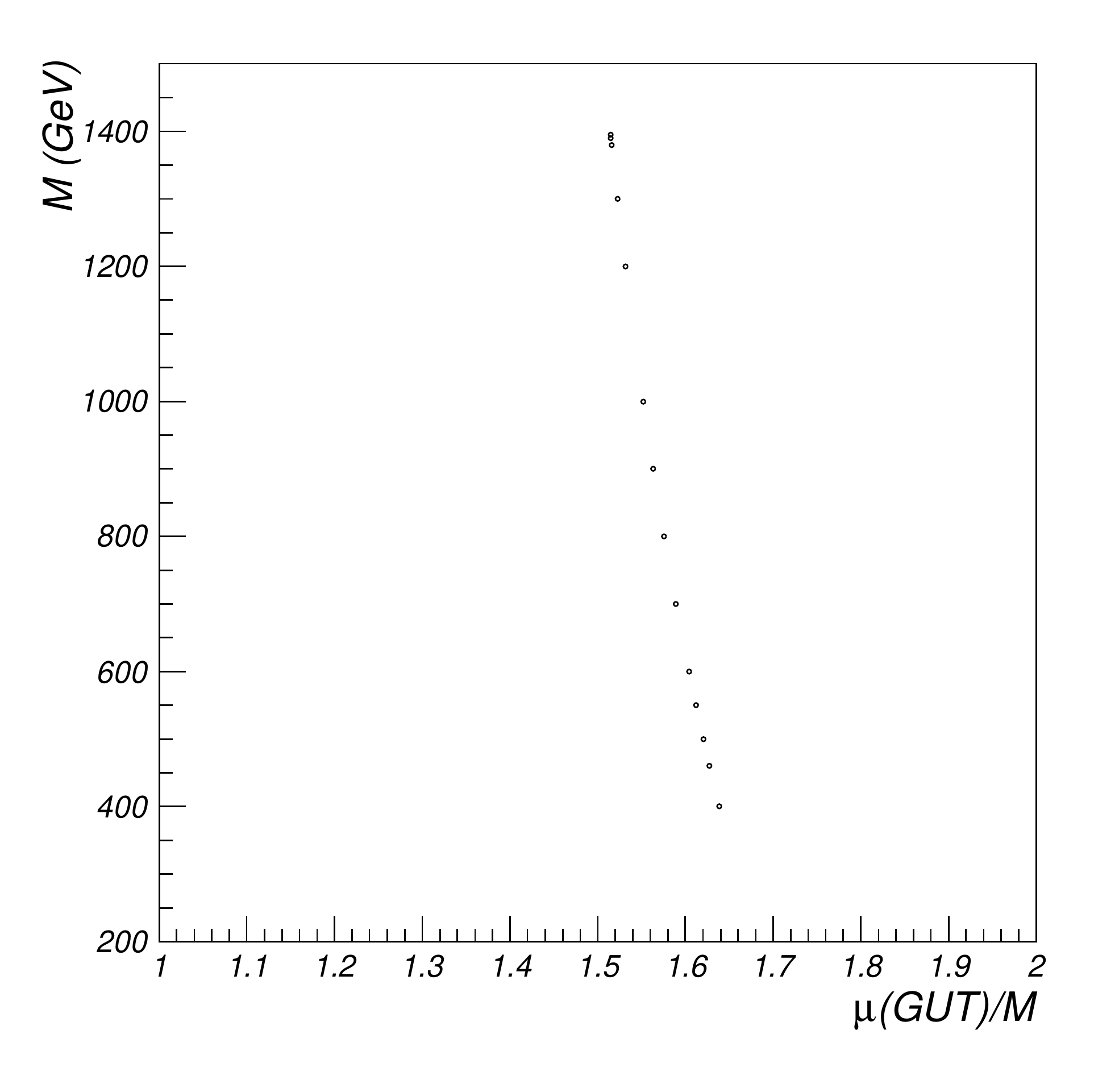}
\caption{Universal gaugino mass as a function of the resulting Higgsino mass parameter, $\mu$ at the unification  scale for those points where both REWSB and viable neutralino dark matter are obtained.}
\label{plotmu}
\end{figure} 
 
As we explained in  the beginning of this chapter, the $\mu$ parameter is computed at the electroweak scale from Eq.\,(\ref{muterm}). 
Using {\tt SPheno\,3.0} we have also computed its value at the unification scale (the effect of the RGEs is not large for this parameter) so that we can compare it with the soft parameters. This might give us an idea of what the possible origin of the $\mu$-term could be
\footnote{In particular, as  noted in Ref.\,\cite{aci},  the Giudice-Masiero  \cite{gm} mechanism would predict in the present model
$\mu=-M/2$ and $B=-3M/2$, which do not lead to consistent REWSB.}.
The results are displayed in Fig.\,\ref{plotmu}, where the ratio $\mu({\rm GUT})/M$ that corresponds for each value of the gaugino mass is plotted. As we can observe, the predicted value for that ratio is approximately constant and satisfies 
$\mu\sim(1.5-1.6)\,M$. At the point of maximal $M$ one has approximately $|\mu|=|A|=3/2|M|$. This could perhaps point towards a higher degree of interdependence among  soft terms.

\subsection{The Higgs mass}

The lightest neutral Higgs, $h$, in the MSSM receives important one-loop corrections to its mass from the top-stop loops. 
The one-loop corrected Higgs mass has an approximate expression of the form
\cite{higgsloops}
\beq
m_h^2\ \simeq \ M_Z^2\cos^22\beta \ +\ 
\frac {3m_t^4}{16\pi^2v^2} \left( \log\frac{m_{\tilde t}^2}{m_t^2} \ +\ 
\frac{X_t^2}{m_{\tilde t}^2}    \left(1-\frac{X_t^2}{12m_{\tilde t}^2}\right)   \right) \ ,
\eeq
where $v^2=v_1^2+v_2^2$, $m_{\tilde t}=(m_{{\tilde t}_L}m_{{\tilde t}_R})^{1/2}$, and $X_t=A_t-\mu \cot\beta$,
all evaluated at the weak scale.
The largest values for the Higgs mass are obtained then for large $\tan\beta$ and large stop masses. 
In particular, the quantity in brackets is maximized for $|X_t|\simeq\, \sqrt{6}m_{\tilde t}$. 
Interestingly enough this maximal value typically correspond to large values for the trilinear soft term 
$A/m\simeq \pm 2$ (see e.g. Ref.\,\cite{Baer:2011ab, Arbey:2011ab}). In our scheme we have  $A/m=-3/\sqrt{2}+\rho_H/\sqrt{2}\simeq -2$
and large values of $\tan\beta=36-41$, so that relatively  {\it large values of the Higgs mass are an automatic
prediction of our scheme}.\\

\begin{figure}[t!]
\hspace*{-0.6cm}
\centering
\includegraphics[width=11.cm, angle=0]{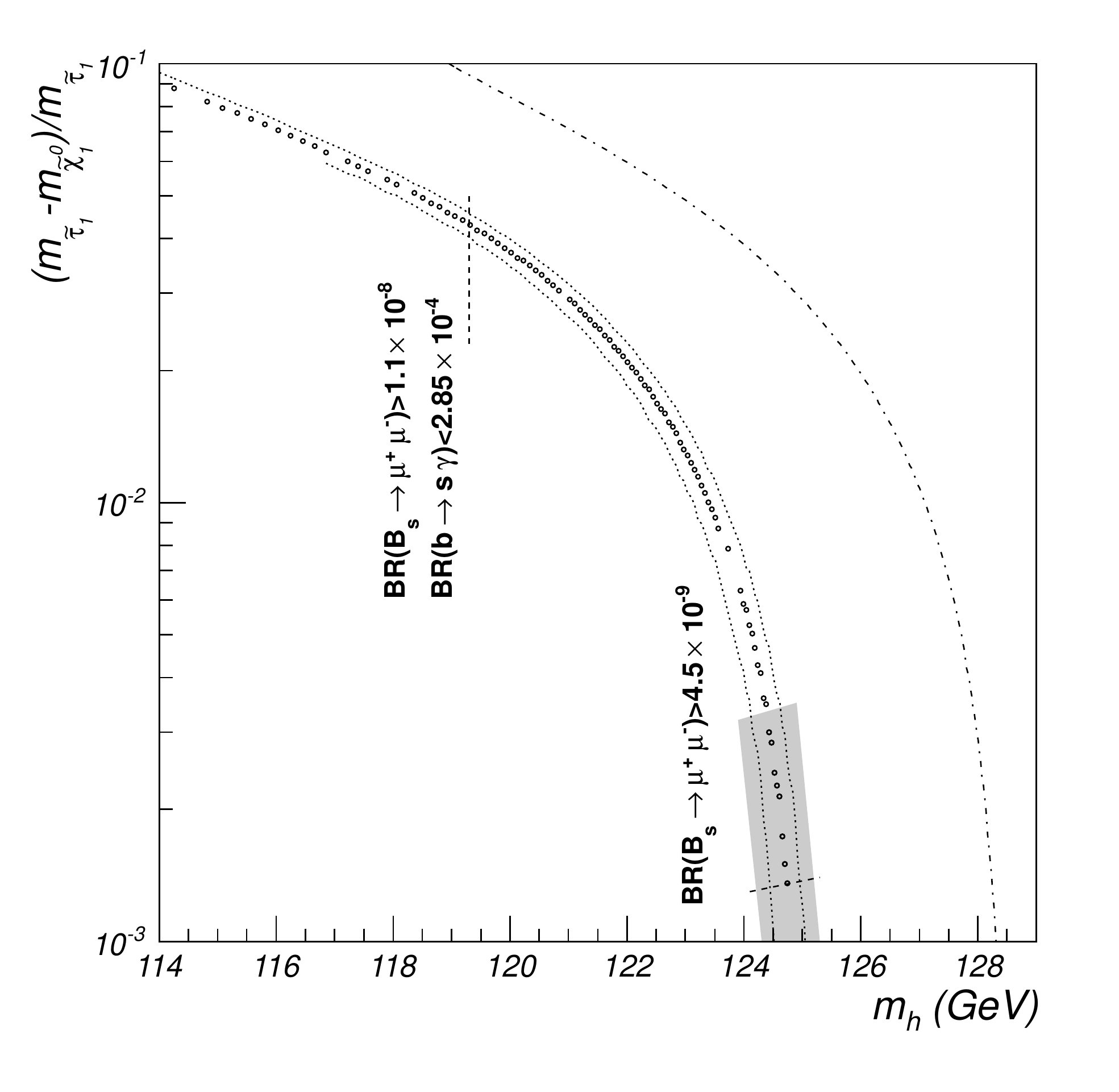}
\caption{The normalized mass difference 
$(m_{{\tilde \tau}_1}-m_{\chi_1^0})/m_{{\tilde \tau}_1}$ as a function of the lightest Higgs mass $m_h$.  Dots correspond to points fulfilling the central value in the result from WMAP for the neutralino relic density and dotted lines denote the upper and lower limits after including the $2\sigma$ uncertainty. 
The dot-dashed line represents points with a critical matter density $\Omega_{matter}=1$. The vertical line corresponds to the 2$\sigma$ limit on BR($b\rightarrow s\gamma$) and the upper bound on BR$(B_s\to\mu^+\mu^-)$ from Ref.\,\cite{cmslhcb} and the recent LHCb result \cite{Aaij:2012ac}. The gray area indicates the points compatible with the latter constraint when the $2\sigma$ error associated to the SM prediction is included.} 
\label{plotHiggs1}
\end{figure}

The 2011 run at LHC has restricted the most likely range for a SM Higgs to the range $115.5-131$~GeV (ATLAS) and 
$114.5-127$~GeV (CMS). Furthermore there is an excess of events in the  $\gamma \gamma$, $ZZ^*\rightarrow 4l$ and 
$WW^*\rightarrow 2l$ channels suggesting the presence of a Higgs boson at a mass around $125$~GeV. 
Although more data are needed to confirm this excess, it is interesting to see whether a Higgs boson in that 
range appears in this construction.  
As we said, our scheme has essentially one free parameter and the 
allowed values for the Higgs mass turn out to be very restricted. We have computed the mass of the Higgs particles
to two-loop order 
using the code {\tt SPheno} linked through the {\tt micrOMEGAs} program
\footnote{We have compared our results with those obtained with {\tt FeynHiggs\,2.8.6} \cite{Heinemeyer:1998yj,Hahn:2005cu}, finding good agreement, within approximately 1~GeV.}.
To show the allowed values for the lightest Higgs mass we display in Fig.\,\ref{plotHiggs1} the ratio
$(m_{{\tilde \tau}_1}-m_{\chi_1^0})/m_{{\tilde \tau}_1}$  versus the value of
the lightest Higgs mass $m_h$. This mass difference is very relevant for the coannihilation effect which is required 
in this scheme to get viable neutralino dark matter.
We also illustrate the variation resulting from the $2\sigma$ uncertainty in the WMAP result. \\

One observes that there is a maximum value of the Higgs mass of order 125 GeV.  For higher values the
neutralino ceases to be viable as a dark matter candidate.  This limit corresponds to the
maximum allowed values $M\simeq 1.4$ TeV and $\tan\beta\simeq 41$ that we discussed above and hence
to a quite massive SUSY spectra, see below. There is also a lower limit coming from the 
lower bound on the constraint BR($b\rightarrow s\gamma)<2.85\times 10^{-4}$ 
which leads to
\beq
119\ {\rm GeV} \leq  \ m_h\ \leq 125\ {\rm GeV}  \ .
\eeq
In the MSSM the bound on BR$(B_s\to\mu^+\mu^-)$ also has an impact on the predicted Higgs mass \cite{Cao:2011sn}.
In our case, if the current LHCb constraint is taken at face value and the SM uncertainty is included in our theoretical predictions, the resulting range for the Higgs mass is reduced to
\beq
124.4\ {\rm GeV} \leq  \ m_h\ \leq 125\ {\rm GeV}  \ .
\eeq

We have to remark at this point that these values are sensitive to the value taken for the top quark mass and the corresponding error.
As we said we have taken the central value in $m_t=173.2\pm 0.9$  \cite{Lancaster:2011wr}.
The value of the Higgs mass is very dependent on the top mass.
As a rule of thumb, one can consider that an increase of $1$~GeV in the top mass leads to an increase of approximately $ 1$~GeV in $m_{h}$ \cite{Heinemeyer:1999zf}.
Note  also that the computation of the Higgs mass includes additional intrinsic errors 
of order 1 GeV, see e.g. Refs.\,\cite{Degrassi:2002fi, Allanach:2004rh}. In any event, it is remarkable that the allowed region in our model
is well within the range allowed by the 2011 LHC data. In particular, generic points in the CMSSM space tend to have a  lighter 
Higgs mass tipically of order 115 GeV or lower. Our particular choice of soft terms plus the constraint of viable neutralino
dark matter force our Higgs mass to be relatively high.\\

\begin{figure}[t!]
\hspace*{-0.6cm}
\centering
\includegraphics[width=8.5cm, angle=0]{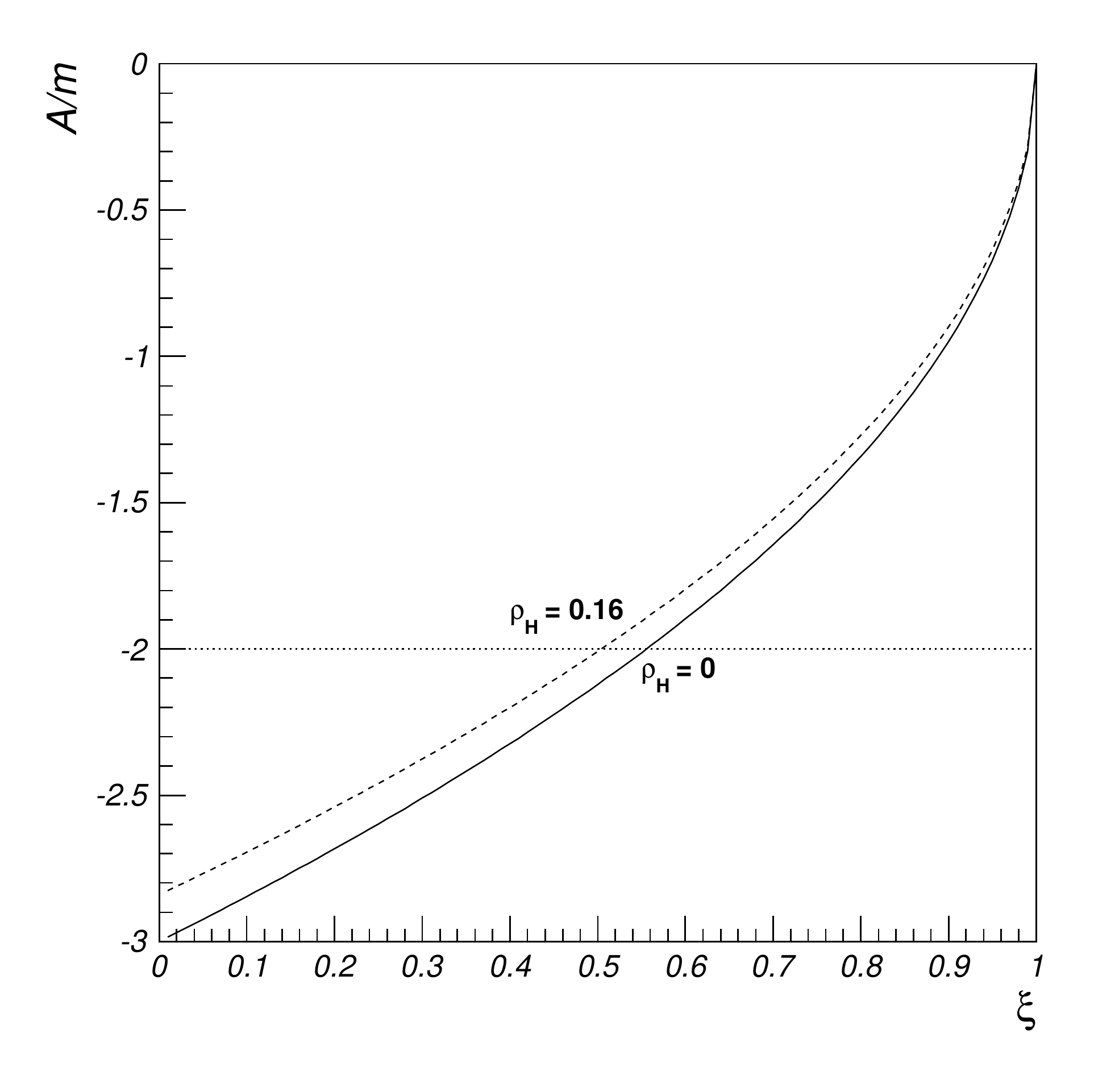}
\caption{Ratio $A/m_f$ at the GUT scale as a function of the modular weight $\xi$ for the case without fluxes (solid line) and when a small flux ($\rho_H=0.16$) is introduced.}
\label{aoverm}
\end{figure} 

It should be pointed out that the regions of the parameter space with larger values of the Higgs mass correspond to a heavy spectrum and therefore predict a small supersymmetric contribution to the muon anomalous magnetic moment, $a_\mu^{\rm SUSY}$. In particular, the points with $m_h>124$~GeV predict $a_\mu^{\rm SUSY}\approx3\times10^{-10}$.
These values show some tension with the observed discrepancy between the experimental value \cite{g-2otr} and the Standard Model predictions using $e^+e^-$ data, which imply $10.1\times10^{-10}<a_\mu^{SUSY}<42.1\times10^{-10}$ at the $2\,\sigma$ confidence level \cite{Hagiwara:2011af} where theoretical and expreimental errors are combined in quadrature (see also Refs.\,\cite{Jegerlehner:2009ry, Davier:2010nc}, which provide similar results). However, if tau data is used this discrepancy is smaller $2.9\times10^{-10}<a_\mu^{SUSY}<36.1\times10^{-10}$ \cite{Davier:2010nc}.\\

As we said, in the context of the CMSSM obtaining a large Higgs mass and not too heavy 
SUSY spectrum requires having $A\simeq -2m$. This may be
considered as a hint of a scheme with all SM localized in intersecting branes and is in fact independent of what the
possible origin of the $\mu$ term is. Indeed, for general (but universal) 
modular weights $\xi$ one has the relation
\beq
A\ = \ -3(1-\xi)^{1/2}\ m \ .
\label{estupenda}
\eeq
For $A/m\simeq -2$  one has $\xi\simeq 0.5$, indicating that indeed large Higgs masses favour 
all SM particles with $\xi\simeq 1/2$ modular weights, which correspond to intersecting branes,
as in our scheme. This is illustrated in Fig.\,\ref{aoverm}.

\subsection{The SUSY spectrum}\label{susyspectrum}

Again, our particular choice of soft terms significantly constrains the spectrum of SUSY particles.
Given that there is only one free parameter, fixing any value for a SUSY particle or Higgs field automatically 
fixes the rest of the spectrum. We give in Table\,\ref{espectromasas} the values of some masses and parameters as we vary 
the universal gaugino mass, $M$.  Let us remember that $\tan\beta$ is not an input, as it is fixed by the boundary conditions on $B$.
\begin{table}[t!] \footnotesize
\renewcommand{\arraystretch}{1.50}
\begin{center}
\begin{tabular}{|c||c|c|c|c|c|c|c|c|c|}
\hline 
M  &  $\tan\beta$ 
 &  ${\tilde g}$    &  ${\tilde Q}_L$
 &   ${\tilde Q}_R$     &    ${\tilde t}_1$  &   $\chi_2^0,\chi_1^+;L_R$   &    $\chi_1^0,{\tilde \tau}_1$        &    $M_A$ &   $m_h$   \\
\hline\hline
 $400$ & $35.3$ &
944   &  $900$  &   $870$    &   $605$    &    314;323  &  $164,175$ &     549 & 116.8 
 \\
 \hline
  $500$ & $37$ &
1160   &  $1107$  &   $1067$    &   $754$    &    397;402  &  $208,219$ & 660 &  118.5
 \\
 \hline
 $600$ & $38.2$ &
1372   &  $1310$  &   $1262$    &   $901$    &    481;482  &  $252,262$ & 769 &  119.7
 \\
 \hline
 $700$ & $39$ &
1583   &  $1511$  &   $1455$    &   $1046$    &    565;561  &  $296,305$ & 875 &  120.8
 \\
 \hline
 $800$ & $39.6$ &
1791   &  $1710$  &   $1644$    &   $1189$    &    649;641  &  $341,349$ & 981 &  121.7
 \\
 \hline
 $900$ & $40.1$ &
1998   &  $1907$  &   $1834$    &   $1330$    &    732;720  &  $386,393$ &  1084 &  122.4
 \\
  \hline
 $1000$ & $40.5$ &
2203   &  $2103$  &   $2020$    &   $1470$    &    816;800  &  $431,436$ &  1187 &  123.1
 \\
  \hline
 $1100$ & $40.8$ &
2424  &  $2314$  &   $2220$    &   $1620$    &    907;886  &  $480,483$ & 1298 & 123.7
 \\
   \hline
 $1200$ & $41.1$ &
2610  &  $2491$  &   $2390$    &   $1746$    &    984;859  &  $521,524$ & 1391 &  124.2
 \\
  \hline
 $1300$ & $41.3$ &
2812  &  $2683$  &   $2575$    &   $1883$    &    1069;1039  &  $567,568$ & 1492 & 124.7
 \\
   \hline
 $1400$ & $41.5$ &
3013 &  $2876$  &   $2760$    &   $2018$    &    1153;1119  &  $612,612$ & 1592 &  125.1
 \\
\hline
\end{tabular}
\end{center} \caption{\small  Sparticle and Higgs masses in GeV and resulting value of $\tan\beta$ as a function of $M$.
Note that there is a maximum value for $M=1.4$~TeV where $\chi_1^0$ becomes degenerate with the lightest  stau,
as the third column from the right shows. At that point the maximum value for the lightest Higgs mass $\simeq 125$~GeV
is obtained.
}
\label{espectromasas}
\end{table}

\begin{figure}[h!]
\hspace*{-0.6cm}
\centering
\includegraphics[width=10.cm, angle=0]{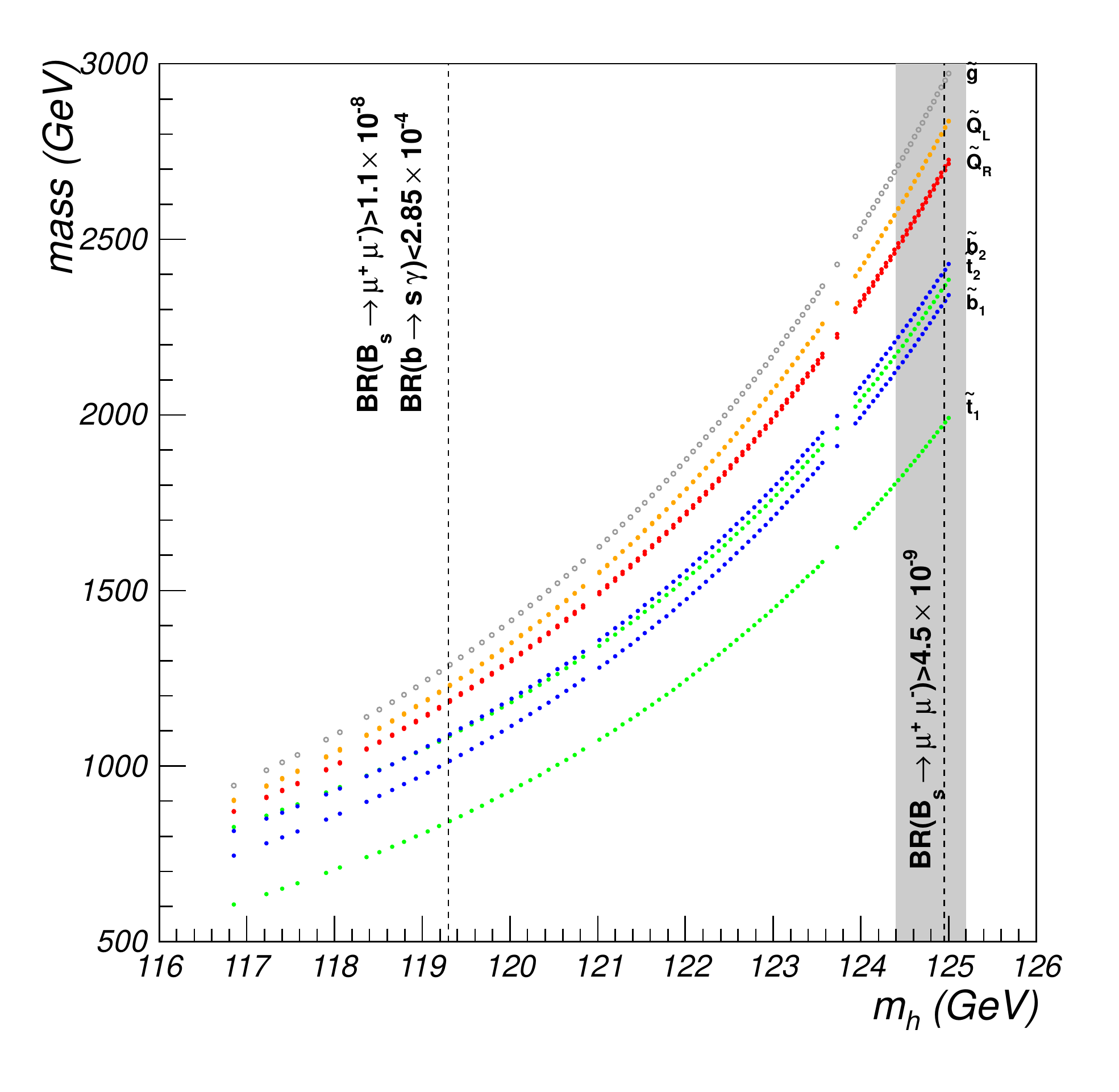} 
\caption{Squark and gluino masses as a function of the Higgs mass. 
The region to the left  of the vertical dashed indicates the constraint $BR(b\to s\gamma)<2.85\times 10^{-4}$ and the upper bound on BR$(B_s\to\mu^+\mu^-)$ from Ref.\,\cite{cmslhcb} and the recent LHCb result \cite{Aaij:2012ac}. The gray area indicates the points compatible with the latter constraint when the $2\sigma$ error associated to the SM prediction is included.}
\label{higgsversusquarks}
\end{figure}

One interesting way of presenting the structure of the SUSY spectrum is in terms of the lightest Higgs mass.
In Fig.\,\ref{higgsversusquarks} we show the masses  of the gluino and the squarks as a function of $m_h$.
The region to the left  of the vertical dashed line is excluded since it leads to  $BR(b\to s\gamma)<2.85\times 10^{-4}$.
Note that this implies that squarks of the first two generations and gluinos in our scheme must be heavier than $\simeq 1.2$~TeV.
This is consistent with LHC  limits obtained with 1\,fb$^{-1}$. For the third generation of squarks, the lightest stop 
has a mass of at least 800 GeV and the heaviest one, along with the sbottoms are heavier than 1 TeV. On the other hand, the recent LHCb limits on BR$(B_s\to\mu^+\mu^-)$ pushes the allowed region to that with a 125 GeV Higgs mass, see \Fig{higgsversusquarks}. \\

If the signal for a Higgs at 125~GeV is real, one expects a quite heavy  spectrum with gluinos of order 3~TeV and squarks of the first two
generations of order 2.8~TeV. The lightest stop would be around 2~TeV and the rest of the squarks at around 2.3~TeV. 
Note however that these values depend strongly on the Higgs mass so that e.g. a Higgs around 124~GeV 
would rather correspond to squarks and gluinos around 2.2~TeV. Given the intrinsic error in the computation of
the Higgs mass, 
this only give us a rough idea of the expected masses for colored particles.
We discuss the testability of such  heavy colored spectra  in the next chapter.\\

\begin{figure}[h!]
\hspace*{-0.6cm}
\centering
\includegraphics[width=10.cm, angle=0]{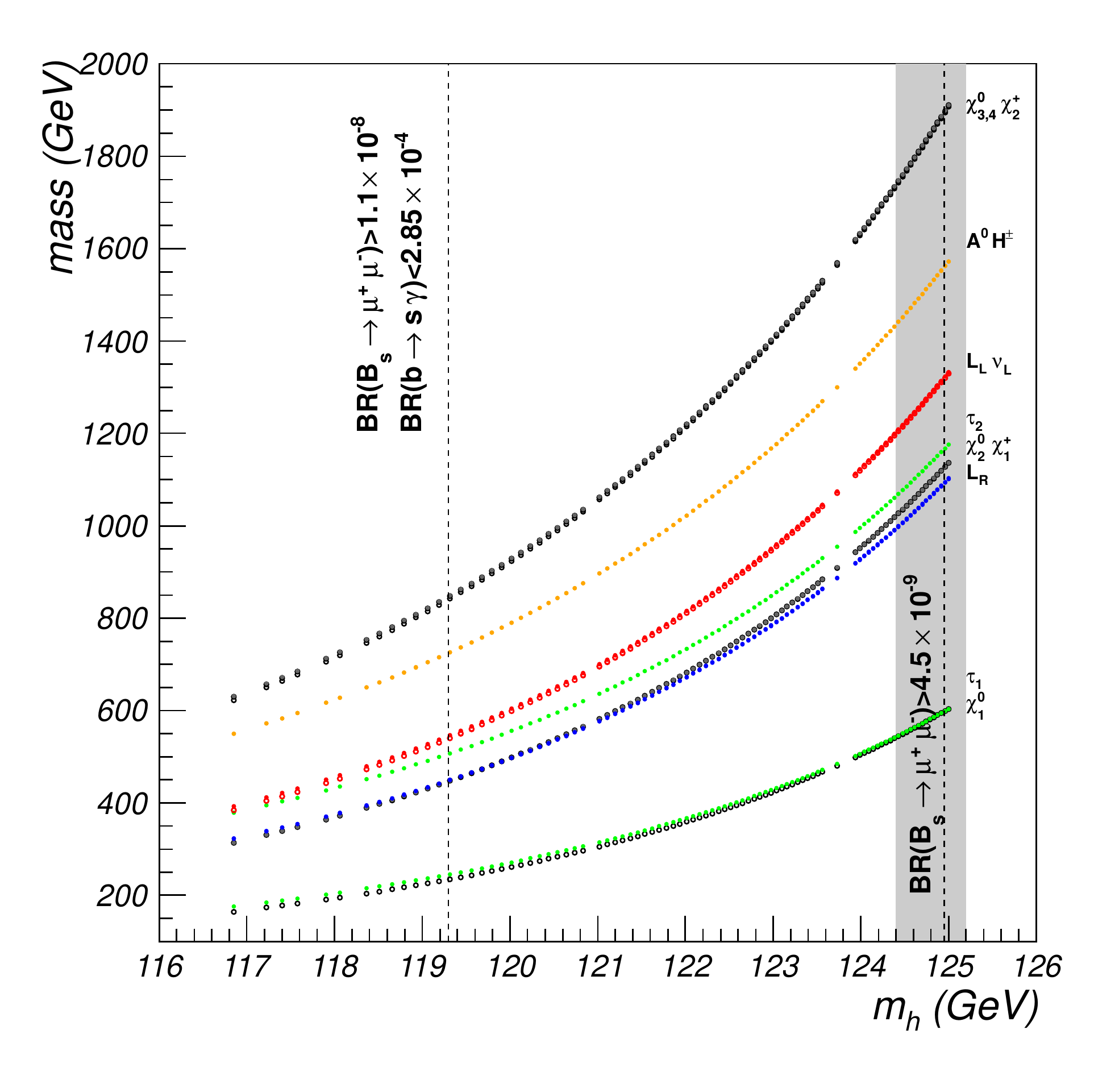}
\caption{Supersymmetric spectrum as a function of the Higgs mass of the slepton sector, together with the masses of the heavy Higgses and the gauginos.  The region to the left  of the vertical dashed indicates the constraint $BR(b\to s\gamma)<2.85\times 10^{-4}$ and the upper bound on BR$(B_s\to\mu^+\mu^-)$ from Ref.\,\cite{cmslhcb} and the recent LHCb result \cite{Aaij:2012ac}. The gray area indicates the points compatible with the latter constraint when the $2\sigma$ error associated to the SM prediction is included.}
\label{higgsversusleptons}
\end{figure} 

In Fig.\,\ref{higgsversusleptons}  we show the spectrum of un-colored particles as a function of the lightest Higgs mass, including neutralinos, charginos, sleptons and 
the rest of the Higgs fields.  The region to the left  of the vertical dashed line is again excluded since  BR$(b\to s\gamma)<2.85\times 10^{-4}$. On the other hand, the $(B_s\to\mu^+\mu^-)$ data forces again the spectrum of uncolored particles to relatively large values. \\

The hierarchy in the sparticle mass pattern is a quick way of classifying a supersymmetric model and understanding the kind of signals it may give rise to in LHC. Several structures have been identified (see Ref.\,\cite{Feldman:2008hs} and references therein) that can originate from the CMSSM or non-universal supergravity scenarios. 
In our case, the model is very close to the CMSSM in the coannihilation region but further constrained. As a consequence, the resulting hierarchy in the supersymmetric spectrum is a very specific one. More specifically, the five lightest supersymmetric particles display the following structure:
\begin{eqnarray} 
\tilde{\chi}_1^0<\tilde{\tau}_1< \tilde{\chi}_2^0\approx\tilde{\chi}_1^\pm <\tilde l_R&{\rm for} & m_h<120\,{\rm GeV}
\,,\nonumber\\
\tilde{\chi}_1^0<\tilde{\tau}_1<\tilde l_R< \tilde{\chi}_2^0\approx\tilde{\chi}_1^\pm &{\rm for} & m_h>120\,{\rm GeV}
\,.\nonumber
\end{eqnarray}
These scenarios are analogous to mSP6 and mSP7, respectively, in Ref.\,\cite{Feldman:2008hs}.
The change of pattern is difficult to appreciate in Fig.\,\ref{higgsversusleptons}, since the mass difference between $\tilde l_R$ and the second-lightest neutralino is small (of order 10~GeV). Also, the mass difference between the second lightest neutralino and the lightest chargino is merely a fraction of a GeV. \\

The almost identical values of the masses of $\chi_2^0$ and $\chi_1^\pm$ is expected since both fields are mostly Winos. 
On the other hand the degeneracy with the ${\tilde l}_R$ fields is a peculiarity of the structure of soft terms in this model. Indeed the 
weak scale masses for these fields have the structure
\beqa
M_{\chi_2^0}^2&  \simeq & \left(\frac{\alpha_2(M_Z)}{\alpha(M_s)}\right)M^2\simeq 0.64\, M^2\,, \\ \nonumber
m_{{\tilde l}_R}^2& \simeq & m^2\ +\ 0.15\,M^2\simeq 0.65\, M^2\,,
\eeqa
where in the second equation the boundary condition $m=M/\sqrt{2}$, characteristic of the present model,  has been used. 
From Fig.\,\ref{higgsversusleptons} we see that the lightest charged sparticle is a  stau, with a mass in between
200 and 550~GeV. The lightest slectrons and charginos are in the region 400 to 1000~GeV.  The remaining Higgs fields will
be heavy, in the 700 to 1600~GeV range. Thus there is a good chance to produce weakly interacting charged sparticles 
in a linear collider.\\

\begin{figure}[t!]
\hspace*{-0.6cm}
\centering
\includegraphics[width=11.cm, angle=0]{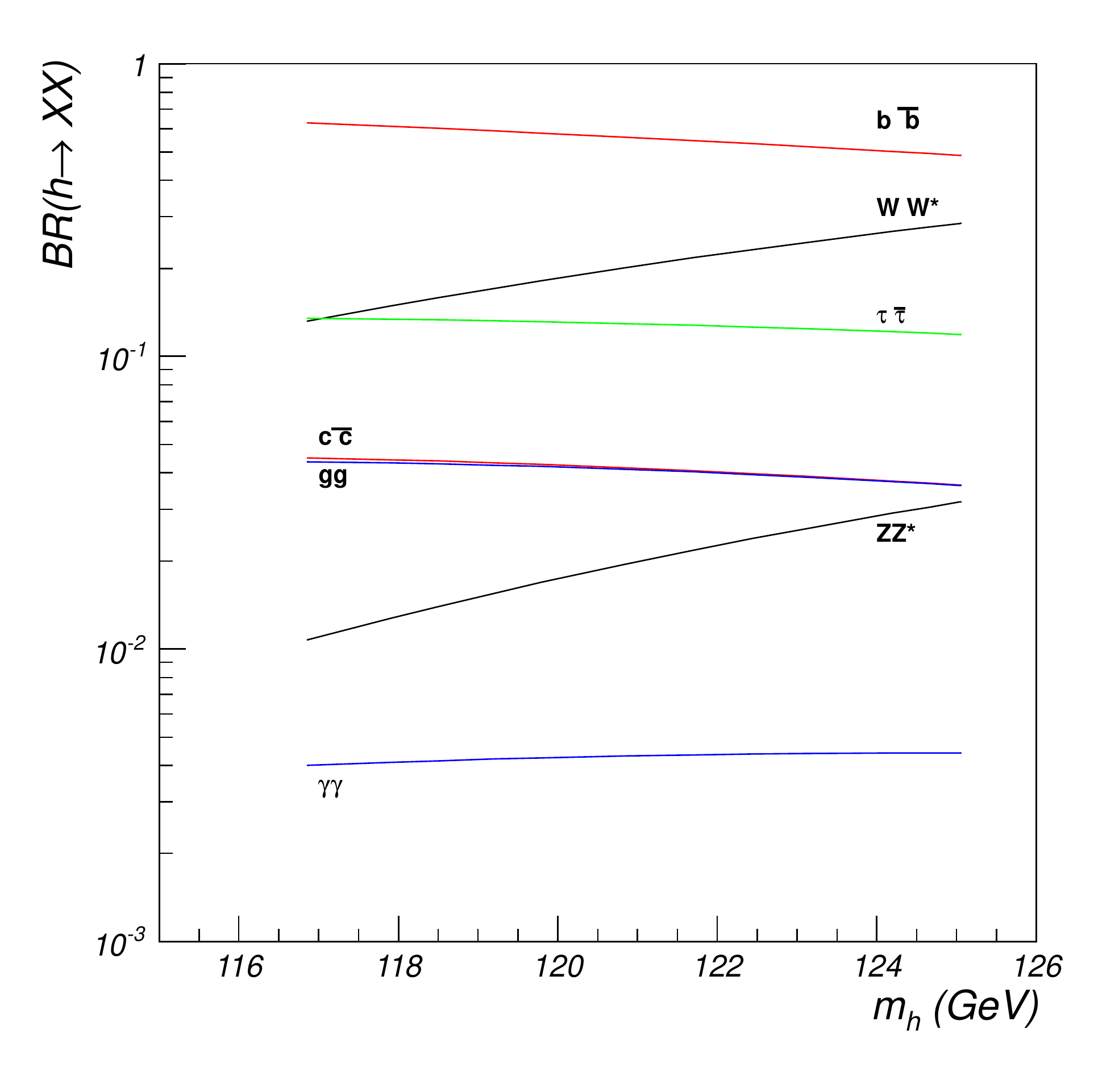}
\caption{Branching fractions for the decay of the lightest CP-even Higgs as a function of its mass.}
\label{plotHiggsbr}
\end{figure} 

For completeness we also show in Fig.\,\ref{plotHiggsbr} the branching ratios of the different decay modes of the lightest CP-even Higgs, computed using code {\tt SPheno\,3.0}, as a function of its mass in this construction. The composition of the lightest Higgs is very similar to that in the CMSSM and therefore these results are quite standard. The leading decay mode is $b\bar b$ although the contribution from $WW$ is almost comparable for large Higgs masses. \\

\begin{figure}[t!]
\hspace*{-0.6cm}
\centering
\includegraphics[width=11.cm, angle=0]{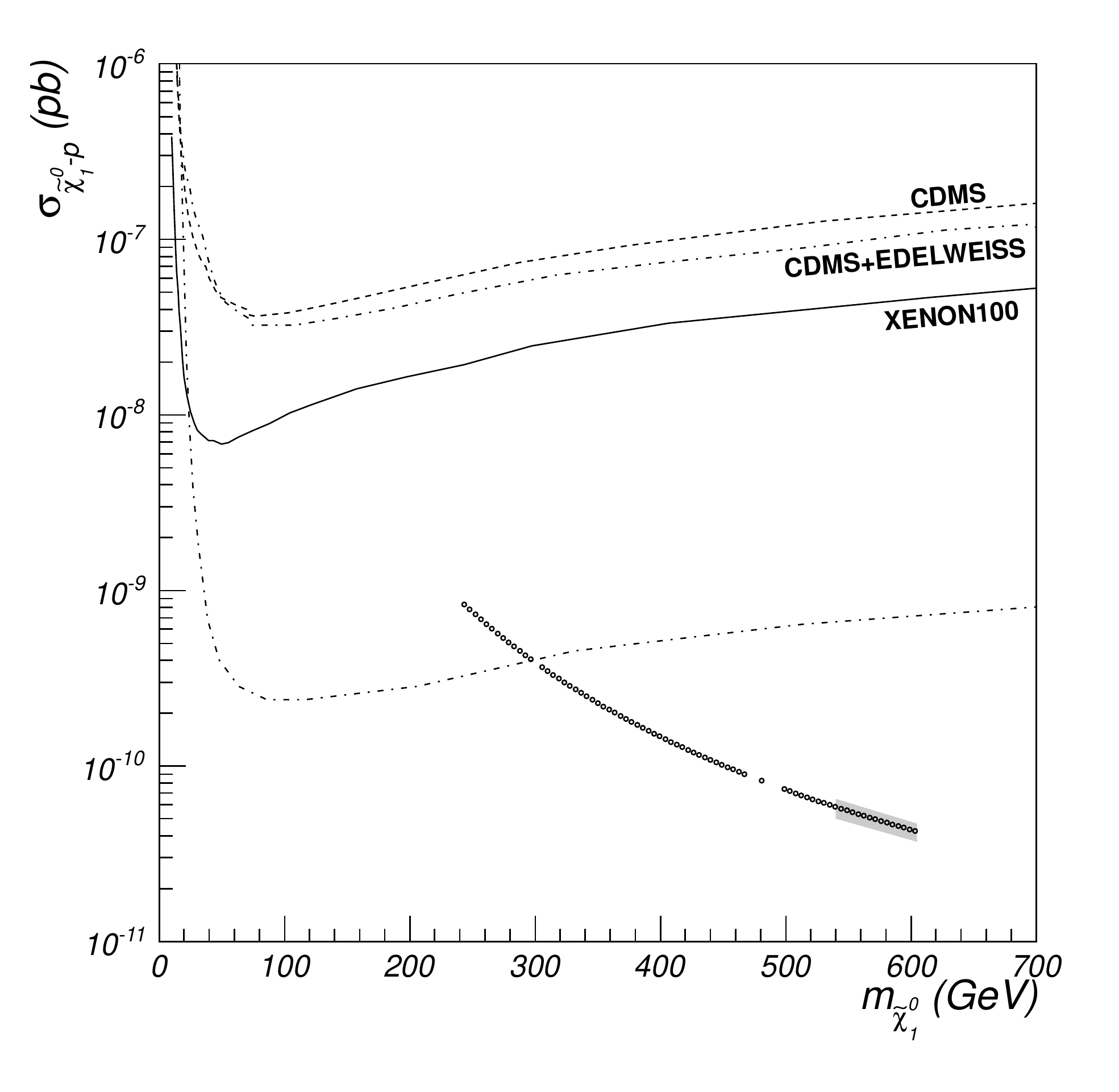}
\caption{Spin-independent part of the neutralino-proton cross section as a function of the neutralino mass for points reproducing the WMAP relic abundance and in agreement with all the experimental constraints. The current sensitivities of the CDMS \cite{Ahmed:2009zw}, CDMS combined with EDELWEISS \cite{Ahmed:2011gh} and XENON100 \cite{Aprile:2011hi} experiments are displayed by means of dashed, dot-dashed and solid lines, respectively. The dotted line represents the expected reach of a 1 ton experiment. The gray area indicates the points compatible with the LHCb constraint when the $2\sigma$ error associated to the SM prediction is included.}
\label{sigsi}
\end{figure}

Let us finally address the direct detectability of dark matter neutralinos in this construction. We show in Fig.\,\ref{sigsi} the theoretical predictions for the spin-independent contribution to its elastic scattering cross section off protons, $\crosssec$, as a function of the neutralino mass, together with current experimental sensitivities from the CDMS (showing also the combination of its data with those from EDELWEISS) and XENON detectors.  
After imposing all the experimental constraints, this scenario predicts $10^{-9}$~pb$\gsim\crosssec\gsim5\times10^{-11}$~pb. This is far from the reach of current experiments. 
Next generation experiments with targets of order 1 ton would be able to probe only 
a portion of the parameter space, corresponding to neutralino masses lighter than 300~GeV (and therefore to Higgses as heavy as approximately 121~GeV). This was to be expected, as these results are typical of the CMSSM in the coannihilation region. 

\section{Detectability at the LHC}

\subsection{Jets and missing transverse energy}

Having already described the SUSY spectrum, let us now address the detectability of this construction at the LHC. In the light of the current and predicted status of the collider, we will consider three possible configurations, with energies of $\sqrt{s}=7,\, 8$ and 14~TeV. 
Our goal is to determine the potential reach of the LHC as a function of the luminosity. In order to do so we have performed a Monte Carlo simulation for the different points in the viable parameter space.\\

As we commented in the previous sections, the SUSY spectrum is calculated for each point using a modified version of {\tt SPheno\, 3.0}. The output, written in Les Houches Accord format, is directly linked to a Monte Carlo event generator. We have used {\tt PYTHIA 6.400} \cite{Sjostrand:2006za} to this aim, linked with {\tt PGS 4} \cite{PGS}, which simulates the response of the detector and uses {\tt TAUOLA} \cite{Jadach:1993hs} for the calculation of tau branching fractions.\\

We include the main sources for SM background, taking into account the production of $t\bar t$ and $WW/ZZ/WZ$ pairs, as well as $W/Z$+jets. The latter give the main contribution \cite{cmshadro, Aad:2011ib} to the background at the relevant energies. 
The production cross sections for these different processes are summarised in Table\,\ref{nlosigma}.
For the production of $W/Z$+jets we have taken the results provided by {\tt PYTHIA}.
\footnote{Calculations of this quantity at the NLO can be found in e.g., Refs.\,\cite{Berger:2009ep, Melnikov:2009wh}. The uncertainty of the result using {\tt PYTHIA} compared with current data and other simulators can be found in Ref.\,\cite{:2010pg}.}\\

\begin{table}[h!] \footnotesize
\renewcommand{\arraystretch}{1.50}
\begin{center}
\begin{tabular}{|c||c|c|c|}
\hline 
& 7~GeV & 8~TeV & 14~TeV\\
\hline
$\sigma_{t\bar t}^{NLO}$	&$152_{-19-9}^{+16+8}$~pb 		&$250$~pb $(^*)$				&$852_{-93-33}^{+91+30}$~pb\\
\hline
$\sigma_{WW}^{NLO}$ 	&$47.04_{-3.2\%}^{+4.3\%}$~pb	&$57.25_{-2.8\%}^{+4.1\%}$~pb	&$124.31_{-2.0\%}^{+2.8\%}$~pb\\
$\sigma_{W^+Z}^{NLO}$	&$11.88_{-4.2\%}^{+5.5\%}$~pb	&$14.48_{-4.0\%}^{+5.2\%}$~pb	&$31.50_{-3.0\%}^{+3.9\%}$~pb\\
$\sigma_{W^-Z}^{NLO}$ 	&$6.69_{-4.23\%}^{+5.6\%}$~pb	&$8.40_{-4.1\%}^{+5.4\%}$~pb		&$20.32_{-3.1\%}^{+3.9\%}$~pb\\
$\sigma_{ZZ}^{NLO}$ 	&$6.46_{-3.3\%}^{+4.7\%}$~pb		&$7.92_{-3.0\%}^{+4.7\%}$~pb		&$17.72_{-2.5\%}^{+3.5\%}$~pb\\
\hline
$\sigma_{W+jets}^{LO}$	&$1.46\times10^5$~pb			&$1.74\times10^5$~pb			&$3.50\times10^5$~pb\\
$\sigma_{Z+jets}^{LO}$	&$6.76\times10^4$~pb			&$7.98\times10^4$~pb			&$1.57\times10^5$~pb\\
 \hline
\end{tabular} 
\caption{\small  Cross sections for the production of $t\bar t$ \cite{Kidonakis:2010dk} and $WW/ZZ/WZ$ \cite{Campbell:2011bn} pairs, as well as $W/Z$+jets. $(^*)$ Rough estimate obtained from the data of Ref.\,\cite{Kidonakis:2010dk}.}
\label{nlosigma}
\end{center}
\end{table}

The production cross section of Supersymmetric particles has been computed using {\tt Prospino 2.1} \cite{Beenakker:1996ed}, which provides the result at NLO. The leading contributions obviously comes from the production of coloured sparticles, $\tilde g\tilde g,\,\tilde g\tilde q,\,\tilde q\tilde q$. The actual values are a function of the gluino and squark masses and have been calculated for each specific case. \\

\begin{figure}[t!]
\hspace*{-0.6cm}
\centering
\includegraphics[width=7.7cm, angle=0]{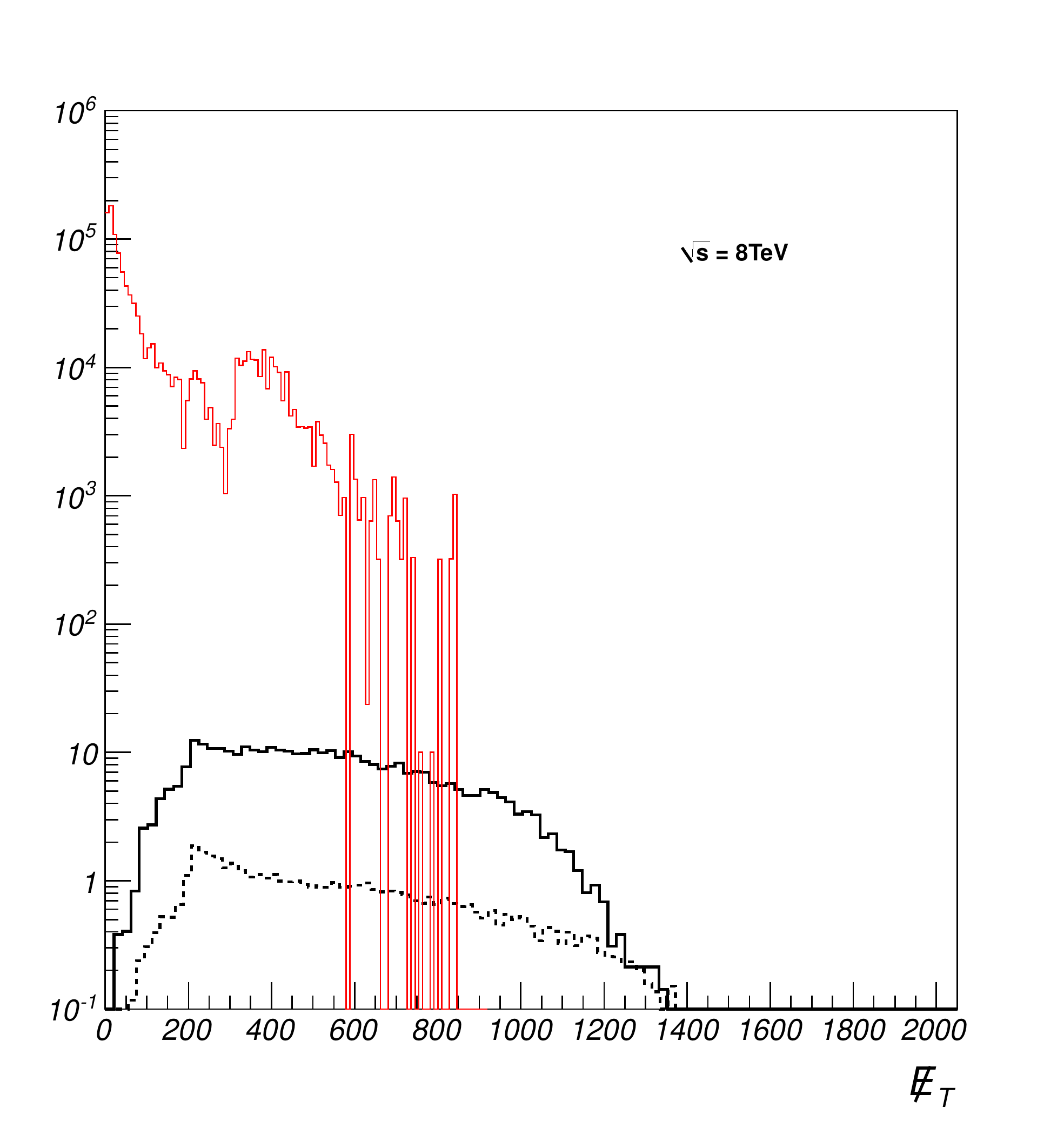}
\includegraphics[width=7.7cm, angle=0]{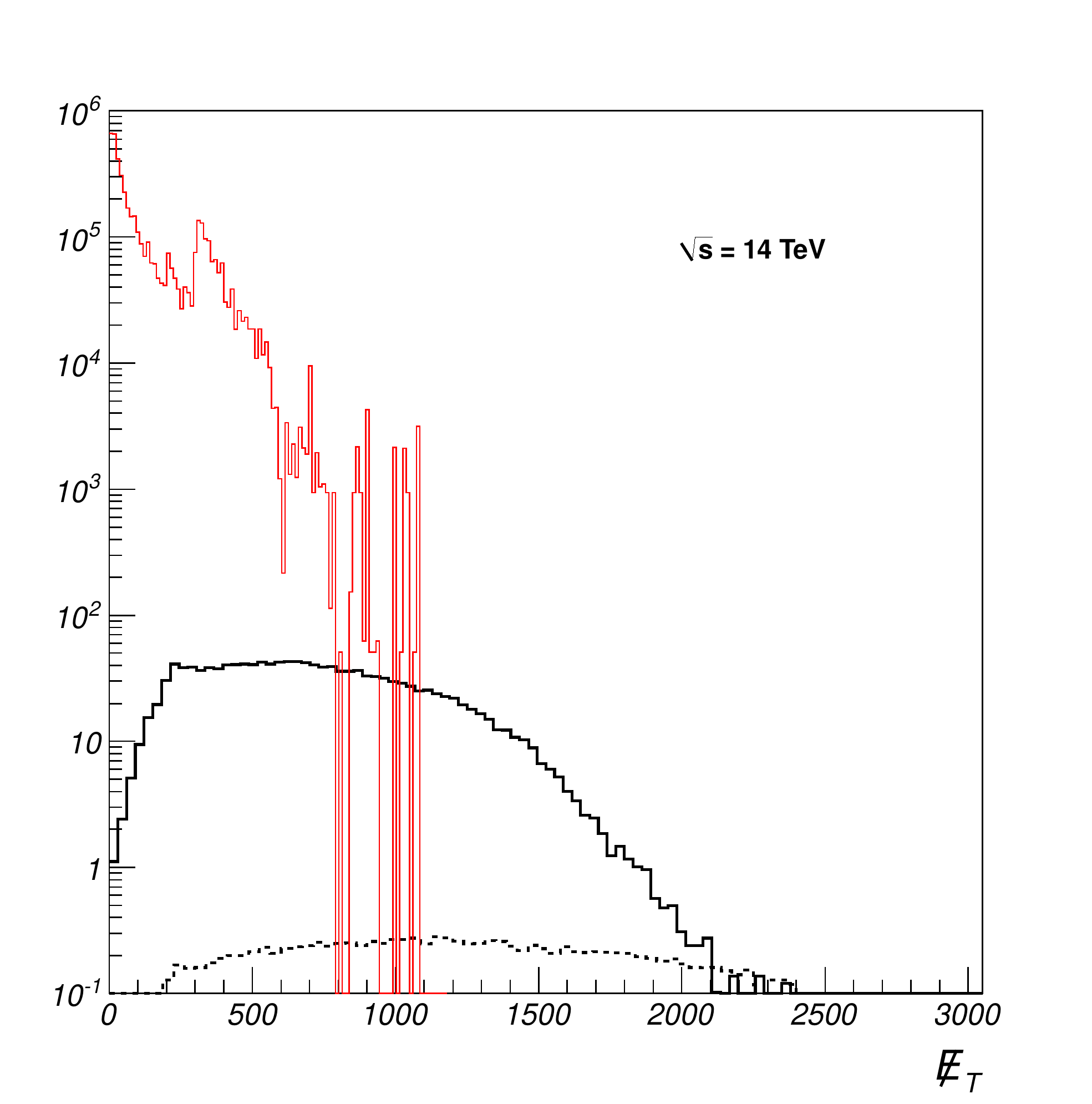}
\caption{Missing energy histogram for the SM background (in red) and SUSY signal in the model. On the left hand-side we assume $\sqrt{s}=8$~TeV and a luminosity of $20$~fb$^{-1}$ and simulate the signal for $M=570$~GeV  (solid line) and $M=700$~GeV (dashed line).
On the right hand-side we assume $\sqrt{s}=14$~TeV and a luminosity of $30$~fb$^{-1}$ and simulate the signal for $M=800$~GeV  (solid line) and $M=1400$~GeV (dashed line).}
\label{plotmet}
\end{figure}

In order to determine the LHC discovery potential we have studied the simplest signal, consisting on the observation of missing transverse energy, $\met$, accompanied by a number ($n\ge 3$) of jets. 
We have used Level 2 triggers in PGS, but supplemented these with additional conditions on the eligible events. Namely, we have implemented the following selection cuts, mimicking those used by the ATLAS Collaboration:
\begin{itemize}\itemsep=0.4ex
 \item[-] Leading jet $P_T$  $>$ 130 GeV,
 \item[-] Second jet $P_T$ $>$ 40 GeV,
 \item[-] Third jet $P_T$ $>$ 40 GeV, 
 \item[-] $m_{eff}$ $>$ 1000 GeV,
\end{itemize}
where $m_{eff}\,=\,E_{T}^{miss} + P_{T}^{jets}$ is calculated from the three leading jets defining the region.
Fig.\,\ref{plotmet} shows a series of histograms for the missing energy resulting from the SM background (red line) and the expected signal events for several examples in the parameter space. In particular, choosing $\sqrt{s}=8$~TeV and a luminosity of 20~fb$^{-1}$ we display the expected signal in our model when $M=570$~GeV and $700$~GeV. Similarly, for  $\sqrt{s}=14$~TeV and a luminosity of 30~fb$^{-1}$ the predictions for the cases $M=800$~GeV and $1400$~GeV are shown.\\

As we can see, the signal dominates over the background above a given value of the missing energy
with a slight dependence on $M$. The actual number of events obviously depends on the luminosity. 
Given a number of signal events $N_s$ and background events $N_b$ that satisfy our series of cuts, a  statistical
 condition for observability may be defined as
\beq
\frac {N_s}{\sqrt{N_b}}>4 \ ,\  \frac {N_s}{N_b}>0.1 \ ,\ N_s>5\,.
\label{statcondition}
\eeq
It is customary to set a fixed cut for the missing energy in order to determine these numbers, however we have implemented an adaptive method which estimates the optimal value for the cut in $\met$ {\em for each value of $M$}. The idea is to maximize the signal-to-background ratio while guaranteeing that the number of signal events is enough ($N_s>5$).
In particular, if the spectrum is heavy and the signal is expected to be centered around a larger $\met$ then the cut in $\met$ can generally be increased so as to reduce the number of background events as long as the number of signal events is above critical. The latter obviously depends on the luminosity. \\

\begin{figure}[t!]
\hspace*{-0.6cm}
\centering
\includegraphics[width=7.7cm, angle=0]{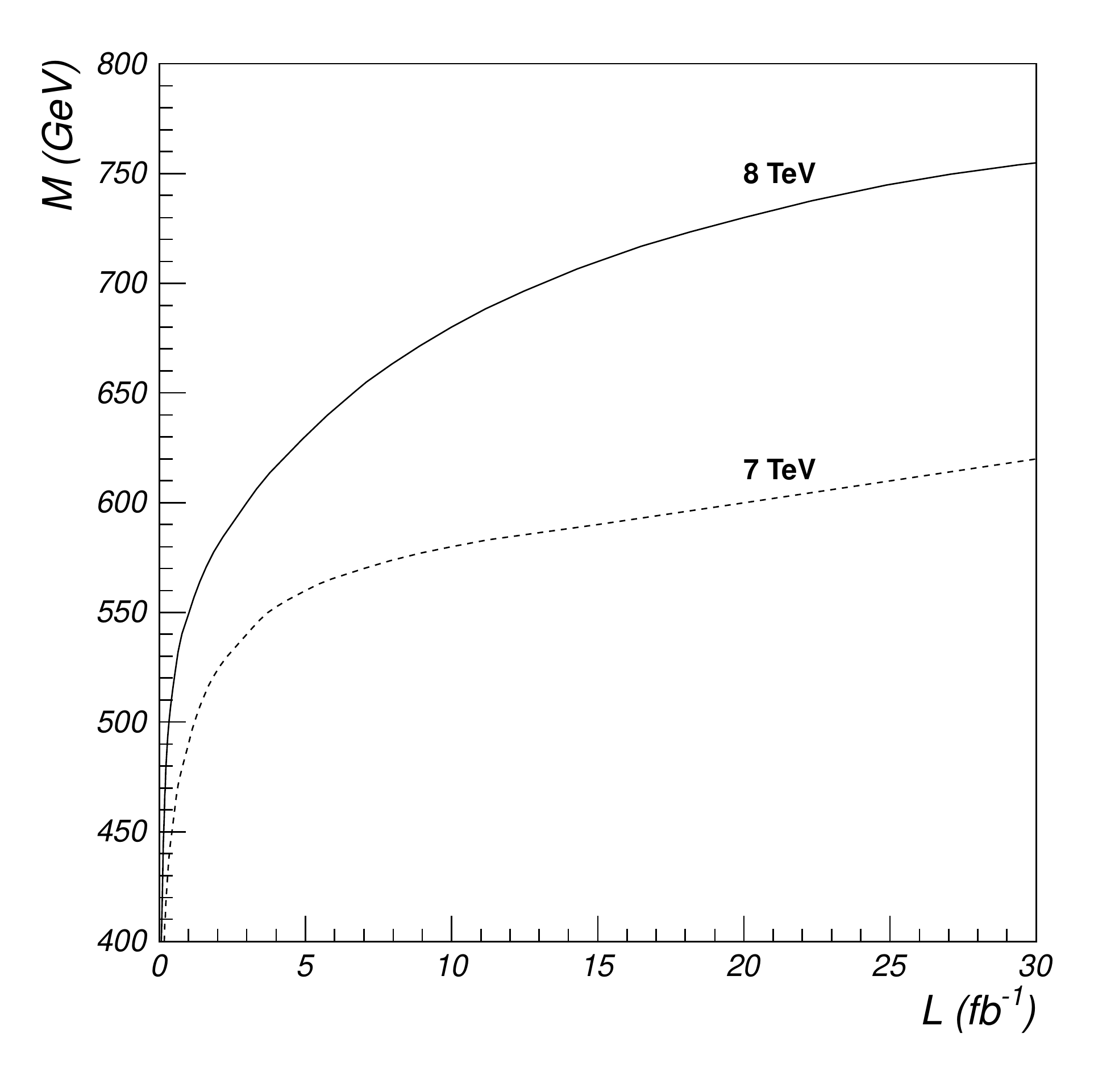}
\includegraphics[width=7.7cm, angle=0]{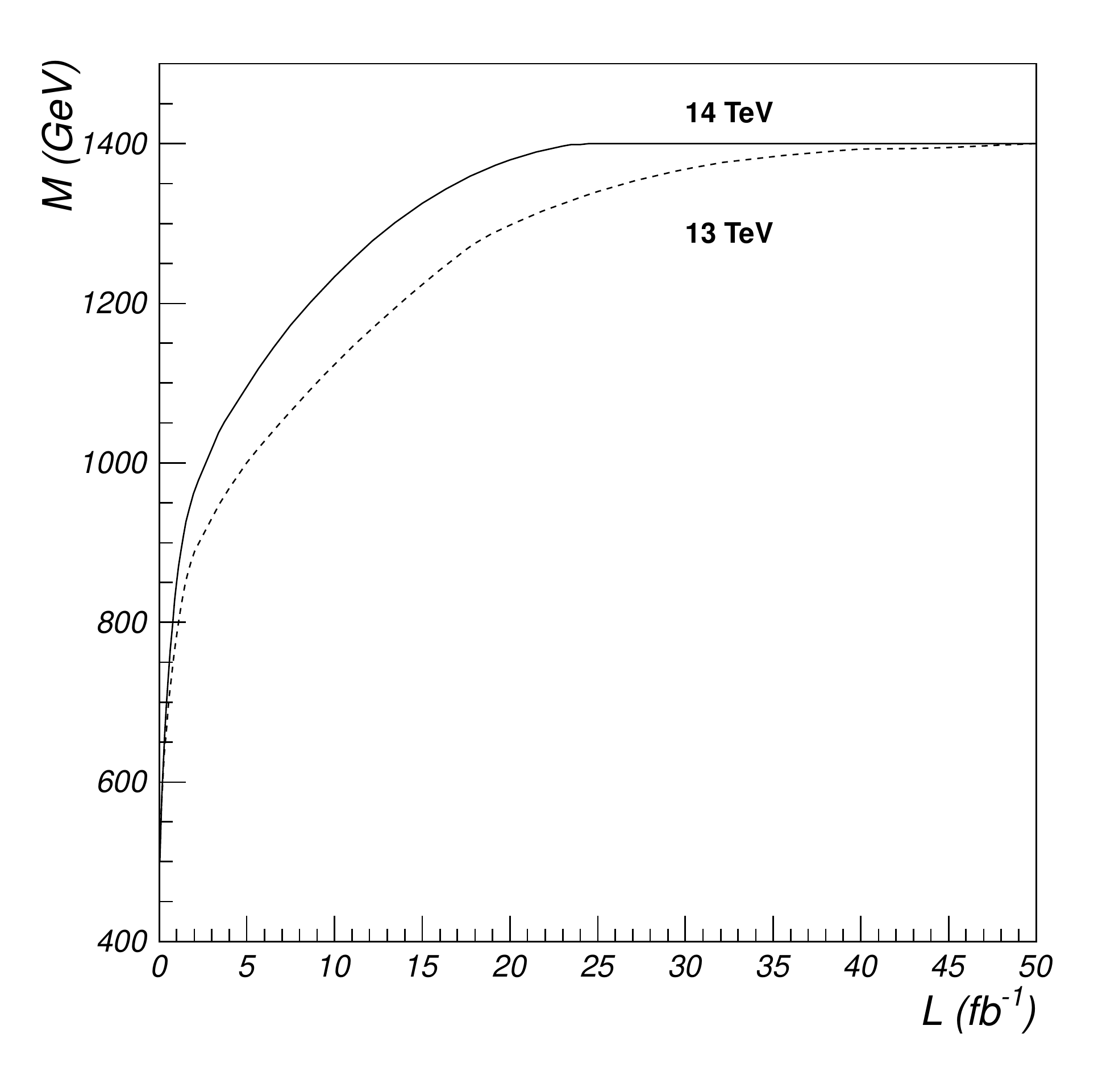}
\caption{Maximum value of $M$ that can be explored at the LHC with $\sqrt{s}=7,\,8$~TeV (left hand-side) and $\sqrt{s}=13,\,14$~TeV (right hand-side) as a function of the luminosity.}
\label{plotluminosity}
\end{figure} 

Using this "adaptive cut" in $\met$ we have determined, for each given value of the luminosity (and for each LHC energy configuration), the maximum value of $M$ for which the number of signal events satisfies condition (\ref{statcondition}). This is, we have calculated the detectability potential of LHC for this specific model. 
The results are displayed in Fig.\,\ref{plotluminosity}, where the maximum value of $M$ is plotted as a function of the luminosity. Operating at $\sqrt{s}=7,\,8$ and $13,\,14$~TeV, LHC will be able to test this scenario up to $M\approx  600,\,750$ and $1400$~GeV, respectively, with a luminosity of $20,\,30$ and $30,\,50$~fb$^{-1}$. 
In fact, the LHC at 14~TeV would be able to explore regions of the parameter space with a larger $M$ than the one displayed in the plot. However, as shown in the previous chapters, there is actually no point of the parameter space above that value for which REWSB and dark matter conditions are fulfilled, and for that reason the line flattens at $M=1400$~GeV.\\

In order to check the validity of our "adaptive cut" in $\met$ we have applied it to the CMSSM and compared the resulting predicted reach with those obtained by the ATLAS \cite{Aad:2011ib} and CMS \cite{cmshadro} collaborations for the same signal. We have obtained a similar reach. Remember in this sense that ATLAS and CMS use a given value for the cut in $\met$ at low masses and a larger value for heavier masses.

\subsection{Other signatures}

As we described in chapter 3, the viable regions of the model correspond to the coannihilation region in which the lightest neutralino and ligtest stau mass are almost degenerate. This class of scenarios has received a lot of attention in the literature \cite{Arnowitt:2001yh, Arnowitt:2006jq, Feldman:2007zn}, since they can give rise to very characteristic signals.
In particular, the following decay chain is dominant for the second-ligest neutralino, $\tilde\chi_2^0\to\tau\tilde\tau_1\to\tau\tau\tilde\chi_1^0$, leading to signals characterised by multiple low energy tau leptons \cite{Arnowitt:2006jq}. In particular, one can search for pairs of opposite sign taus, accompanied by a number of jets, which would be relatively abundant, compared to other characteristic SUSY signals \cite{Feldman:2007zn}.\\

Finally, as we can observe in Fig.\,\ref{plotHiggs1}, the region with larger values of the Higgs mass is precisely that with a smaller mass-splitting between the stau and the lightest neutralino. 
In fact, for Higgs masses above $m_h>124.5$~GeV, for which the recent LHCb constraint on BR($B_s \to \mu^+ \mu^-$) is satisfied,
one finds $m_{\tilde\tau_1}-m_{\tilde\chi_1^0}<1.7$~GeV. This implies that the two body decay $\tilde\tau_1\to\tilde\chi_1^0\tau$ is no longer kinematically allowed and the stau has to undergo three or four body decays ($\tilde\tau_1\to\tilde\chi_1^0\nu_\tau\pi$ or $\tilde\tau_1\to\tilde\chi_1^0\mu\nu_\mu\nu_\tau$). This increases significantly its lifetime which is now larger than $10^{-7}$~s \cite{Jittoh:2005pq}. 
The presence of long-lived staus in the Early Universe has appealing implications for Big Bang Nucleosynthesis (BBN). The stau can form a bound state with nuclei leading to a catalytic enhancement of certain processes (in particular, $^6$Li production) \cite{Pospelov:2006sc}. 

This long-lived stau's provides an interesting possibility, the observation of a stable charged particle in the LHC (due to its lifetime, the stau would decay already outside the detector) \cite{Khachatryan:2011ts, Aad:2011hz}.
Notice that staus in these regions have a mass of order 600~GeV, therefore satisfying the
current bounds for long-lived charged particles obtained in ATLAS (at $\sqrt{s} = 7$~TeV and with a luminosity of 37~pb$^{-1}$), which impose $m_{\tilde\tau_1}>135$~GeV at 95\% CL \cite{Aad:2011hz}. \\

%% file: Yukawas.tex
\chapter{Flux and instanton corrections to Yukawa couplings in local F-theory models}\label{capyuk}

In this chapter we turn to a different question in the phenomenology of Type IIB / F-theory compactifications. We try to adress the issue of hierarchical fermion masses in these large classes of compactifications \cite{afim}.

\section{The problem of rank one Yukawa matrices}
As we said, Yukawa couplings in Type IIB string theory are given by the the integral of the producct of the zero modes of the Dirac equations of the wave functions in extra dimensions. The computation of Yukawa couplings is done in two steps. First we compactify the $10$-dimensional theory down to $d=4$ such that we can express the wavefunctions as a product of the $4d$-wavefunctions and the extra-dimension $6d$-wavefunctions. The second step is to compute the overlap integrals of the three different wavefunctions over the compactified dimensions. As a result we will obtain a parameter (or a matrix of parameters) multiplying the product of the 4-dimensional wavefunctions corresponding to Higgs and fermion fields.\\

In general those wavefunctions are only known in some simple examples, like toroidal orientifolds. For instance, in ${\bf T^6}/{\bf Z_2}\times {\bf Z_2}$ orientifolds  with magnetized D7-branes one can construct semirealistic models  \cite{ms04} in which the Yukawas can be explicitly computed by integration of three overlapping wavefunctions. In these examples the resulting mass matrices have rank one, corresponding to a single massive quark/lepton generation \cite{cim2}. This implies that we need further effects like string instantons in order to yield masses for the lightest generations \cite{ag06}.\\

In the context of intersecting D7-brane models and their non-perturbative extension in F-theory, bifundamental matter fields reside at pairs of 7-branes intersecting in Riemann curves, and Yukawa couplings appear locally at those points where three of these curves intersect. For the case  of local F-theory GUT models that we reviewed in \Sec{fthguts}, in which there is a 4-cycle $S$ on which the GUT 7-branes wrap, the matter fields reside at matter curves $\Sigma_i \subset S$ at which the GUT symmetry is enhanced. In addition, these curves intersect at points at which the symmetry is further enhanced, and which give rise to Yukawa couplings among fields in the matter curves (see \Fig{flocal1}). \\

Since the wavefunctions corresponding to matter fields are localized along the matter curves, the internal 6d wavefunctions of the zero modes are peaked on these curves. Therefore the Yukawa coupling at the intersecting point is expected to depend only on the data around the neighborhood of this intersection point, and not much on the full structure of the compact space. That allows us to compute the Yukawa couplings in the F-theory context in the same way as in Type IIB case mentioned above, i.e. in terms of the overlapping integral over $S$ of the three internal wavefunctions corresponding to three intersecting matter curves \cite{bhv1, bhv2} \cite{fi1, hv08, hktw, fi09, cp09, cchv09, hktw2}\footnote{See \Ref{Leontaris:2010zd} and \Ref{Dudas:2009hu} for other F-theory approaches.}.\\

If we take the $SU(5)$ GUT model, the $SU(5)$ symmetry corresponding to the a 7-brane wrapping the divisor $S$ is enhanced to $SU(6)$ at curves with 5-plets and to $SO(10)$ at curves with 10-plets, as is ilustrated in \Fig{flocal2}. Moreover, assuming that there is only one intersecting point among matter curves for ${\bf 10}\times {\bf {\overline {5}}}\times {\bf {\overline 5}_H}$ and ${\bf 10}\times {\bf { {10}}}\times {\bf { 5}_H}$ $SU(5)$ Yukawa couplings, the resulting Yukawa matrices have rank equal to one \cite{hv08} as we will see later.
Hence only one generation gets massive, and we get the same situation as in the toroidal orientifolds case above mentioned. The first attempt to solve the problem was through the dependence of the Yukawa couplings on the worldvolume fluxes on the 7-branes (i.e., those required for obtaining chirality from these settings). It was thought that they could correct this result and give masses to the rest of quarks and leptons. However these open string fluxes are not enough, since they do not modify the rank of the Yukawa matrices \cite{cchv09},\cite{cp09},\cite{fi09}.  In particular, one can see that the F-term zero mode equations become independent of the worldvolume fluxes in a certain {\it holomorphic gauge} \cite{fi09}  and that, as a consequence, the holomorphic Yukawa couplings remain flux independent.\\

\begin{figure}[ht]
\begin{center}
\begin{tabular}{lcr}
\includegraphics[width=7.5cm]{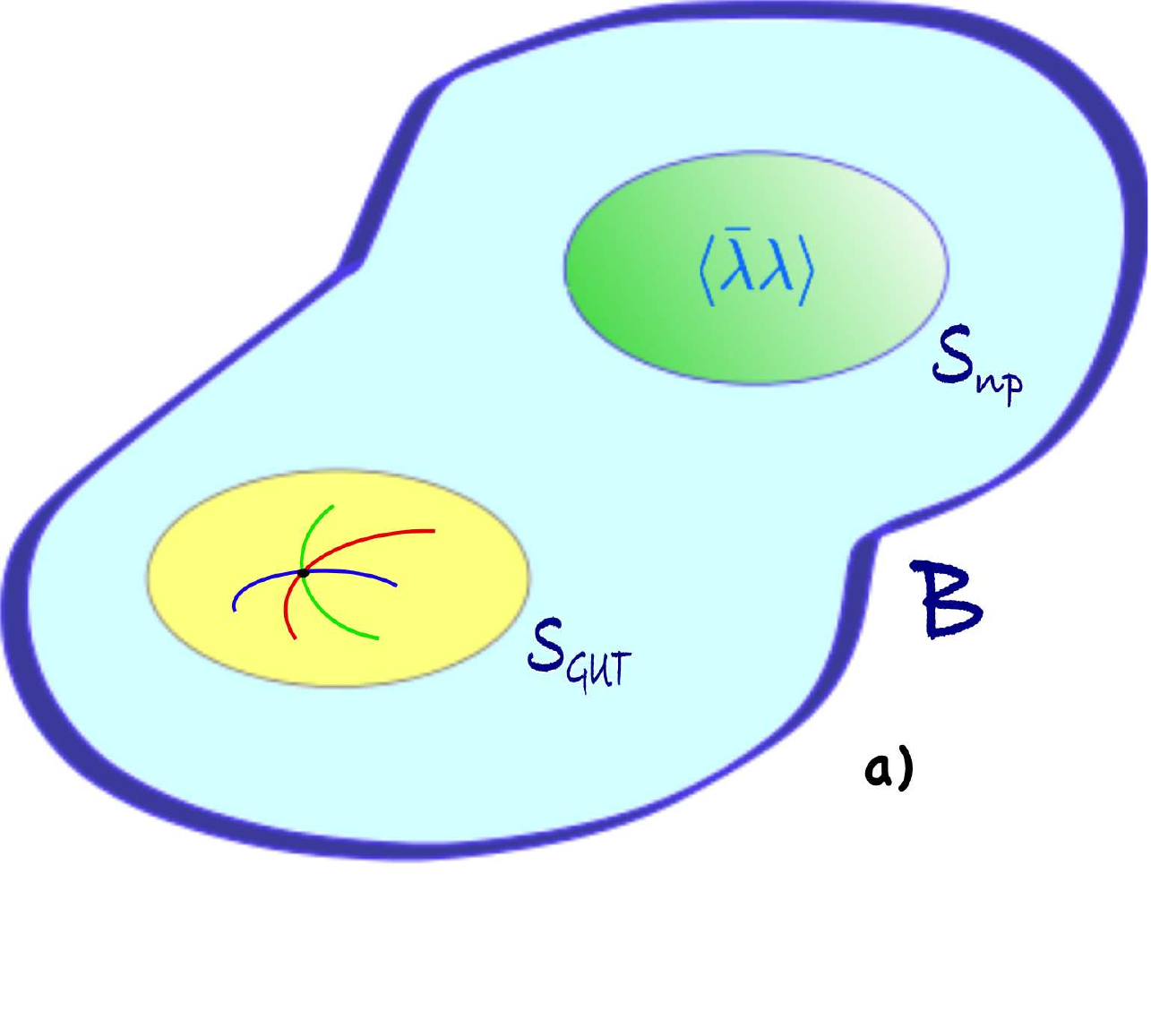}
& \quad &
\includegraphics[width=7.5cm]{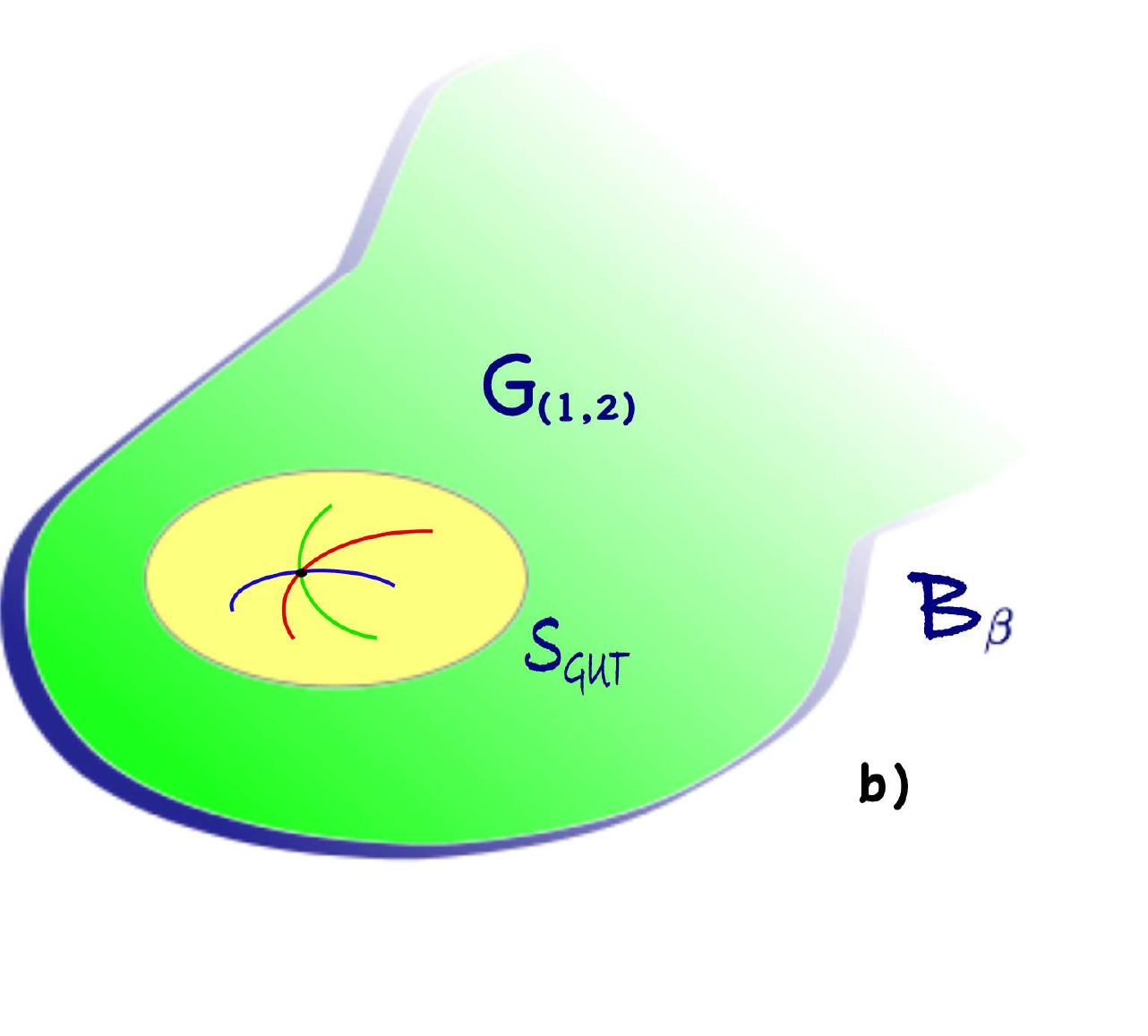}
\end{tabular}
\end{center}
\caption{\small{Sources of corrections to 7-brane Yukawas. Figure $a)$ represents the setup considered in \cite{mm09}, where the Yukawas on a 7-brane stack wrapping the four-cycle $S_{GUT}$ are modified by the gaugino condensate on  7-branes on the distant four-cycle $S_{\rm np}$. Following \cite{km07}, one may identify this setup with the one in figure $b)$, where the non-perturbative sector has been replaced by a $\beta$-deformation of the previous background. This new background contains IASD (1,2) background fluxes that induce a non-commutative deformation on $S_{GUT}$, in the sense of \cite{cchv09}.}}
\label{backr}
\end{figure}

There are however two other possible sources of corrections to the holomorphic Yukawas. The first one corresponds to a non-commutative deformation of the 7-brane gauge theory that can induce corrections to the Yukawas such that the rank of the mass matrix is modified \cite{cchv09}. Such deformation can be generated by placing D7-branes on type IIB backgrounds with closed string IASD fluxes of the $(1,2)$-type, often referred to as $\b$-deformed backgrounds. 
\\

The second possible source is the influence of non-perturbative (instanton or gaugino condensate) effects on distant 4-cycles in the compact manifold \cite{mm09}, see figure \ref{backr}. Although these two proposals look quite different they lead to similar physics and it has been pointed out that they should be equivalent because instanton and gaugino condensate effects generate IASD $(1,2)$ fluxes on the theory \cite{bdkkm10} (see also \cite{km07}). 

\section{Wavefunctions and Yukawa couplings in local F-theory models \label{general}}

In this section we describe the standard computation of wavefunctions and Yukawas for local F-theory models in the absence of any non-perturbative or non-commutative deformation, illustrating the general computation by means of an explicit $U(3)$ toy model. 

\subsection{Local F-theory models from intersecting 7-branes}

As explained in \Sec{ftheory} In local F-theory models 7-branes wrap on a compact divisor $S$ of the threefold base $B$ of an elliptically-fibered Calabi-Yau fourfold 
. The gauge group $G_S$ (e.g. $SU(5)$) on such 7-branes is specified by the singularity type of the elliptic fiber on top of the 4-cycle $S$. Such singularity may be enhanced to a higher type on certain complex submanifolds $\Sigma \subset S$ where $\Sigma = S \cap S'$, often called matter curve, is the intersection locus of $S$ with another divisor $S'$ of $B$ where a different set of 7-branes is wrapped. If we associate $G_{S'}$ and $G_{\Sigma}$ gauge groups to $S'$ and $\Sigma$ respectively, then it is easy to see that $G_S \times G_{S'} \subset G_{\Sigma}$. As in the case of intersecting D7-branes matter fields charged under the gauge group $G_S \times G_{S'}$ are located at the intersection locus $\Sigma = S \cap S'$. Similarly, an enhancement on a point $p \in S$ happens whenever $p$ is the intersection locus of two or more of these matter curves. In particular, Yukawa couplings between three matter fields charged under $G_S$ would be realized at the triple intersections of matter curves, and $G_S$ would be identified with the GUT gauge group, see \Fig{flocal2}.\\

The dynamics governing the above construction can be encoded in the 8d effective action described in \cite{bhv1} which, upon dimensional reduction on the 4-cycle $S$, provides the dynamics of the 4d degrees of freedom.
In particular, the Yukawa couplings between 4d chiral fields arise from the superpotential
\be
W\, =\, m_*^4 \int_{S} \tr \left( F \wedge \Phi \right)
\label{supo7}
\ee
where $m_*^4$ is the F-theory characteristic scale, $F = dA - i A \wedge A$ is the field strength of the gauge vector boson $A$ arising from a stack of 7-branes, and $\Phi$ is a (2,0)-form on the 4-cycle $S$ describing its transverse geometrical deformations. Locally, we can take both $A$ and $\Phi$ to transform in the adjoint of the non-Abelian gauge group $G_p \supset G_S$ associated to the enhanced singularity at the Yukawa point $p$. This initial gauge group is broken by the fact that $\Phi$ and $A$ have a non-trivial profile, and so the actual gauge group is the commutant of $H$ in $G_p$, with $H$ the subgroup generated by $\langle \Phi \rangle$ and $\langle A \rangle$.\\

We can describe separately the effect of $\langle \Phi \rangle$ and $\langle A \rangle$ by assuming that $[\langle \Phi \rangle, \langle A \rangle] = 0$ so that  $H = H_{\Phi} \times H_F$. On one hand, the effect of $\langle \Phi \rangle$ is to describe the system of intersecting divisors considered above, so that $G_\Phi = [H_{\Phi}, G_p] = G_S \times \prod_i G_i$, with $G_i$ the gauge groups associated to 7-branes wrapping the divisors $S_i$ intersecting $S$ on $\Sigma_i$. In particular, for a generic point of $S$ the rank of $\langle \Phi \rangle$ is given by ${\rm rank\, } \langle \Phi \rangle = {\rm rank\, } G_p - {\rm rank\, } G_S$, while it decreases to ${\rm rank\, } G_p - {\rm rank\, } G_{\Sigma_i}$  on top of the matter curve $\Sigma_i$ and vanishes at $p$. On the other hand, the effect of $\langle A \rangle$ is to provide a 4d chiral spectrum and to further break the GUT gauge group $G_S$ down to the subgroup $[H_F, G_S]$, as it is usual in  compactifications with magnetized D-branes \cite{cim2, japan, ConlonWF, DiVecchia, cm09}. Hence, one may obtain a 4d MSSM spectrum from the above construction by first engineering the appropriate $SU(5)$ GUT 4d chiral spectrum via $\langle \Phi \rangle$ and an $\langle A \rangle$ which commutes with $G_S$, and then turn on an extra component of $\langle A \rangle$ along the hypercharge generator in order to break $G_S \raw G_{MSSM}$ \cite{bhv2}.

\subsection{Zero and massive modes at the intersection}\label{zerozero}

The dynamics corresponding to these construction is given by the spectrum of 4d zero modes as a set of internal wavefunctions along $S$, and the couplings between these 4d modes in terms of overlapping integrals of such wavefunctions. All that information is encoded in the superpotential (\ref{supo7}) together with the D-term for $S$,
\be
D\, =\, \int_S \omega \wedge F + \frac{1}{2}  [\Phi, \bar{\Phi}]
\label{FI7}
\ee
where $\omega$ is the fundamental form of $S$.\\

\noindent By variating $A$ and $\Phi$ in the superpotential (\ref{supo7}), one obtains the F-term equations
\begin{subequations}
\label{Fterm7}
\begin{align}
\label{Fterm7A}
\bar{\p}_A \Phi  & =  0\\
\label{Fterm7phi}
F^{(0,2)}  & =  0
\end{align}
\end{subequations}
where $\bar{\p}_A = (\p_{\bar{x}} - i A_{\bar{x}})\, d\bar{x} + (\p_{\bar{y}} - i A_{\bar{y}})\, d\bar{y}$ is the anti-holomorphic piece of the covariant derivative operator $D_A= \p_A + \bar{\p}_A$ on the 4-cycle $S$, of local complex coordinates $(x,y)$. In addition, from (\ref{FI7}) we obtain the D-term equation 
\be
\omega \wedge F + \frac{1}{2} [\Phi, \bar{\Phi}]\, =\, 0
\label{Dterm7}
\ee
where in this local coordinate system $\omega$ can be described as
\be
\omega = \frac{i}{2}\left( dx\wedge d\bar{x} +  dy\wedge d\bar{y}\right).
\label{kahlerform}
\ee


On the other hand, one can also obtain the equation of motion for the 7-brane bosonic fluctuations from the above BPS equations. Indeed, by defining
\be
\Phi_{xy}\, =\, \langle \Phi_{xy} \rangle + \varphi_{xy} \quad \quad A_{\bar{m}}\, =\, \langle A_{\bar{m}} \rangle + a_{\bar{m}}
\label{fluctuation}
\ee
As $[\langle \Phi \rangle, \langle A \rangle] = 0$, \Eq{Fterm7phi} implies $\langle \Phi_{xy} \rangle$ is holomorphic in $(x,y)$. If we expand Eqs.(\ref{Fterm7}) and (\ref{Dterm7}) to first order in the fluctuations $(\varphi, a_{\bar{x}}, a_{\bar{y}})$ one obtains
\begin{subequations}
\label{fluct7}
\begin{align}
\label{Ffluct7A}
\bar{\p}_{\langle A\rangle} \varphi + i [\langle \Phi\rangle , a]   & =  0\\
\label{Ffluct7phi}
\bar{\p}_{\langle A \rangle} a  & =  0 \\
\label{Dfluct}
\omega \wedge \p_{\langle A \rangle} a - \frac{1}{2} [\langle \bar{\Phi} \rangle, \varphi]  & = 0
\end{align}
\end{subequations}
where $a = a_{\bar{x}} d\bar{x} + a_{\bar{y}} d\bar{y}$ and $\varphi = \varphi_{xy} dx \wedge dy$. These are indeed the zero mode equations of motion for the bosonic fluctuations as obtained from the 8d action derived in \cite{bhv1}, and which pair up with the zero mode fermionic fluctuations in 4d $\CN=1$ chiral multiplets as $(a_{\bar{m}}, \psi_{\bar{m}})$ and $(\varphi_{xy}, \chi_{xy})$. The equation of motion for the latter degrees of freedom can be obtained from the part of the 8d action bilinear in fermions, and read \cite{bhv1,fi09}
\begin{subequations}
\label{ferm7}
\begin{align}
\label{Fferm7A}
\bar{\p}_{A} \chi + i [\Phi , \psi]  -2i\sqrt{2}\, \omega \wedge \p_A \eta  & =  0\\
\label{Fferm7phi}
\bar{\p}_{A} \psi - i\sqrt{2}\, [\bar{\Phi}, \eta] & =  0 \\
\label{Dferm}
\omega \wedge \p_{A} \psi - \frac{1}{2} [\bar{\Phi}, \chi]  & = 0
\end{align}
\end{subequations}
where for simplicity we have replaced $\langle \Phi \rangle \raw \Phi$ and $\langle A \rangle \raw A$, and have included the fermionic degree of freedom within the gauge multiplet $(A_\mu, \eta)$. The latter set of equations can be expressed in matrix notation as
\be
{\bf D_A} \Psi\, =\, 0
\label{Dirac9}
\ee
where 
\be
{\bf D_A}\, =\, 
\left(
\begin{array}{cccc}
0 & D_x & D_y & D_z \\
-D_x & 0 & -D_{\bar{z}} & D_{\bar{y}} \\
-D_y & D_{\bar{z}} & 0 & -D_{\bar{x}} \\
-D_z & -D_{\bar{y}} & D_{\bar{x}} & 0
\end{array}
\right)
\quad \quad
\Psi\, =\, \left(
\begin{array}{c}
\psi_{\bar{0}} \\ \psi_{\bar{x}} \\ \psi_{\bar{y}} \\ \psi_{\bar{z}}
\end{array}
\right) \, \equiv\, 
\left(
\begin{array}{c}
- \sqrt{2}\, \eta \\ \psi_{\bar{x}} \\ \psi_{\bar{y}} \\ \chi_{xy}
\end{array}
\right)\label{matrixDirac}
\ee
where $D_m = \p_m - i [A_m, \cdot]$, $m=x,y,z$ is the covariant derivative. In order to define $D_{\bar{z}}$ we are identifying $A_{\bar{z}} =   \Phi_{xy}$ and imposing that all fields are $z$-independent, so that ${D}_{\bar{z}} = -i [\Phi_{xy}, \cdot]$.
These identifications arise from relating a system of intersecting D7-branes with a system of magnetized D9-branes by T-duality. In such D9-brane picture \Eq{Dirac9} is nothing but the standard Dirac equation for the fermionic zero modes, ${\bf D_A}$ being the usual Dirac operator. That allows to write down the eigenmode equation for the 7-brane massive modes in a rather simple way. Analogously we can write the eigenmode equation for the 7-brane massive modes with 
\be
{\bf D_A}^\dag {\bf D_A}\, \Psi\, =\, |m_\rho|^2 \Psi
\label{eigenferm}
\ee
where $m_\rho$ is the mass of fermionic mode and   ${\bf D_A}^\dag$ is given by 
\be
{\bf D_A}^\dag\, =\, 
\left(
\begin{array}{cccc}
0 & D_{\bar{x}} & D_{\bar{y}} & D_{\bar{z}} \\
-D_{\bar{x}} & 0 & -D_z & D_y \\
-D_{\bar{y}} & D_z & 0 & -D_x \\
-D_{\bar{z}} & -D_y & D_x & 0
\end{array}
\right)
\ee

\subsection{A $U(3)$ toy model \label{toy}}

We are going to consider now a simple toy model made up of three intersecting D7-branes as an example of the F-theory local model building features shown above. In particular, we are going to consider a $U(3)$ gauge theory on a four-cycle $S$ of local holomorphic coordinates $(x,y)$, and such that the transverse position field $\Phi$ has the vev
\be
\langle \Phi_{xy} \rangle\, =\, 
\frac{m^2_{\Phi_z}}{3}
\left(
\begin{array}{ccc}
1 \\ & 1 \\ & & 1
\end{array}
\right)
{\Phi_0}
+
\frac{m^2_{\Phi_x}}{3}
\left(
\begin{array}{ccc}
-2 \\ & 1 \\ & & 1
\end{array}
\right)
x
+
\frac{m^2_{\Phi_y}}{3}
\left(
\begin{array}{ccc}
1 \\ & 1 \\ & & -2
\end{array}
\right)
y
\label{vevPhi}
\ee
where $m_{\Phi_x}$, $m_{\Phi_y}$ and $m_{\Phi_z}$ are mass scales introduced so that $\Phi_{xy}$ has the usual dimensions of $L^{-1}$. In the following we will assume for simplicity that 
\be
m_{\Phi_x}^2 = m_{\Phi_y}^2 = m_{\Phi_z}^2 \equiv m_{\Phi}^2
\label{mPhi}
\ee
%
 
From (\ref{vevPhi}) it is easy to see that the initial gauge group is broken as $U(3) \raw U(1)^3$ by the effect of $\langle \Phi \rangle$ alone, and there is then a rank two enhancement $U(1) \raw U(3)$ at the point $p = \{(x,y,z) = (0,0,\Phi_0/3)\}$ where the three D7-branes intersect. 
It can be interpreted geometrically as the fact that the three D7-branes of this model wraps a different four-cycle, algebraically specified by
\begin{subequations}
\label{4cycles}
\begin{align}
S_\a \ :  &\quad 3z + 2x - y - \Phi_0 \, =\, 0\\
S_\b \ :  &\quad 3z - x - y - \Phi_0 \, =\, 0\\
S_\g \ :  &\quad 3z - x + 2y - \Phi_0 \, =\, 0
\end{align}
\end{subequations}
that intersect in the following two-cycles of $S =  \{z = \Phi_0/3\}$
\begin{subequations}
\label{2cycles}
\begin{align}
S_\a \cap S_\b \ :  &\quad \Sigma_a \, =\, \{x=0\} \\
S_\b \cap S_\g \ :  &\quad \Sigma_b \, =\, \{y=0\} \\
S_\g \cap S_\a \ :  &\quad \Sigma_c \, =\, \{x=y\}
\end{align}
\end{subequations}

Each of these curves represent a different sector for the fluctuations of a $U(3)$ adjoint field around the intersecting point where $U(3)$ remains unbroken. These fluctuations come from both bosonic fields $(\phi, a_{\bar{x}}, a_{\bar{y}})$ and fermionic fields in the vector $\Psi$ in (\ref{matrixDirac}). In particular, left-handed 4d chiral fermions in the bifundamental will arise from $U(3)$ off-diagonal fluctuations of $\Psi$, that we label as 
\be
{\psi}_{\bar{m}}\,=\, 
\left(
\begin{array}{ccc}
0 & a^+_{\bar{m}} & c^-_{\bar{m}} \\
a^-_{\bar{m}} & 0 & b^+_{\bar{m}} \\
c^+_{\bar{m}} & b^-_{\bar{m}} & 0
\end{array}
\right) \quad \quad \bar{m} = \bar{0}, \bar{x}, \bar{y}, \bar{z}
\label{sectors}
\ee
while their CPT conjugates will be contained in the off-diagonal entries of ${\psi}_m$.\\

If we assume that $\langle \Phi_{xy} \rangle$ lives in the Cartan subalgebra of $G_p$ then $[\langle \Phi \rangle, \langle \bar{\Phi} \rangle ] = 0$ and so eqs.(\ref{Fterm7phi}) and (\ref{Dterm7}) imply that $\langle F \rangle$ is a primitive $(1,1)$-form on $S$. On the other hand, this $\langle F \rangle$ is the non-trivial magnetic flux that we need in order to obtain chirality. As it happened with $\langle A \rangle$, it is choosen such that $[\langle \Phi_{xy} \rangle, \langle F \rangle] = 0$. More precisely, in  our $U(3)$ model, a convenient choice is given by 
\be
\langle F\rangle \, =\,  i \left( M_x\, dx \wedge d\bar{x} + M_y\, dy \wedge d\bar{y}\right) 
\frac{1}{3}
\left(
\begin{array}{ccc}
1 \\ & -2 \\ & & 1
\end{array}
\right)
\label{fluxD7}
\ee
so that the background D-term equation (\ref{Dterm7}) is satisfied by imposing  $M_y + M_x = 0$ pointwise. For simplicity, in the following we will assume that $M_x$ and $M_y$ are constant.\\

In order to derive the chiral spectrum wavefunctions of this toy model let us consider \Eq{eigenferm}. In general we have that
\be
{\bf D_A}^\dag {\bf D_A}\, =\, - \Delta \Id_4\, - i 
\left(
\begin{array}{cccc}
\sig_{+++} & 0 & 0 & 0\\
0 & \sig_{+--} & F_{y\bar{x}} & F_{z\bar{x}} \\
0 & F_{x\bar{y}} & \sig_{-+-} &  F_{z\bar{y}} \\
0 & F_{x\bar{z}} & F_{y\bar{z}} & \sig_{--+}
\end{array}
\right) 
\label{Lapfer}
\ee
where we have defined  $F_{n\bar{z}} \equiv  D_n \Phi_{xy} $ and\footnote{In our conventions the anticommutator is given by $\{A, B\} \equiv \oh (AB + BA)$.} 
\be
\Delta\, =\, \{D_x, D_{\bar{x}}\} + \{D_y, D_{\bar{y}}\} + \{D_z, D_{\bar{z}}\}
\label{Lapsym}
\ee
\be
\sig_{\eps_x\eps_y\eps_z}\, =\, \oh \left( \eps_x F_{x\bar{x}} + \eps_y F_{y\bar{y}} + \eps_z F_{z\bar{z}}\right)
\label{Lapsig}
\ee

Finally, $F_{n\bar{m}} \sim [F_{n\bar{m}}, \cdot]$ acts in the adjoint on the $U(3)$ gauge indices of $\Psi$, which implies that the worldvolume fluxes $F_{n\bar{m}}$ are felt differently by each matter curve. Indeed, for the $a^{\pm}$ sector in (\ref{sectors}) we have
\be
{\bf D_A}^\dag {\bf D_A}\, =\, -(\Delta_{a^\pm} \pm  M_{xy})\Id_4\, \pm 
\left(
\begin{array}{cccc}
2M_{xy} & 0 & 0 & 0\\
0 & M_x & 0 & -im^2_\Phi\\
0 & 0 & M_y &  0 \\
0 & im_\Phi^2 & 0 & 0
\end{array}
\right)
\label{Lapa}
\ee
where ${M}_{xy} \equiv \oh({M}_x+{M}_y)$. For the $b^\pm$ sector we have instead
\be
{\bf D_A}^\dag {\bf D_A}\, =\, -(\Delta_{b^\pm}\mp M_{xy}) \Id_4\, \pm 
\left(
\begin{array}{cccc}
-2M_{xy} & 0 & 0 & 0\\
0 & -M_x & 0 & 0 \\
0 & 0 & -M_y &  im_\Phi^2 \\
0 & 0 & -im_\Phi^2 & 0
\end{array}
\right)
\label{Lapb}
\ee
and, finally, for the $c^\pm$ sector we have
\be
{\bf D_A}^\dag {\bf D_A}\, =\, -\Delta_{c^\pm} \Id_4\, \pm  m_\Phi^2
\left(
\begin{array}{cccc}
0 & 0 & 0 & 0\\
0 & 0 & 0 & i  \\
0 & 0 & 0 &  -i \\
0 & -i & i & 0
\end{array}
\right)
\label{Lapc}
\ee

Given these expressions and the fact that $M_x$, $M_y$ and $m_\Phi$ are constant it is easy to find the spectrum of eigenvectors of ${\bf D_A}^\dag {\bf D_A}$ in terms of the eigenfunctions $-\Delta \psi_\rho = \rho^2 \psi_\rho$ of the Laplacian. Indeed, in the case of the sector $a^\pm$ we find that the eigenvalues and eigenvectors of the squared Dirac operator are given by
\begin{subequations}
\label{speca}
\begin{align}
\label{speca1}
 & \small{|m_\rho|^2 = \rho^2 \pm  M_{xy}, \,
\Psi =
\left(
\begin{array}{c}
1 \\ 0 \\ 0 \\ 0
\end{array}
\right) \psi_\rho}\ ; \quad
\small{|m_\rho|^2 = \rho^2 \pm  (M_y - M_{xy}), \,
\Psi =
\left(
\begin{array}{c}
0 \\ 0 \\ 1 \\ 0
\end{array}
\right) \psi_\rho} \\
\label{speca2}
& \small{|m_\rho|^2 = \rho^2 \pm  (\lam_a^+ - M_{xy}), \,
\Psi =
\left(
\begin{array}{c}
0\\ \frac{\lam_a^+}{m_\Phi^2}  \\ 0 \\ i
\end{array}
\right) \psi_\rho}\ ; 
\quad 
\small{|m_\rho|^2 = \rho^2 \pm  (\lam_a^- - M_{xy}), \,
\Psi =
\left(
\begin{array}{c}
0\\ \frac{\lam_a^- }{m_\Phi^2} \\ 0 \\ i
\end{array}
\right) \psi_\rho}
\end{align}
\end{subequations}
where
\be
\lam_a^{\pm}\, =\, \frac{M_x}{2} \pm \sqrt{\left(\frac{M_x}{2}\right)^2 + m_\Phi^4}
\label{eigenlam}
\ee
The precise expression for $\psi_\rho$ does in principle depend on which sector we consider, as the Laplacian (\ref{Lapsym}) depends non-trivially on the gauge potential $A$, which acts differently on $a^\pm$.\\

%
%
%
%

It is convenient to express the zero mode wavefunctions in the holomorphic gauge introduced in \cite{fi09}, in which only the holomorphic components of $\langle A \rangle$ are non-vanishing. In the model at hand, such gauge reads
\be
\langle A\rangle^{\rm hol} \, =\, \left( \bar{A_x}dx + \bar{A_y}dy \right)
\frac{1}{3}
\left(
\begin{array}{ccc}
1 \\ & -2 \\ & & 1
\end{array}
\right) \quad\quad
\begin{array}{c}
\bar{A_x}\, =\, -i M_x \bar{x} \\
\bar{A_y}\, =\, -i M_y \bar{y}
\end{array}
\label{holg}
\ee
and it is easy to see that the modes satisfying the zero mode equation (\ref{Dirac9}) for the sectors $a^\pm$ are given by
\be
\small{\Psi_{0, a^\pm}
\, =\, 
\left(
\begin{array}{c}
0\\ -i\frac{\lam_a^\mp}{m_\Phi^2} \\ 0 \\ 1
\end{array}
\right) \psi_{0, a^\pm}, \quad \quad \psi_{0, a^\pm}\, =\, e^{\pm \lam_a^{\mp}  |x|^2}  f_{a^\pm}(y)}
\label{zmhola}
\ee
with $f_{a^{\pm}}$ an arbitrary holomorphic function on the intersection coordinate $y$. In fact, from the Laplace eigenfunction $\psi_{0, a^\pm}$ one may easily construct all the other eigenfunctions of the Laplace operator $\Delta_{a^\pm}$, and so the full spectrum of massive modes in this sector. 
%
%

A similar discussion can be carried out for the sectors $b^{\pm}$ and $c^{\pm}$. We obtain that the zero mode wavefunctions in the holomorphic gauge for these sectors are 
%
\be
\small{\Psi_{0, b^\pm}
\, =\, 
\left(
\begin{array}{c}
0\\ 0 \\ i\frac{\lam_b^\mp}{m_\Phi^2} \\ 1
\end{array}
\right) \psi_{0, b^\pm}, \quad \quad \psi_{0, b^\pm}\, =\, e^{\pm  \lam_b^{\mp}  |y|^2}  f_{b^\pm}(x)}
\label{zmholb}
\ee
and
\be
\small{\Psi_{0, c^\pm}
\, =\, 
\left(
\begin{array}{c}
0\\ i\frac{\lam_c}{m_\Phi^2} \\ -i\frac{\lam_c}{m_\Phi^2} \\ 1
\end{array}
\right) \psi_{0, c^\pm}, \quad \quad \psi_{0, c^\pm}\, =\, \g_c m_* e^{  \lam_c  |x-y|^2} }
\label{zmholc}
\ee
where
\be
\lam_b^\pm\, =\, - \frac{{M}_{y}}{2} \pm \sqrt{\left(\frac{{M}_{y}}{2}\right)^2 + m_\Phi^{4}} \quad ;\quad 
\lam_c\, =\, - \frac{m_\Phi^2}{\sqrt2} 
\label{simlams}
\ee
%
In addition, we are going to assume that $M_x < 0 < M_y$, so that the sectors of interest for computing zero mode Yukawa couplings are $a^+$, $b^+$ and $c^+$. Finally, in (\ref{zmholc}) we have introduced a normalization factor to be fixed later.\\

\subsection{Yukawa couplings \label{yuky}}

Substituting $F = dA - i A \wedge A$ in the superpotential \Eq{supo7}, one obtains a trilinear term of the form 
\be
W_{\rm Yuk}\, =\, - im_*^4 \int_{S} \tr \left( A \wedge A \wedge \Phi \right)
\label{supy}
\ee
which is interpreted as a Yukawa coupling among the zero modes of $A$ and $\Phi$. Following the idea of local F-theory models, charged massless matter resides at curves where 7-branes intersect and therefore the Yukawa couplings $Y_{abc}^{ijk}$ are generated at the intersection of three matter curves $\Sigma_a$, $\Sigma_b$ and $\Sigma_c$, whose zero modes are respectively indexed by $i,j,k$.\\
 
The zero modes we are dealing with in the $U(3)$ model are $\psi_{\bar{m}}$, which are the superpartners of the fluctuations $a_{\bar{m}}$ of $A$, $\chi_{xy}$, that belongs to the same multiplet as the fluctuations $\varphi_{xy}$ of $\Phi$ and the fermion $\eta$ in the gauge multiplet. However, in (\ref{supy}) $\eta$ does not contribute to the Yukawa couplings.
Hence, it is useful to define the vector
\be
\vec{\psi}
\, =\,
\left(
\begin{array}{c}
\psi_{\bar{x}} \\ \psi_{\bar{y}} \\ \chi_{xy}
\end{array}
\right)\, =\, \vec{\psi}_\a \mathfrak{t}_\a
\label{defvec}
\ee
which is a subvector of $\Psi$ in (\ref{matrixDirac}) in order to describe the Yukawa couplings. Here $\mathfrak{t}_\a$ is a generator of the Lie algebra $\mathfrak{g}_p$ of the enhanced group $G_p$ at the Yukawa point $p$, with the normalization $ \tr\, \mathfrak{t}_\a \mathfrak{t}_\b^\dag =  \delta_{\a\b}$. More precisely, $\mathfrak{t}_\a$ is the generator associated to a root $\a$ of $\mathfrak{g}_p$, which  in turn corresponds to a matter curve $\Sigma_\a$ going through that point. The components of $\vec{\psi}_\a$ are scalar wavefunctions describing localized modes at such curve, and in particular its zero modes. As each curve may host several zero modes, we will label each zero mode vector by $\vec{\psi}_\a^{\, i}$, $i$ being the family index.\\

Inserting the zero modes in $W_{\rm Yuk}$ gives the couplings
\be
Y_{abc}^{ijk}\, =\, m_* f_{abc} \int_{S}\, {\rm det\, } (\vec{\psi}_a^{\, i} , \vec{\psi}_b^{\, j}, \vec{\psi}_c^{\, k}) 
\, {\rm d }{\rm vol}_S
\label{yukawa7}
\ee
where $f_{abc} = -i \tr\, ([\mathfrak{t}_a, \mathfrak{t}_b] \mathfrak{t}_c)$ and the integration measure is given by ${\rm d vol}_S = 2 \om^2 = \, {\rm d} x \wedge {\rm d}y \wedge {\rm d}  \bar x \wedge {\rm d} \bar y$.\\

In order to calculate the Yukawa couplings $Y_{abc}^{ijk}$ for the $U(3)$ toy model we are going to use the holomorphic gauge. Since the couplings are gauge invariant we can make this choice in which the zero modes take a simpler form. Turning on 7-brane fluxes $M_x < 0 < M_y$, there will be normalizable zero modes in the $a^+$ and $b^+$ sectors, which couple to those in the $c^+$ sector. Indeed, given the $U(3)$ structure displayed in \Eq{sectors} we see that
\be
\mathfrak{t}_{a^+}\, =\,
\left(
\begin{array}{ccc}
 0 & 1 & 0 \\
 0 & 0 & 0 \\
 0 & 0 & 0 
\end{array}
\right) \quad \quad 
\mathfrak{t}_{b^+}\, =\,
\left(
\begin{array}{ccc}
 0 & 0 & 0 \\
 0 & 0 & 1 \\
 0 & 0 & 0 
\end{array}
\right) \quad \quad 
\mathfrak{t}_{c^+}\, =\,
\left(
\begin{array}{ccc}
 0 & 0 & 0 \\
 0 & 0 & 0 \\
 1 & 0 & 0 
\end{array}
\right)
\label{tmats}
\ee
and so $\tr([\mathfrak{t}_{a^+}, \mathfrak{t}_{b^+}] \mathfrak{t}_{c^+}) =1$. The {\it Higgs} will be placed on matter curve $\Sigma_c$ which corresponds to the non-chiral sector, while the chiral families will arise from the curves $\Sigma_a$ and $\Sigma_b$, and will be indexed by $i$ and $j$ respectively so that the Yukawa couplings will be denoted by $Y^{ij}$.\\
 
We can arrange the the vectors (\ref{defvec}) of \Sec{toy} such that
\be
\vec{\psi}_{a^+}^i
\, =\,
\left(\!\!
\begin{array}{c}
-\displaystyle{\frac{i\lam_a}{m_\Phi^2}} \\ 0 \\ 1
\end{array}
\!\! \right)\chi_{a^+}^i
\qquad ; \qquad
\vec{\psi}_{b^+}^j
\, =\,
\left(\!\!
\begin{array}{c}
0\\ \displaystyle{\frac{i\lam_b}{m_\Phi^2}} \\ 1
\end{array}
\!\! \right)\chi_{b^+}^j
\qquad ; \qquad
\vec{\psi}_{c^+}
\, =\,
\left(\!\!
\begin{array}{c}
\displaystyle{\frac{i\lam_c}{m_\Phi^2}} \\[3mm] -\displaystyle{\frac{i\lam_c}{m_\Phi^2}}  \\ 1
\end{array}
\!\! \right)\chi_{c^+}
\label{simvecs}
\ee
where $\lam_a = \lam_a^-$, $\lam_b = \lam_b^-$ and $\lam_c$ are defined in (\ref{eigenlam}) and (\ref{simlams}), and the scalar wavefunctions $\chi$ are given by
\be
\chi_{a^+}^i\, =\, e^{ \lam_a  |x|^2}  f_i(y) \qquad ; \qquad
\chi_{b^+}^j\, =\, e^{ \lam_b  |y|^2}  g_j(x) \qquad ; \qquad
\chi_{c^+}\, =\, \g_c m_* e^{ \lam_c  |x-y|^2}  
\label{simchis}
\ee

Following \cite{hv08}  we are going to use a basis in which $f_i(y)= \g_{ai} m_*^{4-i} y^{3-i}$ and $g_j(x)= \g_{bj} m_*^{4-j} x^{3-j}$, with $i,j=1,2,3$ for the different zero modes, mimicking the physical case with three families of quarks and leptons. The normalization factors $\g_{ai}$ and $\g_{bj}$ will be fixed later.\\

\noindent Substituting in \Eq{yukawa7} readily gives the couplings
\be
Y^{ij}= - i \g_c \frac{m_*^2}{m_\Phi^4} [\lam_a \lam_b + \lam_c(\lam_a + \lam_b)]
\int_S e^{ \lam_a  |x|^2 + \lam_b  |y|^2 + \lam_c  |x-y|^2}  f_i(y) g_j(x) \, {\rm d}{\rm vol}_S
\label{simy}
\ee
Even though we are working with a local model for $S$, to evaluate the integral in (\ref{simy}) we extend $|x|$ and $|y|$ to infinite radius. This is justified because the  exponentials are localized on the matter curves and the error due to extending the Gaussian integrals is negligible.\\

Notice that the exponential and the measure of the integral are invariant under the diagonal $U(1)$ rotation $x \to e^{i\a} x$ and $y \to e^{i\a} y$. Therefore, the only non-vanishing coupling is $Y^{33}$ because $f_3$ and $g_3$ are constant. Then, we only need to integrate the Gaussian part of the integral of \Eq{simy}
\be
\int_S e^{\lam_a  |x|^2 + \lam_b  |y|^2 + \lam_c  |x-y|^2}  \, {\rm d}{\rm vol}_S =
\pi^2 [\lam_a \lam_b + \lam_c(\lam_a + \lam_b)]^{-1}
\label{exres}
\ee

\noindent Hence, the only non-vanishing Yukawa is given by
\be
Y^{33}= -i\pi^2 \frac{m_*^4 }{m_\Phi^4} \g_{a3} \g_{b3} \g_c 
\label{y33}
\ee
With normalization $\g_{a3}= \g_{b3}= \g_c=1$, the coupling is completely independent of the worldvolume flux, and hence it agrees with the fact that we are working in the holomorphic gauge, as we mentioned at the beginnig of the chapter.  

\section{Non-perturbative effects on intersecting 7-branes\label{npei7}}

We have just shown that for our $U(3)$ toy model, the Yukawa couplings do not depend on 7-brane worldvolume fluxes, where these fluxes satisfy the F-term BPS conditions (\ref{Fterm7phi}) at the level of background. Indeed this result is more general than the case of this $U(3)$ model \cite{cchv09} and has very important consequences from the viewpoint of the fermion mass matrices. In particular, , if all Yukawa couplings arise from a single triple intersection, the Yukawa matrices derived from (\ref{yukawa7}) will have rank one for any choice of worldvolume flux, and so only one family of quarks and leptons will receive a non-trivial mass in such F-theory construction \cite{cchv09}. As in the case of toroidal orientifolds one could think that this a good starting point to generate the observed hierarchical structure of fermion masses. Still this means that we need an extra ingredient beyond the intersecting 7-brane setup in order to avoid this rank-one Yukawas result.\\

In \cite{mm09} it was proposed that such extra contribution to the Yukawa couplings will in general arise from non-perturbative effects on a 7-brane far away from the GUT 4-cycle $S$, as was illustrated in \Fig{backr}. Indeed, if we consider a distant 7-brane whose 4d gauge theory undergoes a gaugino condensation, then a non-perturbative superpotential will be generated for the GUT 7-brane fields, perturbing the previous tree-level superpotential. In particular, there will a non-trivial contribution to the tree-level Yukawa couplings, so that we will instead have
\be
Y^{ijk}_{\rm total}\, =\, Y^{ijk}_{\rm tree} + Y^{ijk}_{\rm np}
\label{totalyuk}
\ee
where $Y^{ijk}_{\rm tree}$ corresponds to the tree-level contribution (\ref{yukawa7}), while $Y^{ijk}_{\rm np}$ stands for the new set of Yukawa couplings that arise at the non-perturbative level. 
In general $|Y^{ijk}_{\rm np}| << |Y^{ijk}_{\rm tree}|$, and the non-perturbative couplings will provide a slight deviation from the tree-level rank-one result. On the other hand there exists an equivalent scenario in which instead of a gaugino condensate on a 7-brane one considers the effect of an Euclidean 3-brane on the same 4-cycle.\\

In fact, as it has been shown in \cite{mm09}, there is not only one, but rather several approaches that one may use in order to compute (\ref{totalyuk}) and it can be computed rather precisely in the case of intersecting 7-branes. The purpose of this section is to introduce two of these approaches. The first approach consists in computing the non-perturbative effect at the level of the 4d effective action, in terms of a non-perturbative superpotential $W_{\rm np}$ generated for the 4d massless and massive fields of the GUT 7-brane. In the second approach the non-perturbative effect is now seen as a non-commutative deformation of the functional (\ref{supo7}), in the sense of \cite{cchv09}.\\

\subsection{4d approach\label{4dapp}}

In general, when computing non-perturbative effects in a string compactification, one does so at the level of the 4d effective theory. In particular, for the 7-brane setup considered above one would first compute the gauge kinetic function $f_{7_{\rm np}}$ of the stack of $n$ 7-branes undergoing a gaugino condensation, and then use the standard 4d expression
\be
W_{\rm np}^{\rm 4d}\, =\, \mu^3 \, e^{-f_{7_{\rm np}}/n}
\label{4dsuponp}
\ee
to compute the gaugino condensate contribution to the 4d effective superpotential. Here $\mu \sim m_*$ is the UV scale at which $f_{7{\rm np}}$ is defined. From the IR viewpoint, $f_{7{\rm np}}$ should be understood as a  holomorphic function of the 4d chiral multiplets of the theory, which arise either from the bulk or from the 7-brane sectors of the compactification. More precisely, such 7-brane kinetic function is of the form
\be
f_{7_{\rm np}}\, =\, T_{\rm np} + \, 
f^{\rm 1-loop}_{7_{\rm np}}\left( B_i, C_j \right)
\ee
where the first contribution amounts to the gauge kinetic function $f_{7_{\rm np}}$ computed at tree-level, and is given by  the complexified K\"ahler modulus  $T_{\rm np} = {\rm Vol\, }({S_{\rm np}}) + i \int_{S_{\rm np}} C_4$ corresponding to the 4-cycle $S_{\rm np}$ wrapped by the gaugino condensing 7-branes. The second contribution arises from threshold effects, and is given by a holomorphic function $f^{\rm 1-loop}_{7_{\rm np}}$ of the bulk/closed string fields $\{B_i\}$, and of the 4d fields $\{C_j\}$ arising from the remaining 7-brane sectors of the compactification.\\

From this 4d viewpoint, the main problem is to find $f^{\rm 1-loop}_{7_{\rm np}}$ as a function of massless and massive 4d fields. This is however implicit in the expression
\be
f^{\rm 1-loop}_{7_{\rm np}}\, =\, - n\, {\rm log \, } \ca \, - \frac{1}{8\pi^2} \int_S \str ({\rm log\, } h\, F \wedge F)
\label{f1loopa}
\ee
derived in \cite{mm09}. Here $h$ is the divisor function of the 4-cycle $S_{\rm np} = \{h = 0\}$ where the non-perturbative effect is taking place, and ${\cal A}$ is a function of the bulk/closed string fields $B_i$ which will not play any role in the following discussion and can be replaced by their vev $\langle B_i \rangle$. While ${\rm log\, } h$ is a scalar bulk quantity, when plugged into the expression (\ref{f1loopa}) one should follow the prescription of \cite{Myers} and consider its non-Abelian pull-back into $S$. That is
\be
{\rm log\, } h  \, =\, {\rm log\, } h|_{S} + m_\Phi^{-2}\, \Phi^m [\CL_m {\rm log\, } h]_S  +\, m_\Phi^{-4}\, \Phi^m \Phi^n [\CL_m \CL_n {\rm log\, } h]_S  + \dots  
\label{napb}
\ee
with $\CL_m \equiv \CL_{X_m}$ the Lie derivative along a vector $X_m$ transverse to $S$. Since $h$ is holomorphic so will be $X_m$ and so, in the local coordinate system used above, we should take $X_m = z$. Also, if as we assume that $S_{\rm np}$ is distant from our GUT 4-cycle $S$, and in particular that they do not intersect, then $h|_S$ will be a holomorphic function of $S$ with no zeroes or poles, hence a constant. This implies that
\be
- f^{\rm 1-loop}_{7_{\rm np}}\, =\, n\, {\rm log \, } \ca \, + N_{\rm D3}\, {\rm log\, } h|_{S} +  \frac{m_\Phi^{-2}}{8\pi^2} \int_S [\p_z{\rm log\, } h]_{S} \, \tr\, (\Phi^z\, F \wedge F)  +\dots
\label{f1loopb}
\ee
where $N_{D3} = (8\pi^2)^{-1} \int_S \tr (F \wedge F) \in \IN$ stands for the D3-brane charge induced by the presence of $F$ and we are not displaying higher orders in $m_\Phi^{-2}$. Clearly, the dependence of $f^{\rm 1-loop}_{7_{\rm np}}$ on the 7-brane fields $\{C_j\}$ arises only from the third term of the rhs of (\ref{f1loopb}), and is still implicit in the integral over the GUT 4-cycle $S$. In order to extract such dependence one must insert the internal wavefunctions for the fields $\{C_j\}$ in the term $\tr (\Phi^z\, F \wedge F)$, and then perform the integral over $S$ in order to obtain the different 4d couplings.\\

Once done so, it is straightforward to compute the non-perturbative contribution to the full 4d superpotential. Indeed, inserting (\ref{f1loopb}) into (\ref{4dsuponp}) we obtain
\bea\nonumber
W_{\rm np}^{\rm 4d} & = & \mu^3   (\ca e^{-T_{\rm np}/n} h^{N_{\rm D3}/n}|_{S} )\, {\rm exp } \left[ \frac{(m_\Phi^2 n)^{-1}}{8\pi^2} \int_S [\p_z{\rm log\, } h]_{S} \, \tr\, (\Phi^z\, F \wedge F)  +\dots\right] \\
& = & \mu^3 \eps \left(1 +  \frac{(m_\Phi^2 n)^{-1}}{8\pi^2} \int_S [\p_z{\rm log\, } h]_{S} \, \tr\, (\Phi^z\, F \wedge F)  +\dots \right)
\eea
where we have defined $\eps = \ca\, e^{-T_{\rm np}/n} h |_{S} ^{N_{\rm D3}/n}$. Upon further defining $\theta = \frac{\mu^3/4\pi^2n}{m_*^4 m_\Phi^2} \p_z {\rm log\, } h |_S$ and up to a constant term we have
\be
W_{\rm np}^{\rm 4d}\, =\, m_*^4 \frac{\eps}{2} \int_S {\theta} \,  \tr \left( \Phi_{xy} F \wedge F\right)
\label{finalsuponp4d}
\ee
where we have identified $\Phi_{xy} = \Phi^z$ 
. We can then approximate the total 4d superpotential by
\be
W_{\rm total}^{\rm 4d}\, =\, W_{\rm tree}^{\rm 4d} + W_{\rm np}^{\rm 4d}\, =\, 
m_*^4\left[  \int_S \tr (\Phi_{xy} F) \wedge dx \wedge dy + \frac{\eps}{2} \int_S {\theta} \,  \tr \left( \Phi_{xy} F \wedge F\right)\right]
\label{totalsupo4d}
\ee

Notice that in this approach the total 4d superpotential is obtained by inserting the zero mode wavefunctions computed at tree level (i.e., the ones of section \ref{general}) into (\ref{totalsupo4d}) and then performing the appropriate integral. That is, we are dimensionally reducing (\ref{totalsupo4d}) with tree level wavefunctions and background values in order to obtain new 4d couplings generated non-perturbatively, and from there performing a 4d analysis. This is in contrast with the non conmutative philosophy applied in the next subsection, where new internal wavefunctions need to be computed from the very beginning.\\

\subsection{Non-commutative approach\label{ncapp}}

In the 4d analysis of \Sec{4dapp} the field theory is obtained from dimensional reduction of the 8d fields living on the worldvolume of a stack of 7-branes. In this sense the superpotential at tree level $W_{\rm tree}^{\rm 4d}$ is obtained from reducing the functional (\ref{supo7}) that depends on the 8d fields $(A_{\bar{m}}, \Phi_{xy})$. Concerning $W_{\rm np}^{\rm 4d}$, it has a different origin, it arises at the level of the 4d effective action through the expression (\ref{4dsuponp}). Nevertheless, as it is clear from \Eq{totalsupo4d}, both superpotentials may be expressed as a sum of two functionals that depend on the 8d fields $(A_{\bar{m}}, \Phi_{xy})$.\\

In this subsection we are going to analize the $W_{\rm total}$ from the non-conmutative 8d point of view. The non-commutative superpotential (\ref{ncsupo}) was proposed in \cite{cchv09} as a way to overcome the Yukawa rank one problem discussed in section \ref{general}. For this case the superpotential is given by
\be
\hat W \, =\, m_*^4 \int_S \tr \left( \hat\Phi\circledast \hat F \right)
\label{ncsupo}
\ee
In this formalism, $7$-brane fields $\hat{A}$ and $\hat{\Phi}$ should be multiplied according to a non-commutative version of the usual scalar and wedge products. More precisely, two scalar functions $f$ and $g$ will be multiplied by the holomorphic Moyal product, which reads
\be
f * g\, =\, fg + \frac{i}{2} \eps\,  \theta^{ij} \p_if \p_jg + \CO(\e^2)  \quad \quad \quad \theta^{yx} = - \theta^{xy} = \theta
\label{Moyal}
\ee
whenever $\theta$ is a constant. For non-constant $\theta  = \theta(x, y)$ this definition has to be modified, as explained in appendix B of \cite{cchv09}, where also the non-commutative version $\circledast$ of the ordinary wedge product was discussed for this case.\\

In what follows we will not dwell into the intrincate aspects of non-conmutative field theory but rather use the non-conmutative formalism of \cite{cchv09} as an efficient method to obtain the local wave functions required to compute the Yukawa couplings.\\

Just like for its commutative counterpart (\ref{supo7}) we may compute the F-term equations for (\ref{ncsupo}). Following \cite{cchv09} they read
\begin{subequations}
\label{ncFterm7}
\begin{align}
\label{ncFterm7A}
\bar{\p}_{A\, \circledast} \hat{\Phi}  & =  \bar{\p} \hat{\Phi} - i [\hat{A}, \hat{\Phi}]_* \, =\, 0\\
\label{ncFterm7phi}
\hat{F}^{(0,2)}  & =  \bar{\p} \hat{A} - i \hat{A} \circledast \hat{A}\, =\, 0 
\end{align}
\end{subequations}
where $\hat{A} = \hat{A}_{\bar{x}} d\bar{x} + \hat{A}_{\bar{y}} d\bar{y}$ is a $(0,1)$-form. As in section \ref{general}, these equations greatly simplify if we take a non-commutative version of the holomorphic gauge of \cite{fi09}, namely setting $\langle \hat{A}_{\bar{m}} \rangle = 0$ for $\bar{m} = \bar{x},\bar{y}$. Indeed, we then have that at the level of the background they amount to set $\langle \hat{\Phi}_{xy} \rangle$ holomorphic. In addition, defining the non-commutative fields fluctuations as
\be
\hat{\Phi}_{xy}\, =\, \langle \hat{\Phi}_{xy} \rangle + \hat{\varphi}_{xy} \quad \quad \hat{A}_{\bar{m}}\, =\, \langle \hat{A}_{\bar{m}} \rangle + \hat{a}_{\bar{m}}
\label{ncfluctuation}
\ee
and expanding (\ref{ncFterm7}) to first order in fluctuations we find the wavefunction equations
\begin{subequations}
\label{ncfluct7}
\begin{align}
\label{ncFfluct7A}
\bar{\p}_{\bar{m}} \hat{\varphi}_{xy} - i [\hat{a}_{\bar{m}}, \langle \hat{\Phi}_{xy} \rangle]   + \eps \sum_{ij} \{\p_i \hat{a}_{\bar{m}}, \p_j (\theta^{ij} \langle \hat{\Phi}_{xy} \rangle) \} & =  \CO(\e^2)\\
\label{ncFfluct7phi}
\bar{\p}_{\bar{x}} \hat{a}_{\bar{y}} -  \bar{\p}_{\bar{y}} \hat{a}_{\bar{x}}  & =  \CO(\e^2)
\end{align}
\end{subequations}
where now all products are commutative, $i,j = x,y$ and again $\theta^{yx} = - \theta^{xy} = \theta$.\\

Besides modifying the wavefunctions, the non-commutative superpotential (\ref{ncsupo}) also induces $\theta$ depending corrections to the Yukawa
couplings. In particular, $\hat W$ includes the trilinear term
\be
\hat W_{\rm Yuk}\, =\, - im_*^4 \int_{S} \tr \left(\hat A \circledast \hat A \circledast \hat \Phi \right)
\label{ncsupy}
\ee
which has the $\eps$ expansion
\be
\hat W_{\rm Yuk} = \hat W_0 + \eps \hat W_1 +  \CO(\e^2)\ .
\label{exphatw}
\ee
To  zeroth order in $\e$ such trilinear term reads
\be
\hat W_0=-im_*^4 \int_{S} \tr \left(\hat A \wedge \hat A \wedge \hat \Phi \right)
\ee
while the first order correction turns out to be
\be
\hat W_1\, =\, m_*^4 d_{abc} \int_{S} \th \, \p \hat A_a \wedge \p \hat A_b  \, \hat \Phi_{c xy} 
+ {\rm cyclic \ permutations \ in \ } a, \, b, \, c
\label{ncsupy3}
\ee 
where a surface term has been dropped. One can check that the corrected superpotential in \Eq{ncsupy3} is equivalent to that in \Eq{totalsupo4d} \cite{afim}.

\section{Wavefunctions and Yukawas in the non-commutative formalism\label{ncyuk}}

The non-commutative approach has a practical advantage for the computation of wavefunctions and Yukawa couplings. That is the reason why we are going to use this formalism for determining the non-perturbative $\theta$ corrections to the Yukawa couplings in our $U(3)$ toy model. For this purpose  we are going to solve the equations of motion that follow from the non-perturbative superpotential $\hat W$ in (\ref{ncsupo}) in order to use the zero mode solutions to determine the $\theta$ corrections corresponding to the trilinear term in $\hat W$.\\

Let us recall that in the holomorphic gauge the F-term equations for the fluctuations are given in equations (\ref{ncfluct7}). In addition there is a D-term which is the non-commutative extension of (\ref{Dterm7}) \cite{cchv09}. Expanding the non-commutative wedge product according to the prescription in \cite{cchv09}, and defining the fluctuations as in  (\ref{ncfluctuation}), yields the D-term equation    
\be
\omega \wedge \p_{\langle \hat A \rangle} \hat a - \frac12 [\langle \hat{\bar{\Phi}} \rangle, \hat\varphi] = 0
\label{Dnc}
\ee
Notice that this equation does not receive $\CO(\e)$ corrections. Now we are going to use this set of F and D equations fordeterminig the zero modes and then for calculating the Yukawa couplings in the non-conmutative formalism.\\

\subsection{Zero modes in the $U(3)$ toy model \label{ncu3}}

We proceed now to obtain the zero modes in each sector.  We will first obtain the case of $\th$ constant in more detail and in \Sec{ncthvar} we will briefly discuss an example with $\th$ depending on the coordinates. Following the steps of \Sec{toy} we are going to choose a background like the one of \Eq{vevPhi} in order to break $U(3)$ to $U(1)^3$ by giving a non zero vev $\langle \hat\Phi_{xy}\rangle$. We will also take the full enhancement at the intersection point of matter curves $\Sigma_a =\{x=0\}$, $\Sigma_b =\{y=0\}$, and $\Sigma_c =\{x=y\}$ defined as the intersection 2-cycles \Eq{2cycles} of the non-compact 4-cycles \Eq{4cycles} where the 7-branes are wrapping.\\ 

As in \Sec{yuky} we are going to define a vector with the components of the fields included in the Yukawa couplings representing the different fluctuations of the backgrounds in each matter curve. Hnce as in \Eq{defvec} we define 
\be
\hat{\vec{\psi}}_{(\a)} = \left(\!\!
\begin{array}{c}
\hat\psi_{\a\bar x} \\ \hat\psi_{\a\bar y} \\ \hat\chi_{\a}
\end{array}
\!\! \right){\mathfrak t}_\a = \hat{\vec{\psi}}_\a {\mathfrak t}_\a  
\label{hatvec}
\ee 
where for simplicity we write $\chi = \chi_{xy}$. We are using the fermionic fluctuations which satisfy the same equations as their supersymmetric partners $(\hat a_{\a\bar x}, \hat a_{\a\bar y}, \hat \varphi_{\a xy})$.\\

To obtain a 4d chiral model we turn on a worldvolume flux of the form (\ref{fluxD7}). We then choose the holomorphic gauge in which 
\be
\langle \hat A \rangle = -\frac{i}{3} (M_x\bar x  dx + M_y \bar y dy){\rm diag} (1, -2, 1)
\label{ahol}
\ee 
exactly as in  (\ref{holg}). As in the commutative case we take $M_x < 0 < M_y$, so that the normalizable
zero modes appear in the $a^+$, $b^+$ and $c^+$ sectors. The corresponding ${\mathfrak t}_\a$ generators
for each curve are those given in (\ref{tmats}). \\

The F-term equation (\ref{ncFfluct7phi}) arising from $\hat F^{(0,2)}=0$ takes the same form in all sectors and will not be written separately. To first order in $\eps$ it implies that $\bar\partial_{\bar x} \hat  \psi_{\a\bar y}= \bar\partial_{\bar y} \hat \psi_{\a\bar x}$, which
will indeed be satisfied by the zero modes determined below. \\

\subsubsection*{Sector $a^+$ \label{nca}}

Given the generator ${\mathfrak t}_{a^+}$ one can evaluate the various commutators and anticommutators in the F-term equations (\ref{ncFfluct7A}). For generic $\th$ we find
\bea
\bar\p_{\bar m} \hat\chi \ & - &im_{\Phi}^2 \, x \hat \psi_{\bar m} 
- \frac{\eps m_{\Phi}^2}{6}[\th + \p_x\th(x-2y-2\Phi_0)]\p_y \hat\psi_{\bar m} \nonumber \\
& - &  \frac{\eps m_{\Phi}^2}{6}[2\th - \p_y\th(x-2y-2\Phi_0)] \p_x \hat\psi_{\bar m}     =  \CO(\e^2)
\label{ancfer}
\eea
for $\bar m = \bar x, \bar y$.We have dropped the subindex $\a=a^+$ that will be reinserted at the end.\\

The D-term equation derived from (\ref{Dnc}) turns out to be
\be
(\p_x-  M_x \bar x) \hat \psi_{\bar x} +  (\p_y- M_y \bar y) \hat \psi_{\bar y} 
+ i m_{\Phi}^2 \bar x \hat \chi = 0
\label{ancferD}
\ee
Notice that the worldvolume fluxes $M_x$ and $M_y$ only appear in the D-term equation.
That feature makes clear from the beginning that the Yukawa couplings, which only depend on the F-terms, will turn out to be independent of fluxes \cite{cchv09}.\\

To find the zero modes we make an Ansatz motivated by the form of the solutions when $\theta=0$ collected in \Eq{simvecs}. In particular we set $ \hat \psi_{\bar y} =0$, which then implies $\bar\partial_{\bar y}\hat \psi_{\bar x} =  \bar\partial_{\bar y} \hat\chi=0$.
When $\th$ is constant we further impose
\be
 \hat \psi_{\bar x} = -\frac{i\lam_a}{m_\Phi^2} \hat \chi \quad ; \quad
\hat \chi= e^{\lam_a  |x|^2} G_a(y, \bar x)
\label{aans}
\ee
In this way the D-term equation is satisfied with $\lam_a = \lam_a^-$, defined in (\ref{eigenlam}). The root $\lam_a^+$ is discarded because it yields zero modes that are not localized on the curve $x=0$. It remains to solve the F-term (\ref{ancfer}). Inserting the Ansatz for $\hat \psi_{\bar x}$ and $\hat \chi$ gives an equation for $G_a(\bar x, y)$ that is easily solved to first order in $\eps$. The final result can be written as 
\be
\hat{\vec{\psi}}_{a^+}^i
\, =\,
\left(\!\!
\begin{array}{c}
-\displaystyle{\frac{i\lam_a}{m_\Phi^2}} \\ 0 \\ 1
\end{array}
\!\! \right)\hat\chi_{a^+}^i
\quad ; \quad
\hat\chi_{a^+}^i\, =\, e^{\lam_a  |x|^2}
\left(f_i(y) - \frac{i\eps\theta}{6}\lam_a^2 \bar x^2 f_i(y)
 -  \frac{i\eps\th}{6} \lam_a \bar x f_i^\prime(y) \right) 
\label{achinc}
\ee
where $f_i(y)$ is an arbitrary holomorphic function, and $i$ labels the different zero modes.\\

\subsubsection*{Sector $b^+$ \label{ncb}}

In this sector the F-term equations (\ref{ncFfluct7A}) reduce to
\bea
\bar\partial_{\bar m} \hat\chi \ & + & i m_{\Phi}^2 \, y \hat \psi_{\bar m} 
+ \frac{\eps m_{\Phi}^2}{6}[\th - \p_y\th(2x-y+2\Phi_0)] \partial_x \hat\psi_{\bar m} \nonumber \\ 
& + & \frac{\eps m_{\Phi}^2}{6}[2\th + \p_x\th(2x-y+2\Phi_0)] \partial_y \hat\psi_{\bar m}     =  \CO(\e^2)
\label{bncfer}
\eea
for $\bar m = \bar x, \bar y$. The D-term equation is given by
\be
(\p_x+ M_x \bar x) \hat \psi_{\bar x} +  (\p_y+  M_y \bar y) \hat \psi_{\bar y} 
- i  m_{\Phi}^2 \bar y \hat \chi = 0
\label{bncferD}
\ee
The subindex $\a=b^+$ will be omitted until the final result.
Now it is consistent to set $\hat \psi_{\bar x} =0$, which then requires   
$\bar\partial_{\bar x}\hat \psi_{\bar y} =  \bar\partial_{\bar x} \hat\chi=0$.\\

Our previous results for $\th$ constant suggest the Ansatz
\be
 \hat \psi_{\bar y} = \frac{i\lam_b}{m_\Phi^2} \hat \chi \quad ; \quad
\hat \chi= e^{\lam_b  |y|^2} G_b(x,\bar y)
\label{bans}
\ee
The D-term equation is then verified with $\lam_b = \lam_b^-$, defined in (\ref{simlams}).
The auxiliary function $G_b$ is determined to first order in $\eps$ substituting the Ansatz in the 
F-term (\ref{bncfer}). In the end we obtain 
\be
\hat{\vec{\psi}}_{b^+}^j
\, =\,
\left(\!\!
\begin{array}{c}
0\\ \displaystyle{\frac{i\lam_b}{m_\Phi^2}} \\ 1 
\end{array}
\!\! \right)\hat\chi_{b^+}^j
\quad ; \quad
\hat\chi_{b^+}^j\, =\, e^{\lam_b  |y|^2}
\left(g_j(x) - \frac{i\eps\theta}{6} \lam_b^2 \bar y^2 g_j(x)
 -  \frac{i\eps\th}{6} \lam_b \bar y g_j^\prime(x) \right) 
\label{bchinc}
\ee
with $j$ indexing different zero modes.\\

\subsubsection*{Sector $c^+$ \label{ncc}}

Substituting the fluctuations and the vevs in (\ref{ncFfluct7A}) yields the F-term equations
\bea
\bar\p_{\bar m} \hat\chi \ & + & i m_{\Phi}^2 \, (x-y) \hat \psi_{\bar m} 
+ \frac{\eps  m_{\Phi}^2}{6}[\th + \p_y\th(x+y-2\Phi_0)] \p_x \hat\psi_{\bar m} \nonumber \\
& - & \frac{\eps  m_{\Phi}^2}{6}[\th + \p_x\th(x+y-2\Phi_0)] \p_y \hat\psi_{\bar m}     =  \CO(\e^2)
\label{cncfer}
\eea
for $\bar m = \bar x, \bar y$.
As before the subindex $\a=c^+$ is dropped.
The D-term equation takes the simple form
\be
\p_x  \hat \psi_{\bar x} +  \p_y \hat \psi_{\bar y} 
- i  m_{\Phi}^2 (\bar x - \bar y) \hat \chi = 0
\label{cncferD}
\ee
In this case the condition $\hat \psi_{\bar x}= -\hat \psi_{\bar y}$ can be imposed consistently with (\ref{ncFfluct7phi}).\\

For $\th$ constant the Ansatz inspired by known results is now  
\be
 \hat \psi_{\bar x}= -\hat \psi_{\bar y} = \frac{i\lam_c}{m_\Phi^2} \hat \chi \quad ; \quad
\hat \chi= e^{\lam_c  |x-y|^2} G_c(\bar x, \bar y)
\label{cans}
\ee
Inserting  the Ansatz in the equations shows that it works with $\lam_c$ given in (\ref{simlams}).
Summarizing we obtain
\be
\hat{\vec{\psi}}_{c^+}
\, =\,
\left(\!\!
\begin{array}{c}
 \displaystyle{\frac{i\lam_c}{m_\Phi^2}}\\[2mm]
 -\displaystyle{\frac{i\lam_c}{m_\Phi^2}} \\ 1 
\end{array}
\!\! \right)\hat\chi_{c^+}
\quad ; \quad
\hat\chi_{c^+}\, =\, \g_c m_* \, e^{\lam_c  |x-y|^2}
\left(1 - \frac{i\eps\theta}{6} \lam_c^2 (\bar x - \bar y)^2 \right) 
\label{cchinc}
\ee
To determine $G_c$ we assumed that it is constant at lowest order in $\eps$. The reason is that
at $\Sigma_c$ there is only one set of zero modes corresponding to the Higgs. The constant is taken to 
be $\g_c m_*$, where $\g_c$ is an adimensional normalization factor.\\

\subsubsection{Zero modes with $\th$ coordinate dependent \label{ncthvar}}

If we use a $\th$ constant it turns out it is not possible to generate a realistic pattern for the Yukawa couplings. This motivates us to consider a more general case with $\th$ coordinate dependent. We have been able to find the zero modes in closed form when $\th$ is the linear function
\be
\th= \th_0 + \th_1 x + \th_2 y
\label{thfun}
\ee
where the $\th_\ell$ are constants. In this case it is no longer consistent to make an Ansatz such as (\ref{cans}) in which the non-zero $\hat \psi_{\a\bar m}$ are proportional to $\hat \chi_\a$.\\

To illustrate the strategy that works for the linear $\th$ let us focus in the $a^+$ sector. Again it is allowed to take $\hat \psi_{a^+\bar y}=0$. The new Ansatz for the non-trivialfluctuations consists of first setting
\be
\hat \psi_{a^+\bar x}=  -\frac{i\lam_a}{m_\Phi^2} e^{\lam_a  |x|^2} H_a(y,x,\bar x)
\label{newaansA}
\ee 
and then solving for $\hat \chi_{a^+}$ from the D-term equation (\ref{ancferD}).
The constant $\lam_a$ is again equal to the $\lam_a^-$ defined in (\ref{eigenlam}). It thus follows that
\be
\hat \chi_{a^+}=  e^{\lam_a  |x|^2} \left(H_a + \frac{\lam_a}{m_\Phi^4 \bar x} \p_x H_a\right)
\label{newaansB}
\ee 
We also impose the condition that $\hat \chi_{a^+} \to f_i(y)$ when $x \to 0$. To determine the function
$H_a$ we substitute the above $\hat \psi_{a^+\bar x}$ and $\hat \chi_{a^+}$ into the F-term (\ref{ancfer})
and solve to first order in $\eps$. In this way we find
\bea
H_a & = & f_i(y) + \eps\th_0 \bar x \a_0 + \eps\th_1\left(\bar x \a_1 + 
\frac{i\lam_a^2}{3(\lam_a^2 + m_\Phi^4)}f_i^\prime(y)\right) + \eps\th_2 \bar x \a_2 \nonumber \\
\a_0 & = &  - \frac{i}{6} \lam_a^2 \bar x f_i(y) -  \frac{i}{6} \lam_a f_i^\prime(y)
\label{hasol}\\
\a_1 & = & \frac{i\lam_a^2}{3(\lam_a^2 + 2m_\Phi^4)}
(\lam_a - m_\Phi^4 x\bar x )f_i(y) 
-\frac{i\lam_a}{3}\left(\frac{m_\Phi^4}{(\lam_a^2+ m_\Phi^4)}x- y - \Phi_0 \right)f_i^\prime(y) \nonumber \\ 
\a_2 & = &   - \frac{i\lam_a}{6} y f_i^\prime(y)  - \frac{i \lam_a^2}{6}
\left(\frac{\lam_a - m_\Phi^4 x\bar x}{(\lam_a^2+ 2m_\Phi^4)} + 2y\bar x + \Phi_0 \bar x \right)f_i(y)
\nonumber
\eea
Notice that for $\theta$ constant we recover our previous result (\ref{achinc}). We remark that the
F-term equations are satisfied to $\CO(\e)$ for any $\lam_a$. Hence, we expect the $\lam_a$ dependence
to drop out completely in the computation of Yukawa couplings.\\

The wavefunctions in the $b^+$ sector can be found in a similar fashion. We start with 
$\hat \psi_{b^+\bar x}=0$, together with
\be
\hat \psi_{b^+\bar y}=  \frac{i\lam_b}{m_\Phi^2} e^{\lam_b  |y|^2} H_b(x,y,\bar y)
\label{newbans}
\ee 
where $\lam_b$ is equal to $\lam_b^-$ defined in (\ref{simlams}).
The corresponding $\hat \chi_{b^+}$ is such that the D-term equation (\ref{bncferD}) is verified and is
required to satisfy $\hat \chi_{b^+} \to g_j(x)$ when $y \to 0$.
The function $H_b$ is determined from the F-term eq.(\ref{bncfer}). To order $\eps$ it is given by
\be
H_b  =  g_j(x) + \eps\th_0 \bar y \b_0 + \eps\th_1 \bar y \b_1 
+ \eps\th_2 \left(\bar y \b_2 + \frac{i\lam_b^2}{3(\lam_b^2 + m_\Phi^4)}g_j^\prime(x)\right) 
\label{hbsol}
\ee
The $\b_\ell$ are obtained from the $\a_\ell$ in eq.(\ref{hasol}) as $\b_0=\a_0$, $\b_1=\a_2$,
and $\b_2 = \a_1$, upon the exchanges $x \to y$, $\lam_a \to \lam_b$, and $f_i(y) \to g_j(x)$.
The F-term equations do not constrain the value of $\lam_b$.

In the $c^+$ sector we take
\be
\hat \psi_{c^+\bar x}=  \frac{i\lam_c}{m_\Phi^2} e^{\lam_c  |x-y|^2} H_c(x,\bar x, y,\bar y) = 
- \hat \psi_{c^+\bar y}
\label{newcans}
\ee 
where $\lam_c=-m_\Phi^2/\sqrt2$. {} From the D-term equation (\ref{cncferD}) 
$\hat \chi_{c^+}$ is then determined to be
\be
\hat \chi_{c^+}=  e^{\lam_c  |x-y|^2} \left(H_c 
+ \frac{\lam_c}{m_\Phi^4}\frac{ \p_x H_c - \p_y H_c}{\bar x - \bar y}\right)
\label{newcansB}
\ee
For $H_c$ we make the Ansatz
\be
H_c  =  m_* \g_c \left[ 1 + \eps(\bar x-\bar y)(\th_0 \nu_0 + \th_1\nu_1 + \th_2\nu_2)\right] 
\label{hcsol}
\ee
and deduce the $\nu_\ell$ substituting in the F-term eqs.(\ref{cncfer}). This procedure yields
\bea
\nu_0 & = &  - \frac{i}{6} \lam_c^2 (\bar x-\bar y)  
\label{nusols}\\
\nu_1 & = & \frac{i\lam_c^2}{12(\lam_c^2 + m_\Phi^4)} \left\{ 2\lam_c   
-(\bar x-\bar y)\left[(2\lam_c^2 + 3m_\Phi^4)x + (2\lam_c^2 + m_\Phi^4)y -2(\lam_c^2 + m_\Phi^4)\Phi_0\right] \right\}    
\nonumber \\
\nu_2 & = & -\frac{i\lam_c^2}{12(\lam_c^2 + m_\Phi^4)} \left\{ 2\lam_c   
+(\bar x-\bar y)\left[(2\lam_c^2 + 3m_\Phi^4)x + (2\lam_c^2 + m_\Phi^4)y -2(\lam_c^2 + m_\Phi^4)\Phi_0\right] \right\}    
\nonumber
\eea
These solutions can be simplified inserting the actual value of $\lam_c$. However, the F-term equations are satisfied
for generic $\lam_c$.

\subsection{Yukawa couplings in the  $U(3)$ toy model \label{ncyuks}}

In this section we are going to calculate the corrected Yukawa couplings up to $\CO(\e)$ order for our $U(3)$ toy model. For this purpose we are going to use the wavefunctions determined in section \Sec{ncu3}. As was commented before, we take the {\it Higgs} to  arise from the curve $\Sigma_c$, whereas the quark and lepton families come from the curves $\Sigma_a$ and $\Sigma_b$, and are indexed by $i$ and $j$ respectively. \\

The full Yukawa couplings  $Y^{ij}$ have two types of contributions denoted by $Y_0^{ij}$ and $Y_1^{ij}$. Both contributions come respectively from the two terms of the trilinear superpotential $\hat W_{\rm Yuk}$ given in \Eq{exphatw}. Namely, $Y_0$ is given by (\ref{yukawa7}) replacing the zero modes defined in $\vec{\psi}_\a$ by the ones defined in $\hat{\vec{\psi}}_\a$. On the other hand, from $\hat W_1$ in (\ref{ncsupy3})
we obtain a contribution
\bea
(Y_1)_{abc}^{ijk} \!\! & = & \!\!\! -m_* d_{abc} 
\int_{S} \th  \left(\p_x \hat \psi_{a\bar x}^i \,\p_y \hat \psi_{b \bar y}^j 
-\p_y \hat \psi_{a\bar x}^i \,\p_x \hat \psi_{b \bar y}^j  
- \p_x \hat \psi_{a\bar y}^i \,\p_y \hat \psi_{b \bar x}^j 
+ \p_y \hat \psi_{a\bar y}^i \,\p_x \hat \psi_{b \bar x}^j \right)\hat\chi_c^k \, d{\rm vol}_S
\nonumber \\
& + & {\rm cyclic \ permutations \ in \ } a, \, b, \, c
\label{ncy1}
\eea 
Both pieces $Y_0$ and $Y_1$ have an $\eps$ expansion since the zero modes can be written as $\hat{\vec{\psi}} = \hat{\vec{\psi}}^{(0)} + \eps \hat{\vec{\psi}}^{(1)} + \CO(\e^2)$. Taking into account the expansion of $\hat W_{\rm Yuk}$ we see that indeed the Yukawa couplings have the schematic structure
\be
Y = Y_0^{(0)} + \eps\left( Y_0^{(1)}+ Y_1^{(0)}\right) +  \CO(\e^2) 
\ee
where we have omitted indices for simplicity. Here $Y_0^{(0)}$ and $Y_0^{(1)}$ both originate from $Y_0$. More concretely,  $Y_0^{(0)}$ is computed from (\ref{yukawa7}) replacing $\vec{\psi}_\a^i$ by $(\vec{\psi}_\a^i)^{(0)}$, whereas  $Y_0^{(1)}$ is given by 
\bea\nonumber
(Y_{{\rm tree}}^{(1)})_{abc}^{ijk} & = & m_* f_{abc} \int_{S}\, \left[ {\rm det\, } \left((\vec{\psi}_a^i)^{(1)} , (\vec{\psi}_b^j)^{(0)}, (\vec{\psi}_c^k)^{(0)} \right) + {\rm det\, } \left((\vec{\psi}_a^i)^{(0)} , (\vec{\psi}_b^j)^{(1)}, (\vec{\psi}_c^k)^{(0)} \right) \right. \\ & & \quad \quad \quad \quad 
\left. +\,  {\rm det\, } \left((\vec{\psi}_a^i)^{(0)} , (\vec{\psi}_b^j)^{(0)}, (\vec{\psi}_c^k)^{(1)} \right)\right] \, d{\rm vol}_S
\label{yukawa7np1}
\eea
with  $\vec{\psi}_\a^i$ replaced by $\hat{\vec{\psi}}_\a^i$. On the other hand, the $\CO(\e)$ contribution $Y_1^{(0)}$ is computed inserting the uncorrected wavefunctions $(\vec{\psi}_\a^i)^{(0)}$ in (\ref{ncy1}).\\

Concerning the holomorphic functions appearing in  the wavefunctions we again adopt the basis of \cite{hv08} in which 
$f_i(y)= \g_{ai} m_*^{4-i} y^{3-i}$ and $g_j(x)= \g_{bj} m_*^{4-j} x^{3-j}$, with $i,j=1,2,3$. The normalization factors $\g_{ai}$ and $\g_{bj}$ will be specified later.

We will first determine the couplings when $\th$ is constant. The calculation simplifies considerably because the wavefunctions $\hat\psi_{\a \bar m}$ are either zero or are proportional to $\hat \chi_\a$ as shown in equations (\ref{aans}), (\ref{bans}) and (\ref{cans}). From section \ref{yuky} we already know that $(Y_0^{(0)})^{ij}$ is zero for $i\not=3, j\not=3$, and $(Y_0^{(0)})^{33}=-i\pi^2 \g_{a3}\g_{b3} \g_c m_*^4/m_\Phi^4$. We will then concentrate on the pieces $Y_0^{(1)}$ and $Y_1^{(0)}$ for which we can write down explicit expressions using the properties of the wavefunctions derived in section \ref{ncu3}.\\

Inserting the known wavefunctions in (\ref{yukawa7np1}) we obtain the  $\CO(\e)$ contribution from $\hat W_0$ 
\bea
(Y_0^{(1)})^{ij} & = & -\frac{\th m_*^2 \g_c}{6 m_\Phi^4}(\lam_a\lam_b + \lam_a\lam_c + \lam_b\lam_c)\, I_0^{ij} 
\nonumber \\
I_0^{ij} & = & \int_S \bigg\{f_i(y) g_j(x)\left[\lam_c^2(\bar x- \bar y)^2 
+ \lam_a^2{\bar x}^2 +\lam_b^2{\bar y}^2\right]  \label{yuk01} \\[2mm]
& + & \left[ \lam_a \bar x f_i^\prime(y) g_j(x) + \lam_b \bar y f_i(y) g_j^\prime(x) \right]\bigg\}
 e^{\lam_a  |x|^2 + \lam_b  |y|^2 + \lam_c  |x-y|^2} \, d{\rm vol}_S
\nonumber
\eea 
Substituting in (\ref{ncy1}) yields the $\CO(\e)$ contribution from $\hat W_1$  
\bea
(Y_1^{(0)})^{ij} & = & -\frac{m_*^2 \g_c}{2 m_\Phi^4}I_1^{ij} 
\nonumber \\
I_1^{ij} & = & \int_S \th \bigg\{ 
\lam_c^2 (\bar x- \bar y)\left[\left(\lam_a^2 \bar x - \lam_b^2 \bar y\right)f_i(y) g_j(x)  
+\lam_a f_i^\prime(y) g_j(x) - \lam_b f_i(y) g_j^\prime(x) \right]
\nonumber \\[2mm]
& + &  \lam_a  \lam_b \left[\lam_a \lam_b \bar x \bar y f_i(y) g_j(x) 
- f_i^\prime(y)  g_j^\prime(x) \right] 
\bigg\} e^{\lam_a  |x|^2 + \lam_b  |y|^2 + \lam_c  |x-y|^2} \, d{\rm vol}_S
\label{yuk10}
\eea 
This formula is also valid when $\th$ is coordinate dependent. Here we have used that in the $U(3)$ model $d_{a^+ b^+ c^+} =  \str(\mathfrak{t}_{a^+},\mathfrak{t}_{b+},\mathfrak{t}_{c^+}) = \frac12$.\\

The coupling $Y^{33}$ does not receive $\CO(\e)$ corrections. Indeed, the integrals appearing in $(Y_0^{(1)})^{33}$ and $(Y_1^{(0)})^{33}$ vanish because the product of wavefunctions in the integrand is not invariant under the diagonal $U(1)$ rotation $x \to e^{i\a} x$ and $y \to e^{i\a} y$. By the same token we also conclude that for constant $\th$ only the couplings $Y^{22}$, $Y^{31}$, and  $Y^{13}$ can be different from zero. \\

To evaluate the integrals we extend $|x|$ and $|y|$ to infinity as we did in absence of $\theta$ corrections obtaining
\be
Y^{22}= \frac{\eps \pi^2 \th m_*^6}{3 m_\Phi^4}\g_{a2}\g_{b2} \g_c \quad ; \quad
Y^{31} = -\frac{\eps \pi^2 \th m_*^6}{3 m_\Phi^4}\g_{a3}\g_{b1} \g_c  \quad ; \quad
Y^{13} = -\frac{\eps \pi^2 \th m_*^6}{3 m_\Phi^4}\g_{a1}\g_{b3} \g_c 
\label{thconstresu}
\ee
Observe that, up to normalization, the couplings are independent of worldvolume flux, in agreement with the general result of \cite{cchv09}.\\

Besides the $\th$ constant case, we have calculated the couplings for perturbations linear in the local coordinates, i.e.
\be
\th  \ = \ 3i(\th_0 \ +\  \th_1 x \  + \ \th_2 y)
\label{lineartheta}
\ee
where the $3i$ factor is added to simplify the results. Inserting the wavefunctions given in section \ref{ncthvar} and evaluating the integrals leads to the Yukawa matrix
\be
\frac{Y}{Y^{33}} = 
\left(
\begin{array}{ccc}
\CO(\e^2)   &   \CO(\e^2)   &   \eps m_*^2 \frac { \g_{a1}}{\g_{a3}}(\th_0+ \theta_1 \Phi_0)\\
\CO(\e^2)   &  \eps m_*^2 \frac { \g_{a2} \g_{b2} } {\g_{a3}\g_{b3}}  [(\theta_1+\theta_2)\Phi_0-\theta_0]  
&  \eps m_* \frac {  \g_{a2}}{\g_{a3}}\theta_2 \\
\eps m_*^2 \frac { \g_{b1}}{\g_{b3}}(\theta_0+\theta_2\Phi_0) &  \eps m_*  \frac { \g_{b2}}{\g_{b3}}\theta_1 &  1
\end{array}
\right) 
\label{yukasmm}
\ee
As expected, up to normalization the couplings are independent of worldvolume fluxes. 
\\

\section{Flux dependence, D-terms and Yukawa couplings\label{norms}}

One interesting feature of the $U(3)$ example developed in last sections, is that for the case of constant fluxes and linearly dependent perturbations displays a hierarchical structure which could be useful in a more realistic local $SU(5)$ F-theory GUT. In particular, setting $\g_{ai}=\g_{bi}=1$, one observes that the Yukawa matrix \Eq{yukasmm} has eigenvalues of order $1,\epsilon , \epsilon^2$ and hence has a promising structure to generate such mass hierarchies. In this section we will discuss  possible  phenomenological implications of the Yukawa structure obtained in the last section.

\subsection{The $Y(D)=Y(L)$ problem}

Despite of the hierarchical structure of  \Eq{yukasmm}, the Yukawa couplings are flux independent and this has an important consequence from the phenomenological point of view. In an $SU(5)$ local GUT the Yukawa couplings of D-quark and leptons come from  couplings ${\bf 10}\times {\bf {\overline 5}}\times {\bf {\overline 5}_H}$ and before  the addition of hypercharge fluxes one has identical Yukawa couplings for D-quarks and leptons of all three generations, i.e., $Y^{ij}(D)=Y^{ij}(L)$. Since \Eq{yukasmm} is independent of  fluxes this equality will persist even after the addition of hypercharge fluxes.  These equalities are however not consistent with the measured
quark and lepton masses of the first two generations.
\\

Actually, this  seems to be a  general problem of Yukawa couplings in  local F-theory GUT models, and is ultimately connected to the existence of the holomorphic gauge discussed in previous sections. Using this holomorphic gauge the fluxes disappear from the local F-term equations and the resulting holomorphic Yukawas are flux independent.

\subsection{Normalization and flux dependence in physical couplings}

As we commented in \Sec{general} the gauge we have been using for calculating wave functions and Yukawas is the holomorphic gauge for practical reasons. However this is not a physical gauge and we need to make a gauge tranformation to the so called real gauge. In this real gauge  
\be
\langle A\rangle^{\rm real}\, =\, \frac{i}{2} \left[ M_x \left(xd\bar{x} -\bar{x}dx\right) + M_y \left(y d\bar{y} - \bar{y}dy\right) \right]
\frac{1}{3}
\left(
\begin{array}{ccc}
1 \\ & -2 \\ & & 1
\end{array}
\right)
\ee
%
Let us consider the $a$ sector and to simplify let us  impose the BPS condition on the fluxes, i.e. $M_x=M$ and $M_y=-M$. We take $M<0$ so that the normalizable zero modes are in the $a^+$ sector. The wave functions in real gauge will be given by 
\be
\vec{\psi}_{a\, i}^{ {\rm real}}
\, =\,
\left(\!\!
\begin{array}{c}
-\displaystyle{\frac{i\lam_a}{m_\Phi^2}} \\ 0 \\ 1
\end{array}
\!\! \right)\
\chi^{{\rm real}}_{a\, i} 
\label{realveca}
\ee
where $\lam_a=\lam_a^-$ is defined in \Eq{eigenlam} and the real scalar wavefunction is 
\be
\chi^{{\rm real}}_{a\, i} = e^{- \sqrt{\left( \frac{M}{2}\right)^2 + m_\Phi^4}\,  |x|^2} e^{-\frac{|M|}{2} |y|^2} f_i(y) \ .
\label{chireala}
\ee

In this gauge it is no so straightforward to see that Yukawas are flux independent as in the holomorphic gauge case because in this case fluxes are present both in F-term and D-term equations. However because of integrands in Yukawa couplings are gauge invariant, the result of the calculation of Yukawas is the same as in \Eq{yukasmm} i.e. flux independent.\\

Nevertheless the wave functions we have been using to compute the Yukawa couplings were not normalized as they should if one is to compare with physical quantities. Notice from \Eq{chireala} that in the real gauge an exponential suppression along the  $y$ coordinate is made explicit. The wave function is only sensitive to local physics and one can  normalize the wave function without the addition  of any volume dependent cut-off.
To normalize the states we perform the integration
\be
||\chi^{{\rm real}}_{a \, i}||^2 = 
 \int_{S}\, |\chi^{{\rm real}}_{a \, i}|^2
\, d{\rm vol}_S\ =\  
|\g_i|^2 \frac{\pi^2 (3-i) !} {2 \sqrt{\left( \frac{M}{2}\right)^2 + m_\Phi^4} \, \big(|M|/m_*^2\big)^{4-i}} \ ,
\label{chinorm}
\ee
where we have extended  the integration along  $|x|$ and $|y|$ to infinite radius as in the calculation of Yukawa couplings. The normalization condition amounts to imposing 
\be
\langle \vec{\psi}_{a \, i}^{\rm real} | \vec{\psi}_{a \, j}^{\rm real} \rangle  =
\,  m_*^2 \int_S \tr \,( \vec{\psi}_{a \, i}^{\rm real} \cdot \vec{\psi}_{a\, j}^{\, \dag\, {\rm real}})\, {\rm d vol}_S\, =
\, \d_{ij}
\label{normcond}
\ee 
where the $\d_{ij}$ structure arises because the exponentials in the wavefunctions, as well as the measure, are invariant under the diagonal $U(1)$ rotation $x \to e^{i\theta} x$ and $y \to e^{i\theta} y$. Upon normalization we get
\be
\g_{ai}^2\ =\ \frac {( |M|/m_*^2)^{3-i}}{(3-i)!} \times  \CN_{a}^{-1}
\ \ ,\ \ 
\CN_{a}\ =\  \frac {\pi^2m_*^4}{2m_{\Phi}^4} (1+\sqrt{1+ \frac {4m_{\Phi}^4}{|M|^2}}) \ ,
\label{gamillas}
\ee
where $\CN_{a}$ is generation independent. Note that in the dilute flux limit $m_*,m_{\Phi}\gg |M|$ one has $ \CN_{a}\simeq  m_*^4 / (m_{\Phi}^2|M|)$.
\\

Similar results are obtained for the matter fields in curve $b$. In the case of the curve $c$ with the flux choice in section 2 there is  no flux along the curve. In this case the exponential damping along the curve  is missing and one has to take a volume dependent cut-off in order to normalize the wave function. However, we will not need to do so because in the Yukawa coupling ratios in \Eq{yukasmm} the normalization of the c curve cancels out. Plugging these values for $\g_{ai},\g_{bi}$ in eq.(\ref{yukasmm}) one obtains a matrix
\be
\frac{Y}{Y^{33}} = 
\left(
\begin{array}{ccc}
\CO(\epsilon ^2)    &   \CO(\epsilon ^2)    &   \frac{ |M| }{\sqrt{2}} \eps (\th_0+ \theta_1 \Phi_0)\\
\CO(\epsilon ^2)    &   |M| \eps [(\theta_1+\theta_2)\Phi_0-\theta_0]  
&  |M|^{1/2}  \eps \theta_2 \\
\frac{|M| }{\sqrt{2}}\eps (\theta_0+\theta_2\Phi_0) &    |M|^{1/2} \eps \theta_1 &  1
\end{array}
\right) \ . 
\label{yukasmf}
\ee
As we can see, the physical Yukawa couplings do now depend on the fluxes. 
Notice that there exist an invariance under the rescalings
\be
(\epsilon\theta_0, \epsilon\theta_{1,2} ,   \Phi_0 , M)\ \rightarrow \   (\lambda ^2\epsilon\theta_0, \lambda \epsilon\theta_{1,2} ,   \lambda\Phi_0 , \lambda ^{-2}M)
\label{scalings}
\ee
and there is  no explicit dependence on the ``stringy'' scales $m_*,m_{\Phi}$. This is as it should since Yukawa couplings in type IIB string compactifications may be computed in the large volume limit just in terms of the compactified 10d field theory, without explicit reference to any string theoretical scale. 
\\

\subsection{A model of quark-lepton hierarchies}

We can take the $U(3)\rightarrow U(1)$ model developed in last sections as a toy model for the  $SO(12)\rightarrow SU(5)$ and $E_6\rightarrow SU(5)$  symmetry structure underlying the ${\bf 10}\times {\bf {\overline 5}}\times {\bf {\overline 5}_H}$ and ${\bf 10}\times {\bf {{10}}}\times {\bf { 5}_H}$ Yukawas in a local $SU(5)$ GUT. In order to achive the breaking of $SU(5)$ GUT down to the SM in local F-theory GUT's it is also necessary to turn on a worldvolume magnetic field $F_Y$ for the hypercharge generator in $SU(5)$. In these local models the matter is localized at the curves and then it is feeling both the flux in the matter curves $\Sigma_a$ coming from the $U(1)_a$ and the hypercharge flux in the 4-cycle $S_{GUT}$.  Although the {\it net}  hypercharge  flux on the matter curves associated to quark and leptons is assumed to vanish (so that the SM family spectrum is not spoiled), the local value of the hypercharge flux density  at the intersection Yukawa point is in general non-vanishing. This effect can then make the physical Yukawa couplings sensitive to the hypercharge flux. \\

One can model out this situation by taking the above wave functions with constant fluxes and making the replacement $M =M_0  +YM_1$ in the physical Yukawa couplings in \Eq{yukasmf}. Here $M_0$, $M_1$ are the curve flux and the bulk hypercharge flux respectively, and $Y=1,-4,2,-3,6$ are the hypercharges  of the SM fields $Q_L,U_R,D_R,L,E_R$ respectively, in standard notation. This may be a good approximation to the extent that the local flux in the vicinity of the Yukawa point may be slowly varying. This might be the case recalling that the physical Yukawa coupling and the normalized wave functions are only sensitive to backgrounds in the vicinity of this Yukawa point. Left- and right-handed fermions in a given coupling have different hypercharges $Y_{L,R}$ so that one finds that the ratio of physical Yukawa couplings have a hypercharge dependence of the form
{ 
\beq
\frac{Y}{Y^{33}} = 
\left(
\begin{array}{ccc}
  \CO(\epsilon^2)   &   \CO(\epsilon^2)  &    (\theta_0+ \theta_1\Phi_0)\frac {\epsilon}{\sqrt{2}}  |M_R|\\
   \CO(\epsilon^2)    &    \epsilon [(\theta_1+\theta_2)\Phi_0-\theta_0]   |M_LM_R|^{1/2}    &  \epsilon \theta_2  |M_R|^{1/2} \\
(\theta_0+\theta_2\Phi_0)  \frac {\epsilon}{\sqrt{2}}  |M_L|\ &  \epsilon \theta_1 |M_L|^{1/2} &  1
\end{array}
\right)  \ ,
\label{yukasmmflux} 
\eeq }
where
\be
M_L \ =\ M_0\ +\ Y_L M_1 \ \ ,\ \ 
M_R\ =\ M_0 \ +\ Y_RM_1  .
\ee

In  $SU(5)$ GUT's the right handed $D$-quarks  live in a ${\bf {\overline 5}}$ matter curve whereas the left-handed quarks $Q_L$ live in a ${\bf { {10}}}$. For the leptons the opposite happens, left-handed leptons $L$ live in the ${\bf {\overline 5}}$ and the right-handed leptons in the ${\bf { {10}}}$. This means that when going from a $D$-quark Yukawa matrix to a lepton matrix we have to interchange $Y_R,M_R\leftrightarrow Y_L,M_L$. In the case of $U$-quark masses Yukawa couplings come from an intersection ${\bf 10}\times {\bf { {10}}}\times {\bf { 5}_H}$, both left and right $U$-quarks are in a ${\bf 10}$ and one has to symmetrize the Yukawa coupling. Note also that in this case, as noted in \cite{bhv2} the 10-plet matter curve must self-pinch  or rather both 10-plet branches must be related by some discrete symmetry \cite{hktw} in order to be able to get eventually rank=3 matrices. We will further assume that there is a single intersection point of matter curves for each of the two types of Yukawa couplings.\\

It is interesting to check numerically whether this kind of structure  is able to describe the observed hierarchy of fermion masses and their mixing.  Since the above matrix is not hermitian it is simpler to compute the eigenvalues and eigenvectors of its product by its adjoint and take the square root. We have looked for values of the six parameters $\epsilon\theta_0, \epsilon\theta_1, \epsilon\theta_2, \Phi_0 , M_0, M_1$  (with all parameters real for simplicity)  able to reproduce the observed fermion hierarchies and mixings. Note that in a realistic setting  these sets of parameters are different for the $D/L$ and the $U$ physical Yukawa couplings since the corresponding intersection points are in general different in a $SU(5)$ local GUT. We will compare the results with the {\it observed } ratios of physical Yukawa couplings all evaluated at a scale of order the electroweak scale. One has for those  (see e.g. \cite{Fritzsch} )
\begin{equation}
\begin{array}{lcl}
\left(\frac{Y_1}{Y_3}\right)_U = (0.5-1.6)\cdot10^{-5} &  \mathrm{and}  & \left(\frac{Y_2}{Y_3}\right)_U = (3-4)\cdot10^{-3}\\
\left(\frac{Y_1}{Y_3}\right)_D = (0.6-1.8)\cdot10^{-3} & \mathrm{and} & \left(\frac{Y_2}{Y_3}\right)_D = (1-3)\cdot10^{-2}\\
\left(\frac{Y_1}{Y_3}\right)_L = (2.8)\cdot10^{-4} & \mathrm{and} & \left(\frac{Y_2}{Y_3}\right)_L = (5.9)\cdot10^{-2}.
\end{array}
\end{equation}
The experimental CKM mixing matrix with 90\%  CL is (see \cite{pdg})
\begin{equation}
|V_{CKM}|  = \left(\begin{array}{ccc}
0.9741-0.9756 & 0.219-0.226  & 0.0025-0.0048 \\
 0.219-0.226 & 0.9732-0.9748  & 0.038-0.044\\
 0.004-0.014 & 0.037-0.044   & 0.9990-0.9993
\end{array}\right) \qquad 
\end{equation}
As an example take  parameters 
\be
(\epsilon\theta_0, \epsilon\theta_1, \epsilon\theta_2,\Phi_0, M_0, M_1)_{D, L} \  =  \  (-0.066, 0.10, -0.27, 0.5,1.41,-0.31) 
\ee
\be
(\epsilon\theta_0, \epsilon\theta_1, \epsilon\theta_2, \Phi_0, M_0, M_1)_U \  =  \  (-0.033, -0.27, -0.33, -0.1, -1.1, -0.47) \ .
\ee
for the backgrounds at the $D/L$ and $U$ Yukawa intersecting points respectively.
One then obtains mass  ratios
\begin{equation}
 \begin{array}{lcl}
  (m_1,\ m_2,\ m_3)_U & = & (6.9\cdot10^{-5},\ 3.8\cdot10^{-3},\ 1)\\
  (m_1,\ m_2,\ m_3)_D & = & (0.65\cdot10^{-3},\ 2.96\cdot10^{-2},\ 1)\\
  (m_1,\ m_2,\ m_3)_L & = & (4.8\cdot10^{-4},\ 5.3\cdot10^{-2},\ 1)
 \end{array}
\end{equation}
with a CKM mixing matrix
\begin{equation}
V_{CKM}  = \left(\begin{array}{ccc}
0.9834 & 0.1812  & 0.0056\\
 0.1809 & 0.9827  & 0.0382\\
0.0125  & 0.0365   & 0.9992
\end{array}\right)
\end{equation}
The agreement is quite good taking into account the simplicity of the model and the uncertainties. In particular the first  (smallest) eigenvalues are less reliable since in the computation of the Yukawa couplings we have neglected effects of order $\epsilon ^2$ which could be relevant for the
masses and mixings of the first generation.  Many other solutions leading to similarly acceptable results exist. Note that due to the scale invariance in eq.(\ref{scalings}) the same numerical results may be obtained with magnetic fluxes a factor $1/\lambda^2$ smaller by compensating taking larger values for the rest of the parameters. Note also that in the above we have computed ratios of physical Yukawa couplings. However, flux effects do also affect the relative size of the third generation physical Yukawa couplings. Indeed, from eq.(\ref{gamillas}) one obtains 
\beq
\frac {Y^{33}(L)}{Y^{33}(D)} \ =\  \frac {(\CN_{D} \CN_{Q})^{1/2} }  {(\CN_{L} \CN_{E})^{1/2} } \ .
\eeq
As we saw, in the dilute flux limit with $m_*,m_{\Phi}\gg |M|$ one has $ \CN_{a}\simeq  m_*^4 / (m_{\Phi}^2|M|)$ so 
\be
\frac {Y^{33}(L)}{Y^{33}(D)} \  \simeq \  \frac {( |M_0+Y_LM_1||M_0+Y_EM_1|)^{1/2} }  {( |M_0+Y_DM_1||M_0+Y_QM_1|)^{1/2} }
\ =\  \frac {( |M_0-3M_1||M_0+6M_1|)^{1/2} }  {( |M_0+2M_1||M_0+M_1|)^{1/2} } \ .
\ee
For the above choice of parameters this leads to $Y^{33}(L)/Y^{33}(D)\simeq 1.13$ , so that the (successful) standard $SU(5)$ prediction $m_b(M_{GUT})=m_{\tau} (M_{GUT})$ is not much distorted.\\

In summary, although the above estimations are based on the results obtained for a simple $U(3)$ toy model with constant magnetic  flux, the lesson seems to be more general.  Non-perturbative effects from distant 7-branes sectors (or, equivalently, local closed string $(1,2)$ fluxes) can give rise to the observed hierarchy of quark and lepton masses. On the other hand  turning on hypercharge fluxes on $SU(5)$ models  seems able to explain the difference in masses between the D-quarks and charged leptons of the lightest generations.

%% file: conclusionsfin.tex
\chapter{Conclusions}

If string theory is indeed the fundamental theory underlying the SM, its low energy physics should be able to describe the different properties and parameters of the SM. In this work we have addressed a couple of phenomenological aspects of a large class of Type IIB/F-theory string compactifications which are able to contain the MSSM as its low-energy limit. In the first part we explore the structure of SUSY-breaking  soft terms induced by modulus-dominated SUSY breaking and their phenomenological consequences, particularly in the context of recent LHC results.
In the second part we address the issue of the origin of hierarchical Yukawa couplings in MSSM-like constructions induced by non-perturbative effects.\\

We start our discussion (\Ch{ch1}) with a brief review of the construction of chiral 4d Type II orientifold compactifications with $N=1$ supersymmetry. These include Type IIA orientifolds with chiral multiplets at intersecting D6-branes and their mirrors  based on Type IIB magnetized intersecting D7-branes. The non-perturbative extensions of the latter are better described in terms of F-theory compactified on complex CY 4-folds, which are also briefly reviewed, particularly in the context of local F-theory GUT's. We describe how in this class of models  there are three large classes of fields, those residing at intersecting  7-branes (I), those corresponding to Wilson lines on the bulk of the 7-branes (A), and those parametrizing the 7-brane positions ($\phi$). The distribution of SM chiral generations and Higgs multiplets in these 3 classes of fields is strongly restricted by the experimental existence of a large Yukawa coupling, that of the top quark. This leaves only three options: (I-I-I), all fermions and Higgses residing at intersecting 7-branes; (I-I-A): same, but Higgses living in the bulk; (A-A-$\phi$): all live in the bulk of the 7-branes. This different distributions lead to very different SUSY breaking spectra when analyzed in \Ch{stdm}. We also describe some essential features of the low energy effective action (gauge kinetic functions and Kahler metrics of chiral matter fields). In particular the {\it modular weights $\xi_\alpha$ } of the three classes of fields  I, A, $\phi$ above are $1\over2$, 1, 0 respectively. In this chapter  we also include some novel formulae estimating the effect  of the magnetic fluxes  in the Kahler metrics and gauge kinetic functions. The structure of this effective action is crucial for the computation of SUSY-breaking soft terms in \Ch{stdm}. We also illustrate these results in a detailed specific MSSM-like model whose effective action is described in \App{appA}.\\

In \Ch{stdm} we present a study of the SUSY-breaking induced under the assumption  of modulus-dominance on Type IIB orientifolds with  matter fields residing on D7-branes and their F-theory extensions. This modulus dominance SUSY-breaking may be understood as originated in the presence of closed string ISD fluxes in the IIB background. Armed with the results of \Ch{ch1}, we compute the SUSY-breaking soft terms in a scheme in which it is assumed that the MSSM gauge group resides at a local set of 7-branes wrapping a 4-cycle inside a CY. This is done within the spirit of the
{\it swiss cheese } type of CY compactifications first introduced in \Ref{conlon1}. The results are summarized in \Tab{opciones reales} and \Tab{correcciones de flujos}.\\

We have studied the phenomenological viability of these soft terms by impossing two conditions: 1) Correct radiative EW symmetry breaking and 2) neutralino relic density consistent with WMAP results.  We also impose a number of experimental conditions from LEP, LHC and weak decays, in particular limits from $b\rightarrow s\gamma$ and $B_s\rightarrow \mu^+\mu^-$. To check for  all these conditions we run the soft terms from the string/GUT scale down to the EW scale and obtain the low-energy SUSY spectra. The scheme has only three parameters, the universal gaugino mass 
$M$, the $\mu$-term and a small parameter $\rho_H$ describing possible small magnetic flux corrections. Imposing the above conditions the low-energy physics is extremely constrained.  Interestingly enough, we find that the only consistent field distribution is (I-I-I) in which all MSSM fields live at intersecting D7-branes. This is in fact the structure favored by local F-theory GUT's in which all matter fields live at matter curves, which may be interpreted as the  intersection of  7-branes.  Consistent results are obtained only in a region with large tan$\beta\simeq 41$ in which correct neutralino relic abundance is obtained thanks to neutralino-stau coannihilation.\\

In view of the latter result we have performed a through analysis of this favored scheme with all fields at intersecting 7-branes. Interestingly enough  the model has the built-in aproximate identity $A/m=-\sqrt{2}(3-\rho_H)\simeq -2$ which induces a large stop mixing. This, in addition to the mentioned large tan$\beta$,  gives rise to a relativly heavy lightest Higgs boson, with a mass in the region 119-125 GeV. In fact the recent  LHCb limit on $BR(B_s\rightarrow \mu^+\mu^-)$ forces $M \geq 1.4$ TeV corresponding to the upper region with $m_h\simeq 125$ GeV. The SUSY spectrum is relatively heavy with gluinos and squarks with a mass around 2.8-3.0 TeV and stops around 2 TeV. The lightest stau is around 600 GeV,
just slightly above the lightest neutralino.  The signatures would be quite similar to those of the CMSSM in the stau coannihilation region, with very characteristic signatures involving multi-tau events. To study its detectability  at LHC we have performed a Monte Carlo  analysis 
of the jets+missing energy signal using PYTHIA 6.400 linked with PGS, which simulates the response of the LHC detectors. We have included the most important sources of SM background. We find that the model may be tested at LHC(14 TeV) with an integrated luminosity of 25 fb$^{-1}$.
Still, forthcoming LHC limits on $BR(B_s\rightarrow \mu^+\mu^-)$ have also the potential of testing the model soon.\\

Finally in \Ch{capyuk} we turn to the study of Yukawa couplings in local F-theory  GUT compactifications. Yukawa couplings in Type IIB/F-theory compactifications are obtained by performing an overlap integral of the three wave functions associated to the participating fields. This is a difficult task since such wave functions are only known for simple cases like toroidal orientifolds. On the other hand in local F-theory GUT's the Yukawa couplings are associated  to points in the base $S$  in which the matter curves intersect. Since the SM fields are localized (with a Gaussian profile) at the matter curves, the Yukawa coupling is only sensitive to the region around the intersection point. This allow us to compute the Yukawa couplings with only a local information about the wave functions.\\

In this chapter we start by solving the local equations of motion for the matter fields in a simple $U(3)$ toy model with three singlet generations. Plugging the resulting wave functions in the triple overlap integral we find  the known result that only one generation gets Yukawa couplings. In addition we argue that non-perturbative gaugino condensation (or instanton) corrections originated on distant 7-branes can induce the required corrections to obtain non-vanishing Yukawa couplings for the lightest generations. We compute the modified wave functions in the presence of such corrections and performing again the overlap integral we find the results in \Eq{yukasmm}, which lead to hierarchical Yukawas. On general grounds one finds also that the holomorphic Yukawa couplings obtained are independent of the magnetic fluxes present. This is potentially problematic since e.g. in a realistic setting with an $SU(5)$ unification  all three down quark Yukawas would be identical to their corresponding charged lepton Yukawas, which is phenomenologically untenable. We argue however that after including  the wave function normalization required to get the physical Yukawas, an explicit dependence on hypercharge appears which can explain the difference betwen D and L Yukawas. We also check using this $U(3)$ toy model that the structure of the non-perturbative corrections obtained, along with the wave function normalizations is sufficient to understand the observed structure of fermion masses and mixings in a more realistic setting. Thus indeed non-perturbative corrections have the potential to explain the observed hierarchies of fermion masses.

%% file: appendixA.tex
\chapter{MSSM-like model}\label{appA}

In this appendix we are going to apply the results in \Ch{ch1} to an explicit MSSM-like example with intersecting D6-branes or their T-dual equivalent with magnetized D7-branes.\\

Toroidal models with $\mathcal{N}=1$ SUSY may be constructed if an additional $Z_2 \times Z_2$ orbifolding is performed \cite{CSU, ms}. In the case of the model we are dealing with, it has its origin in one of the simplest SUSY-quivers with four nodes, namely the SUSY-triangle of  \Ref{CIMQuivers}.\\

\begin{table}[htb]
\renewcommand{\arraystretch}{1.5}
\begin{center}
\begin{tabular}{|c|c|c|c|c|}
\hline 
 \textbf{Brane type} & $N_i$  &  $(n_i^1,m_i^1)$  &  $(n_i^2,m_i^2)$   & $(n_i^3,m_i^3)$ \\ 
\hline
$a_2 = a'_2$ & $N_a=3$ & (1,0) & (3,1) & (3,-1/2)\\
\hline
$b_2$ & $N_b=2$ & (1,1) & (1,0) & (1,-1/2)\\
\hline
$c_2$ & $N_c=1$ & (0,1) & (0,-1) & (2,0)\\
\hline
\end{tabular}
\caption{\small{Wrapping numbers of a three generation MSSM with $\mathcal{N}=1$ SUSY locally}}\label{tabwrap}
\end{center}
\end{table}
  
To get $N=1$ SUSY, it is necessary that this local model is also embedded in a Type IIA $T^6/ Z_2\times Z_2$ orientifold with D6-branes wrapping 3-cycles. Then, one can provide a complete tadpole free version of this model. In this case the branes must be fixed by some element of the $Z_2\times Z_2$ orbifold group. The orbifold action projects the initial $U(N_a)$ Chan-Paton gauge group down to $U(N_a/2)$. Finally, the values of the wrapping numbers are chosen to be the same as in  \Tab{tabwrap}.\\

Due to the fact that there exists a T-duality between D6-branes at angles and magnetized D7-branes, we are going to construct the model in terms of three sets of intersecting Type IIB D7-branes whose magnetized numbers are indicated in \Tab{tabwrap}. For each group of branes the state of magnetization is characterized by the integers $(n_a^i,m_a^i)$, where $m_a^i$ is the wrapping number and $n_a^i$ the unit of magnetic flux in such two-torus. Namely, what we are doing is turning on a constant Abelian world-volume magnetic field $F=dA$, satisfying
\begin{equation}
\label{ecm2.11} \frac{m_a^i}{2\pi}\int_{T^2_i}F_a^i=n_a^i.
\end{equation}

\noindent The T-duality allows us to introduce the angles 
\begin{equation}\label{ecm2.12}
\psi_a^i=\arctan2\pi\alpha'F_a^i=\arctan\frac{\alpha'n_a^i}{m_a^iA_i}
\end{equation}
where $(2\pi)^2 A_i$ is the area of the $T^2_i$. The conditions for preserving $\mathcal{N}=1$ SUSY \cite{BLT,CU} are the following 
\begin{equation}\label{ecm2.13}
\sum_{i=1}^3\psi_a^i=\frac{3\pi}{2}\ \textrm{mod}\ 2\pi.
\end{equation}

Notice that T-duality along the horizontal direction in each $T^2_i$ gives the dual picture of D6-branes at angles, e.g., for $T^2_i$, $A_i=R_{ix}R_{iy}$, and the dual angle is $\vartheta^i_a=\arctan(n_a^iR_{ix}/m_a^iR_{iy})$.\\

We construct the model with three sets of D7-branes and their corresponding images under $\Omega\mathcal{R}$. In \Tab{tabwrap7} we can see the characteristics and the angles defined in \Eq{ecm2.12}, where $\pi\gamma_2=\arctan(3\alpha'/A_2)$, $\pi\gamma_3=\arctan(6\alpha'/A_2)$, $\pi\beta_1=\arctan(\alpha'/A_1)$ and $\pi\beta_3=\arctan(2\alpha'/A_3)$.\\

\begin{table}[ht]
\renewcommand{\arraystretch}{1.5}
\begin{center}
\begin{tabular}{|c|c|c|c|c|c|}
\hline
\textbf{Branes} & $N_i$  &  $(n_i^1,m_i^1)$  &  $(n_i^2,m_i^2)$   & $(n_i^3,m_i^3)$ & $(\psi_a^1,\psi_a^2,\psi_a^3)$\\
\hline
$D7_a$ & 6 + 2 & (1,0) & (3,1) & (3,-1/2) & $(\frac{\pi}{2},\pi\gamma_2, \pi-\pi\gamma_3)$\\
\hline
$D7_b$ & 4 & (1,1) & (1,0) & (1,-1/2) & $(\pi\beta_1,\frac{\pi}{2},\pi-\pi\beta_3)$\\
\hline
$D7_c$ & 2 & (0,1) & (0,-1) & (2,0) & $(0,\pi,\frac{\pi}{2})$\\
\hline
\end{tabular}
\caption{\small{Wrapping numbers of a three generation MSSM with D7-branes}}\label{tabwrap7}
\end{center}
\end{table}

\noindent If we apply the condition  we conclude that in order to get SUSY we need
\begin{eqnarray}
\label{ecm2.14} \gamma_2 = \gamma_3 = \gamma \; \Rightarrow A_3=2A_2\\
\label{ecm2.15} \beta_1 = \beta_3 = \beta \; \Rightarrow A_3=2A_2 
\end{eqnarray}
and using (\ref{ecm2.14}) and (\ref{ecm2.15}) we obtain
\begin{eqnarray}
\label{ecm2.16} A&=&A_1=A_2=\frac{1}{2}A_3\\
\label{ecm2.17} \tan\pi\beta&=&\frac{\alpha'}{A_1}=\frac{\alpha'}{A}\\
\label{ecm2.18} \tan\pi\gamma&=&\frac{3\alpha'}{A_2}=\frac{3\alpha'}{A}
\end{eqnarray}
and from eq. (\ref{ecm2.17}) and (\ref{ecm2.18}) we have
\begin{equation}
\label{ecm2.19} \tan\pi\gamma=3\tan\pi\beta  
\end{equation}
finally, from eq (\ref{ecm2.19}) we see that $\gamma>\beta$.\\

We will also need $D7_{b^\ast}$ and $D7_{c^\ast}$-branes (images of $D7_b$ and $D7_c$ under $\Omega\mathcal{R}$) in order to obtain the spectrum of the model, in Table \ref{tabwrap72} we can see the corresponding $\psi_a^i$.

\begin{table}[ht]
\renewcommand{\arraystretch}{1.5}
\begin{center}
\begin{tabular}{|c|c|c|c|c|}
\hline
\textbf{Branes}  &  $(n_i^1,m_i^1)$  &  $(n_i^2,m_i^2)$   & $(n_i^3,m_i^3)$ & $(\psi_a^1,\psi_a^2,\psi_a^3)$\\
\hline
$D7_{b^\ast}$  & (1,-1) & (1,0) & (1,1/2) & $(\pi-\pi\beta,\frac{\pi}{2},\pi\beta)$\\
\hline
$D7_{c^\ast}$  & (0,-1) & (0,1) & (2,0) & $(\pi,0,\frac{\pi}{2})$\\
\hline
\end{tabular}
\caption{\small{Wrapping numbers for D7-branes images under $\Omega\mathcal{R}$}}\label{tabwrap72}
\end{center}
\end{table}

From wrapping numbers of Tables \ref{tabwrap7} and \ref{tabwrap72} we can calculate the intersection numbers
\begin{equation}
\begin{array}{lcl}
I_{ab}\ =   \ 1, & & I_{ab*}\ =\ 2, \\ 
I_{ac}\ =   \ -3, & & I_{ac*}\ =\ -3, \\ 
I_{bd}\ =   \ -1,  & & I_{bd*}\ =\ 2, \\ 
I_{cd}\ =   \ 3, & & I_{cd*}\ =\ -3, \\ 
I_{bc}\ =   \ -1 , & & I_{bc*}\ =\ -1, 
\end{array}
\end{equation}
which provide us the massless spectrum of the model which is displayed in Table \ref{tabSpectrum} (remember that $a=a'=d$). This is the chiral spectrum of the MSSM enlarged to include right-handed neutrinos $N_R$ and an extra $U(1)_{B-L}$.

\begin{table}[ht]
\renewcommand{\arraystretch}{1.3}
\begin{center}
\begin{tabular}{|c|c|c|c|c|c|c|c|} 
\hline Intersection & 
 Matter fields  &   &  $Q_a$  & $Q_b $ & $Q_c $ & $Q_d$  & $Q_Y$ \\ 
\hline 
\hline $ab$ & $Q_L$ &  $(3, 2)$ & 1  & -1 & 0 & 0 & 1/6 \\ 
\hline $ab^*$  & $q_L$ & $2(3,2)$ &  1  & 1  & 0  & 0  & 1/6 \\ 
\hline $ac$ & $U_R$ &  $3( {\bar 3},1)$ &  -1  & 0  & 1  & 0 & -2/3 \\ 
\hline $ac^*$  & $D_R$   &  $3({\bar 3},1)$ &  -1  & 0  & -1  & 0 & 1/3 \\ 
\hline $bd$ & $ L $ &  $(1,2)$ &  0   & -1 & 0  & 1 & -1/2 \\ 
\hline $bd^*$ & $ l $ &  $2(1,2)$ &  0  & 1   & 0  & 1 & -1/2 \\ 
\hline $cd$ & $N_R$ & $3(1,1)$ &  0  & 0  & 1  & -1  & 0   \\ 
\hline $cd^*$ & $E_R$  & $3(1,1)$ &  0  & 0  & -1  & -1  & 1 \\ 
\hline $bc$ &  $H$ & $(1,2)$ &  0 & -1 & 1  &  0 & -1/2 \\ 
\hline $bc^*$ & $ {\bar H}$ &  $(1,2)$ & 0 & -1 & -1 & 0  & 1/2 \\ 
\hline 
\end{tabular} 
\caption{\small{ Chiral spectrum of the SUSY's SM obtained from the magnetized D7-brane MSSM-like model. The hypercharge generator is defined as $Q_Y = \frac{1}{6}  Q_a - \frac{1}{2} Q_c - \frac{1}{2}  Q_d$.}}\label{tabSpectrum}
\end{center}
\end{table}

\subsection{Kähler potential}\label{section MSSM kaler}

Using what we have learnt in section \Sec{subsection kahler} about Kähler metrics for untwisted fields \Eq{kalunt1} and \Eq{kalunt2} we obtain
\begin{itemize}
\item $D7_c$-$D7_c$
\begin{equation}\label{ecm2.34}
\kappa^2\tilde{K}_{1}=\frac{1}{u_1t_2}\quad;\quad \kappa^2\tilde{K}_{2}=\frac{1}{u_2t_1}\quad;\quad  \kappa^2\tilde{K}_{3}=\frac{1}{u_3s}
\end{equation}

\item $D7_b$-$D7_b$
\begin{equation}\label{ecm2.35}
\kappa^2\tilde{K}_{1}=\frac{1}{u_1t_1}\quad;\quad \kappa^2\tilde{K}_{2}=\frac{1}{u_2t_2s}\left(s+\frac{t_2}{2} \right)\quad;\quad  \kappa^2\tilde{K}_{3}=\frac{1}{u_3t_3}
\end{equation}

\item $D7_a$-$D7_a$
\begin{equation}\label{ecm2.36}
\kappa^2\tilde{K}_{1}=\frac{1}{u_1t_1s}\left(9s+\frac{t_1}{2} \right)\quad;\quad \kappa^2\tilde{K}_{2}=\frac{1}{u_2t_2}\quad;\quad  \kappa^2\tilde{K}_{3}=\frac{1}{u_3t_3}
\end{equation}
\end{itemize}

\noindent And concerning the twisted fields \Eq{kalt1} we have
\begin{itemize}
\item $D7_a$-$D7_b$
\begin{equation}\label{ecm2.44}
\!\!\!\!\!\kappa^2\tilde{K}_{C^{7_a7_b}}=\frac{\sqrt{2}}{(st^3)^{1/4}}\frac{1}{u_1^{1/2-\beta}u_2^{1/2+\gamma}u_3^{1+\beta-\gamma}}\sqrt{\frac{\Gamma(\frac{1}{2}+\beta)\Gamma(\frac{1}{2}-\gamma)\Gamma(\gamma-\beta)}{\Gamma(\frac{1}{2}-\beta)\Gamma(\frac{1}{2}+\gamma)\Gamma(1+\gamma-\beta)}}
\end{equation}

\item $D7_a$-$D7_c$ and $D7_a$-$D7_{c^\ast}$
\begin{equation}\label{ecm2.45}
\kappa^2\tilde{K}_{C^{7_a7_c}}=\frac{\sqrt{2}}{3(stu_1)^{1/2}}\frac{1}{u_2^{1-\gamma}u_3^{1/2+\gamma}}\frac{\Gamma(1/2-\gamma)}{\Gamma(1-\gamma)}
\end{equation}

\item $D7_a$-$D7_{b^\ast}$
\begin{equation}\label{ecm2.46}
\!\!\!\!\!\kappa^2\tilde{K}_{C^{7_a7_b\ast}}=\frac{\sqrt{2}}{(st^3)^{1/4}}\frac{1}{u_1^{1/2+\beta}u_2^{1/2+\gamma}u_3^{1-\beta-\gamma}}\sqrt{\frac{\Gamma(\frac{1}{2}-\beta)\Gamma(\frac{1}{2}-\gamma)\Gamma(\gamma+\beta)}{\Gamma(\frac{1}{2}+\beta)\Gamma(\frac{1}{2}+\gamma)\Gamma(1-\gamma-\beta)}}
\end{equation}

\item $D7_b$-$D7_c$ and $D7_b$-$D7_{c^\ast}$
\begin{equation}\label{ecm2.47}
\kappa^2\tilde{K}_{C^{7_b7_c}}=\frac{\sqrt{2}}{(stu_2)^{1/2}}\frac{1}{u_1^{1-\beta}u_3^{1/2+\beta}}\frac{\Gamma(1/2-\beta)}{\Gamma(1-\beta)}
\end{equation}
\end{itemize}

\noindent To simplify the above results we are going to use the following relations
\begin{equation}\label{ecm2.48}
\frac{\Gamma(\frac{1}{2}-\delta)}{\Gamma(1-\delta)}=\frac{1}{\sqrt{\pi}}B(1/2,\ 1/2-\delta)
\end{equation}
where
\begin{equation}\label{ecm2.49}
B(x,y)=\frac{\Gamma(x)\Gamma(y)}{\Gamma(x+y)}
\end{equation}
is the Beta function and we can make a Taylor expansion
\begin{equation}
\frac{1}{\sqrt{\pi}}B(1/2,\ 1/2-\delta)\approx B(\delta=0)(1+B_0(\delta=0)\delta)
\end{equation}
where $B_0(\delta)=\psi_0(1-\delta)-\psi_0(1/2-\delta)$ and $\psi_0=\Gamma'(z)/\Gamma(z)$ the Psi function. One can check that $B_0(\delta)=2\log2+\pi^{2/3}\delta$, and so
\begin{equation}\label{ecm2.51}
\frac{\Gamma(\frac{1}{2}-\delta)}{\Gamma(1-\delta)}\approx1+2\log2\cdot\delta
\end{equation}

Therefore, using eqs.(\ref{ecm2.49}) and (\ref{ecm2.51}) in eqs.(\ref{ecm2.44})-(\ref{ecm2.47}) we obtain for small $\gamma$, $\beta$
\begin{itemize}
\item $D7_a$-$D7_b$
\begin{equation}\label{ecm2.52}
\kappa^2\tilde{K}_{C^{7_a7_b}}=\frac{\sqrt{2}}{(st^3)^{1/4}}\frac{1}{u_1^{1/2-\beta}u_2^{1/2+3\beta}u_3^{1-2\beta}}\sqrt{\frac{B(1/2+\beta, 2\beta)}{(-2\beta) B(1/2-\beta, -2\beta)}}
\end{equation}

\item $D7_a$-$D7_c$ and $D7_a$-$D7_{c^\ast}$
\begin{equation}\label{ecm2.53}
\kappa^2\tilde{K}_{C^{7_a7_c}}=\frac{\sqrt{2}}{3(stu_1)^{1/2}}\frac{1}{u_2^{1-\gamma}u_3^{1/2+\gamma}}(1+2\log2\cdot\gamma)
\end{equation}

\item $D7_a$-$D7_{b^\ast}$
\begin{equation}\label{ecm2.54}
\kappa^2\tilde{K}_{C^{7_a7_b\ast}}=\frac{\sqrt{2}}{(st^3)^{1/4}}\frac{1}{u_1^{1/2+\beta}u_2^{1/2+3\beta}u_3^{1-4\beta}}\sqrt{\frac{B(1/2-\beta, 4\beta)}{(-4\beta) B(1/2+\beta, -4\beta)}}
\end{equation}

\item $D7_b$-$D7_c$ and $D7_b$-$D7_{c^\ast}$
\begin{equation}\label{ecm2.55}
\kappa^2\tilde{K}_{C^{7_b7_c}}=\frac{\sqrt{2}}{(stu_2)^{1/2}}\frac{1}{u_1^{1-\beta}u_3^{1/2+\beta}}(1+2\log2\cdot\beta)
\end{equation}
\end{itemize}

An example for small $\gamma$ in Eqs. (\ref{ecm2.52}), (\ref{ecm2.54}) is more complicated. Noticed though that the metrics in (\ref{ecm2.53}), (\ref{ecm2.55}) scale like
\begin{equation}
 K \sim \frac{1}{t^{1\over2}} \left( 1 + c \frac{1}{t^{1\over2}} \right)
\end{equation}
in agreement with the result in \Eq{ult}

\subsection{Gauge kinetic functions}
Substituying in \Eq{ecm2.56} the values of \Tab{tabwrap7} we obtain
\begin{equation}\label{ecm2.57}
f_a=T_3+9S=T+9S\quad ; \quad f_b=T_3+S=T+S\quad ; \quad f_c=T_3=T
\end{equation}

From the above expression we see that in principle there is no gauge coupling unification because the three gauge kinetic functions are different. However there is a limit in which we do obtain gauge coupling unification. This is the case when $T\gg S$. We can see from (\ref{ecm2.57}) that in this limit $f_a\approx T$, $f_b\approx T$ and $f_c = T$, thus there will be gauge coupling unification.

\subsection{A tadpole free MSSM-like model}
As we have explained we have to cancell the tadpoles in order to guarantee the consistency of the theory. For the case of orientifolds the cancellation conditions (\ref{eqRR}) are modified in the following way
\begin{equation}
\begin{array}{l}
\sum_a N_a n_a^1n_a^2n_a^3 = 16 \\ 
\sum_a N_a n_a^1m_a^2m_a^3 = -16\beta_2\beta_3 \\ 
\sum_a N_a m_a^1n_a^2m_a^3 =-16\beta_1\beta_3  \\ 
\sum_a N_a m_a^1m_a^2n_a^3 =-16\beta_1\beta_2
\end{array}
\end{equation}
In the model we are dealing with, taking the values of Table \ref{tabwrap7}, the above equalities imply
\begin{equation}\label{tadorienti}
\begin{array}{l}
\sum_a N_a n_a^1n_a^2n_a^3 = 16 = 8\times9 + 4\\ 
\sum_a N_a n_a^1m_a^2m_a^3 = -8 \neq -4 \\ 
\sum_a N_a m_a^1n_a^2m_a^3 =-8 \neq -2  \\ 
\sum_a N_a m_a^1m_a^2n_a^3 =-16 \neq -4
\end{array}
\end{equation}

This means that the above conditions are not fulfilled and then we are not able to cancell the tadpoles. In order to guarantee the consistency of the theory we will add some additional branes (hidden branes)
\begin{equation}\label{hidd}
\begin{array}{l}
4(-2,1)(-3,1)(-3,1/2)\\
6(1,0)(1,0)(2,0)
\end{array}
\end{equation}
and their corresponding images under $\Omega\mathcal{R}$. Then the whole D-brane framework fulfill the tadpole free conditions
\begin{equation}\label{tadorienti2}
\begin{array}{l}
\sum_a N_a n_a^1n_a^2n_a^3 = 16 = 8\times9 + 4 + (-72 + 12)\\ 
\sum_a N_a n_a^1m_a^2m_a^3 = -8 = -4 + (-4) \\ 
\sum_a N_a m_a^1n_a^2m_a^3 =-8 = -2 + (-6)  \\ 
\sum_a N_a m_a^1m_a^2n_a^3 =-16 = -4 + (-12).
\end{array}
\end{equation}

\begin{table}[ht]
\renewcommand{\arraystretch}{1.5}
\begin{center}
\begin{tabular}{|c|c|c|c|c|}
\hline
\textbf{Branes} & $N_i$  &  $(n_i^1,m_i^1)$  &  $(n_i^2,m_i^2)$   & $(n_i^3,m_i^3)$ \\
\hline
$D7_a$ & 6 + 2 & (1,0) & (3,1) & (3,-1/2) \\
\hline
$D7_b$ & 4 & (1,1) & (1,0) & (1,-1/2) \\
\hline
$D7_c$ & 2 & (0,1) & (0,-1) & (2,0) \\
\hline
$D7_x$ & 4 & (-2,1) & (-3,1) & (-3,1/2)\\
\hline
\end{tabular}
\caption{\small{Wrapping numbers of a three generation MSSM with D7-branes and \textit{hidden} branes}}\label{tabwrapexotic}
\end{center}
\end{table}

Finding a MSSM-like model that is tadpole free is not a trivial task. Actually there is only another MSSM-like model \cite{ms04, fi} that fulfils these consistency conditions. However, the fact of adding \textit{hidden} branes implies that there will be exotics in our model. Actually if we compute the intersection between the D7-branes of Table \ref{tabwrapexotic} that contains the stacks of D7-branes (a,b,c) and the stack of \textit{hidden} branes\footnote{Notice that we are only using the first stack of \textit{hidden} branes of (\ref{hidd}) because the second one does not intersect with any D7-brane}, we obtain 
\begin{equation}
\begin{array}{lcl}
I_{xa}\ =   \ 0, & & I_{xa*}\ =\ 0, \\ 
I_{xb}\ =   \ 3, & & I_{xb*}\ =\ 2, \\ 
I_{xc}\ =   \ 6,  & & I_{xc*}\ =\ 6. 
\end{array}
\end{equation}
and therefore the chiral exotic spectrum consist of $3(1, \bar{2}_b, 1; 2_x)$ for the $xb$ intersection, $3(1, 2_b, 1; 2_x)$ for $xb^*$, and $3(1, 1, 1; 2_x)$ both for $xc$ and $xc^*$ intersections.